\documentclass[journal]{IEEEtran}
\IEEEoverridecommandlockouts

\usepackage{cite}
\usepackage{amsmath,amssymb,amsfonts}
\usepackage{algorithmic}
\usepackage{graphicx}
\usepackage{textcomp}
\usepackage{xcolor}
\usepackage{float}
\usepackage{overpic}
\usepackage{subfigure}
\usepackage[colorlinks,
linkcolor=red,
anchorcolor=red,
citecolor=green]{hyperref}
\def\BibTeX{{\rm B\kern-.05em{\sc i\kern-.025em b}\kern-.08em
    T\kern-.1667em\lower.7ex\hbox{E}\kern-.125emX}}

\begin{document}
	%
	\title{BLNet: A Fast Deep Learning Framework for Low-Light Image Enhancement with\\ Noise Removal and Color Restoration}
	%
	%
	%
	
	\author{Xinxu~Wei,
	Xianshi~Zhang\IEEEauthorrefmark{1}, Shisen~Wang, Cheng~Cheng, Yanlin~Huang, 
	Kaifu~Yang,
	\\and Yongjie~Li,~\IEEEmembership{Senior Member,~IEEE}
	\thanks{This work was supported by Guangdong Key R\&D Project (\#2018B030338001) and Natural Science Foundations of China (\#61806041, \#62076055). (Corresponding author: Xian-Shi~Zhang, email: zhangxianshi@uestc.edu.cn)}%
	\thanks{Xinxu~Wei, Xianshi~Zhang, Cheng~Cheng, Yanlin~Huang, Kaifu~Yang, Yongjie~Li are with the MOE Key Lab for Neuroinformation, University of Electronic Science and Technology of China (UESTC), Chengdu 610054, China.}}

	\maketitle


\maketitle

\begin{abstract}
Images obtained in real-world low-light conditions are not only low in brightness, but they also suffer from many other types of degradation, such as color bias, unknown noise, detail loss and halo artifacts. In this paper, we propose a very fast deep learning framework called Bringing the Lightness (denoted as BLNet) that consists of two U-Nets with a series of well-designed loss functions to tackle all of the above degradations. Based on Retinex Theory, the decomposition net in our model can decompose low-light images into reflectance and illumination and remove noise in the reflectance during the decomposition phase. We propose a Noise and Color Bias Control module (NCBC Module) that contains a convolutional neural network and two loss functions (noise loss and color loss). This module is only used to calculate the loss functions during the training phase, so our method is very fast during the test phase. This module can smooth the reflectance to achieve the purpose of noise removal while preserving details and edge information and controlling color bias. We propose a network that can be trained to learn the mapping between low-light and normal-light illumination and enhance the brightness of images taken in low-light illumination. We train and evaluate the performance of our proposed model over the real-world Low-Light (LOL) dataset), and we also test our model over several other frequently used datasets (LIME, DICM and MEF datasets). We conduct extensive experiments to demonstrate that our approach achieves a promising effect with good rubustness and generalization and outperforms many other state-of-the-art methods qualitatively and quantitatively. Our method achieves high speed because we use loss functions instead of introducing additional denoisers for noise removal and color correction. 
The code and model are available at https://github.com/weixinxu666/BLNet.\\
\end{abstract}

\begin{IEEEkeywords}
Low-light image enhancement, Retinex decomposition, Image denoising, Color bias control, Deep learning
\end{IEEEkeywords}

\begin{figure}
	\flushleft
	\subfigure{
		\begin{minipage}{10\linewidth} 
			\includegraphics[width=4.2cm]{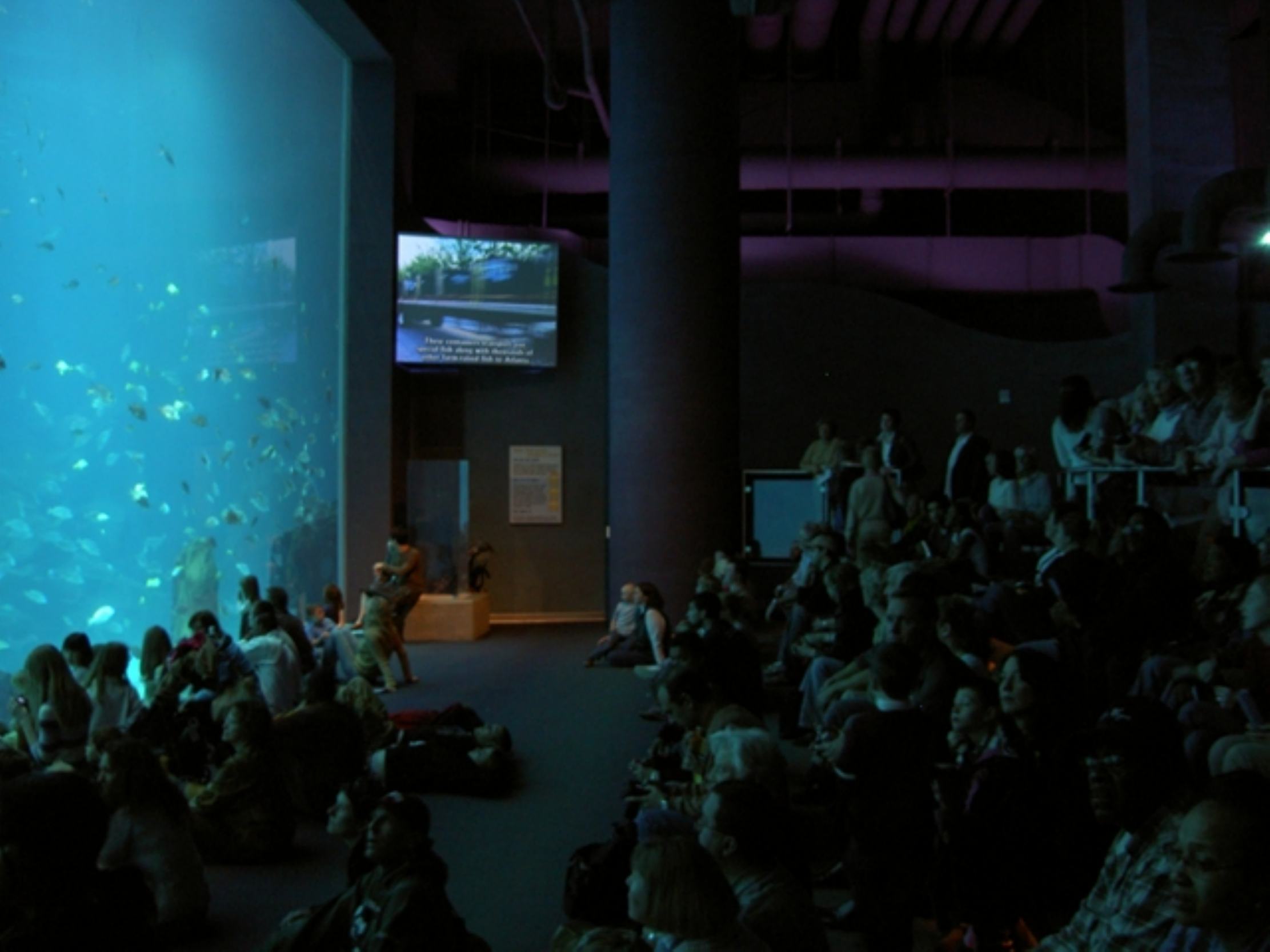}\vspace{-3pt}  
			\includegraphics[width=4.2cm]{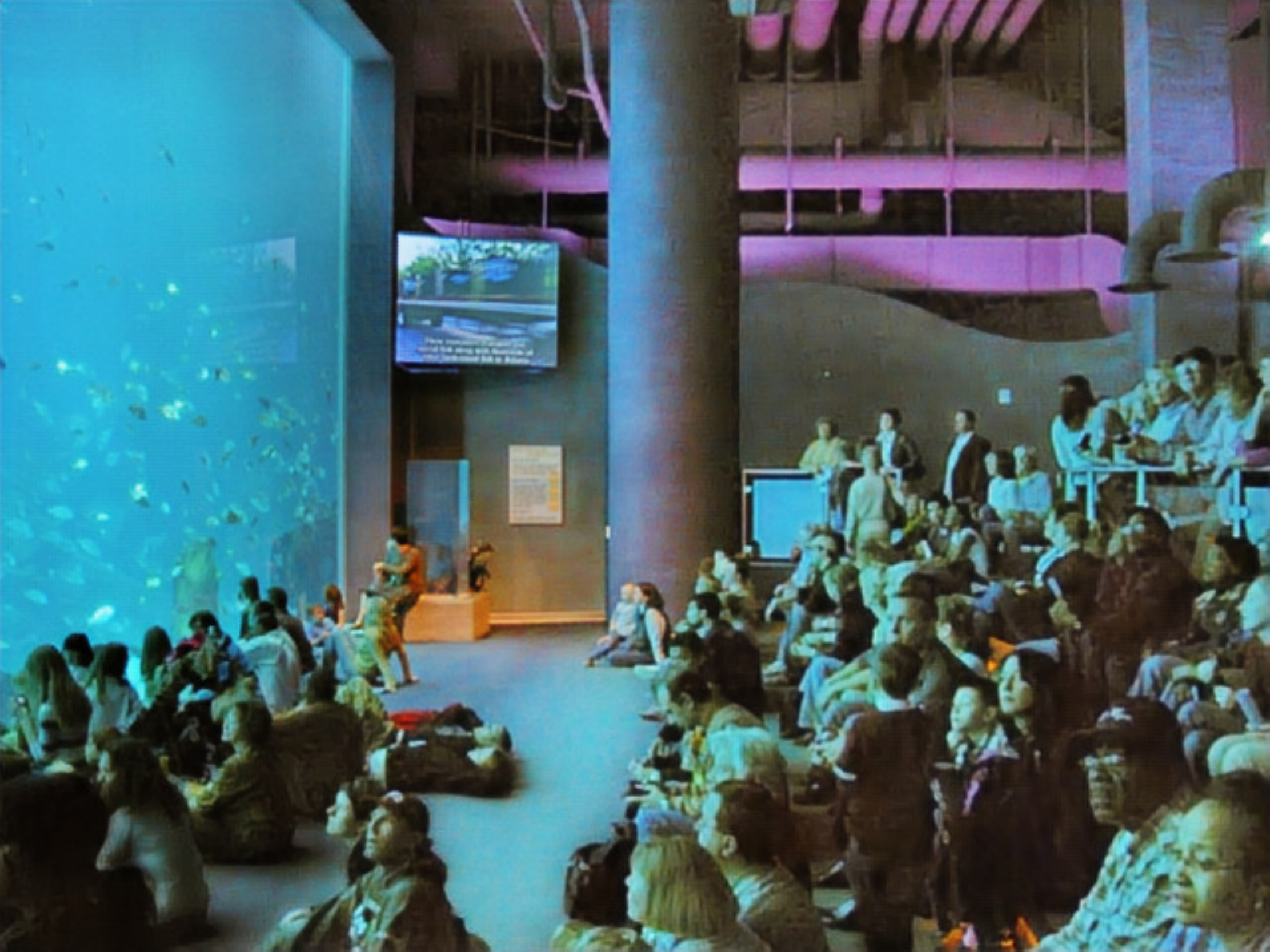}\vspace{-3pt} 
		\end{minipage}
	}
	\quad
	\subfigure{
		\begin{minipage}{10\linewidth}
			\includegraphics[width=4.2cm]{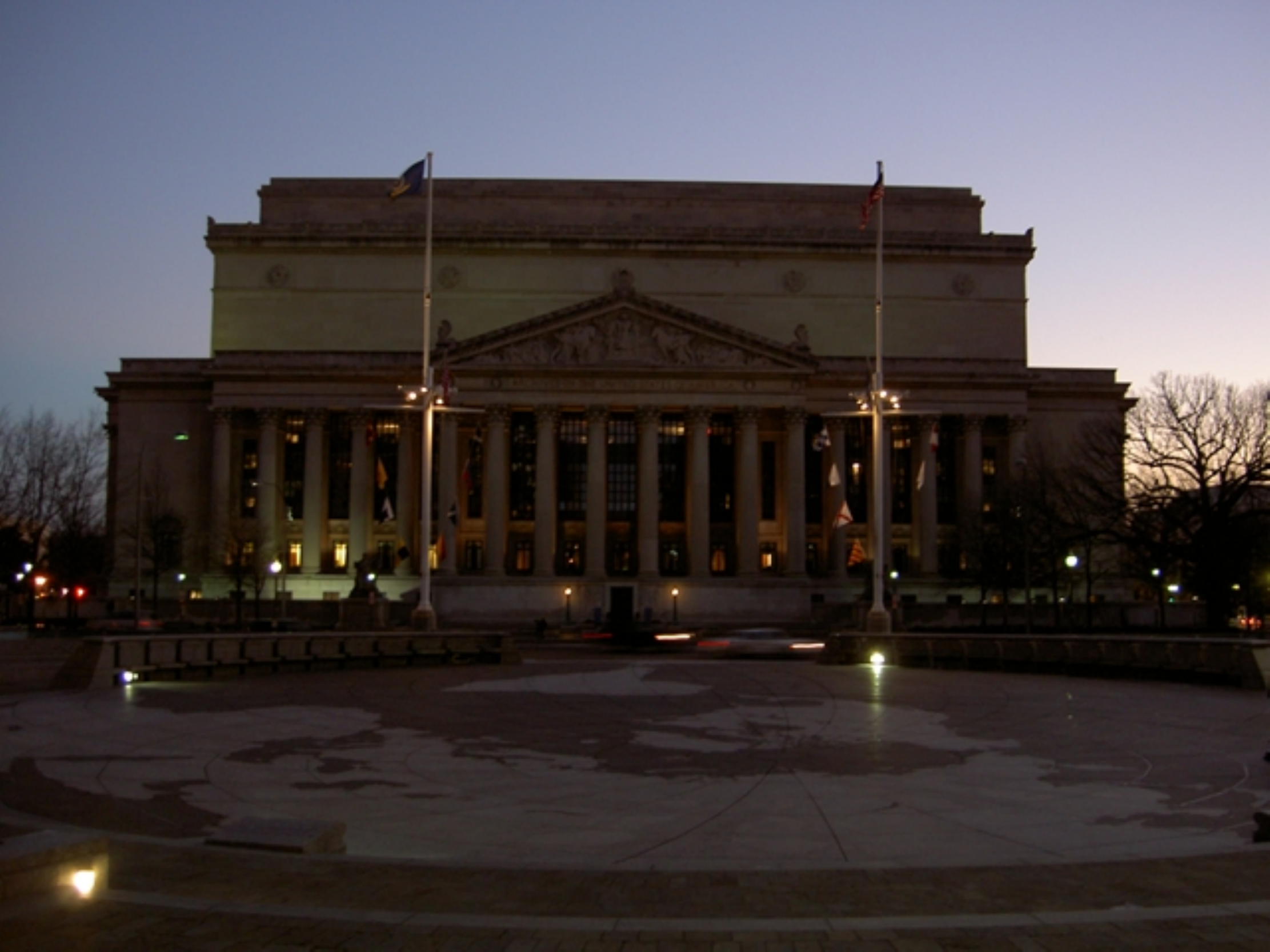}\vspace{-3pt}
			\includegraphics[width=4.2cm]{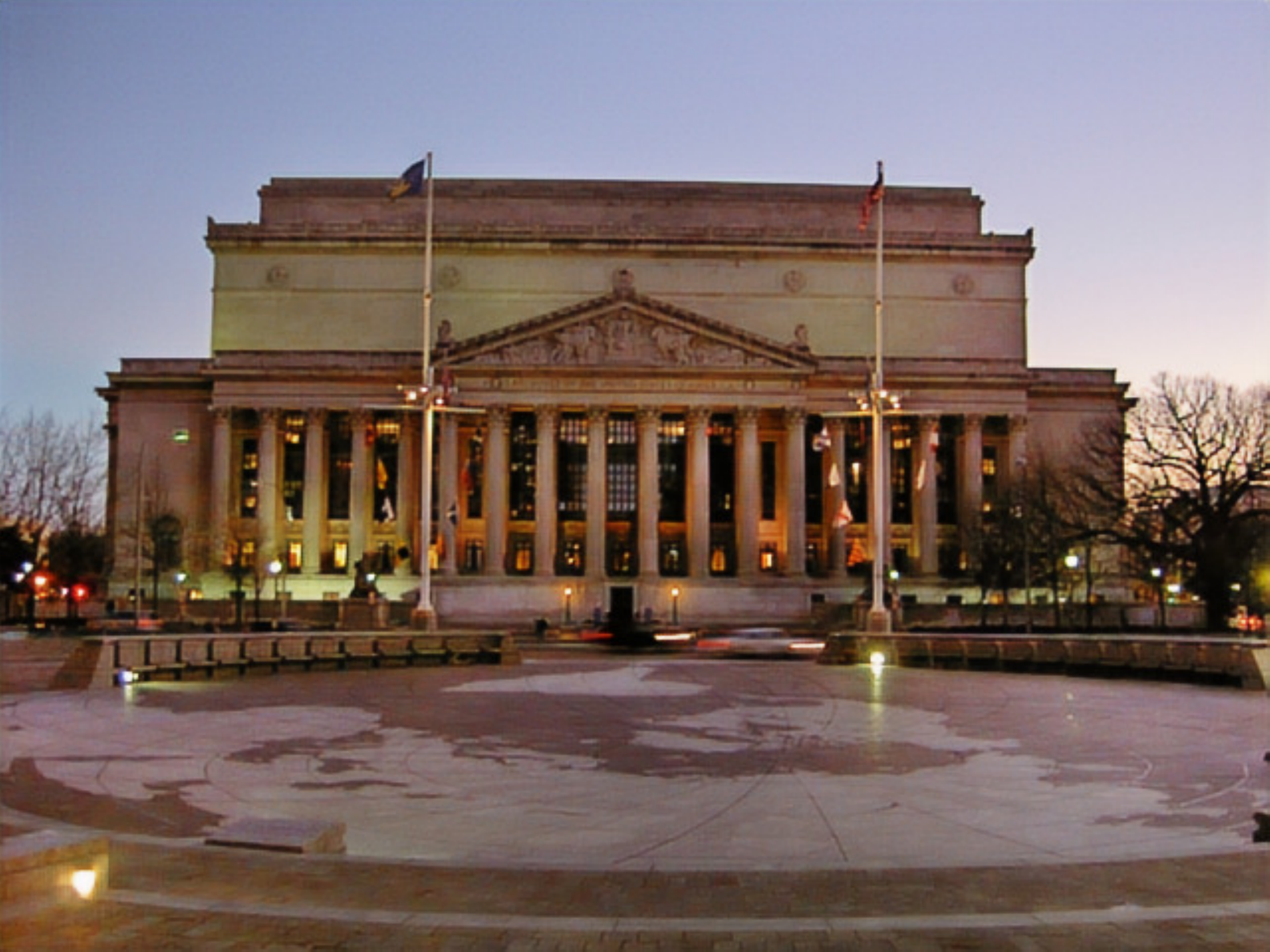}\vspace{-3pt}
		\end{minipage}
	}
	\quad
	\subfigure{
		\begin{minipage}{10\linewidth}
			\includegraphics[width=4.2cm]{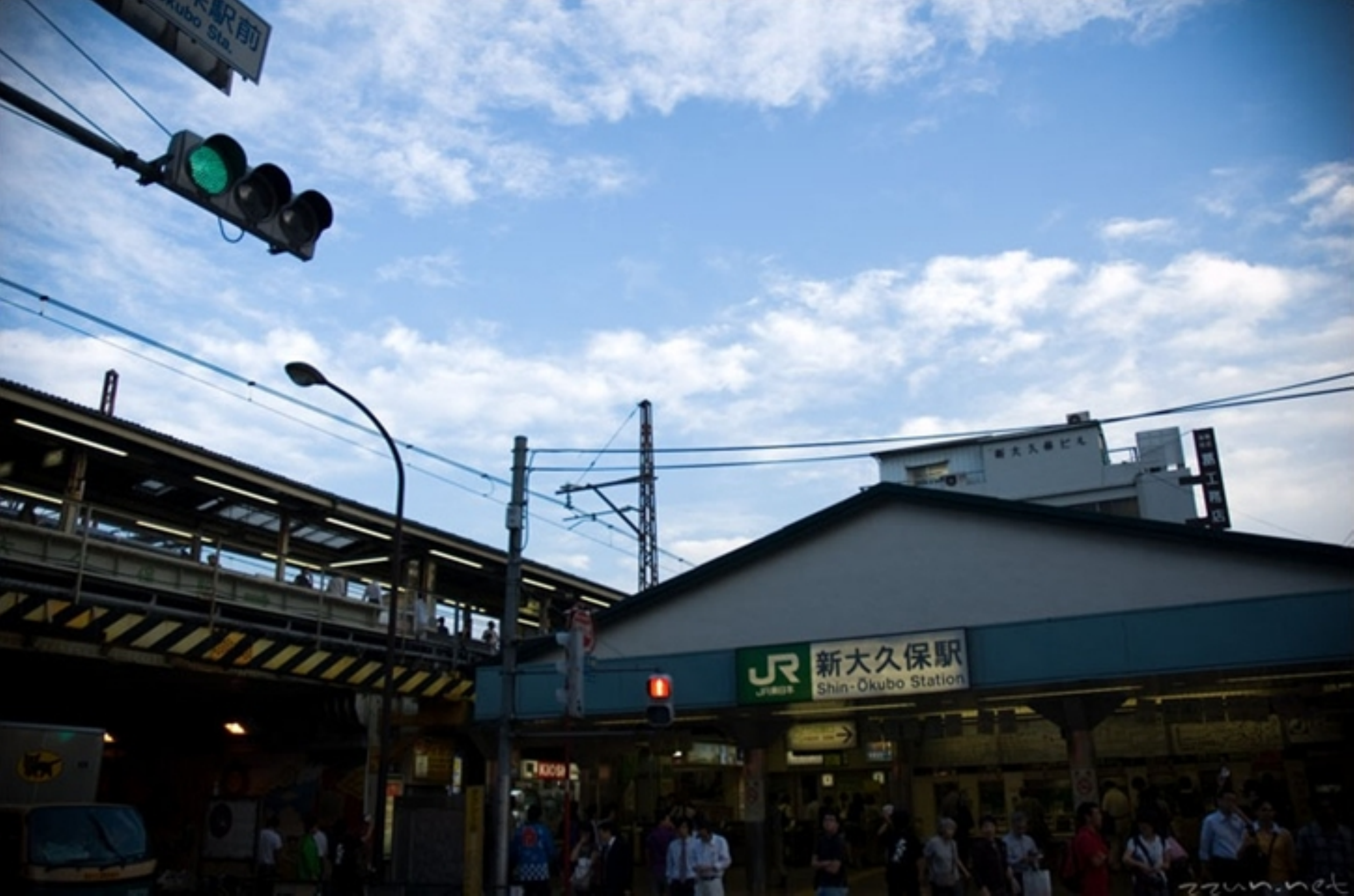}\vspace{-3pt}
			\includegraphics[width=4.2cm]{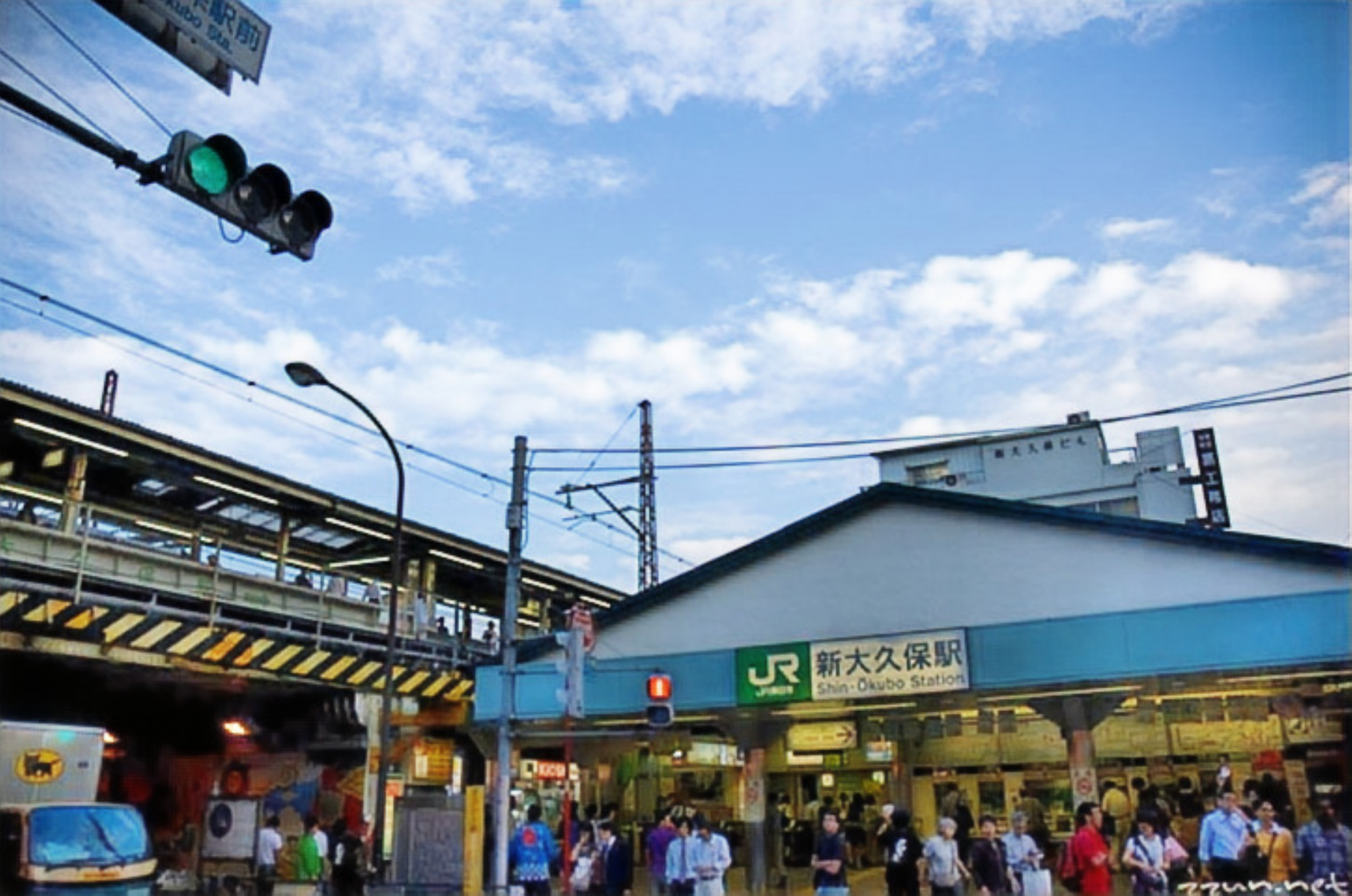}\vspace{3pt}
		\end{minipage}
	}
	\caption{Some low-light image examples and their corresponding enhanced results using our BLNet. Left column: low-light images. Right column: enhanced results using BLNet.}
\end{figure}

\section{Introduction}
\IEEEPARstart{L}{ow-light} image enhancement is a very challenging low-level computer vision task because, while enhancing the brightness, we also need to control the color bias, suppress amplified noise, preserve details and texture information and restore blurred edges. 
Images captured in insufficient lighting conditions often suffer from several types of degradation, such as poor visibility, low contrast, color distortion and severe ISO 
noise, which have negative effects on other computer vision tasks, such as image recognition \cite{lecun1998gradient} \cite{simonyan2014very} \cite{he2016deep}, object detection \cite{redmon2016you} \cite{girshick2014rich} \cite{girshick2015fast} 
and image segmentation \cite{chen2017deeplab} \cite{he2017mask} \cite{zhao2017pyramid}. Therefore, there is a huge demand for low-light image enhancement. 
According to \cite{zhang2019kindling}, although adjusting camera settings (e.g., increasing ISO, extending exposure time and using flash) can enhance the brightness of the image and improve the visibility, it can also bring about specific problems, making the image suffer from degradation to varying degrees. For example, increasing the ISO may introduce additional noise and cause parts of the image to be overexposed. Extending the exposure time may blur the objects in the image. Using flash enhances the brightness of captured images, but it may also lead to an unnatural image with color bias and uneven brightness.
In recent years, a great number of approaches have been proposed and achieved remarkable results in low-light image enhancement, but, to the best of our knowledge, there are few successful methods for simultaneously dealing with all degradations contained in low-light images (such as low brightness, color bias, noise pollution, detail and texture loss, edge blurring, halo artifacts and contrast distortion). \\
It is still a very challenging task to cope with so many problems simultaneously. 
Although many existing methods can improve the brightness of low-light images, some of them cause serious color distortion after increasing the brightness \cite{jobson1997multiscale}\cite{guo2016lime}\cite{wei2018deep}\cite{wang2018gladnet}\cite{jiang2021enlightengan}\cite{zhang2019kindling}. In addition, some methods do not suppress noise when enhancing brightness, resulting in serious noise in the image after enhancement \cite{jobson1997multiscale}\cite{wang2019rdgan}\cite{guo2020zero}. Other methods \cite{lv2018mbllen}\cite{li2018structure}\cite{ren2018joint}, in order to eliminate noise, cause the enhanced and denoised image to be too smooth, resulting in blurred edges, missing details and texture information and color bias. 
For example, KinD \cite{zhang2019kindling} uses a simple U-Net to decompose low-light images. Noise hidden in the dark regions is amplified in reflectance, so a deeper U-Net is used in KinD to remove noise in reflectance, but this introduces color bias.
Degradation can affect not only our visual effects but other visual tasks as well.
Therefore, how to control the color bias during denoising is a problem worth studying.
Based on Retinex Theory \cite{land1977retinex}, a natural image can be decomposed into two parts: reflectance and illumination. The reflectance part contains details, texture and color information of the original image. The illumination part carries information about the intensity and distribution of light in the original image, and it should be smooth enough and not contain any high-frequency information such as details and texture.
The decomposition task concentrates both the high-frequency components and the noise on the reflectance part, whereas the illumination part is mainly composed of low-frequency components, so there is almost no noise in the illumination part and we only need to denoise the reflectance part.
Many previous methods do not consider denoising the enhanced image, such as MSRCR \cite{jobson1997multiscale}, GLADNet \cite{wang2018gladnet} and EnlightenGan \cite{jiang2021enlightengan}. Some methods use other state-of-the-art denoising methods as post-processing operations for enhanced images, such as BM3D \cite{dabov2006image}, DnCNN \cite{zhang2017beyond} and CBDNet \cite{guo2019toward}, which can slow down the speed of the model considerably. In other methods, a denoising network or algorithm is specially designed to make the enhanced image noise-free, such as KinD \cite{zhang2019kindling}, which also slows down the speed of the model significantly. 
In this paper, we improve RetinexNet \cite{wei2018deep} and propose a novel framework based on Retinex Theory \cite{land1977retinex} to achieve brightness enhancement, denoising, color bias correction and detail and texture preservation simultaneously and efficiently. The proposed framework consists of two U-Nets \cite{ronneberger2015u} and a convolutional neural network (CNN). In the decomposition phase, we use a U-Net to decompose low-light images into reflectance and illumination. According to \cite{xiong2020unsupervised}, the up-and-down sampling structure of U-Net can be used as a decomposition network. However, the main problem with using U-Net as the decomposition network is that the decomposition results are accompanied by color distortion and inconsistency. In addition, a large amount of noise will appear in reflectance after decomposition because the noise is originally hidden in the dark regions of the low-light image. 
We propose a Noise and Color Bias Control Module (denoted as NCBC Module), which consists of a CNN and two loss functions, noise loss and color loss, to suppress the noise and correct the color distortion.
Different from previous methods, instead of designing a network specifically to deal with noise, we suppress the noise when conducting decomposition and enhancement to avoid noise amplification. We also use color loss to constrain the color information of the image and recover the correct color from the dark regions, while also removing noise and enhancing brightness. In the enhancement phase, we use a U-Net to reconstruct the illumination and a series of loss functions to constrain the supervised training process.
We highlight the contributions of this paper as follows:
\begin{itemize}
	\item Based on the Retinex theory, we improve RetinexNet \cite{wei2018deep} and propose a very fast deep-learning framework consisting of two U-Nets for low-light image decomposition and enhancement, which can simultaneously address the problems of low brightness, color bias, amplified noise, detail and texture loss, blurred edges and distorted contrast.\\
	\item We propose a Noise and Color Bias Control Module consisting of a convolutional neural network. This module preserves the color information while smoothing the noise in the reflectance, and it reconstructs the high-frequency information.\\
	\item Benefiting from the constraints of a series of well-adjusted loss functions, in the enhancement phase, we propose a very fast enhancement network to enhance the brightness of illumination in an image-to-image translation way.\\
	\item Extensive comparisons and ablation experiments are conducted on commonly used datasets to demonstrate the superiority of our method qualitatively and quantitatively, including its fast computation speed.
\end{itemize}

\section{Related Works}
In recent years, many effective methods have been developed for the tasks of low-light image enhancement and image denoising.

\subsection{Low-Light Image Enhancement Methods.}

According to the algorithm principle, low-light image enhancement methods can be roughly grouped into two categories: traditional methods and deep learning methods.

\subsubsection{Traditional Enhancement Methods}
In traditional methods for low-light image enhancement, histogram equalization (HE)-based approaches are a major category. There are many improved methods based on HE. Dynamic histogram equalization (DHE) divides the histogram of the image into sub-blocks and uses HE to stretch the contrast for each sub-block. Adaptive histogram equalization (AHE) \cite{pizer1987adaptive} changes image contrast by calculating the histogram of multiple local areas of the image and redistributing the brightness. AHE is more suitable for improving the local contrast and details of the image, but the noise is amplified. Contrast limited adaptive histogram equalization (CLAHE) \cite{pizer1990contrast} restricts the histogram of each sub-block and well controls the noise brought by AHE. Brightness preserving dynamic histogram equalization (BPDHE) \cite{ibrahim2007brightness} preserves the same brightness between the enhanced image and the pre-enhanced image. DHECI \cite{nakai2013color} uses differential gray-level histograms for color-image enhancement.\\
Methods based on Retinex Theory \cite{land1977retinex} decompose the image into reflectance and illumination. Such methods maintain the consistency of the reflectance, increase the brightness of the illumination and then take the pixel-wise product to enhance the low-light image. Single Scale Retinex (SSR) \cite{jobson1997properties} aims to restore the brightness of illumination after Retinex decomposition. Multi-Scale Retinex (MSR) \cite{jobson1997multiscale} combines the filtering results of multiple scales based on SSR. MSRCR adds a color recovery factor to tackle the color distortion caused by contrast enhancement in local areas of the image. NPE \cite{wang2013naturalness} strikes a balance between contrast enhancement and naturalness of estimated illumination. MF \cite{fu2016fusion} is a fashion-based method for contrast enhancement of the illumination. SRIE \cite{fu2016weighted} is a weighted variational model to estimate reflectance and illumination. BIMEF \cite{ying2017bio} provides a bio-inspired dual-exposure fusion algorithm to provide accurate contrast and lightness enhancement, which obtains results with less contrast and lightness distortion. LIME \cite{guo2016lime} estimates a structure-aware illumination map with structure prior and uses BM3D \cite{dabov2006image} for the post-processing denoising operation. Dong at el. \cite{dong2011fast} proposed an efficient algorithm by inverting an input low-lighting video and applying an optimized image de-haze algorithm on the inverted video. CRM \cite{ying2017new} uses the camera response model for the enhancement process. RRM \cite{li2018structure} is a Robust Retinex Model that considers noise mapping to improve the enhancement performance of low-light intensity noise images. LECARM \cite{ren2018lecarm} is a novel enhancement framework using the response characteristics of cameras to tackle color and lightness distortions. JED \cite{ren2018joint} is a joint low-light enhancement and denoising strategy. DIE \cite{zhang2019dual} is an automatic exposure correction solution, which can produce high-quality results for images of various exposure conditions. STAR \cite{xu2020star} gives a novel structure and texture aware Retinex model. LR3M \cite{ren2020lr3m} is a low-rank regularized retinex model which can suppress noise in the reflectance map. PRIEN \cite{li2021low} proposes a progressive-recursive image enhancement network.

\subsubsection{Deep Learning Based Enhancement Methods}
With the advent of deep learning, a great number of state-of-the-art methods have been developed for low-light image enhancement. LLNet \cite{lore2017llnet} is a deep auto-encoder model for enhancing lightness and denoising simultaneously. LLCNN \cite{tao2017llcnn} is a CNN-based method utilizing multi-scale feature maps and SSIM loss for low-light image enhancement. MSR-net \cite{shen2017msr} is a feedforward CNN with different Gaussian convolution kernels to simulate the pipeline of MSR for directly learning end-to-end mapping between dark and bright images. GLADNet \cite{wang2018gladnet} is a global illumination-aware and detail-preserving network that calculates global illumination estimation. LightenNet \cite{wang2018gladnet} serves as a trainable CNN by taking a weakly illuminated image as the input and outputting its illumination map. MBLLEN \cite{lv2018mbllen} uses multiple subnets for enhancement and generates the output image through multi-branch fusion. RetinexNet \cite{wei2018deep} decomposes low-light input into reflectance and illumination and enhances the lightness over illumination. EnlightenGan \cite{jiang2021enlightengan} trains an unsupervised generative adversarial network (GAN) without low/normal-light pairs. KinD \cite{zhang2019kindling} first decomposes low-light images into a noisy reflectance and a smooth illumination and then uses a U-Net to recover reflectance from noise and color bias. RDGAN \cite{wang2019rdgan} proposes a Retinex decomposition based GAN for low-light image enhancement. SID \cite{chen2018learning} uses a U-Net to enhance the extremely dark RAW image. RetinexDIP \cite{zhao2021retinexdip} provides a unified deep framework using a novel ”generative” strategy for Retinex decomposition. Zhang et al. \cite{zhang2020self} presented a self-supervised low-light image enhancement network, which is only trained with low-light images. Zero-DCE \cite{guo2020zero} estimates the brightness curve of the input image without any paired or unpaired data during training.

\section{Motivation}
Based on the assumption of Retinex Theory \cite{land1977retinex}, a natural image (S) can be decomposed into two components: reflectance (R) and illumination (L). 
\begin{equation}
	S=R*L
\end{equation}
where * represents a pixel-wise product operator. Reflectance is usually a three-channel image that contains color and the most of the high-frequency components, such as details and texture information. Illumination is a very smooth single-channel image that only contains low-frequency components, such as the intensity and distribution of lumination. In the process of decomposition, illumination is smooth enough to be regarded as noise-free since it only contains low-frequency information. However, noise hidden in the dark is amplified in the reflectance, which results in a very low peak signal-to-noise ratio (PSNR) for the reflectance. Previous approaches, such as LIME \cite{guo2016lime}, RetinexNet \cite{wei2018deep} and KinD\cite{zhang2019kindling}, used additional well-designed denoisers, such as BM3D \cite{dabov2006image}, CBDNet \cite{guo2019toward} or an embedded denoiser, to denoise the reflectance. However, there may be some problems such as color bias and loss of high-frequency details in the reflectance after applying extra denoisers. Furthermore, additional denoisers can significantly reduce the forward inferencing speed of the whole pipeline. In the enhancement phase, the brightness of the illumination is enhanced, but if the intensity of brightness and the distribution of light are not restored correctly, the result will be overexposed or underexposed. The color information of the image depends not only on the reflectance but also on the brightness information of the illumination. Incorrectly predicted illumination maps can also result in color bias. A variety of degradations may arise after enhancing the brightness of the low-light image. Many of the previous methods used multiple sub-methods or sub-networks to tackle some of these problems in numerous steps \cite{guo2016lime}\cite{wei2018deep}\cite{zhang2019kindling}, which can slow down the speed of the pipeline. Unlike previous methods, we aim to enhance the lightness of low-light images without introducing extra networks to deal with real-world noise and color distortion. We use two simple but effective U-Nets and the NCBC Module by carefully adjusting a series of well-designed loss functions rather than designing multiple sub-networks to deal with these problems individually. We are inspired by RetinexNet \cite{wei2018deep} and we improve the shortcomings of RetinexNet.

\begin{figure*}[htbp]
	\centering
	\includegraphics[width=17cm]{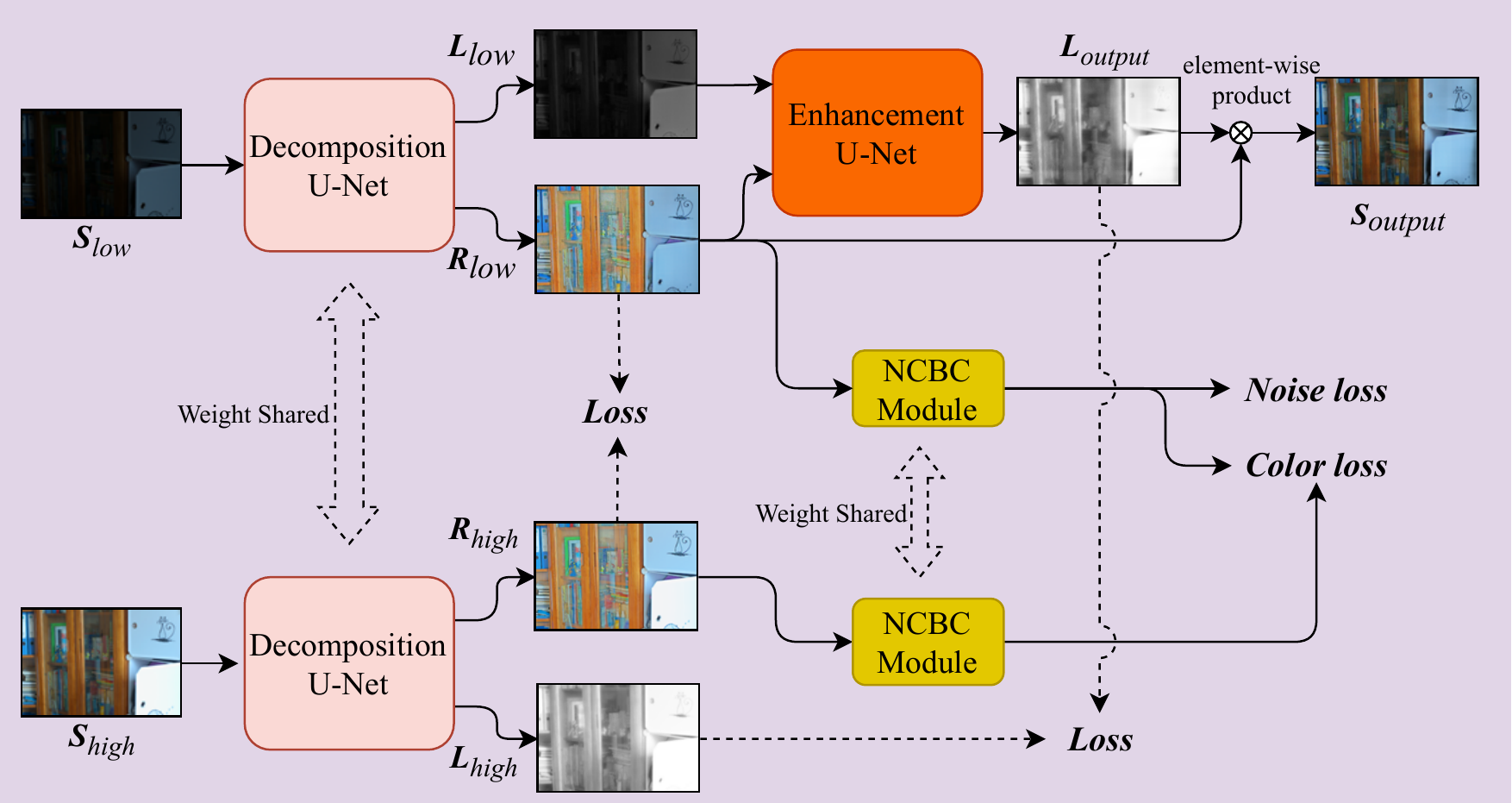}
	\includegraphics[width=18cm]{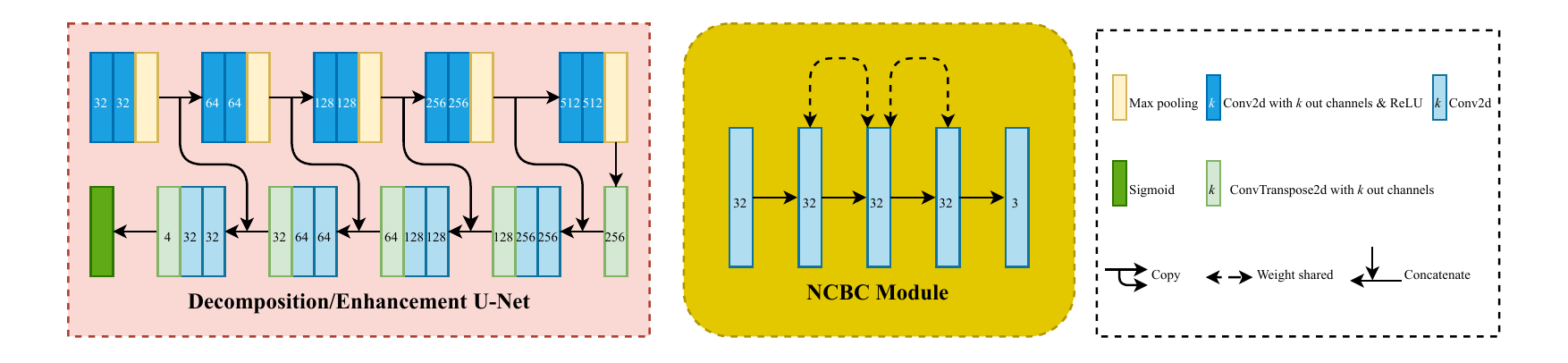}
	\caption{The network architecture of our BLNet. Both the Decomposition Net and Enhancement Net adopt the U-Net structure. Except for the difference in the number of channels in their input and output, the structure and parameters settings of the intermediate network are the same. The NCBC Module is a CNN that consists of several stacked convolutional blocks, and the middle three convolutional layers are weight sharing.}
	\label{model}
\end{figure*}

\section{Methodology}
As shown in Fig. \ref{model}, inspired by Retinex Theory, our proposed method adopts two U-Nets as decomposition and enhancement networks. In the decomposition phase, we use a U-Net to decompose the low-light image into two components (i.e., reflectance and illumination) and introduce a CNN as the Noise and Color Bias Control Module (NCBC Module) for suppressing the noise in reflectance while controlling color bias. In the enhancement phase, we also use a U-Net as our enhancement net, whose kernel settings and network structure are nearly the same as the first one, except for the number of channels of input and output images. By carefully adjusting the well-designed loss functions, we can use only these two U-Nets to enhance brightness, remove noise, restore color and preserve details simultaneously. In the following sections, we will explain our motivation, network architecture and loss functions in detail.

\subsection{Decomposition Network}
RetinexNet \cite{wei2018deep} uses a plain CNN as the decomposition net to decouple the low-light image into the reflectance and illumination, but this CNN can amplify the noise and introduce color distortion. KinD \cite{zhang2019kindling} uses a shallow U-Net as the decomposition net, but the noise is still amplified in the reflectance and there is also a serious color distortion. Therefore, inspired by \cite{zhang2019kindling}, we use a deeper U-Net as our decomposition net because the up-and-down sampling structure of U-Net has the function of denoising and will keep the noise from being amplified as much as possible. However, unpleasant noise and color distortion still exist in the reflectance, so we introduce the NCBC Module to tackle the noise and color bias in the reflectance. In the decomposition, inspired by RetinexNet \cite{wei2018deep} and KinD \cite{zhang2019kindling}, we also use the reconstruction loss $L_{rc}$ proposed in RetinexNet \cite{wei2018deep} to constrain the process of decomposition.
\begin{equation}
L_{rc}=\sum_{i}\sum_{j}\lambda_{ij}\left \|R_{i}\ast L_{j} - S_{j}  \right \|_{1}
\end{equation}
where $*$ denotes the pixel-wise product and $i,j$ denote the low- and high-frequency components, respectively. $R_{high}$, $R_{low}$ and $L_{high}$, $L_{low}$ represent the reflectance and the illumination decomposed by GroundTruth and low-light input, respectively. When $i=j$, $\lambda_{ij}=1$; otherwise, $\lambda_{ij}=0.001$.\\
We use the equal loss $L_{equal}$ proposed in \cite{wei2018deep} to constrain the reflectance to be as similar as possible to that decomposed by GroundTruth. 
\begin{equation}
L_{equal}=\left \| R_{low} - R_{high} \right \|_{1}
\end{equation}
Previous Retinex-based methods \cite{wei2018deep}\cite{zhang2019kindling} use the smoothness loss to maintain the spatial smoothness of illumination so that the illumination contains only low-frequency information, such as intensity and distribution of brightness, and all of the high-frequency information, such as details and texture, are retained in the reflectance map. We also use the smoothness loss $L_{smooth}$ proposed in RetinexNet \cite{wei2018deep} to keep our illumination spatially smooth.
\begin{equation}
L_{smooth}=\sum_{i}\left \| \triangledown _{h}L_{i}\ast e^{-\lambda \ast \triangledown_{h}R_{i}] } \right \|+\sum_{i}\left \| \triangledown _{v}L_{i}\ast e^{-\lambda \ast \triangledown_{v}R_{i}] } \right \|
\end{equation}
where $i=low,high$ and $\triangledown_{h}$, $\triangledown_{v}$ denote the gradients in the horizontal and vertical directions. According to RetinexNet \cite{wei2018deep}, we also set the weight coefficient $\lambda$ to 10.\\
The total decomposition loss is as follows: 
\begin{equation}
L_{decom}=1.0L_{rc} + 0.1L_{smooth} +  0.01L_{equal}
\end{equation}
According to the previous methods \cite{guo2016lime}\cite{wei2018deep}\cite{wang2019rdgan}\cite{zhang2019kindling}, the illumination map is smooth enough and contains only low-frequency information, and all of the noise after decomposition appears in the reflectance map. However, in our opinion, the illumination map can contain some high-frequency details and texture information, which can be restored by our enhancement net.

\subsection{Noise and Color Bias Control Module}
Through experiments, we found that directly using TV loss to smooth the reflectance is not good for denoising and may lead to image distortion. In order to denoise the reflectance map without introducing additional networks or denoisers, which may slow down the processing of the pipeline and cause color distortion in the reflectance map, we propose the NCBC Module to suppress the noise and control the color bias of the reflectance. At the same time, the details smoothed out by the TV loss along with the noise in the reflectance are transferred and reconstructed into the illumination. Therefore, unlike other methods, as shown in Fig. \ref{ablation ncbc}, our illumination also contains some high-frequency information such as details and texture. In this way, we can save the details in the illumination map, which means that the details can be restored and reconstructed by the enhancement net. This allows us to filter out the noise in the reflectance map while preserving the details and texture information.\\
We use a stacked CNN to extract the feature map of the reflectance as the estimated noise map, which contains noise level and color distribution information of the reflectance. The color and texture of the image belong to global information. Even though the convolution operation is a local operator, we can construct a CNN by stacking multiple convolutional layers to enlarge the receptive field so as to obtain the global feature information of the image, such as color and texture. Hence, the output of our NCBC Module has the ability to capture the feature map of global noise level and color information of the image.\\
Our NCBC Module consists of a plain CNN, whose architecture is shown in Fig. \ref{model}, and two loss functions: noise loss and color loss. The input images of our NCBC Module are: (1) the reflectance with noise and color distortion, which is decomposed from the low-light input images and the reflectance without noise; and (2) the color distortion, which is decomposed from normal-light GroundTruth. We apply TV loss to the output of the reflectance with noise and color distortion to smooth the $R_{low}$. 
Total variance loss (TV loss) \cite{guo2019toward} is as follows:
\begin{equation}
	L_{TV}^{low}=\left \| \triangledown _{h}\phi (R_{low})] \right \|_{2}^{2}+\left \| \triangledown _{v}\phi (R_{low})] \right \|_{2}^{2}
\end{equation}
where $\phi$ means the output feature map of our NCBC Module, and $\triangledown_{h}$ and $\triangledown_{v}$ denote the gradients in the horizontal and vertical directions.
The noise level of GroundTruth is zero by default. According to \cite{zhao2016loss}, we use MSE loss to denoise and constrain the similarity of $R_{low}$ and $R_{high}$ while using TV loss to smooth the reflectance. 
\begin{equation}
L_{MSE}=\left \| R_{low} - R_{high}\right \|_{2}^{2}
\end{equation}
So our noise loss is as follows:
\begin{equation}
	L_{noise}=0.05L_{TV}^{low} + 1.0L_{MSE}
\end{equation}
In addition, we use color loss $L_{color}$ to recover the reflectance from the color distortion caused by the decomposition U-Net in the process of decomposition, and we apply TV loss on the reflectance map of low-light images for denoising. Inspired by low-lightGan \cite{kim2019low}, the color loss is as follows:
\begin{equation}
	L_{color}=\left \| \phi (R_{low})- \phi (R_{high})\right \|_{1}
\end{equation}
where $\phi$ is the output feature map of our NCBC Module.
Low-lightGan \cite{kim2019low} uses a Gaussian blur function to remove local details of images while preserving global information, such as color and texture, and then it adopts color loss to alleviate the color distortion problem by forcing the global information of the two images to be as close as possible.\\
The total NCBC loss is as follows: 
\begin{equation}
	L_{NCBC}=0.2L_{noise} +  0.1L_{color}
\end{equation}

\begin{figure}
	\flushleft
	\subfigure[RetinexNet\cite{wei2018deep}(17.78/0.49/0.83)]{
		\begin{minipage}[b]{0.48\linewidth}
			\flushleft
			\includegraphics[width=4.25cm]{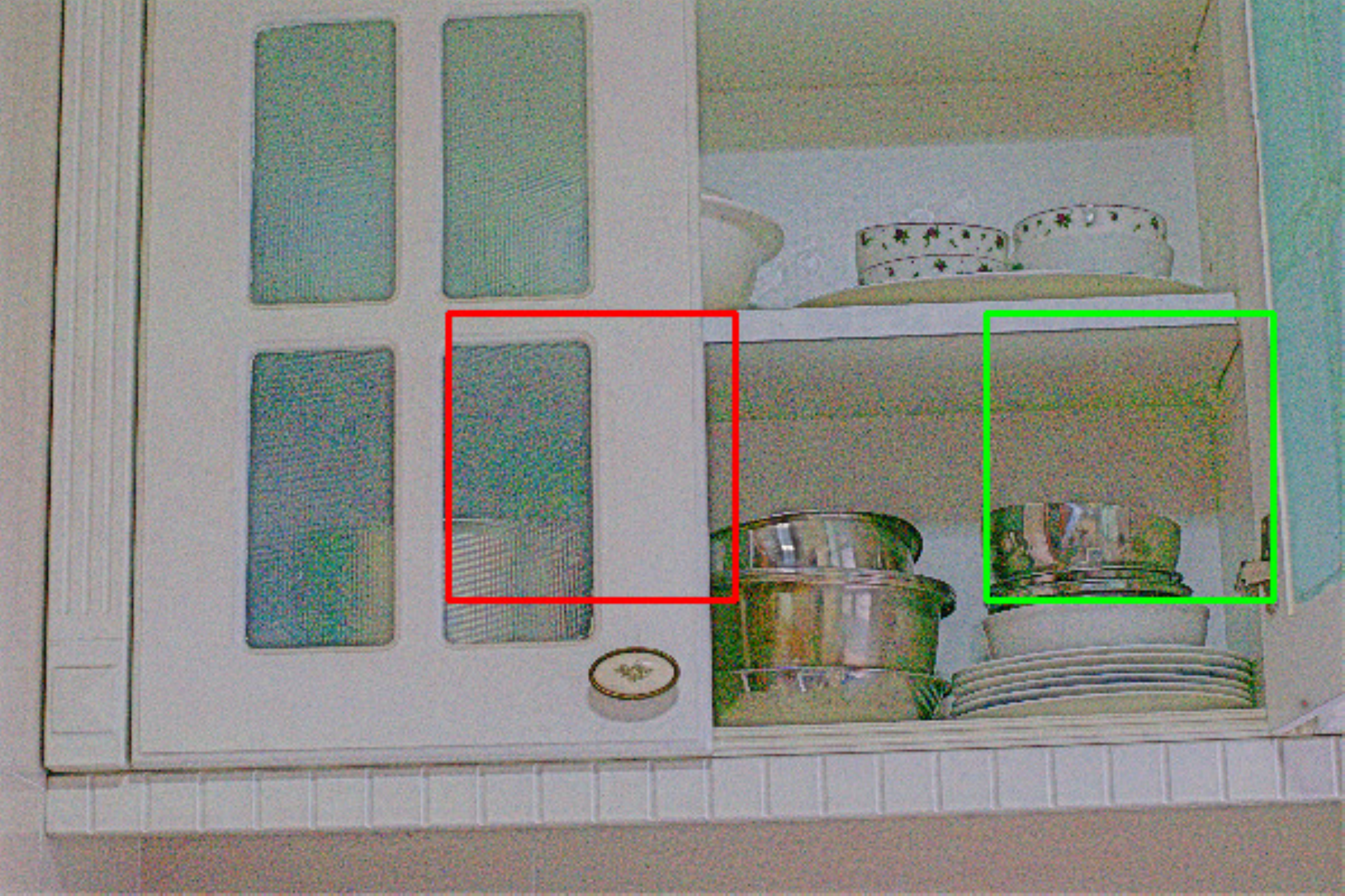}\vspace{1pt}
			\begin{minipage}[b]{0.48\linewidth}
				\flushleft
				\includegraphics[width=2.1cm]{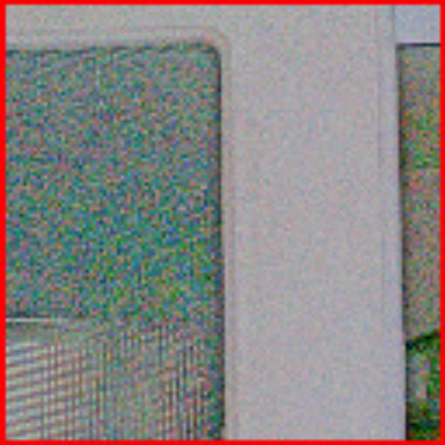}
			\end{minipage}
			\begin{minipage}[b]{0.48\linewidth}
				\flushleft
				\includegraphics[width=2.1cm]{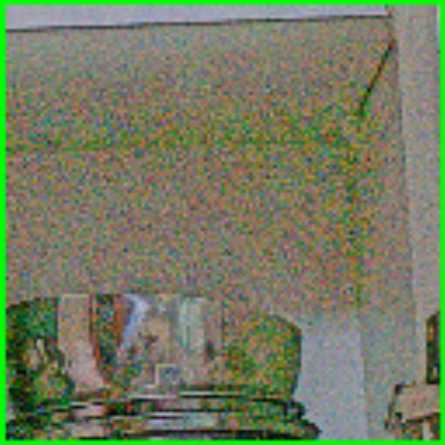}
			\end{minipage}
	\end{minipage}}
	\subfigure[Ours w/o NCBC(17.70/0.83/0.94)]{
		\begin{minipage}[b]{0.48\linewidth}
			\flushleft
			\includegraphics[width=4.25cm]{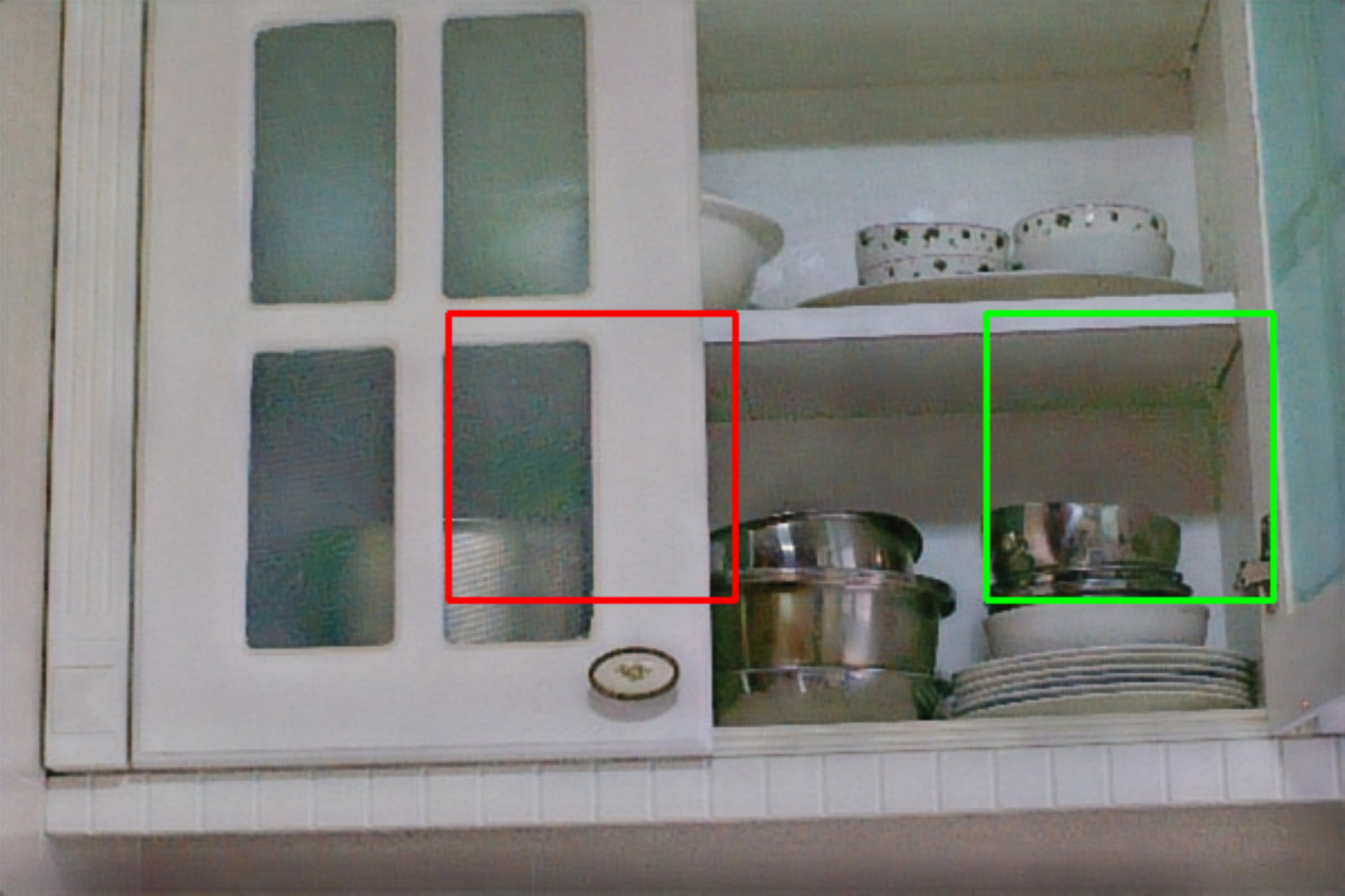}\vspace{1pt}
			\begin{minipage}[b]{0.48\linewidth}
				\flushleft
				\includegraphics[width=2.1cm]{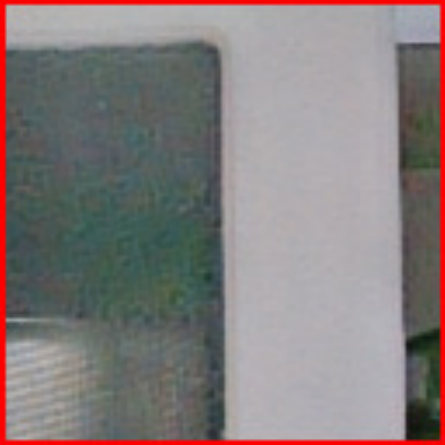}
			\end{minipage}
			\begin{minipage}[b]{0.48\linewidth}
				\flushleft
				\includegraphics[width=2.1cm]{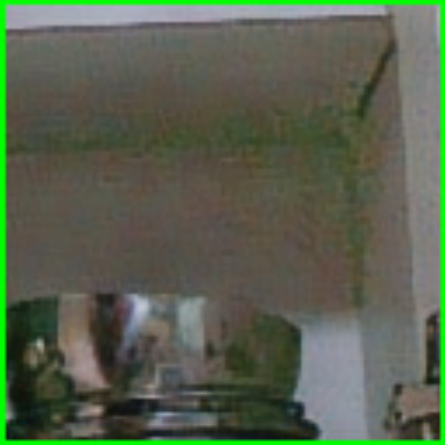}
			\end{minipage}
	\end{minipage}}
	\subfigure[Ours with NCBC(22.06/0.87/0.62)]{
		\begin{minipage}[b]{0.48\linewidth}
			\flushleft
			\includegraphics[width=4.25cm]{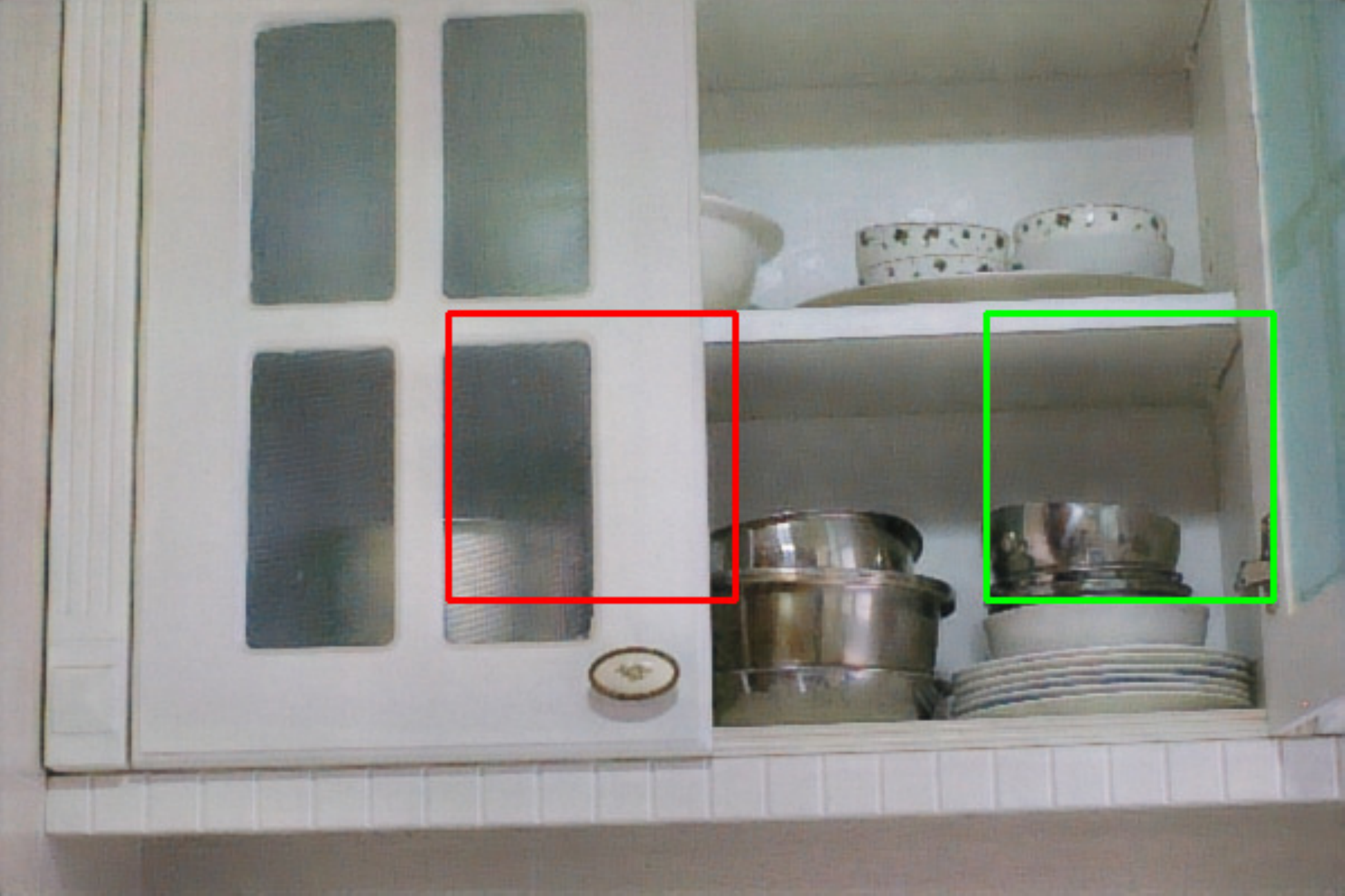}\vspace{1pt}
			\begin{minipage}[b]{0.48\linewidth}
				\flushleft
				\includegraphics[width=2.1cm]{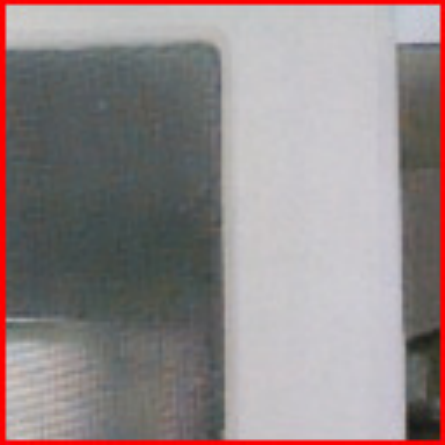}
			\end{minipage}
			\begin{minipage}[b]{0.48\linewidth}
				\flushleft
				\includegraphics[width=2.1cm]{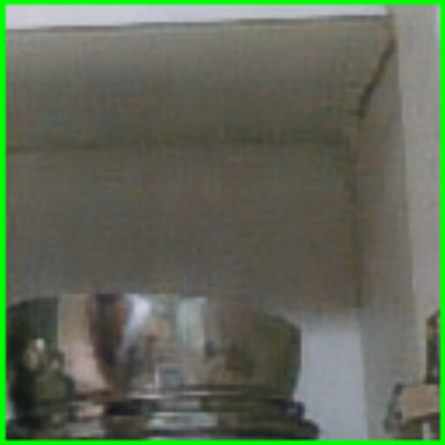}
			\end{minipage}
	\end{minipage}}
	\subfigure[GroundTruth]{
		\begin{minipage}[b]{0.48\linewidth}
			\flushleft
			\includegraphics[width=4.25cm]{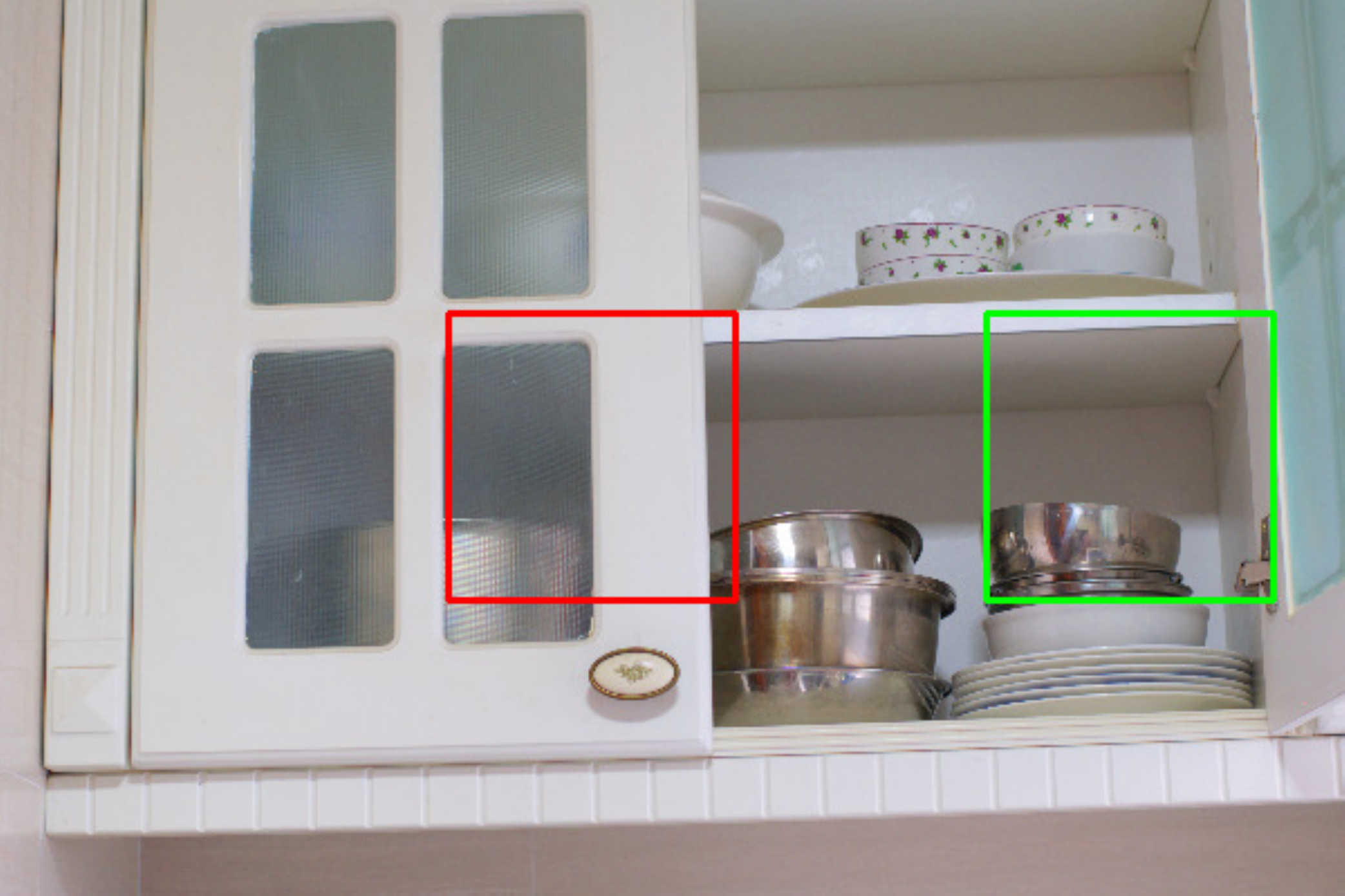}\vspace{1pt}
			\begin{minipage}[b]{0.48\linewidth}
				\flushleft
				\includegraphics[width=2.1cm]{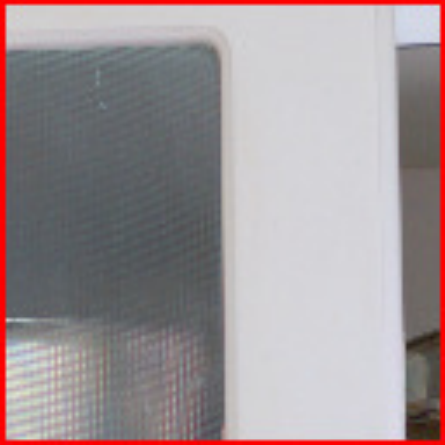}
			\end{minipage}
			\begin{minipage}[b]{0.48\linewidth}
				\flushleft
				\includegraphics[width=2.1cm]{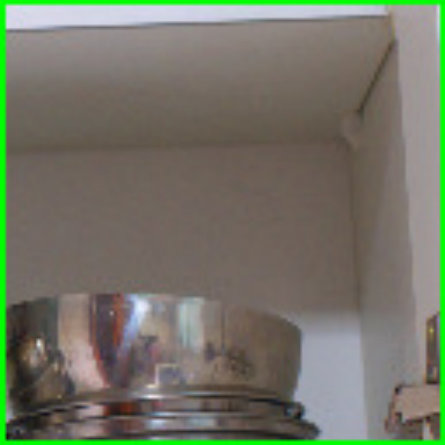}
			\end{minipage}
	\end{minipage}}
	\caption{Ablation study on the LOL dataset (PSNR/SSIM/Color Bias Average).}
	\label{abla1}
\end{figure}

\subsection{Enhancement Network}
The network structure and kernel settings of our enhancement network are the same as those of the decomposition network. In particular, we use a series of loss functions for the brightness enhancement in order to deal with noise removal, color restoration and detail and texture preservation simultaneously. The effectiveness of our method depends primarily on the contribution of our enhancement loss function. The enhancement loss function in the enhancement phase is as follows:
\begin{equation}
L_{eh}=1.0 L_{rc} + 1.0 L_{bri} + 1.0 L_{per}  + 1.0 L_{grad}
\end{equation}
We use reconstruction loss $L_{rc}$ to constrain the final output after the pixel-wise product to be as close as possible to the GroundTruth. The reconstruction loss is as follows:
\begin{equation}
L_{rc}=\left \|R_{low}\ast L_{output} - S_{high}  \right \|_{1}
\end{equation}
where $R_{low}$, $L_{output}$ and $S_{high}$ respectively represent the reflectance decomposed from the low-light image, the brightness-improved illumination by our enhancement net and the normal-light input GroundTruth.\\ 
We use $L1$ loss as our brighten loss function $L_{bri}$ to enhance the brightness of the illumination decomposed from low-light images and constrain the output brightness-improved illumination to be as similar as possible to its counterpart from GroundTruth. The brighten loss function is as follows:
\begin{equation}
L_{bri}=\left \| L_{output} - L_{high} \right \|_{1}
\end{equation}
where $L_{output}$ and $L_{high}$ denote the brightness-improved output of the enhancement net and the decomposed illumination of the normal-light GroundTruth, respectively.\\
Perceptual loss is widely used in super-resolution and style tranfer \cite{johnson2016perceptual}. Because we transfer some of the image details from reflectance to illumination, we use perceptual loss to preserve the texture information during illumination reconstruction in the enhancement network. In addition, perceptual loss can make our output as similar to the normal-light GroundTruth as possible in perception. 
Perceptual loss $L_{per}$ is as follows:
\begin{equation}
L_{per}=\frac{1}{CHW}\left \| \phi (R_{low}\ast L_{output})-\phi (S_{high}) \right \|_{2}^{2}
\end{equation}
where $\phi$ is the 31st feature map obtained by the VGG16 \cite{simonyan2014very} network pre-trained on ImageNet database. C,H,W represents the number of channels, the height and the width of the input image, respectively.\\
We find that the obtained output is rather rough, and the details and edges information are still lost partly. Therefore, we use gradient loss $L_{grad}$ to maintain the balance of the sharpness and smoothness of the output.
\begin{equation}
\begin{aligned}
L_{grad}=\left \| \triangledown_{h}(R_{low}\ast L_{output})-\triangledown_{h}(S_{high})  \right \|_{1}
\\+ \left \| \triangledown_{v}(R_{low}\ast L_{output})-\triangledown_{v}(S_{high})  \right \|_{1}
\end{aligned}
\end{equation}
where $\triangledown_{h}$, $\triangledown_{v}$ denote the gradients in the horizontal and vertical directions, respectively. $S_{high}$ is the normal-light GroundTruth.\\
The channel number of the enhancement net is four. Like RetinexNet \cite{wei2018deep}, we also concatenate the illumination and reflectance of the low-light image decomposed by the decomposition net as the four-channel input of the enhancement net. 
Because both the illumination and the reflectance contain high-frequency details and texture information after decomposition, concatenating the illumination and reflectance as the input of our enhancement net can help the enhancement net reconstruct the details and texture information. 
In addition, the output of the enhancement net is the corresponding brightness-improved illumination with a single channel. Finally, we multiply the output illuminantion with the noise-removed reflectance at the pixel-wise level and obtain the final output, which is noise-free, detail-preserved, color-restored and brightness-improved.

\section{Experimental Evaluation}
\subsection{Implementation details}
We train our model on the LOL \cite{wei2018deep} training dataset and evaluate it on the LOL validation dataset. In addition, we test our model on three popular test datasets: DICM \cite{lee2013contrast} dataset, LIME \cite{guo2016lime} dataset and MEF \cite{ma2015perceptual} dataset. The LOL dataset consists of two parts: the real-world dataset and the synthetic dataset. Because the LOL synthetic dataset cannot simulate the degradation of real-world images very well, we deprecate the synthetic dataset where low-light images are synthesized from normal-light images and only use the real-world dataset, which contains 500 low-light/normal-light image pairs captured by adjusting the exposure time and ISO of the camera. We use 485 image pairs as the training set and the remaining 15 image pairs as the validation set. We use the Adam \cite{kingma2014adam} optimizer to optimize the training of the model and set the training batch-size to four and the patch-size of random crop to 320x320. We use the PyTorch framework to build our model on a PC with an Nvidia TITAN XP GPU and an Intel Core i7-9700 3.00GHz CPU.\\
We evaluate our method on the LOL \cite{wei2018deep} validation dataset and test it on several widely used datasets, including DICM \cite{lee2013contrast} dataset, LIME \cite{guo2016lime} dataset and MEF \cite{ma2015perceptual} dataset. We adopt PSNR, SSIM \cite{wang2004image}, LPIPS \cite{zhang2018unreasonable}, FSIM \cite{zhang2011fsim} and UQI \cite{wang2002universal} as the quantative metrics to measure the performance of our method. In addition, we use Angular Error \cite{hordley2004re} and DeltaE\cite{sharma2005ciede2000} as the metrics of color distortion to calculate the color bias between our results and GroundTruth. The Angular Error is as follows:
\begin{equation}
	Angular Error=arcos(\frac{<S_{output},S_{high}>}{\left \| S_{output} \right \| \cdot \left \| S_{high} \right \|})
\end{equation}
where $S_{output}$ and $S_{high}$ denote the final output of our method and the normal-light input GroundTruth, respectively.

\subsection{Ablation Study}
\begin{figure}
	\flushleft
	\subfigure[Our model without NCBC (1.16/1.14/1.15)]{
		\begin{minipage}[b]{0.48\linewidth}
			\flushleft
			\includegraphics[width=4.25cm]{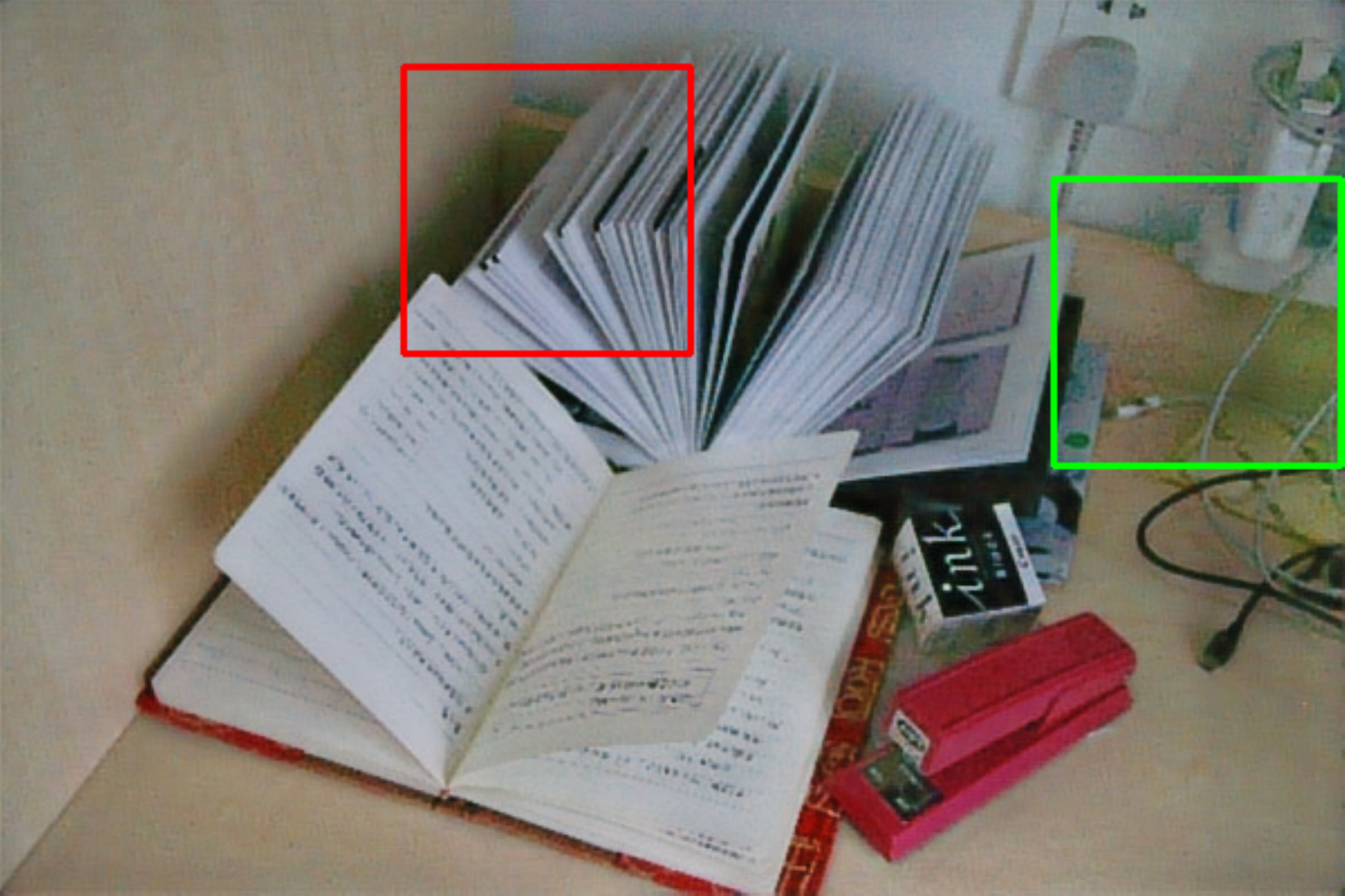}\vspace{1pt}
			\begin{minipage}[b]{0.48\linewidth}
				\flushleft
				\includegraphics[width=2.1cm]{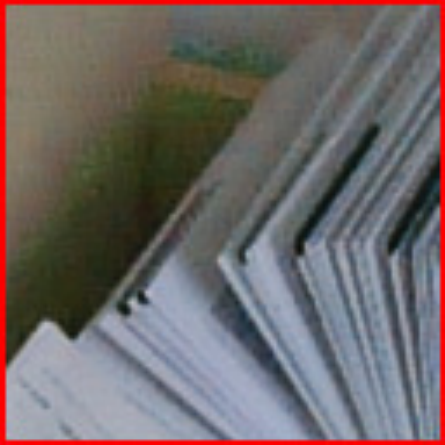}
			\end{minipage}
			\begin{minipage}[b]{0.48\linewidth}
				\flushleft
				\includegraphics[width=2.1cm]{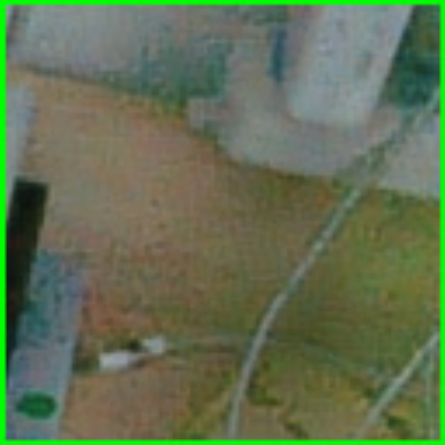}
			\end{minipage}
	\end{minipage}}
	\subfigure[Our model with NCBC (1.03/0.99/1.01)]{
		\begin{minipage}[b]{0.48\linewidth}
			\flushleft
			\includegraphics[width=4.25cm]{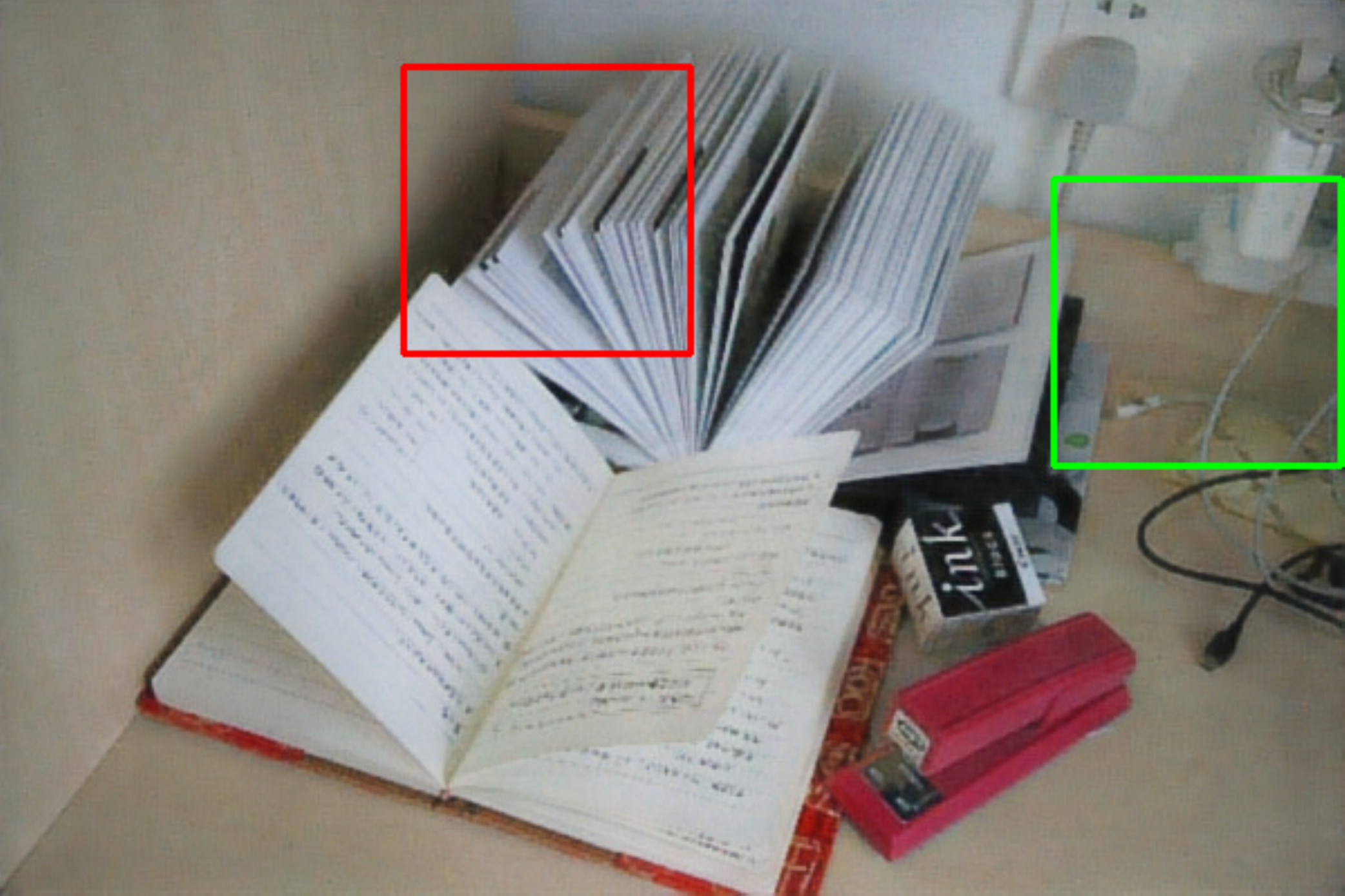}\vspace{1pt}
			\begin{minipage}[b]{0.48\linewidth}
				\flushleft
				\includegraphics[width=2.1cm]{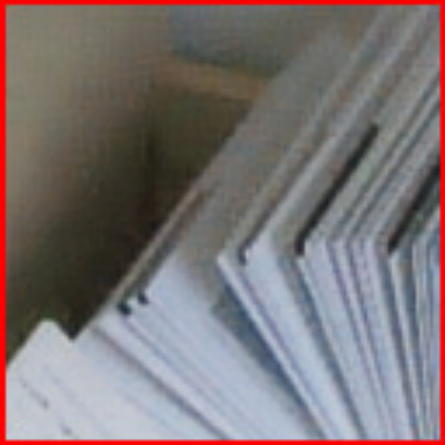}
			\end{minipage}
			\begin{minipage}[b]{0.48\linewidth}
				\flushleft
				\includegraphics[width=2.1cm]{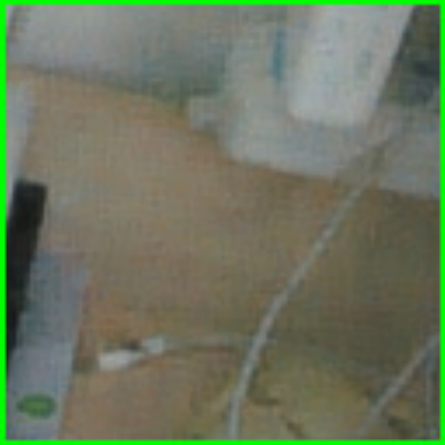}
			\end{minipage}
	\end{minipage}}
	\subfigure[Our model without NCBC (1.73/1.87/1.80)]{
		\begin{minipage}[b]{0.48\linewidth}
			\flushleft
			\includegraphics[width=4.25cm]{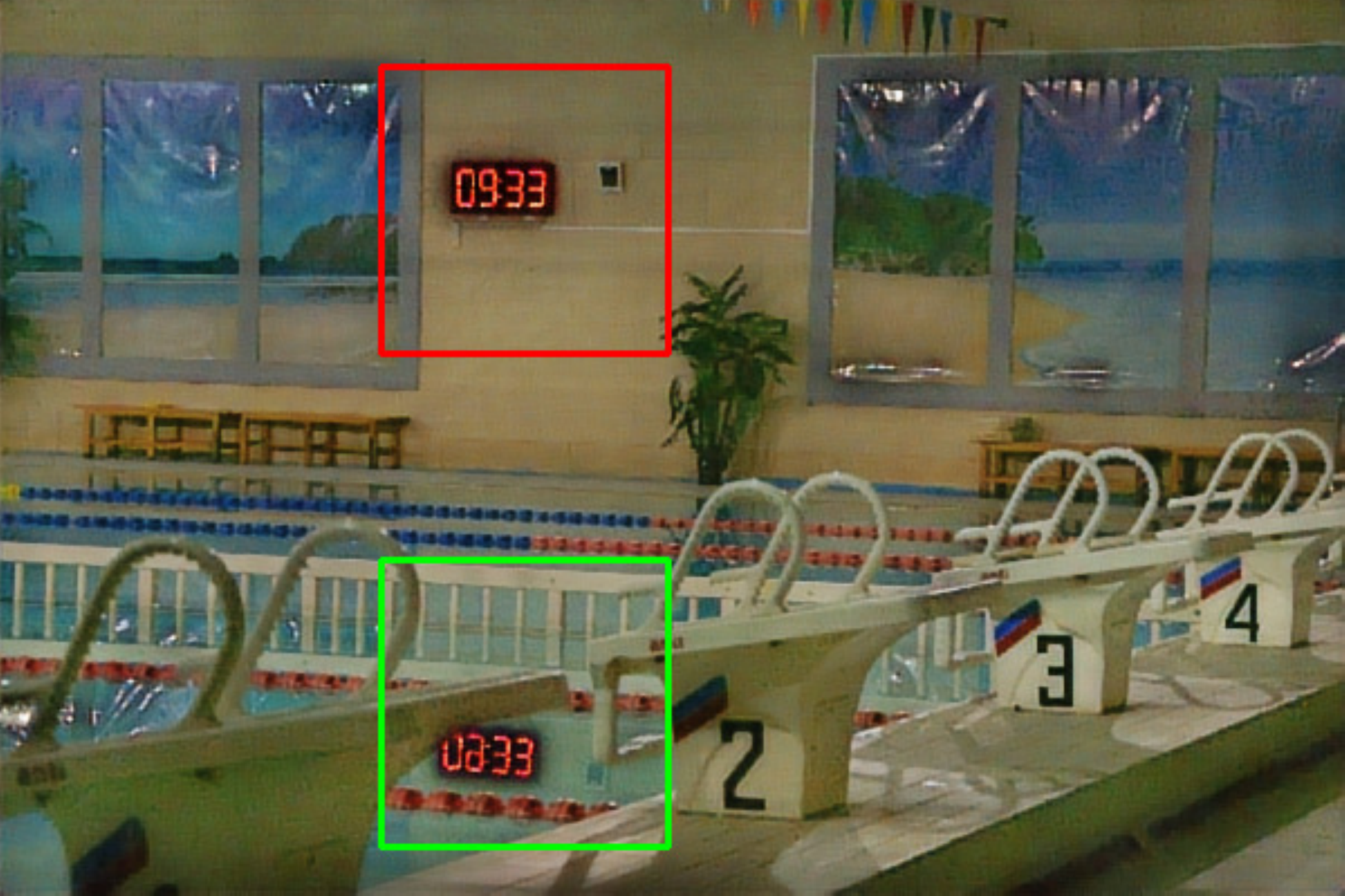}\vspace{1pt}
			\begin{minipage}[b]{0.48\linewidth}
				\flushleft
				\includegraphics[width=2.1cm]{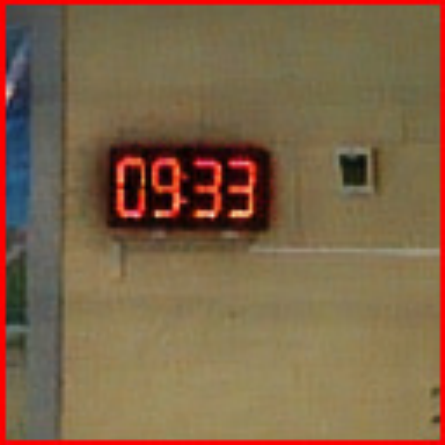}
			\end{minipage}
			\begin{minipage}[b]{0.48\linewidth}
				\flushleft
				\includegraphics[width=2.1cm]{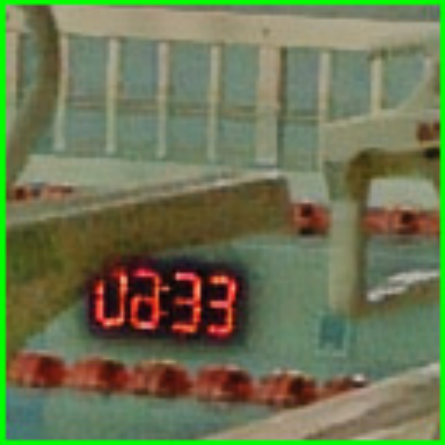}
			\end{minipage}
	\end{minipage}}
	\subfigure[Our model with NCBC (0.86/1.24/1.05)]{
		\begin{minipage}[b]{0.48\linewidth}
			\flushleft
			\includegraphics[width=4.25cm]{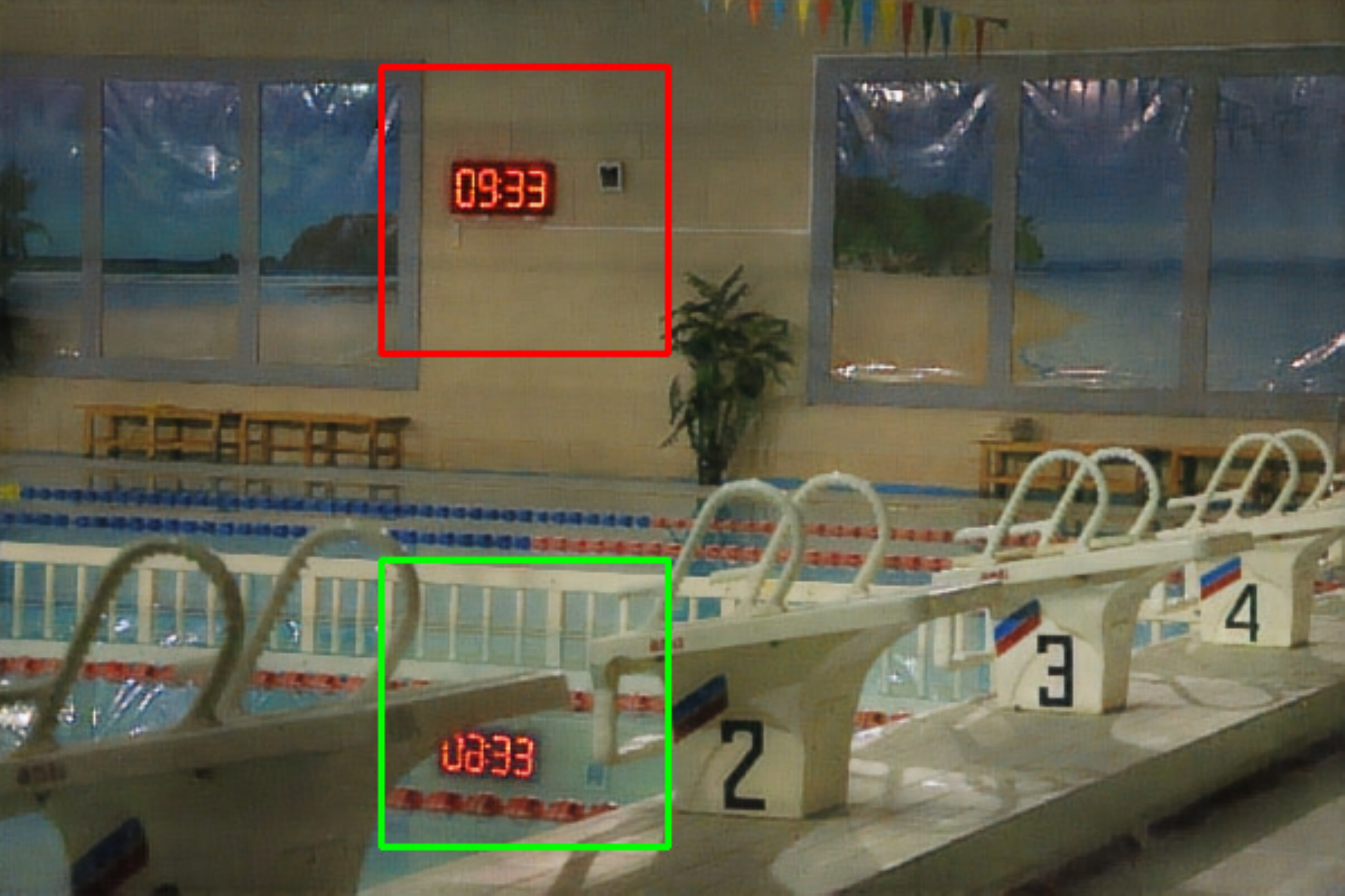}\vspace{1pt}
			\begin{minipage}[b]{0.48\linewidth}
				\flushleft
				\includegraphics[width=2.1cm]{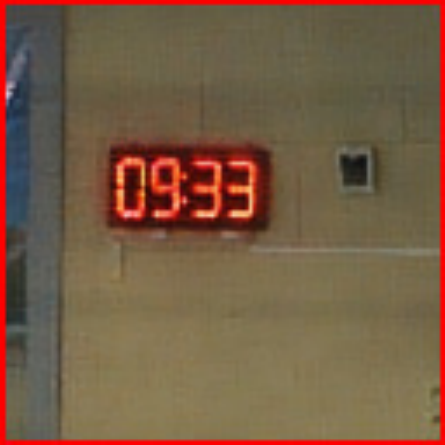}
			\end{minipage}
			\begin{minipage}[b]{0.48\linewidth}
				\flushleft
				\includegraphics[width=2.1cm]{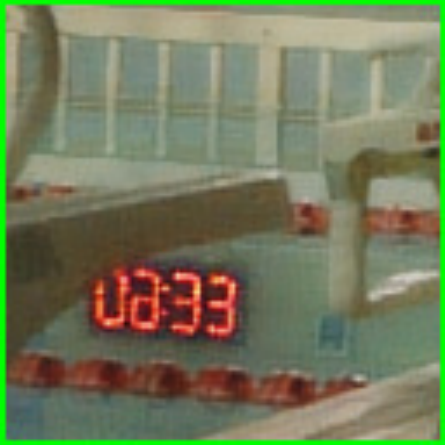}
			\end{minipage}
	\end{minipage}}
	\caption{Ablation study on LOL dataset (color bias mean/color bias median/color bias average).}
	\label{abla2}
\end{figure}

\begin{figure*}
	\flushleft
	\subfigure{
		\begin{minipage}{10\linewidth} 
			\includegraphics[width=4.4cm]{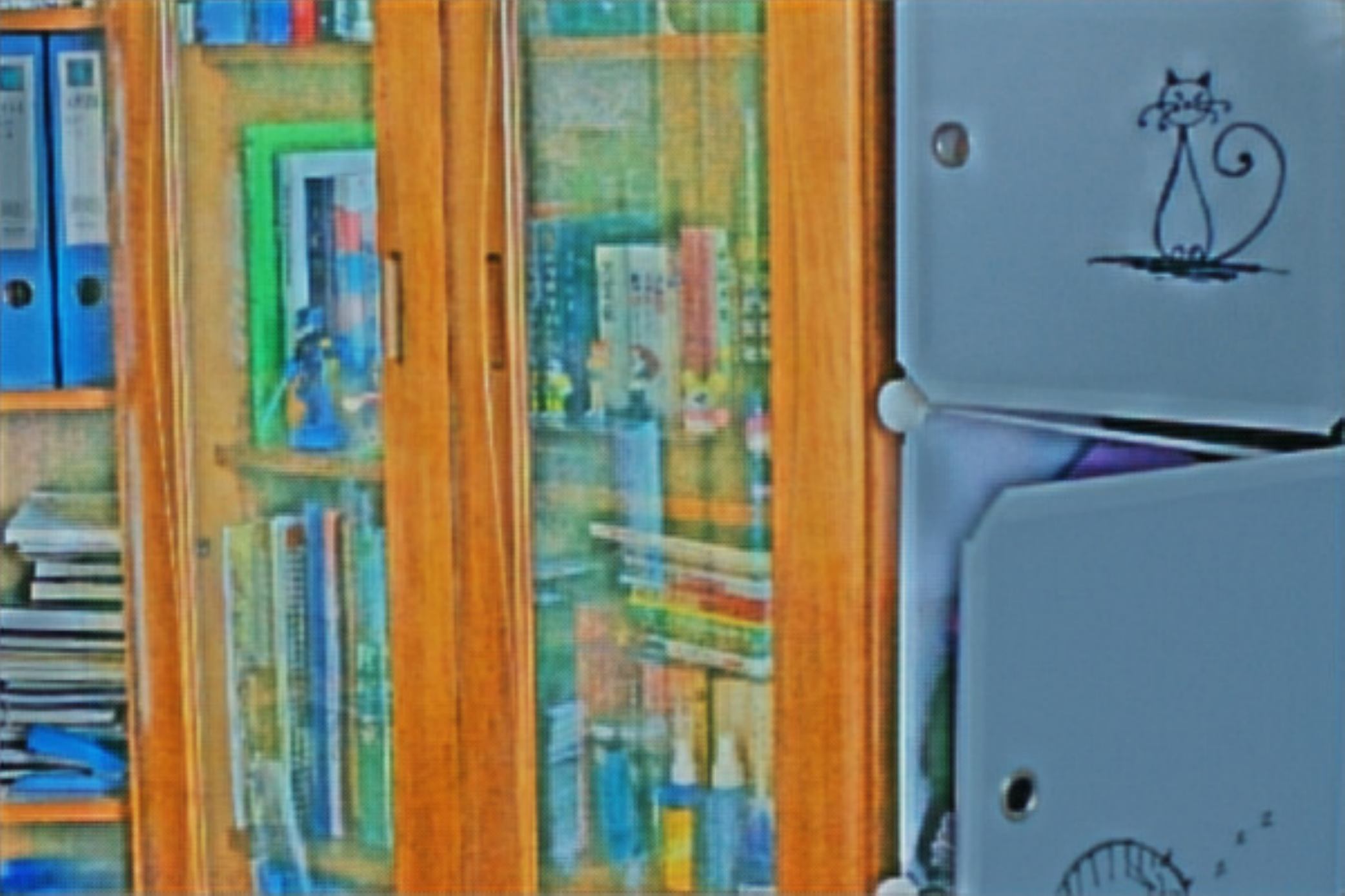}\vspace{-1pt}  
			\includegraphics[width=4.4cm]{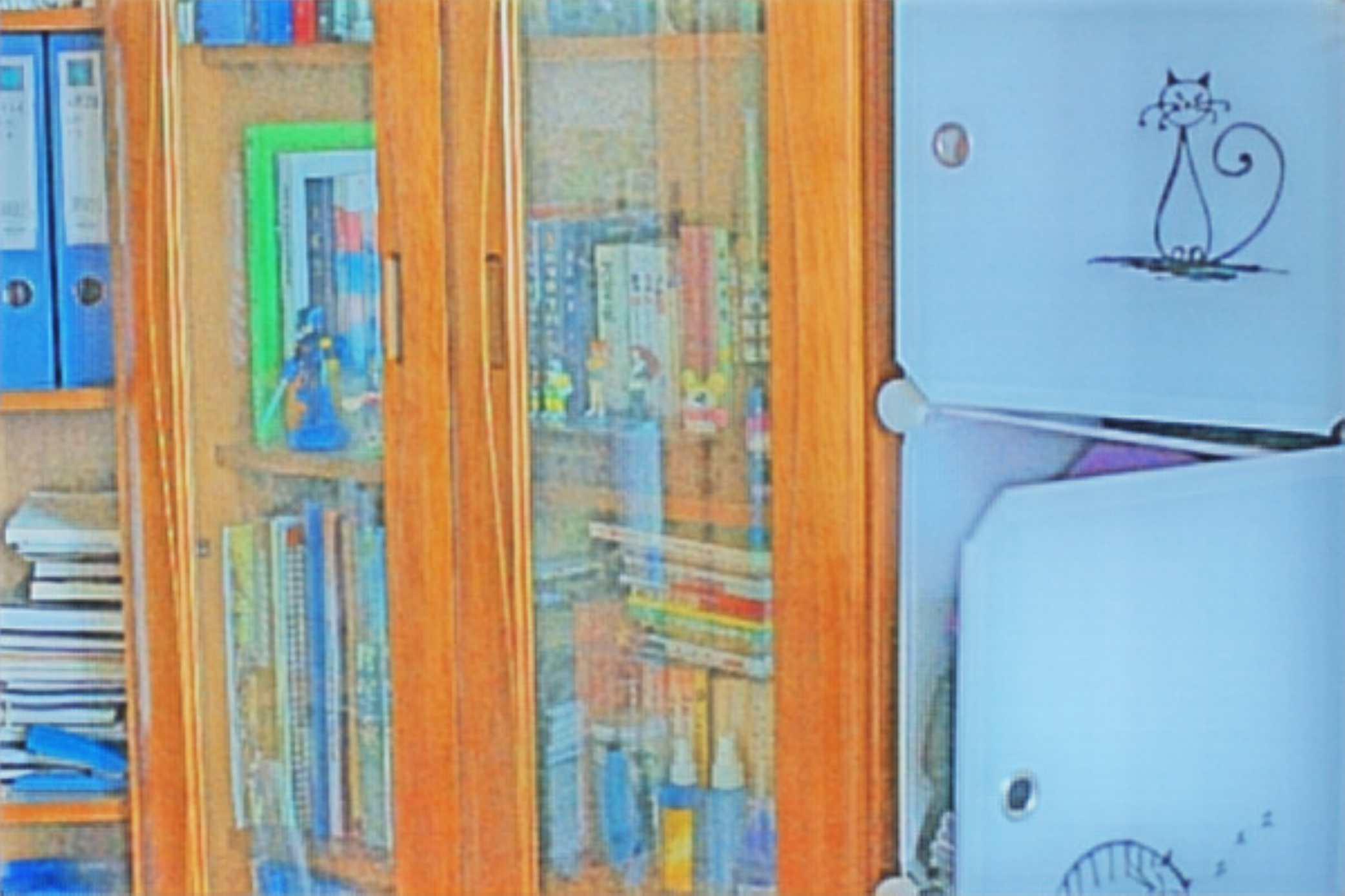}\vspace{-1pt} 
			\includegraphics[width=4.4cm]{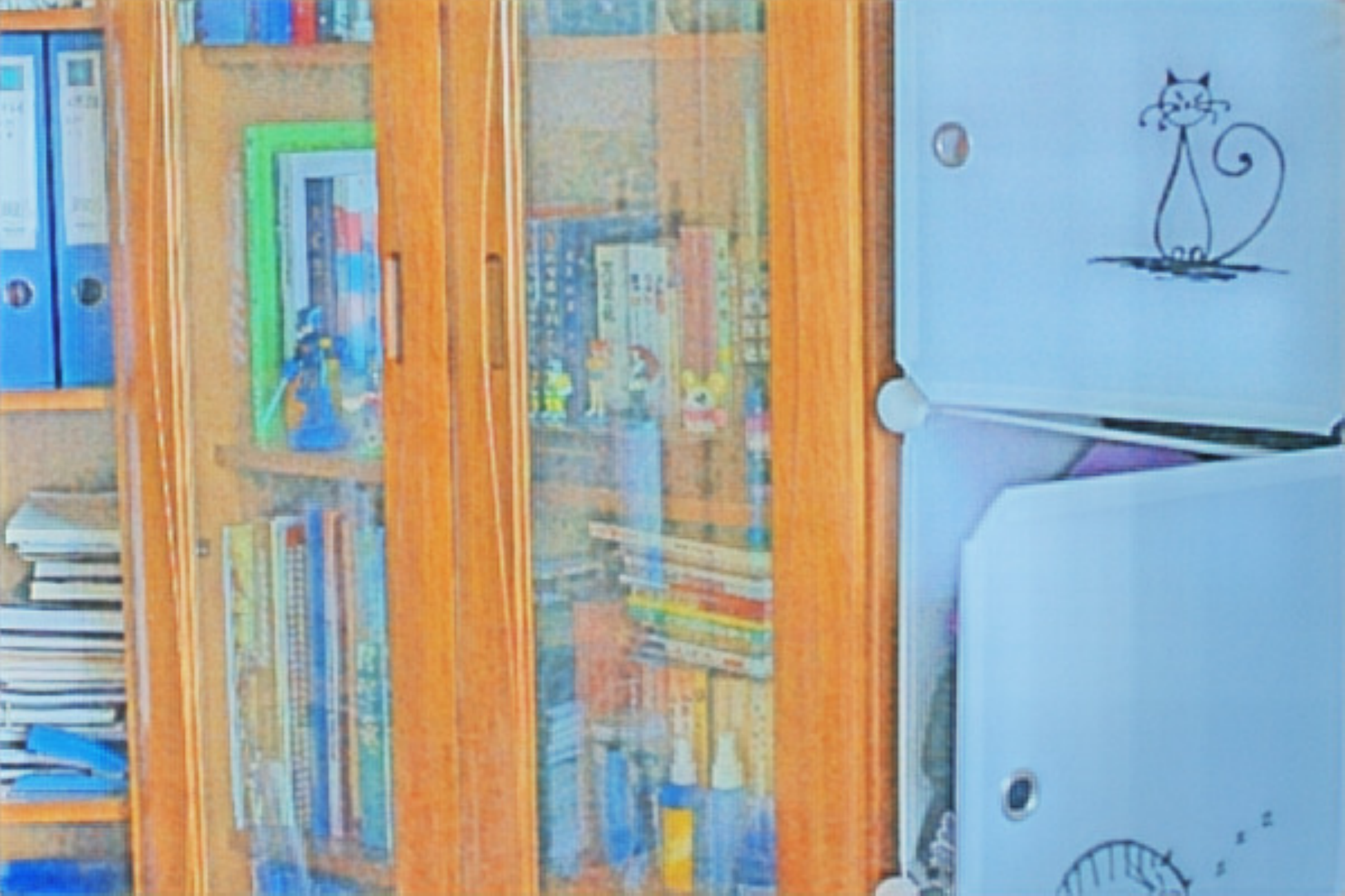}\vspace{-1pt} 
			\includegraphics[width=4.4cm]{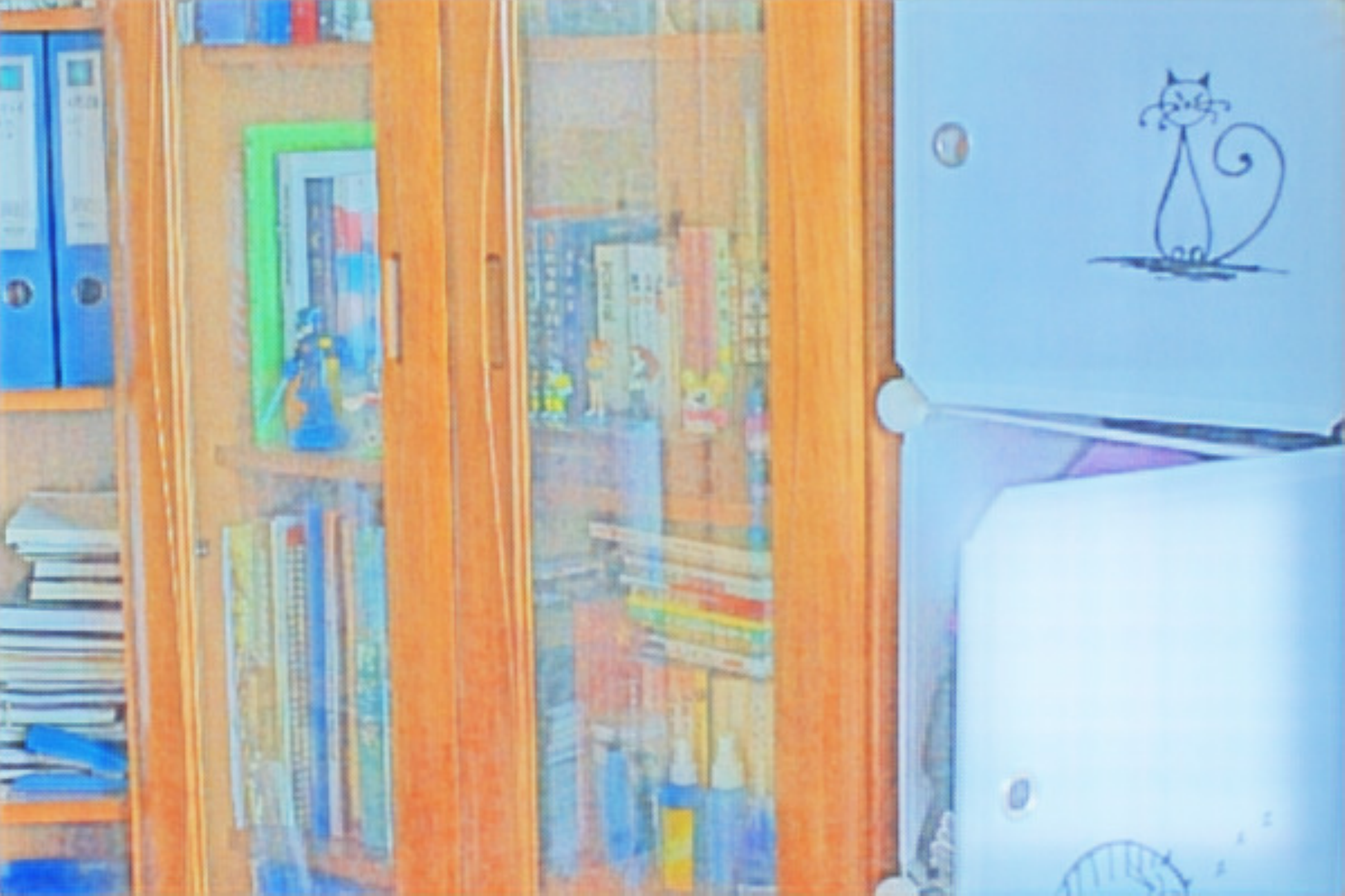}\vspace{-1pt} 
		\end{minipage}
	}
	\quad
	\subfigure{
		\begin{minipage}{10\linewidth}
			\includegraphics[width=4.4cm]{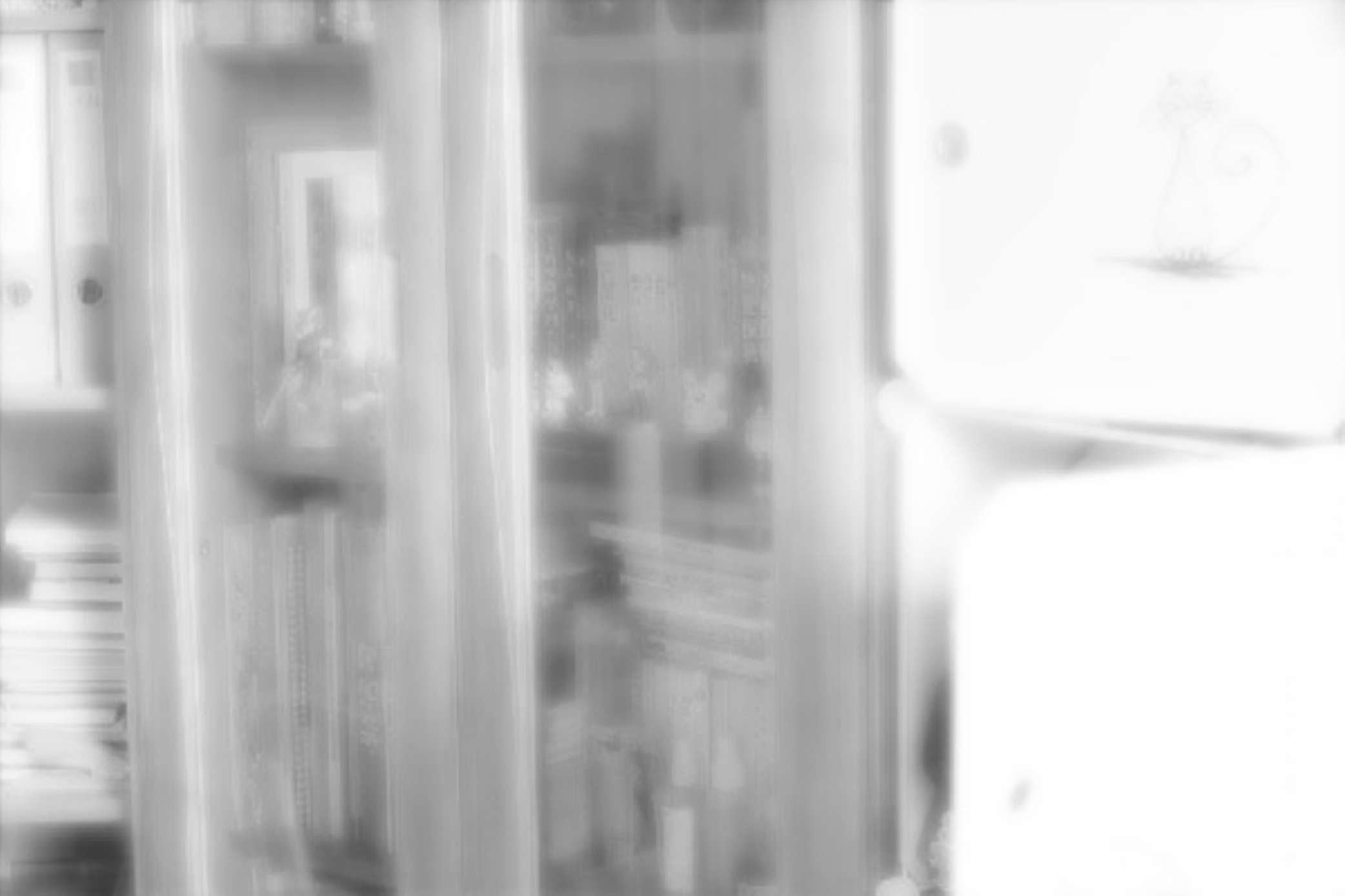}\vspace{-1pt}  
			\includegraphics[width=4.4cm]{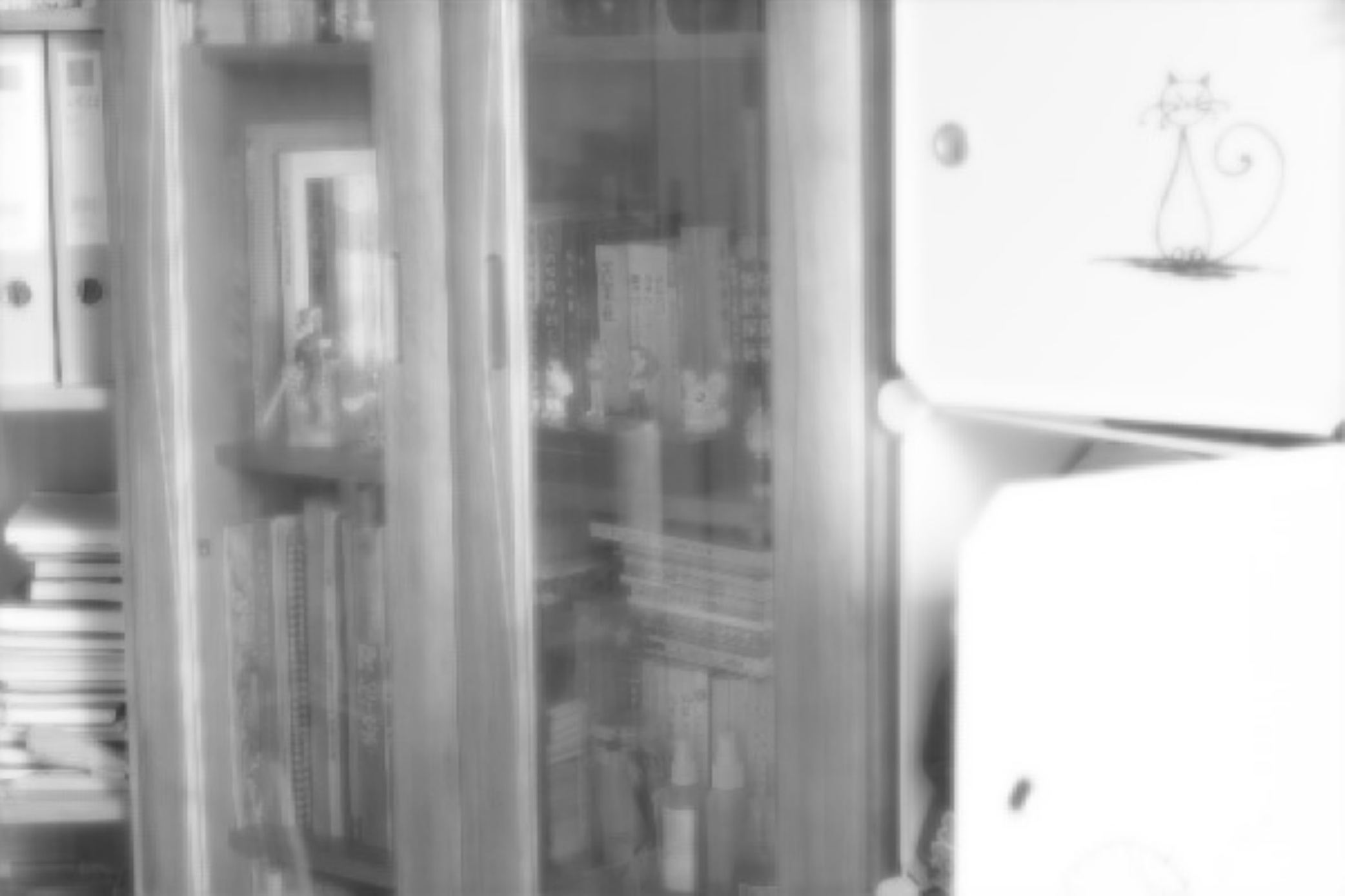}\vspace{-1pt} 
			\includegraphics[width=4.4cm]{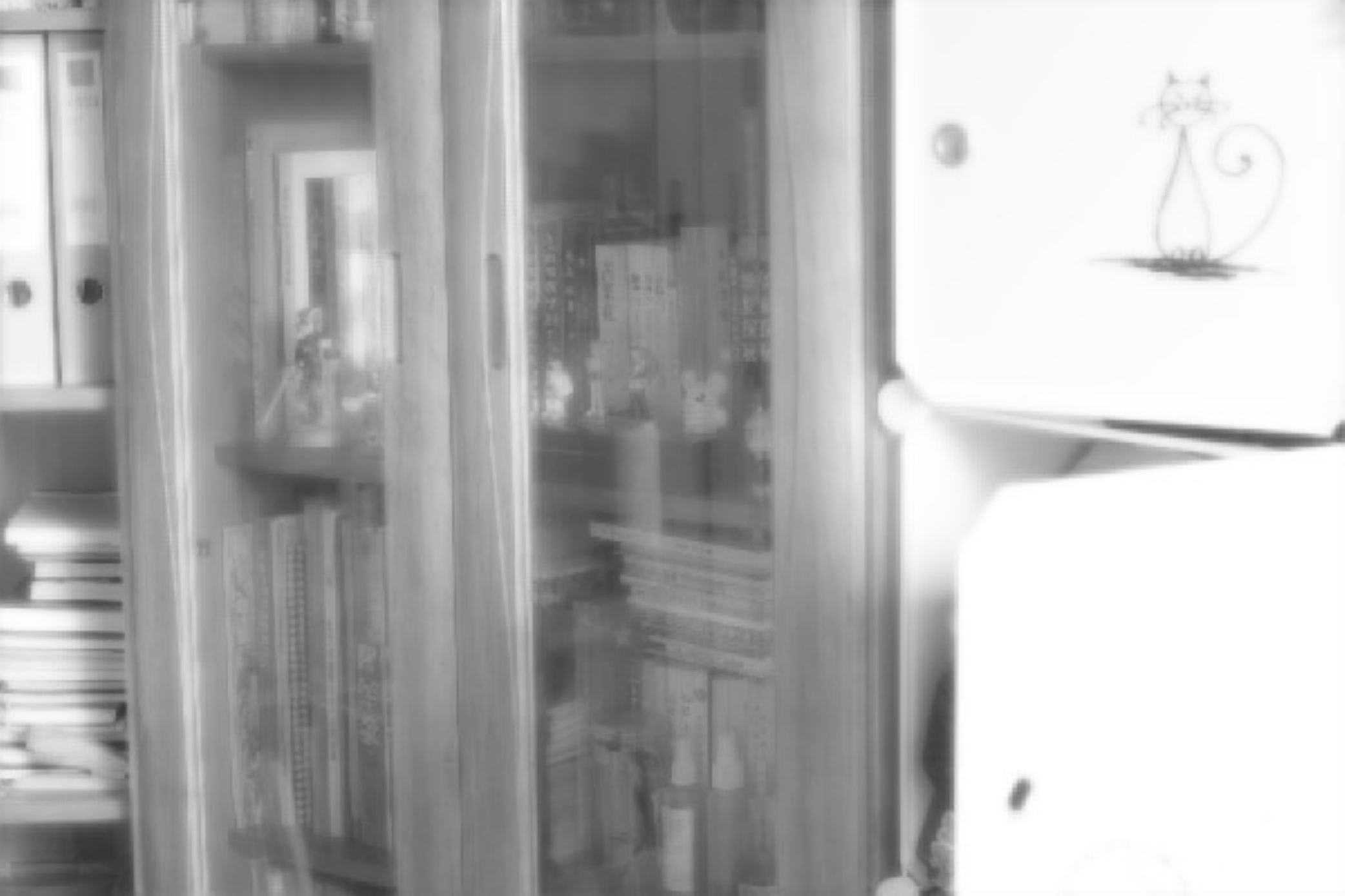}\vspace{-1pt} 
			\includegraphics[width=4.4cm]{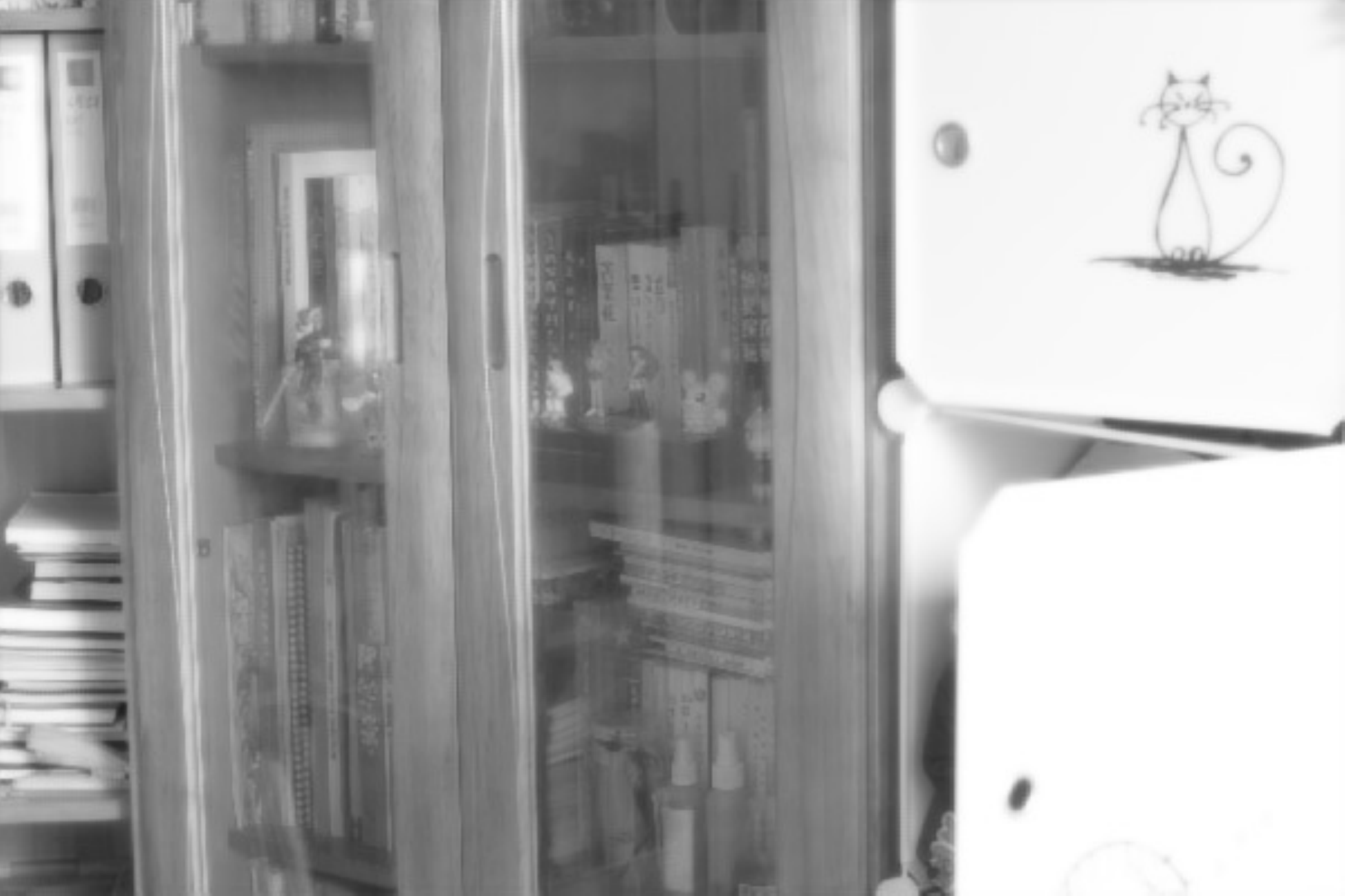}\vspace{0pt} 
		\end{minipage}
	}
	\caption{The images in the first and second row from left to right represent the low-light image's reflectance and GroundTruth's illumination (low-light image's illumination is too dark to compare) decomposed by our decomposition net without the NCBC Module, using noise loss with weight coefficients of 0.2, 0.4 and 0.7.}
	\label{ablation ncbc}
\end{figure*}

As shown in Fig. \ref{abla1}, obvious noise and serious color bias exist in the results generated by RetinexNet \cite{wei2018deep}. Our model without the NCBC Module is able to enhance the brightness well but cannot deal with noise, color bias and unpleasant artifacts. As shown in Fig. \ref{abla2}, the results generated by our method without the NCBC Module are relatively low in terms of PSNR and SSIM, with slight color distortion and unpleasant artifacts. In contrast, our model with the NCBC Module can not only enhance the brightness but also suppress noise, correct color bias and eliminate halo artifacts well.
As shown in Fig. \ref{ablation ncbc}, when we remove the NCBC Module from the training of the decomposition network, the result is full of noise, artifacts and color distortion. In contrast, when we add the NCBC Module and use the noise loss to suppress noise, most of the noise in the reflectance is removed. By increasing the coefficient of noise loss, we can increase the degree of noise suppression in the reflectance. The greater the coefficient of noise loss we set, the better the denoising effect we can achieve. We set the coefficient of noise loss to 0.2, which enables our model to achieve the best results in noise reduction and color bias correction.

\begin{figure*}[htbp]
	\flushleft
	\subfigure[Input]{
		\begin{overpic}[scale=.205]{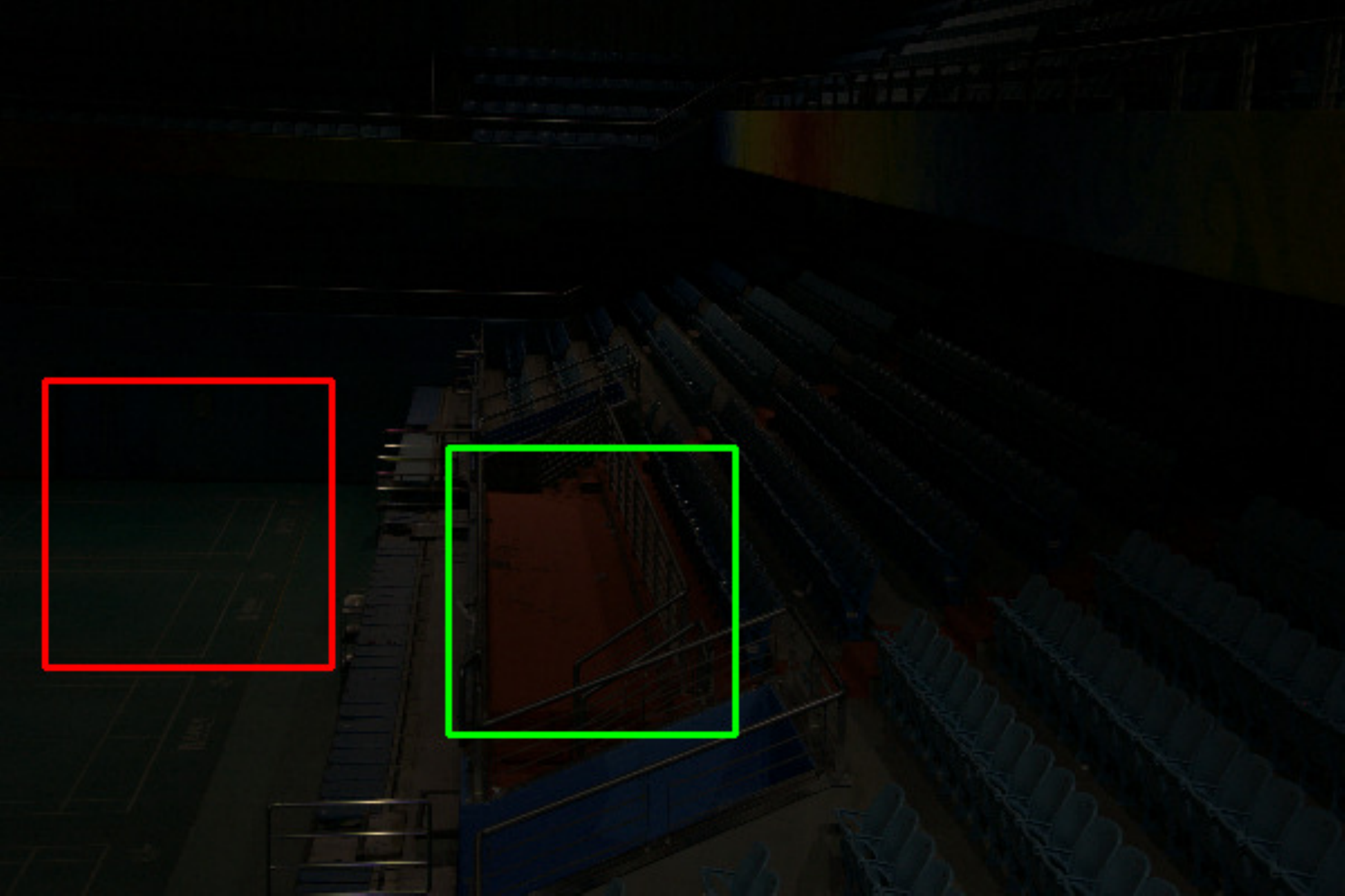}   		\put(70,0){\includegraphics[scale=.28]%
		{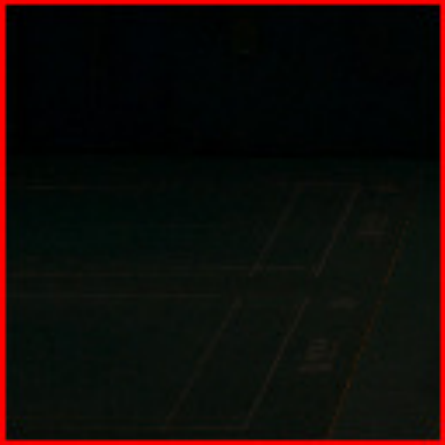}} 
		\end{overpic}
	}%
	\subfigure[MSRCR \cite{jobson1997multiscale}]{
		\begin{overpic}[scale=.205]{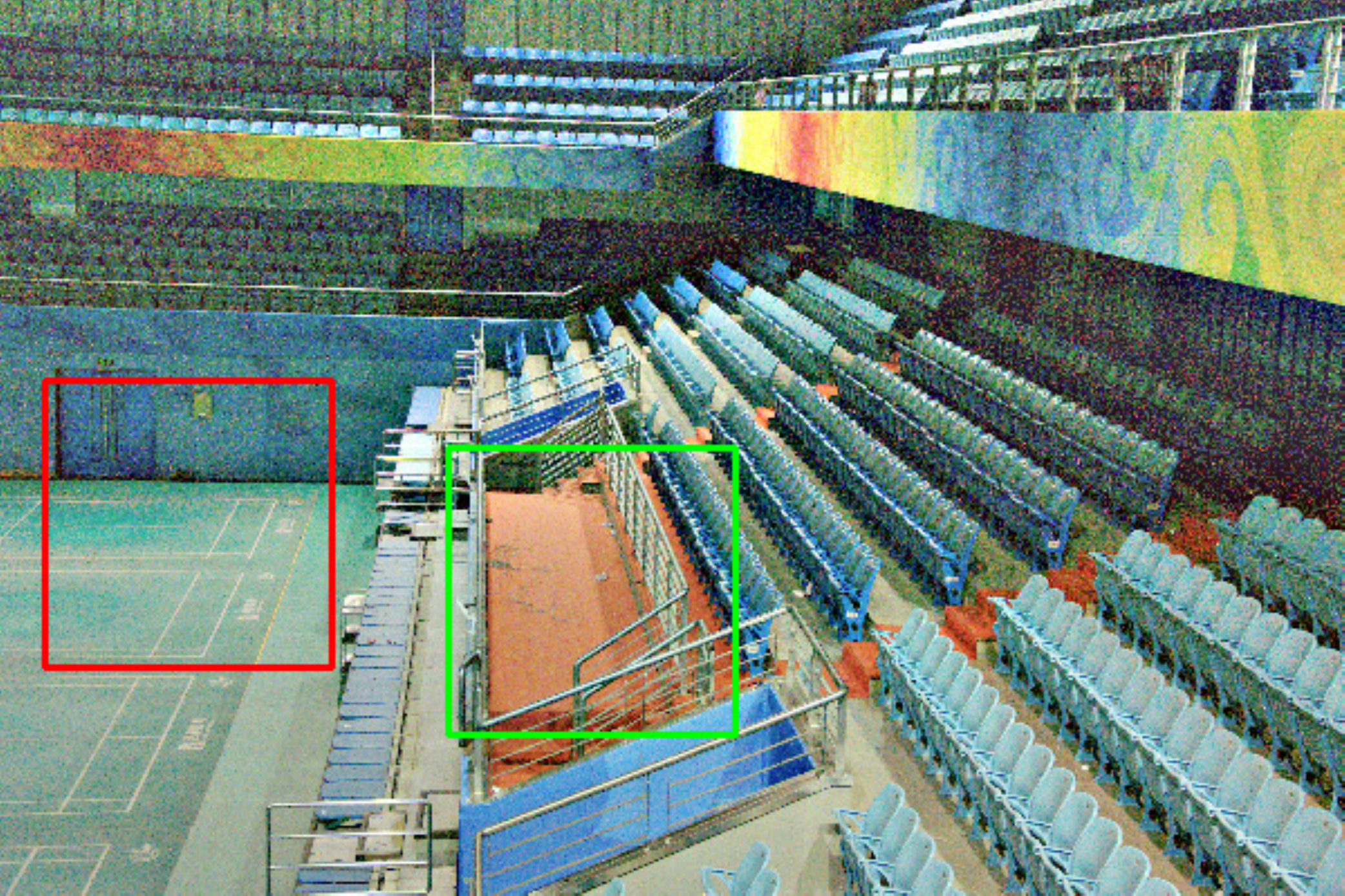}   		\put(70,0){\includegraphics[scale=.28]%
		{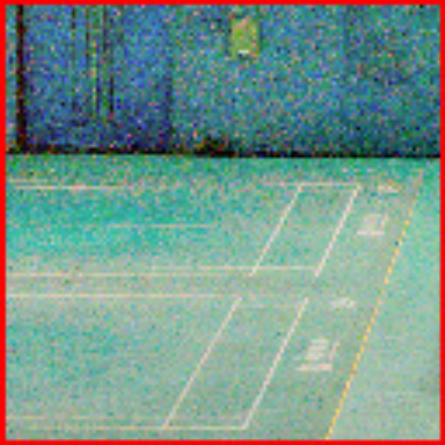}} 
		\end{overpic}
	}%
	\subfigure[BIMEF \cite{ying2017bio}]{
		\begin{overpic}[scale=.205]{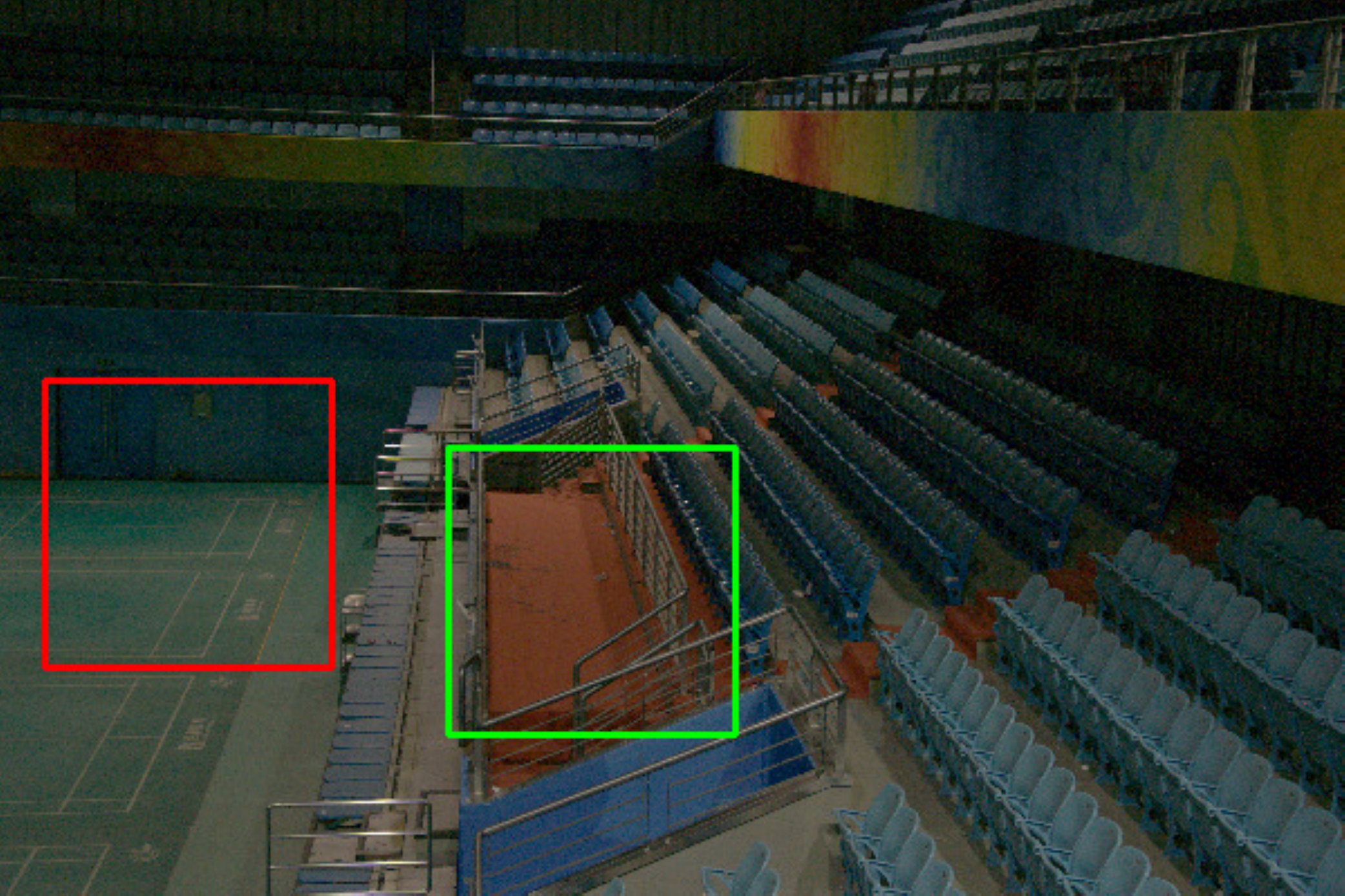}   		\put(70,0){\includegraphics[scale=.28]%
		{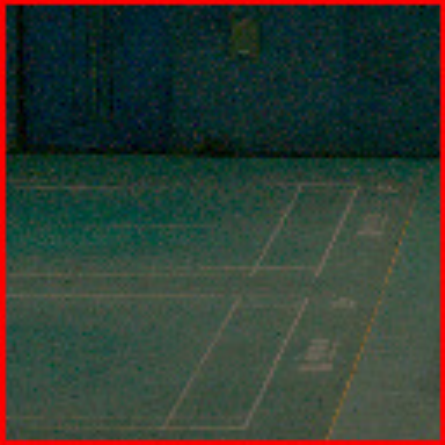}} 
		\end{overpic}
	}%
	\subfigure[LIME \cite{guo2016lime}]{
		\begin{overpic}[scale=.205]{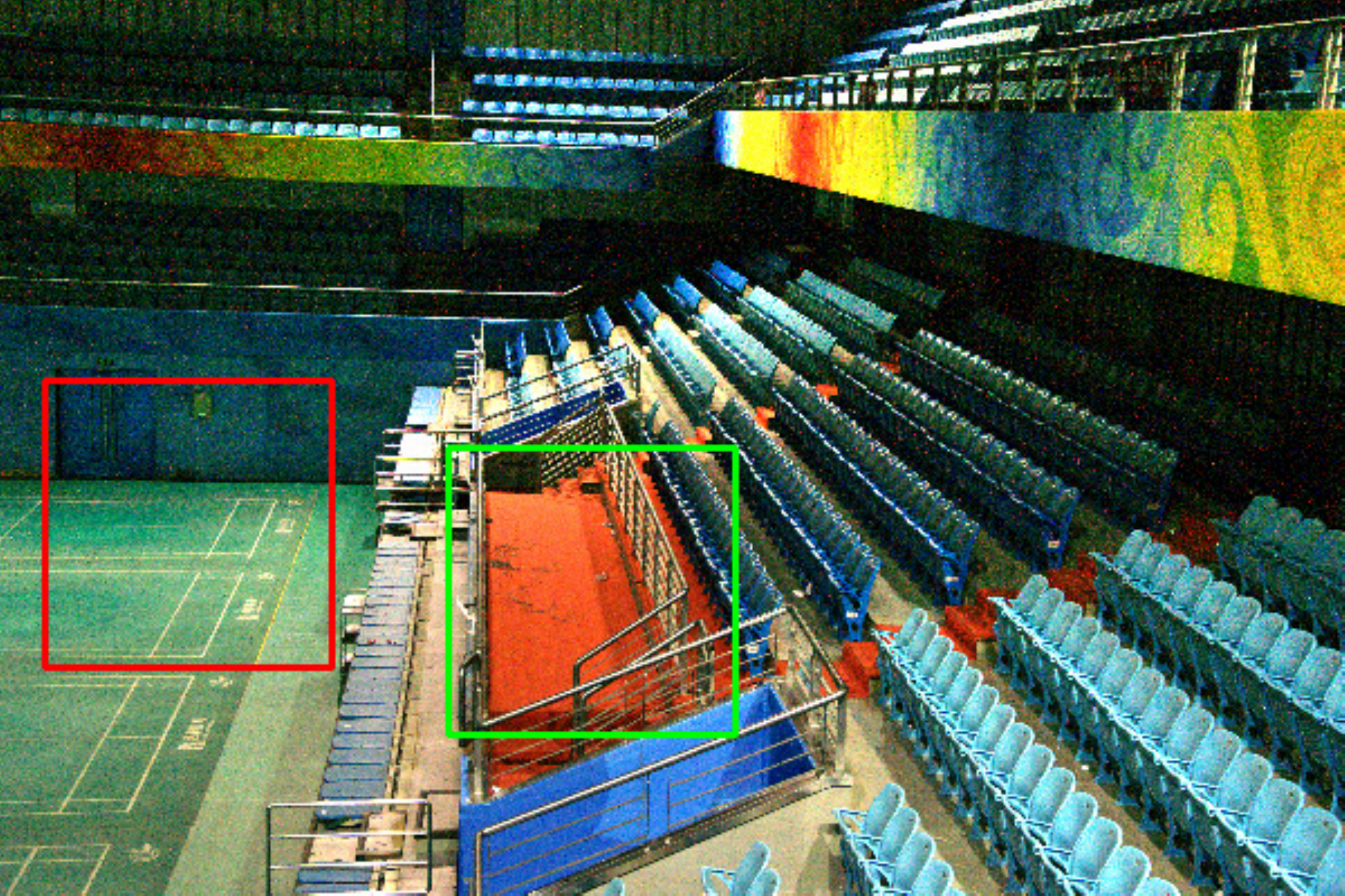}   		\put(70,0){\includegraphics[scale=.28]%
		{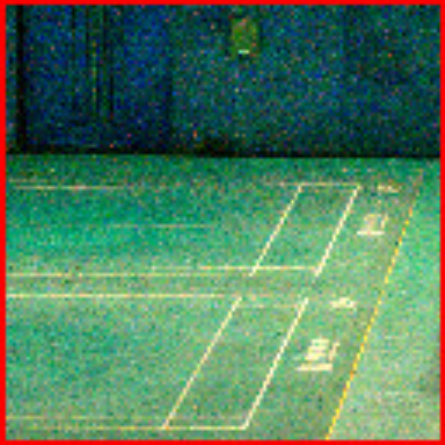}} 
		\end{overpic}
	}%
	
	\subfigure[Dong \cite{dong2011fast}]{
		\begin{overpic}[scale=.205]{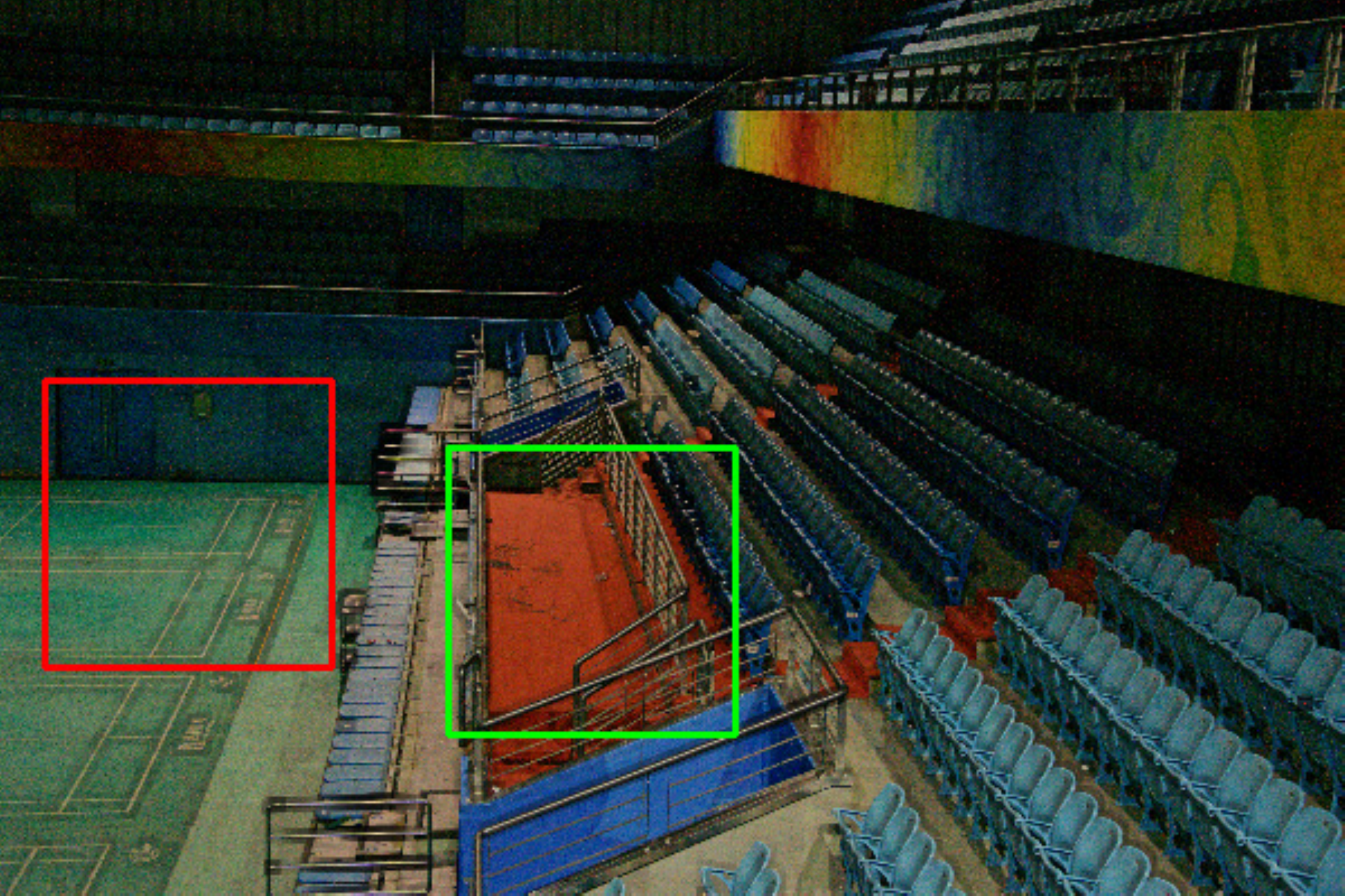}   		\put(70,0){\includegraphics[scale=.28]%
		{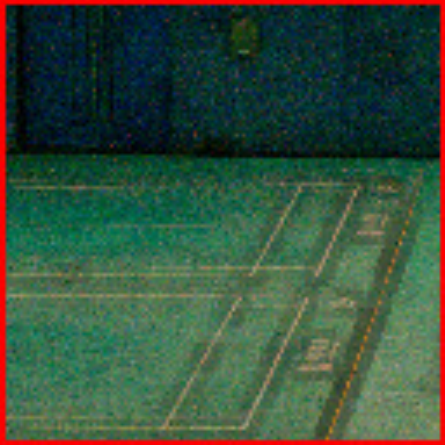}} 
		\end{overpic}
	}%
	\subfigure[SRIE \cite{fu2016weighted}]{
		\begin{overpic}[scale=.205]{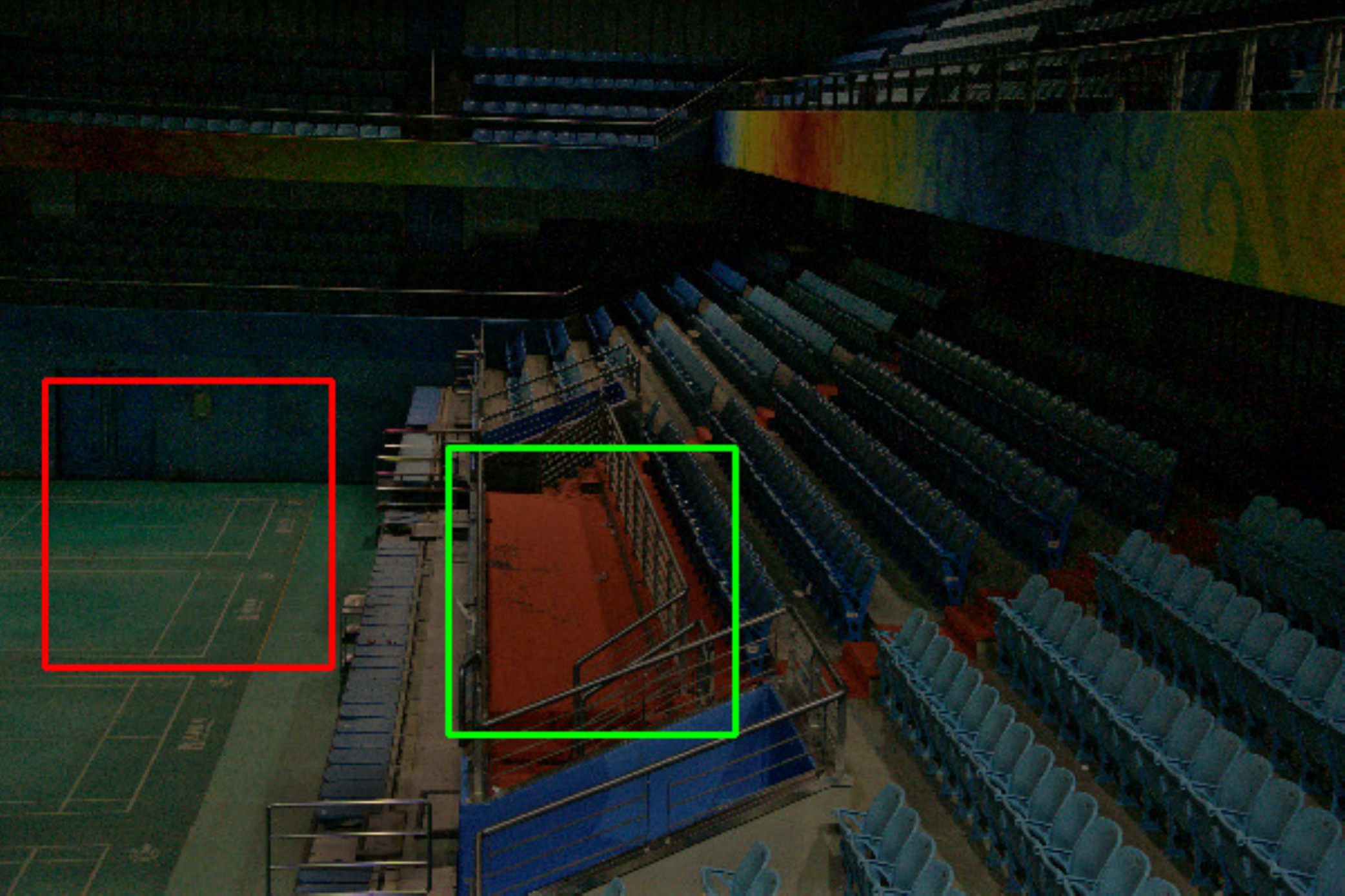}   		\put(70,0){\includegraphics[scale=.28]%
		{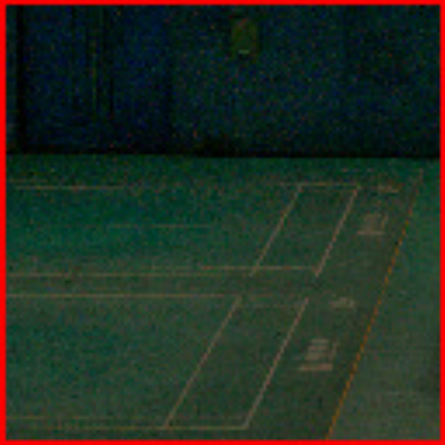}} 
		\end{overpic}
	}%
	\subfigure[MF \cite{fu2016fusion}]{
		\begin{overpic}[scale=.205]{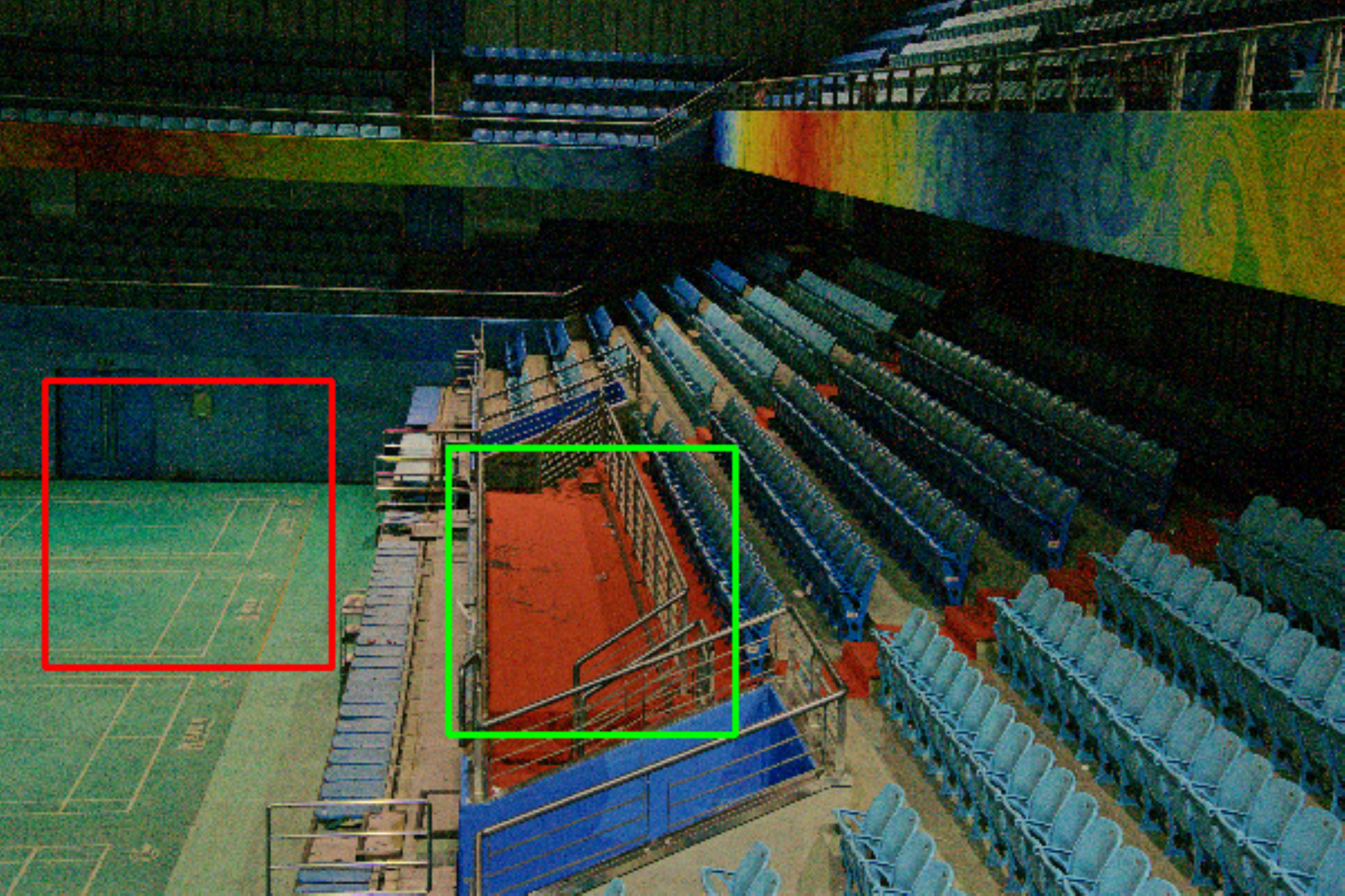}   		\put(70,0){\includegraphics[scale=.28]%
		{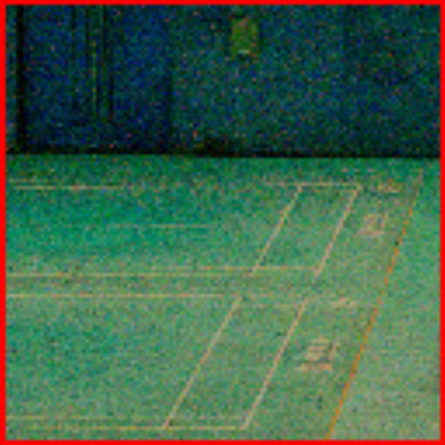}} 
		\end{overpic}
	}%
	\subfigure[NPE \cite{wang2013naturalness}]{
		\begin{overpic}[scale=.205]{pic/778/Dong-eps-converted-to.pdf}   		\put(70,0){\includegraphics[scale=.28]%
		{pic/778/Dong_0-eps-converted-to.pdf}} 
		\end{overpic}
	}%
	
	\subfigure[RRM \cite{li2018structure}]{
		\begin{overpic}[scale=.205]{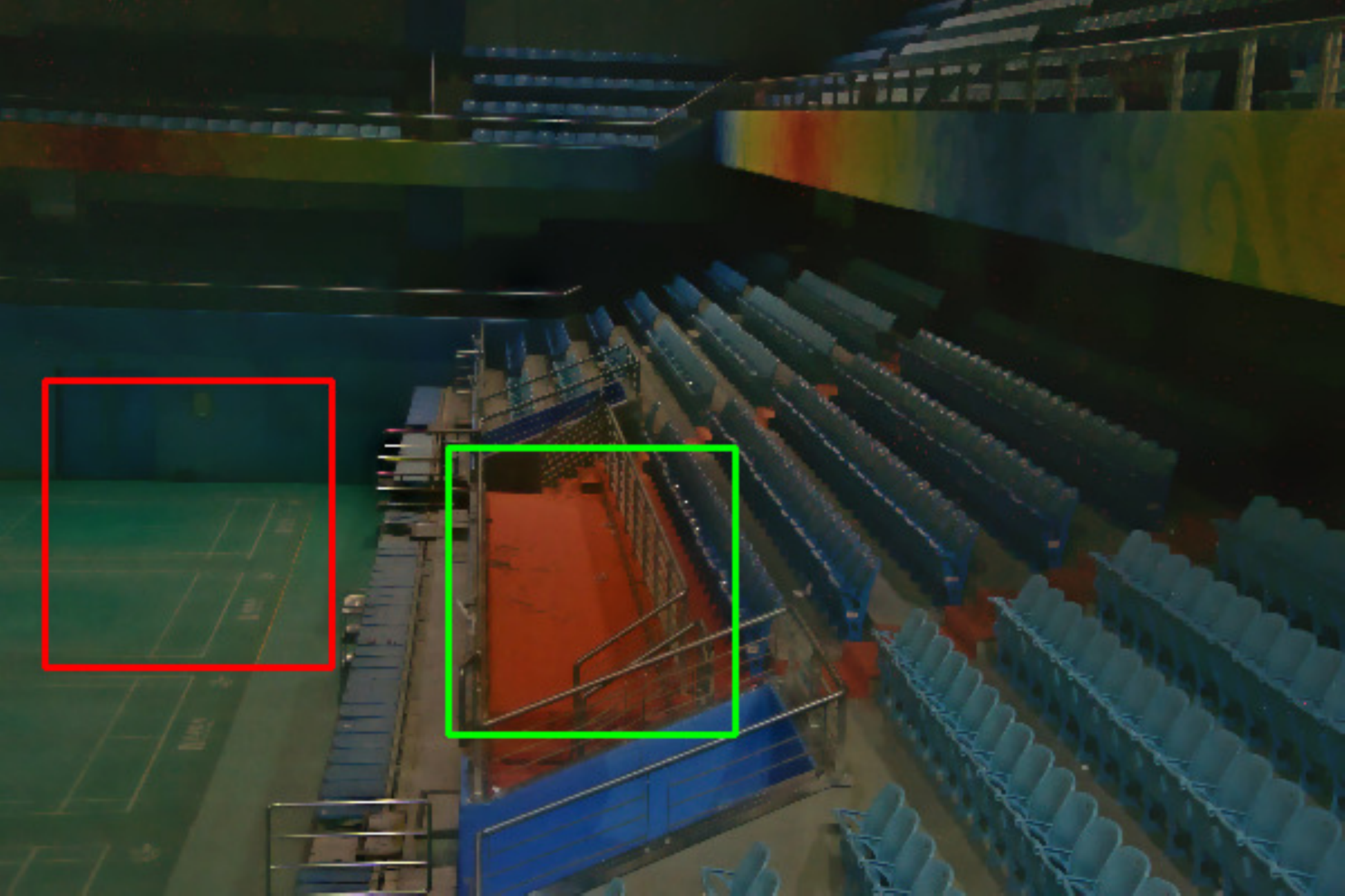}   		\put(70,0){\includegraphics[scale=.28]%
		{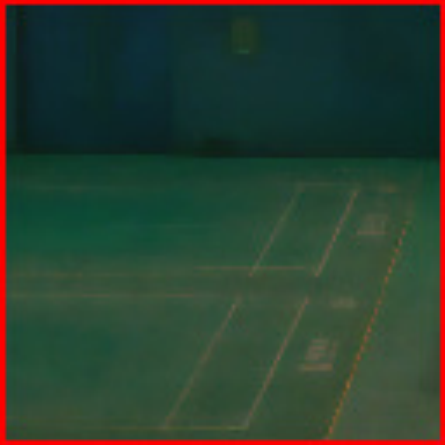}} 
		\end{overpic}
	}%
	\subfigure[MBLLEN \cite{lv2018mbllen}]{
		\begin{overpic}[scale=.205]{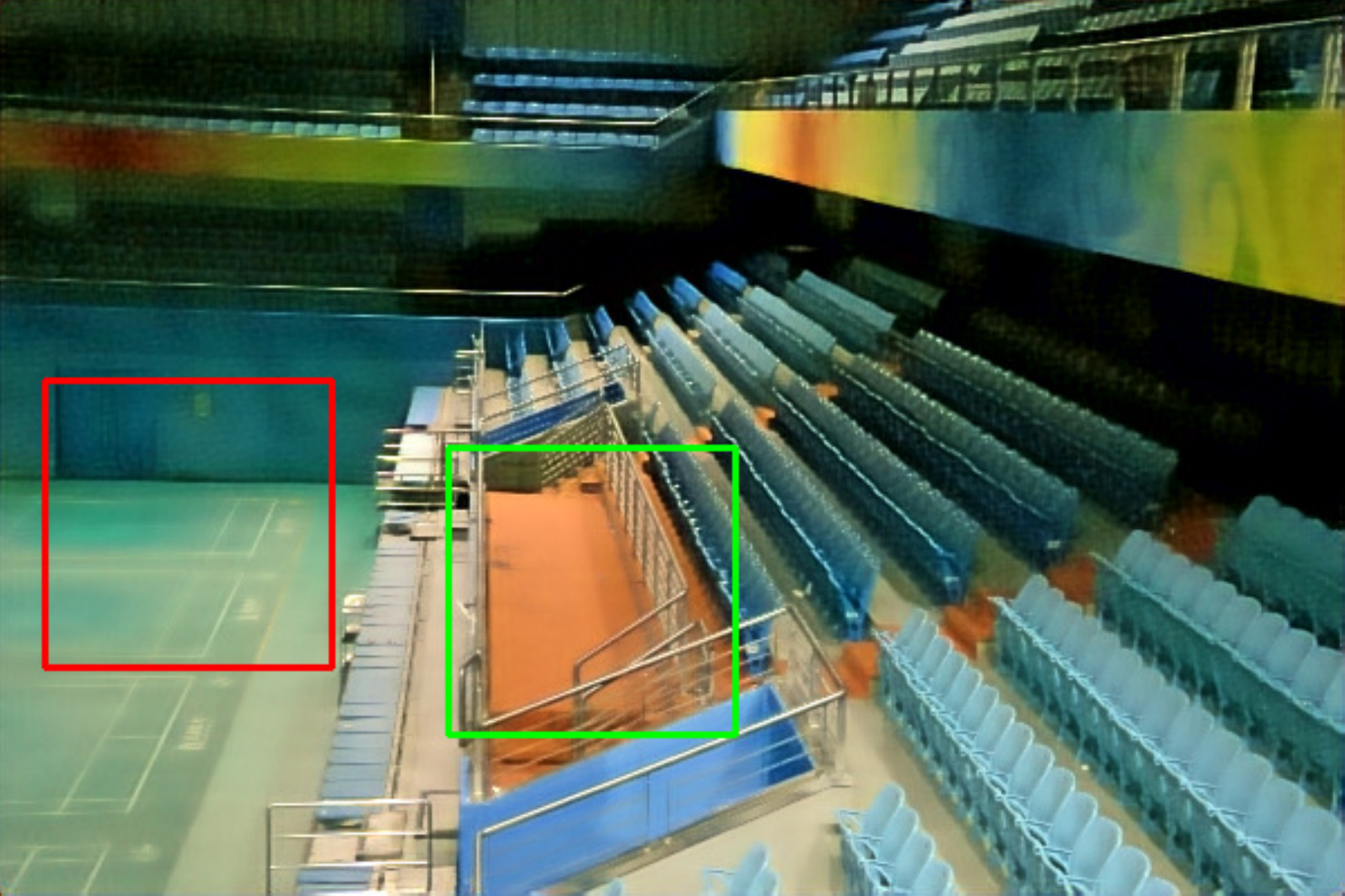}   		\put(70,0){\includegraphics[scale=.28]%
		{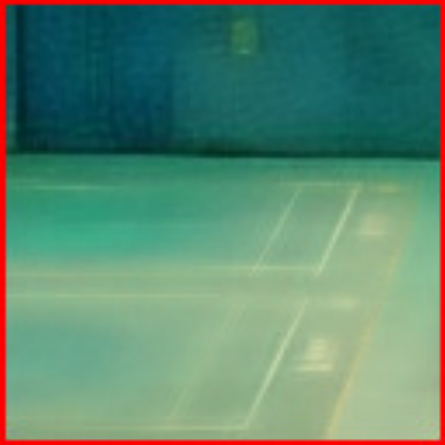}} 
		\end{overpic}
	}%
	\subfigure[RetinexNet \cite{wei2018deep}]{
		\begin{overpic}[scale=.205]{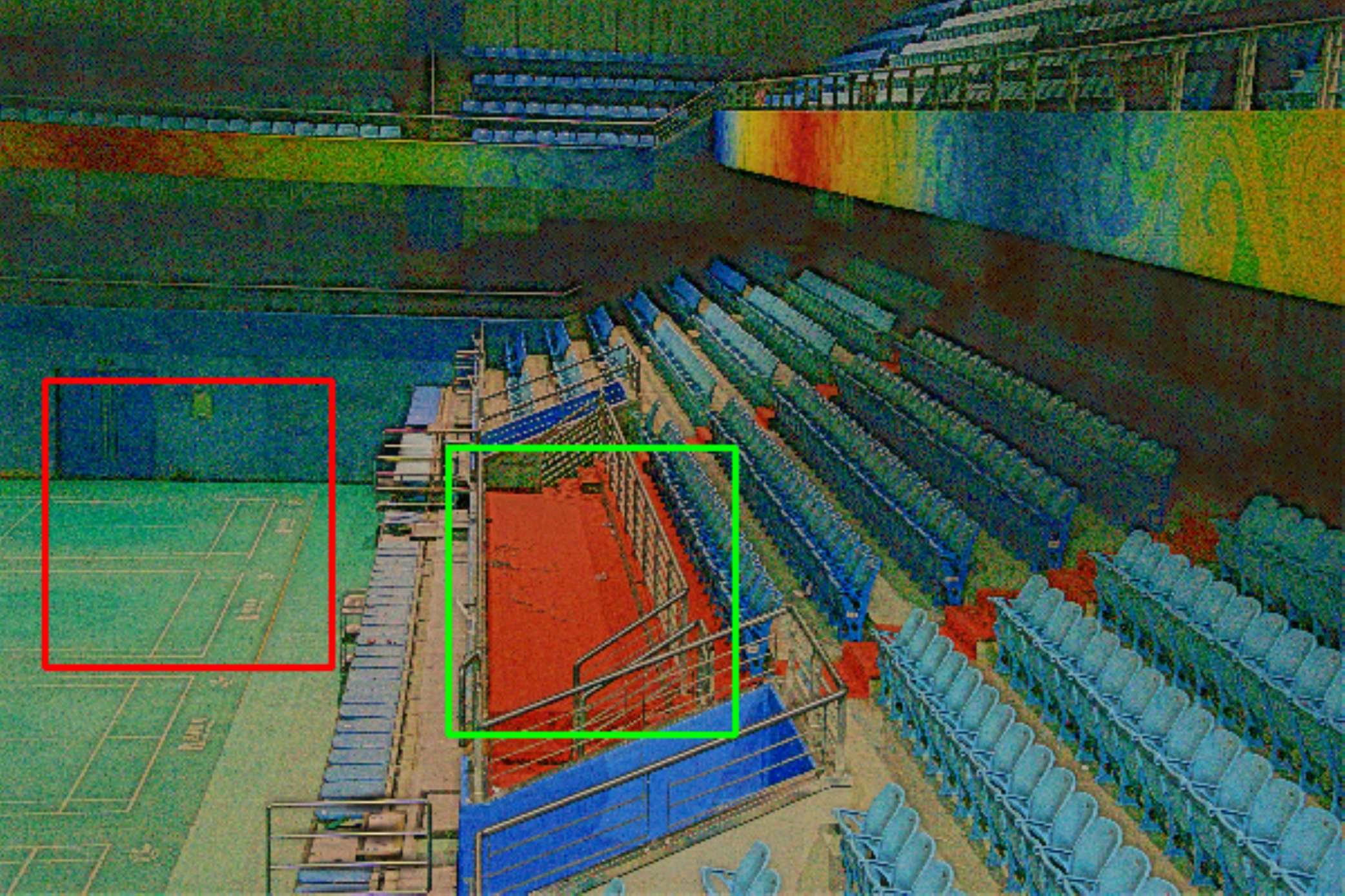}   		\put(70,0){\includegraphics[scale=.28]%
		{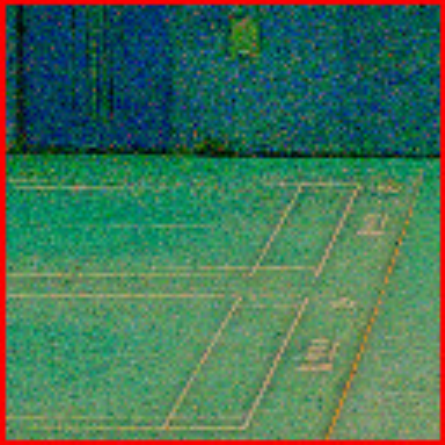}} 
		\end{overpic}
	}%
	\subfigure[GLAD \cite{wang2018gladnet}]{
		\begin{overpic}[scale=.205]{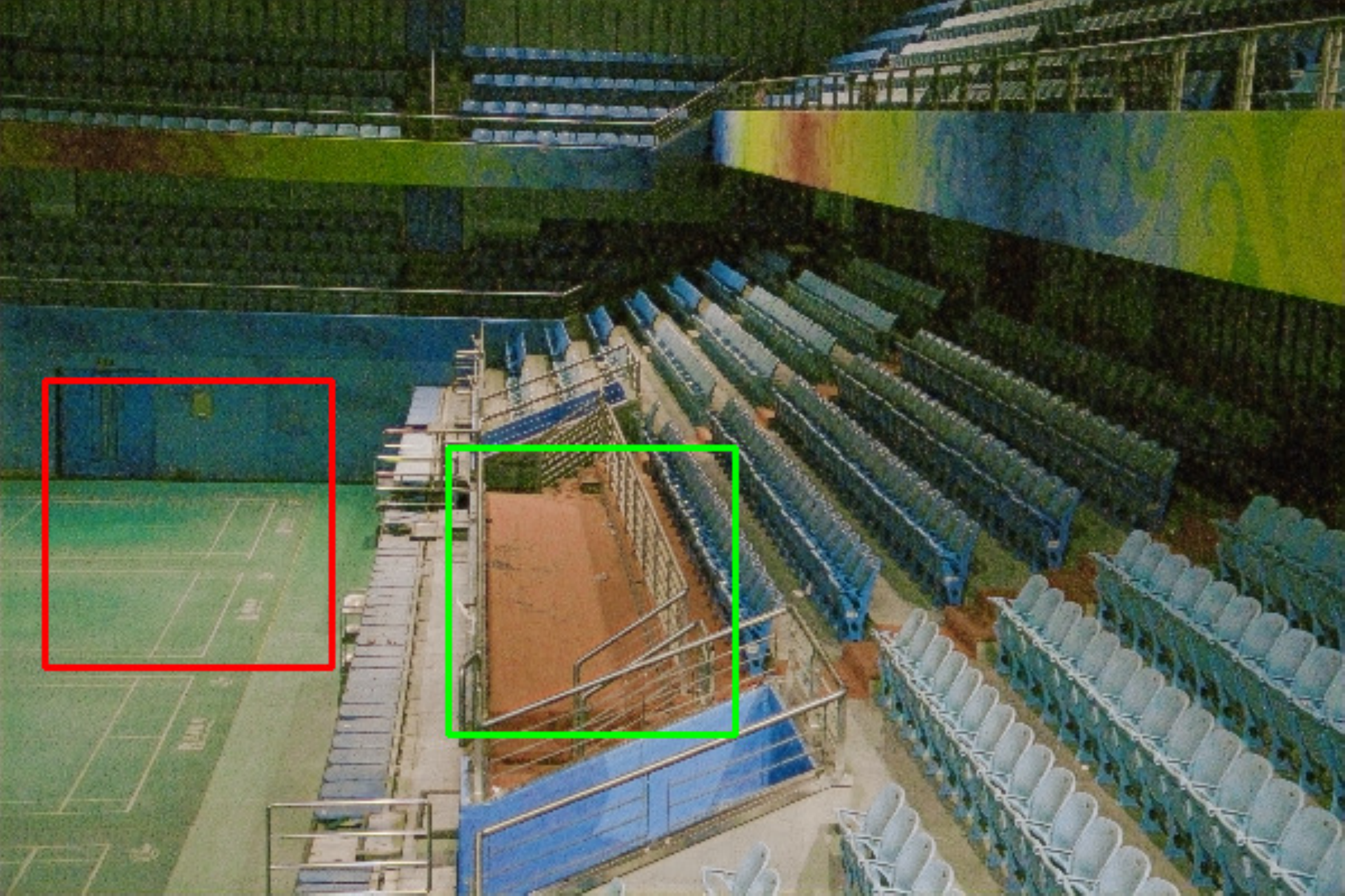}   		\put(70,0){\includegraphics[scale=.28]%
		{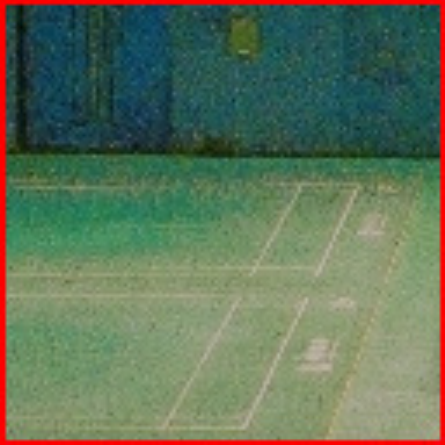}} 
		\end{overpic}
	}%
	
	\subfigure[EnlightenGan \cite{jiang2021enlightengan}]{
		\begin{overpic}[scale=.205]{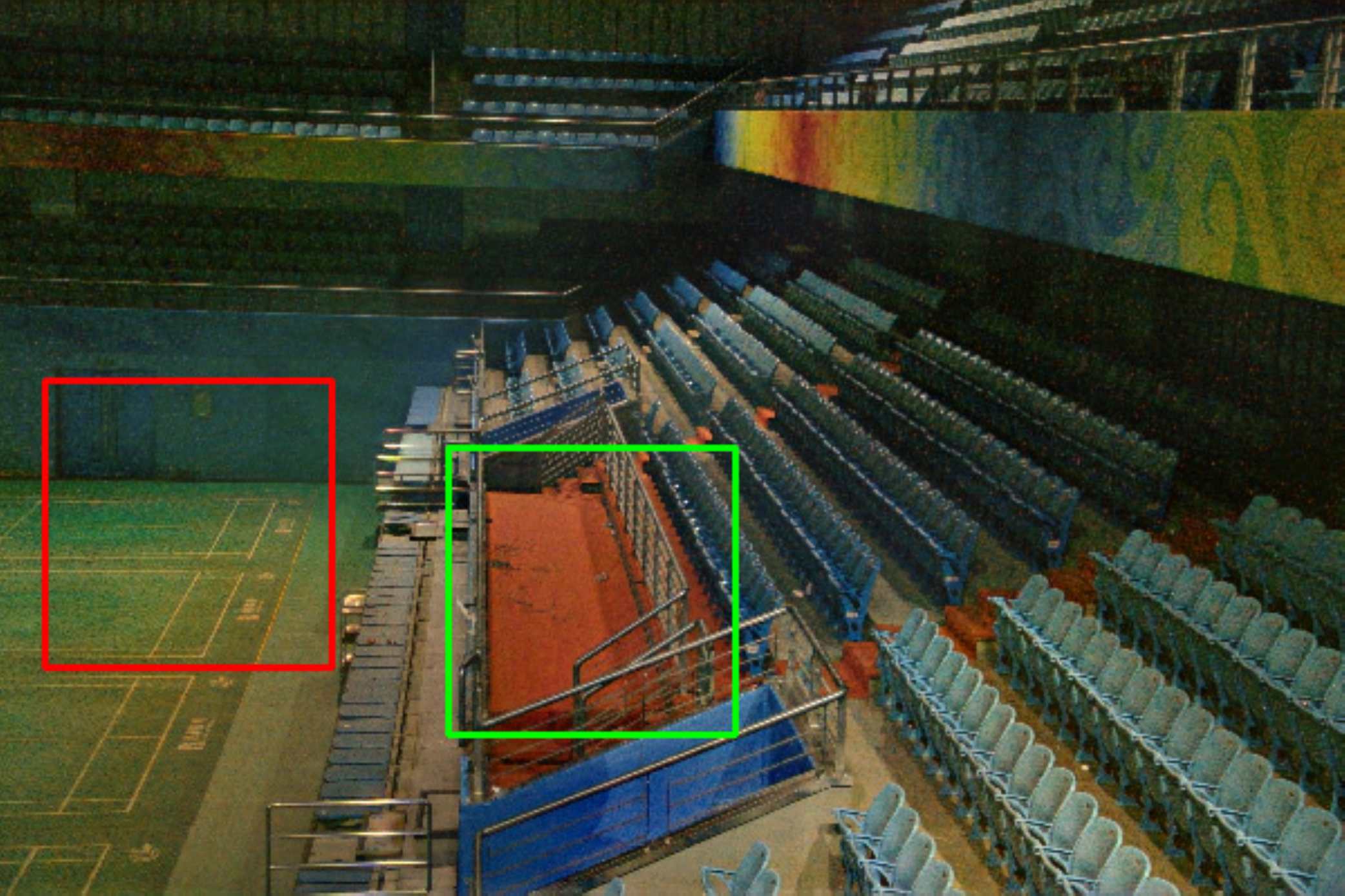}   		\put(70,0){\includegraphics[scale=.28]%
		{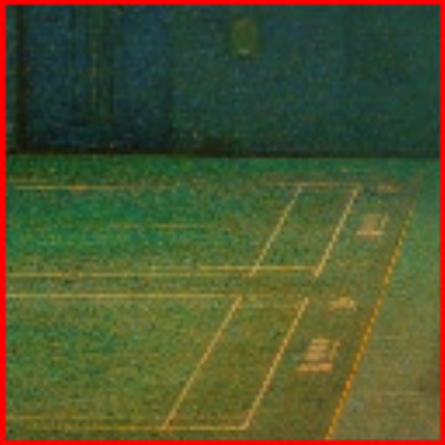}} 
		\end{overpic}
	}%
	\subfigure[Zero-DCE \cite{guo2020zero}]{
		\begin{overpic}[scale=.205]{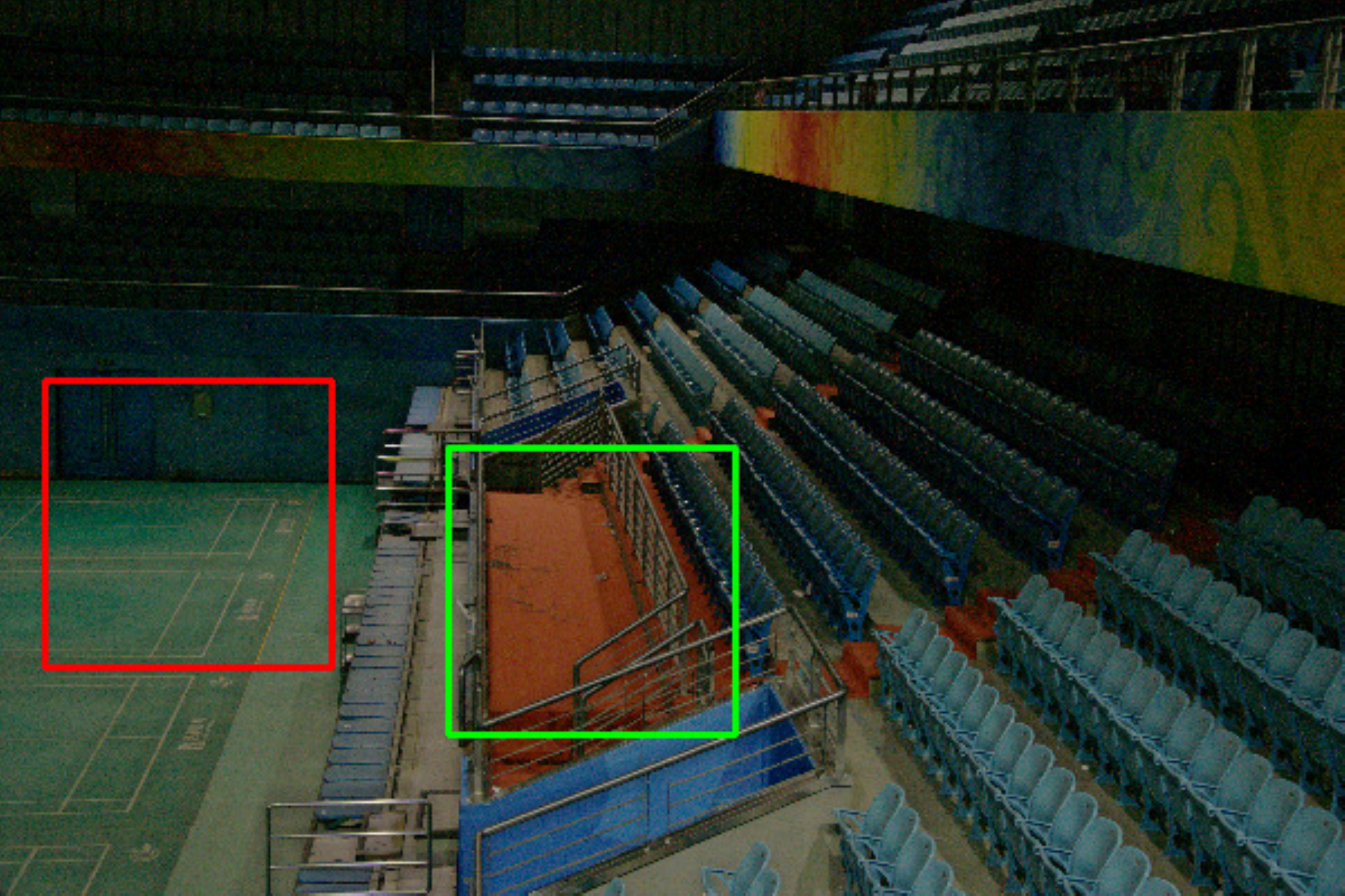}   		\put(70,0){\includegraphics[scale=.28]%
		{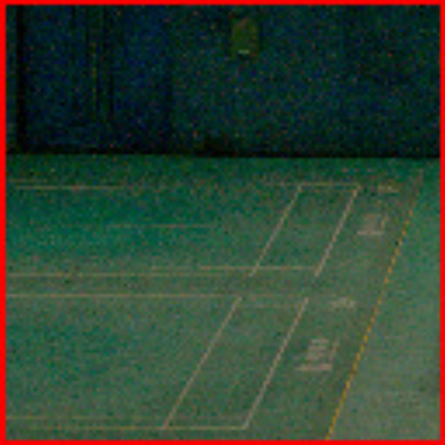}} 
		\end{overpic}
	}%
	\subfigure[Ours]{
		\begin{overpic}[scale=.205]{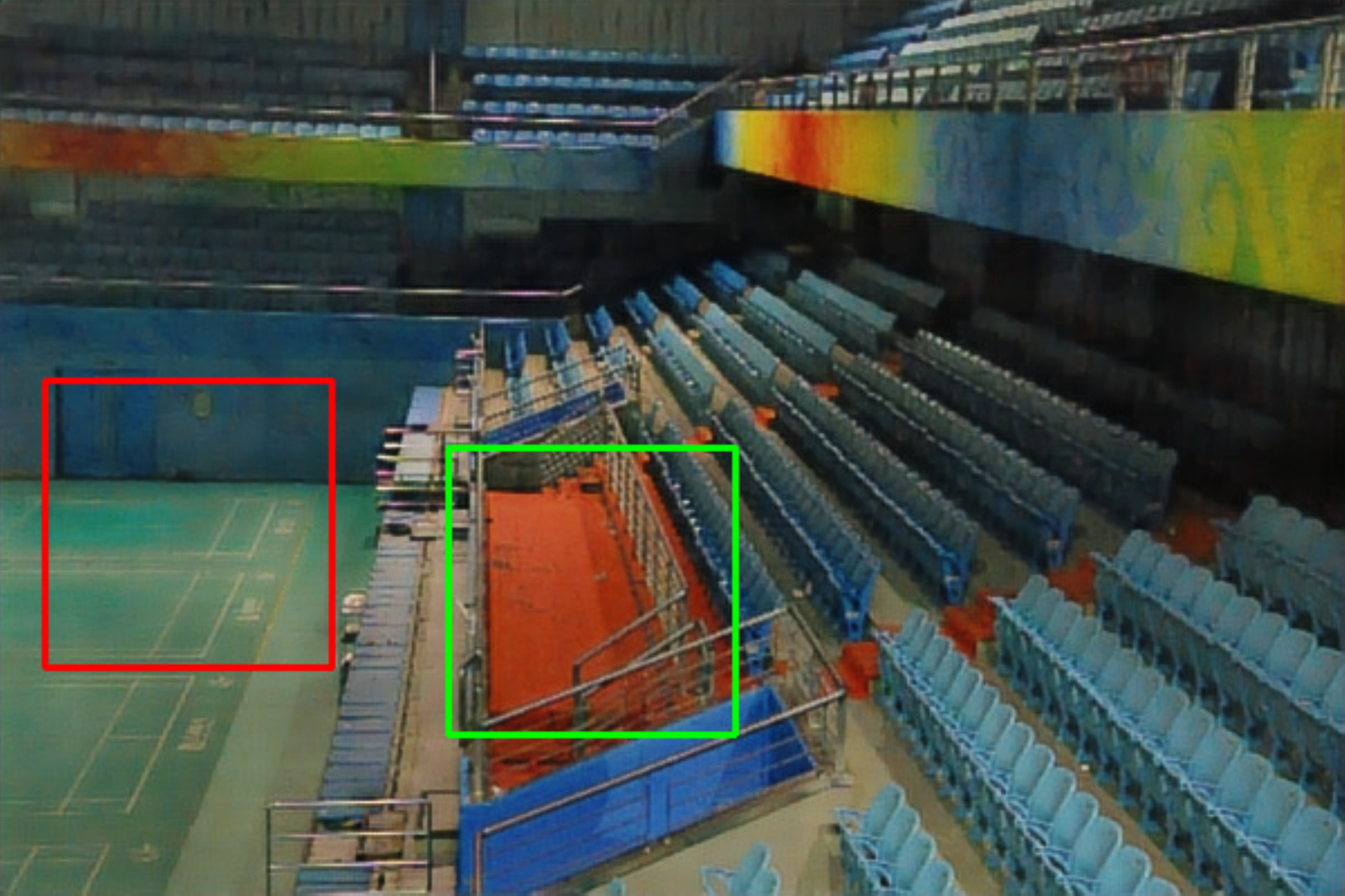}   		\put(70,0){\includegraphics[scale=.28]%
		{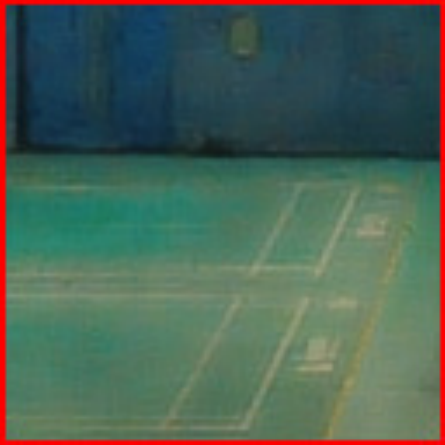}} 
		\end{overpic}
	}%
	\subfigure[GroundTruth]{
		\begin{overpic}[scale=.205]{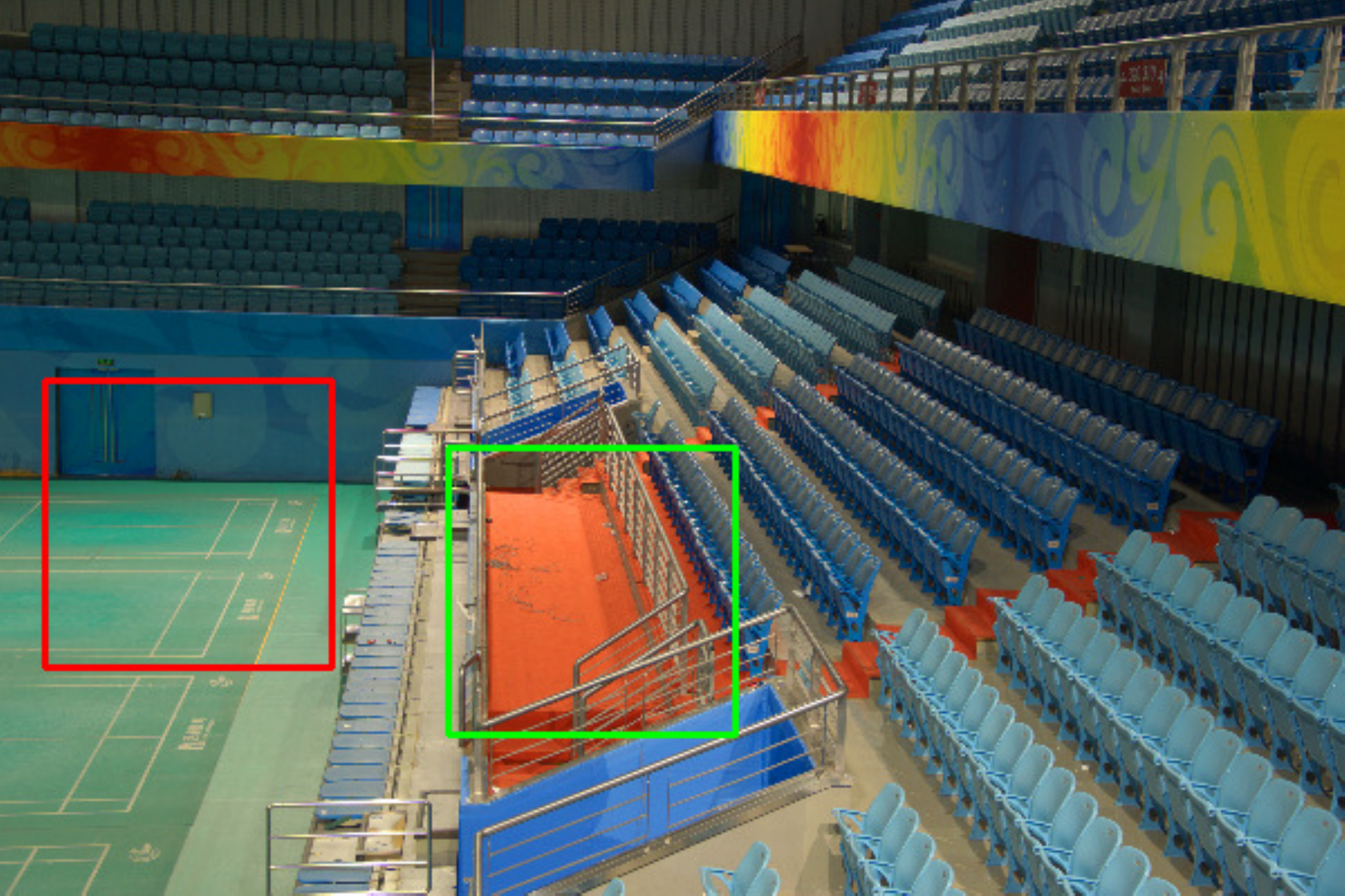}   		\put(70,0){\includegraphics[scale=.28]%
		{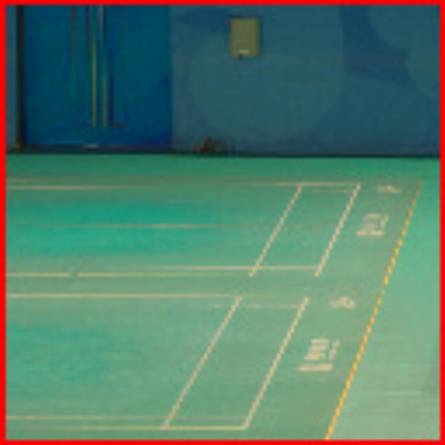}} 
		\end{overpic}
	}%
	
	\flushleft
	\caption{Visual comparison with other state-of-the-art methods on LOL real-world validation dataset, where the degradation is hidden in darkness.}
	\label{court}
\end{figure*}

\begin{table}[htbp] 
	\centering \caption{Quantitative comparison of several metrics between our method
		and other state-of-the-art methods on LOL dataset. “↑” indicates the higher the better, “↓” indicates the lower the	better. \color{red}Red: the best, \color{blue}Blue: the second best.}
	\begin{tabular}{  c | c  c  c  c  c  c  c }
		\hline  Traditional Methods  & PSNR↑     & SSIM↑   & LPIPS↓  &FSIM↑    & UQI↑  \\
		\hline
		\hline  Input & 7.7733 & 0.1914 & 0.4173 &0.7190   &0.0622 \\
		MSRCR \cite{jobson1997multiscale}  & 13.1728 & 0.4615 & 0.4404 &0.8450   &0.7884 \\
		BIMEF \cite{ying2017bio}  & 13.8752 & 0.5949 & 0.3673 &0.9263   &0.7088 \\
		LIME \cite{guo2016lime}  & 16.7586 & 0.4449 & 0.4183 &0.8549   &0.8805 \\
		Dong \cite{dong2011fast} & 16.7165 & 0.4783 & 0.4226 &0.8886   & 0.8078\\
		SRIE \cite{fu2016weighted}  & 11.8552 & 0.4954 & 0.3657 &0.9085   &0.5033 \\
		MF \cite{fu2016fusion}  & 16.9662 & 0.5075 & 0.4092 &0.9236   &0.8572 \\
		NPE \cite{wang2013naturalness}  & 16.9697 & 0.4839 & 0.4156 &0.8964   &0.8943 \\
		RRM \cite{li2018structure}  & 13.8765 & 0.6636 & \color{blue}0.3476 &0.8821   &0.7275 \\
		LECARM \cite{ren2018lecarm}  & 14.4099 & 0.5448 & 0.3687 &0.9288   &0.6406 \\
		JED \cite{ren2018joint}  & 13.6857 & 0.6509 & 0.3549 &0.8812   &0.7143 \\
		PLM \cite{yu2017low}  & 16.2620 & 0.4617 & 0.4284 &0.8265   &0.8892 \\
		DIE \cite{zhang2019dual}  & 14.0181 & 0.5188 & 0.3910 &0.9172   &0.7027 \\
		\hline
		\hline  DL Methods  & PSNR↑     & SSIM↑   & LPIPS↓  &FSIM↑    & UQI↑  \\
		\hline
		\hline
		MBLLEN \cite{lv2018mbllen}  & 17.8583 & 0.7247 & 0.3672 &0.9262   &0.8261 \\
		RetinexNet \cite{wei2018deep}  & 16.7740 & 0.4249 & 0.4670 &0.8642   &0.9110 \\
		GLAD \cite{wang2018gladnet}  & \color{blue}19.7182 & 0.6820 & 0.3994 & \color{blue}0.9329   &\color{blue}0.9204 \\
		RDGAN \cite{wang2019rdgan}  & 15.9363 & 0.6357 & 0.3985 &0.9276   &0.8296 \\
		Zero-DCE \cite{guo2020zero}  & 14.8671 & 0.5623 & 0.3852 &0.9276   &0.7205 \\
		Zhang \cite{zhang2020self}  & 19.4968 & \color{blue}0.7003 & 0.3911 &0.8514   &0.8521 \\
		EnlightenGan \cite{jiang2021enlightengan}  & 17.4828 & 0.6515 & 0.3903 &0.9226  &0.8499 \\
		\hline 
		\hline 
		Our model w/o NCBC & 18.4446 & 0.7605 & 0.3514 &0.9261   &0.9259 \\
		Our model  & \color{red}20.1447 & \color{red}0.7918 & \color{red}0.3126 &\color{red}0.9454   &\color{red}0.9371 \\
		\hline\end{tabular}\vspace{0cm}
	\label{lol_eval}
\end{table}

\begin{table}[htbp] 
	\centering \caption{Quantitative comparison of color distortion between our method
		and other state-of-the-art methods on LOL dataset.“↓” indicates the lower the better. \color{red}Red: the best, \color{blue}Blue: the second best.}
	\begin{tabular}{  c | c  c | c | c  c  }
		\hline  Traditional Methods  & Mean↓     & Median↓   &Average↓  &DeltaE↓ \\
		\hline
		\hline  
		MSRCR \cite{jobson1997multiscale}  & 3.7421 & 4.5877 & 4.1649 & 27.4496\\
		BIMEF \cite{ying2017bio}  & 3.4004 & \color{blue}3.5187 & \color{blue}3.4595 & 33.8820\\
		LIME \cite{guo2016lime}  & \color{blue}3.2096 & 4.0825 & 3.6460  & 21.1816\\
		Dong \cite{dong2011fast} & 3.3499 & 4.1481 & 3.7490 & 25.3349 \\
		SRIE \cite{fu2016weighted}  & 3.4488 & 4.0751 & 3.7620 & 44.3194\\
		MF \cite{fu2016fusion}  & 3.3277 & 3.9810 & 3.6544 & 24.5488\\
		NPE \cite{wang2013naturalness}  & 3.5588 & 4.2505 & 3.9046 & 22.6374\\
		RRM \cite{li2018structure}  & 3.3745 & 3.5821 & 3.4784 & 32.9843\\
		LECARM \cite{ren2018lecarm}  & 3.4091 & 3.9668 & 3.6979 & 34.5143\\
		JED \cite{ren2018joint}  & 3.4064 & 3.8651 & 3.6357 & 33.8342\\
		PLM \cite{yu2017low}  & 3.4274 & 3.7085 & 3.4829 & 22.0553\\
		DIE \cite{zhang2019dual}  & 3.4597 & 4.1164 & 3.7880 & 34.1583\\
		\hline
		\hline  Deep Learning Methods  & Mean↓     & Median↓   &Average↓  &DeltaE↓\\
		\hline
		\hline
		MBLLEN \cite{lv2018mbllen}  & 3.2716 & 4.4620 & 3.8669 & 21.5774\\
		RetinexNet \cite{wei2018deep}  & 3.7501 & 4.4975 & 4.3589 & 21.3550\\
		GLAD \cite{wang2018gladnet}  & 3.3110 & 3.8021 & 3.5565 & \color{blue}16.0393\\
		RDGAN \cite{wang2019rdgan}  & 4.3899 & 5.3027 & 4.8463 & 26.3796\\
		Zero-DCE \cite{guo2020zero}  & 4.1051 & 4.6860 & 4.3955 & 31.4451\\
		Zhang \cite{zhang2020self}  & 3.3744 & 3.9104 & 3.6424 & 17.6652\\
		EnlightenGan \cite{jiang2021enlightengan}  & 4.5296 & 5.2536 & 4.8916 & 21.9113\\
		\hline  
		\hline 
		Our model w/o NCBC   & 2.8645 & 3.9083 & 3.3864 & 16.0604\\
		Our model  & \color{red}2.1785 & \color{red}2.3870 & \color{red}2.3024 & \color{red}13.5264\\
		\hline\end{tabular}\vspace{0cm}
	\label{lol_eval_color}
\end{table}

\begin{table}[htbp] 
	\centering \caption{Quantitative comparison of the entropy and inference speed between our method and other state-of-the-art deep learning methods on LOL dataset.“↓” indicates the lower the	better. \color{red}{Red: the best, \color{blue}Blue: the second best.} 
		\color{black}*All of the deep learning models are run on a PC with Nvidia TiTan XP GPU.}
	\begin{tabular}{  c | c  c | c  c  c  }
		\hline  Deep Learning Methods  & GE↑     & CE↑   &Time cost↓ \\
		\hline
		\hline  
		Low-Light Input   & 4.7432 & 14.3837 & - \\
		GroundTruth   & 7.0400 & 21.3163 & - \\
		\hline  
		\hline 
		MBLLEN \cite{lv2018mbllen}  &\color{blue} 7.1421 & 21.3664 & 80 ms \\
		RetinexNet \cite{wei2018deep}  & 6.8346 & 21.1266 & 20 ms \\
		GLAD \cite{wang2018gladnet}  & 7.1141 & \color{blue}21.5237 & 25 ms \\
		RDGAN \cite{wang2019rdgan}  & 6.6327 & 20.1053 & 30 ms \\
		Zero-DCE \cite{guo2020zero}  & 6.5964 & 18.9072 & \color{red} 2 ms \\
		Zhang \cite{zhang2020self}  & 7.0673 & 21.4085 & 20 ms \\
		EnlightenGan \cite{jiang2021enlightengan}  & 7.0664 & 21.3476 & 20 ms \\
		\hline  
		\hline 
		Our model w/o NCBC   & 7.0336 & 21.3651 & 7 ms \\
		Our model  & \color{red}7.1450 & \color{red}21.5882 & \color{blue}7 ms \\
		\hline\end{tabular}\vspace{0cm}
	\label{lol_eval_entropy}
\end{table}

\begin{figure*}[htbp]
	\flushleft
	\subfigure[Input]{
		\begin{overpic}[scale=.205]{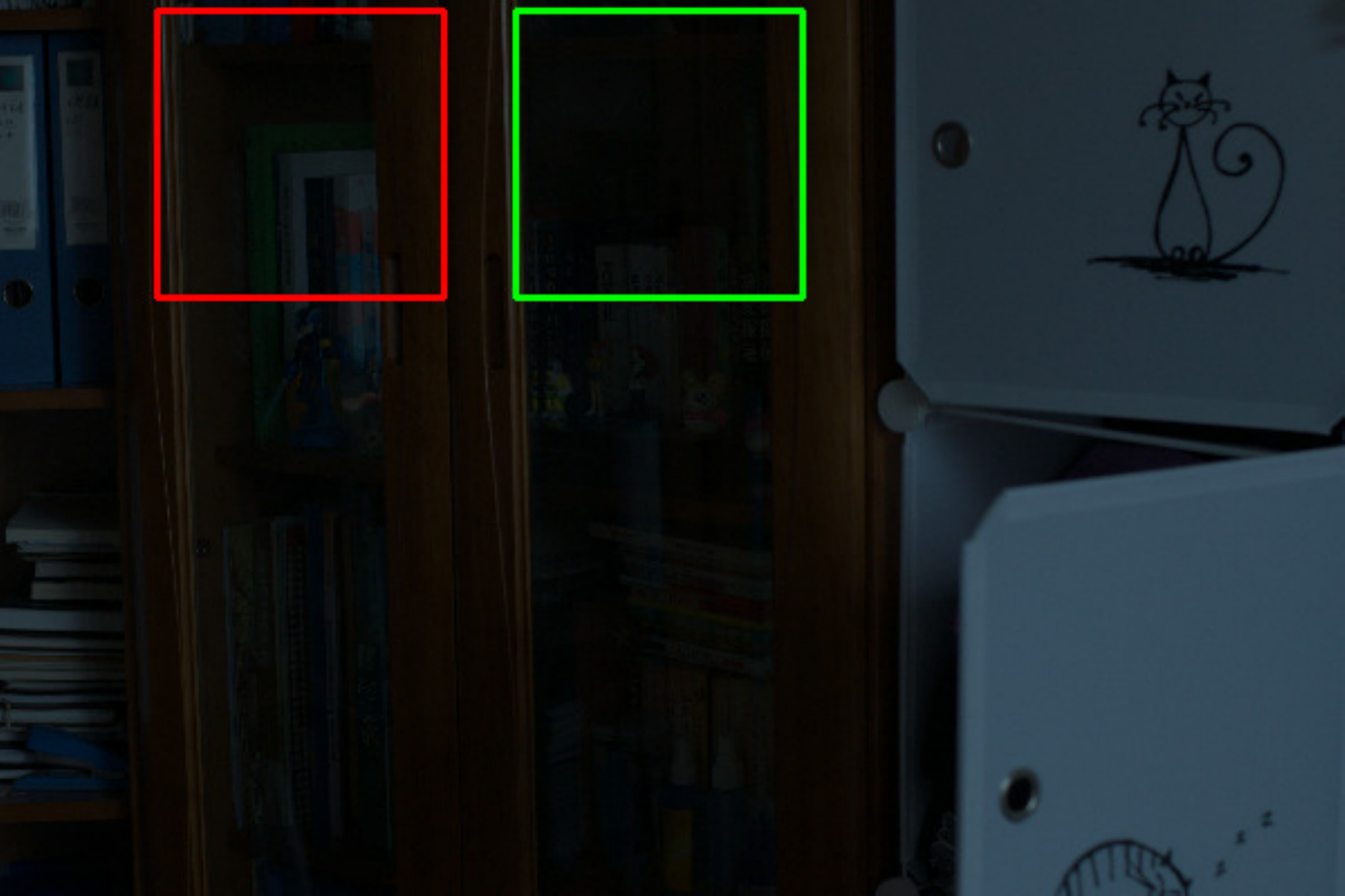}   		\put(70,0){\includegraphics[scale=.28]%
				{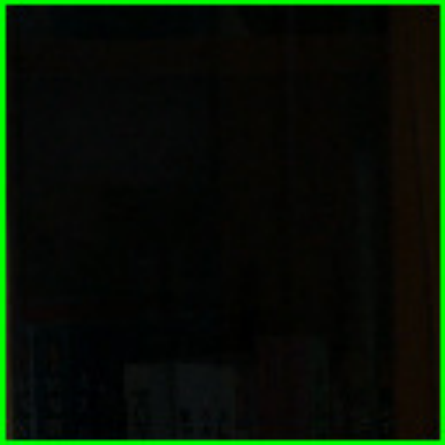}} 
		\end{overpic}
	}%
	\subfigure[MSRCR \cite{jobson1997multiscale}]{
		\begin{overpic}[scale=.205]{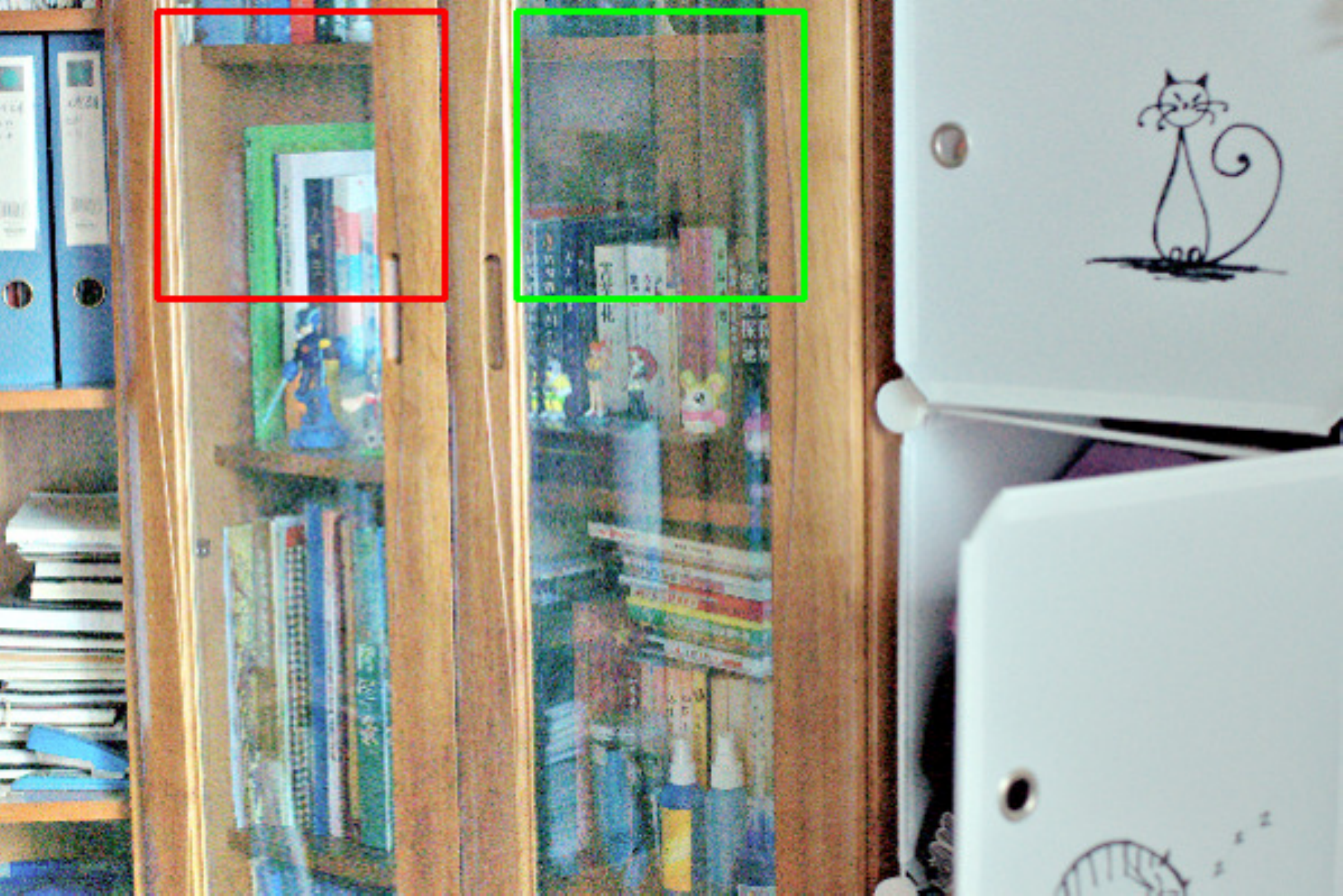}   		\put(70,0){\includegraphics[scale=.28]%
				{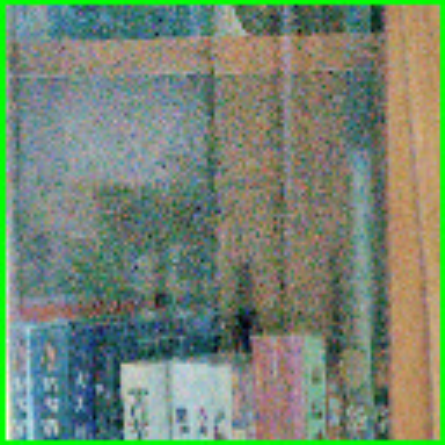}} 
		\end{overpic}
	}%
	\subfigure[BIMEF \cite{ying2017bio}]{
		\begin{overpic}[scale=.205]{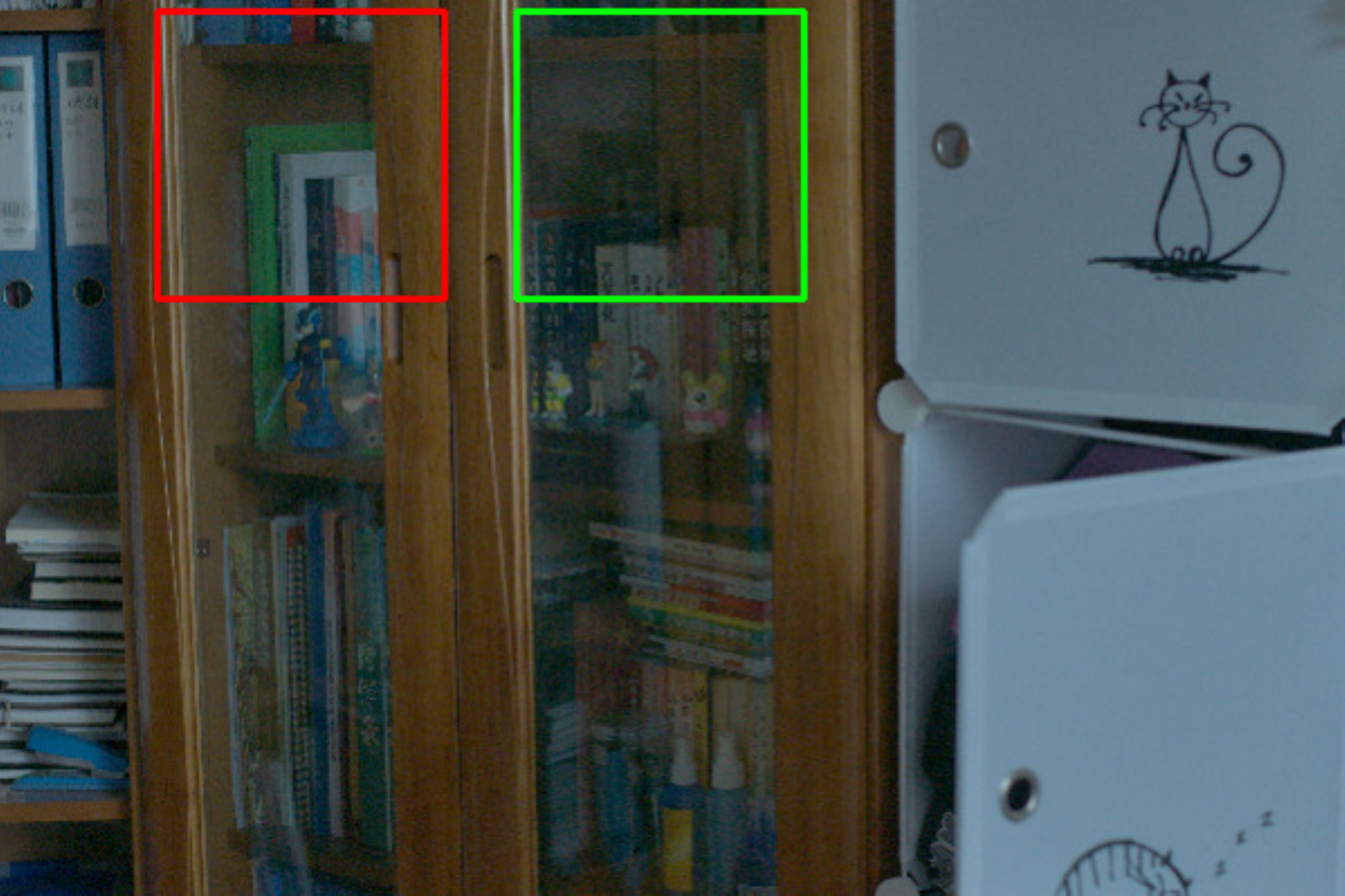}   		\put(70,0){\includegraphics[scale=.28]%
				{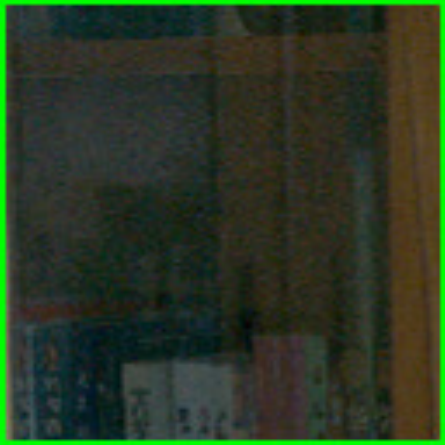}} 
		\end{overpic}
	}%
	\subfigure[LIME \cite{guo2016lime}]{
		\begin{overpic}[scale=.205]{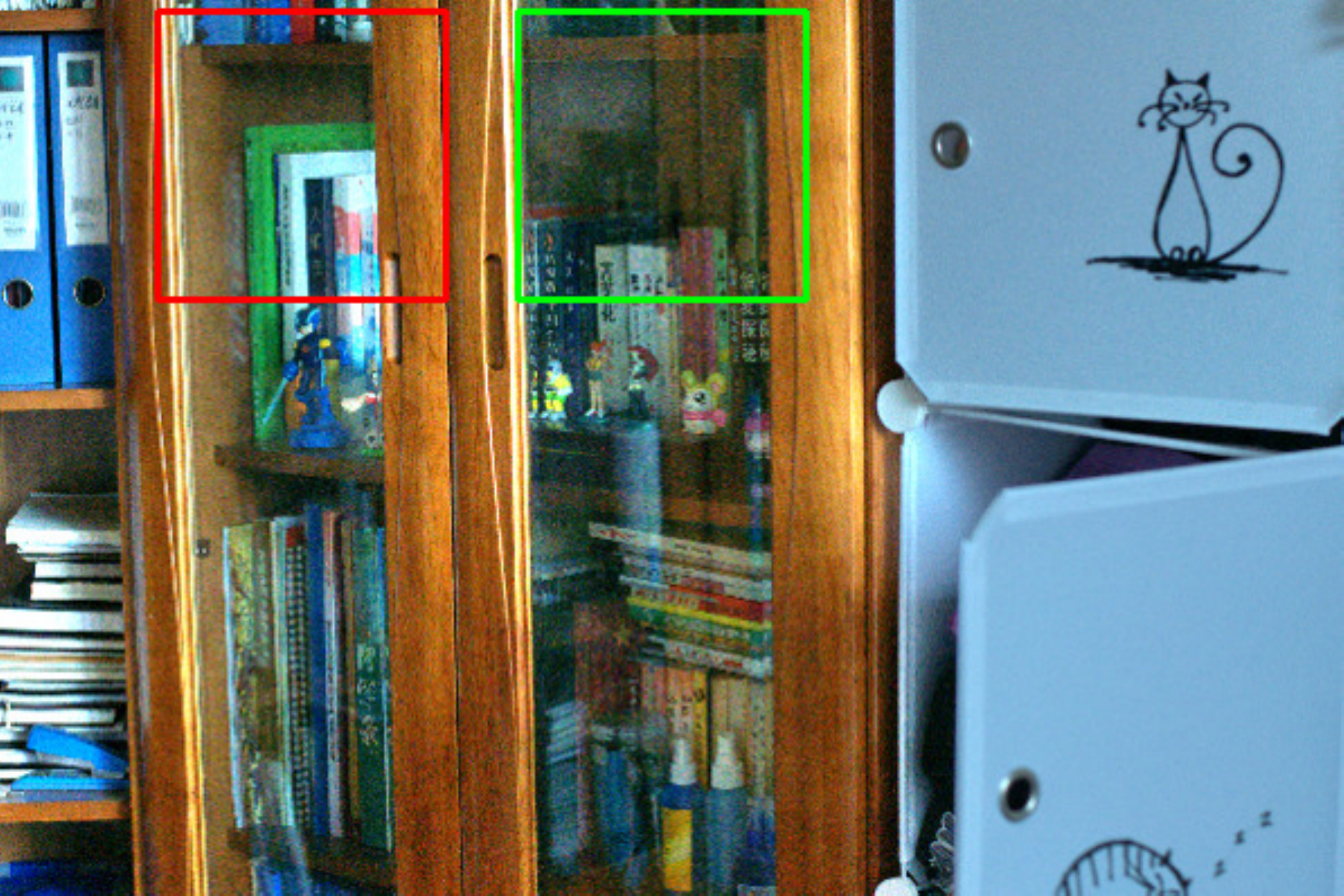}   		\put(70,0){\includegraphics[scale=.28]%
				{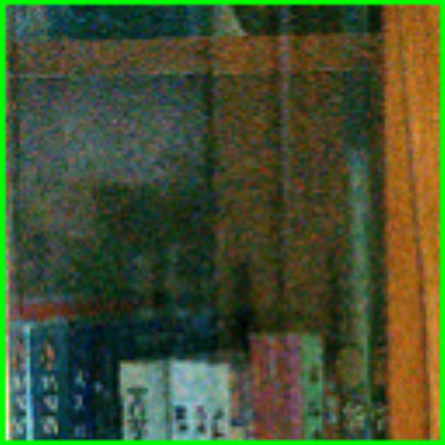}} 
		\end{overpic}
	}%
	
	\subfigure[Dong \cite{dong2011fast}]{
		\begin{overpic}[scale=.205]{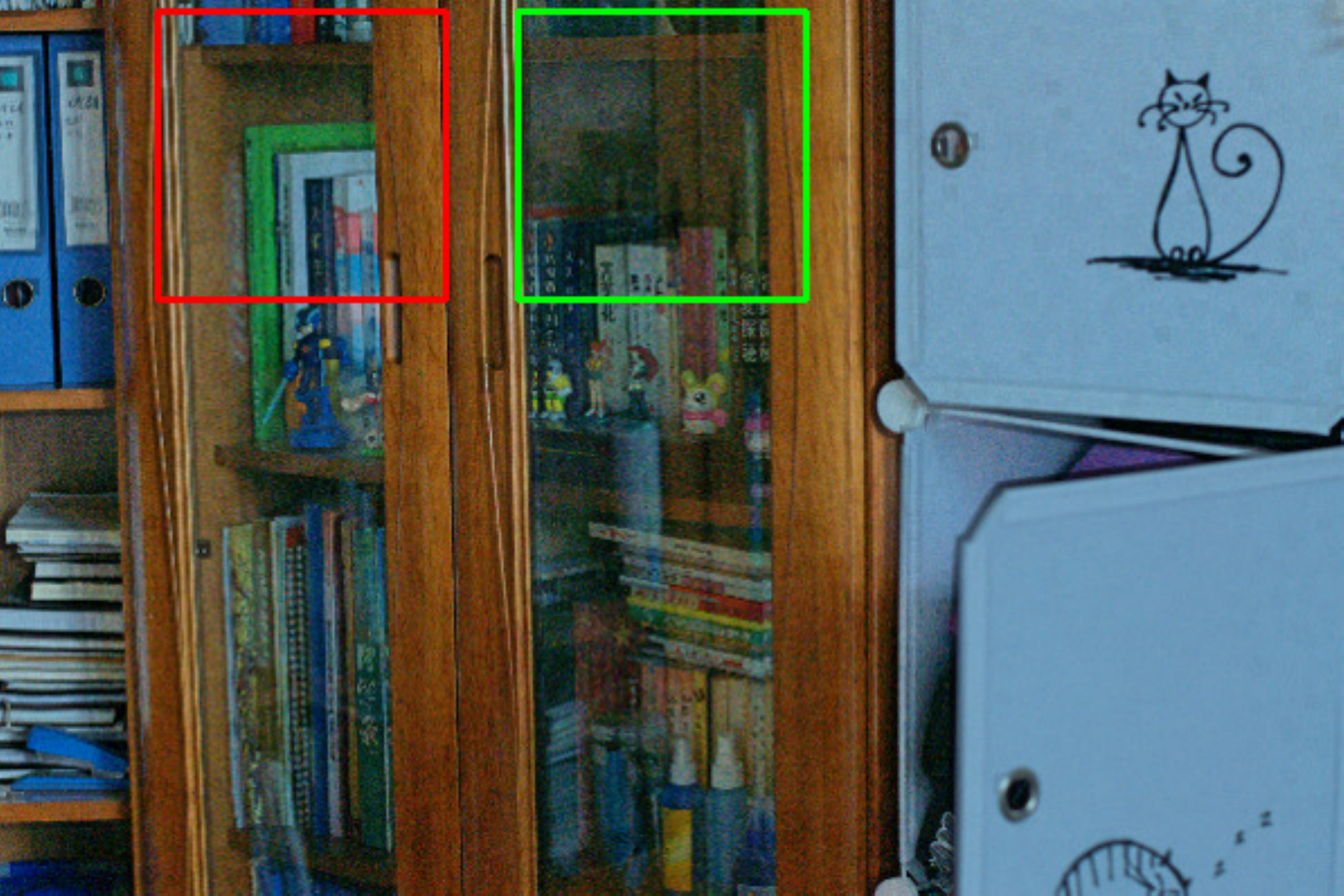}   		\put(70,0){\includegraphics[scale=.28]%
				{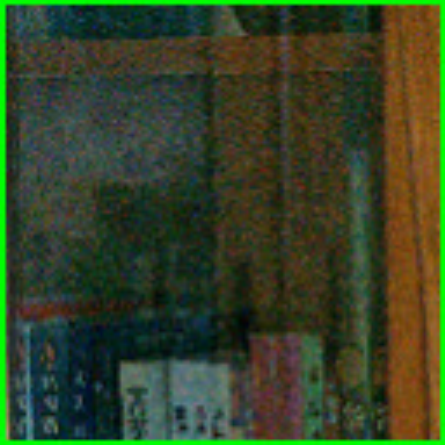}} 
		\end{overpic}
	}%
	\subfigure[SRIE \cite{fu2016weighted}]{
		\begin{overpic}[scale=.205]{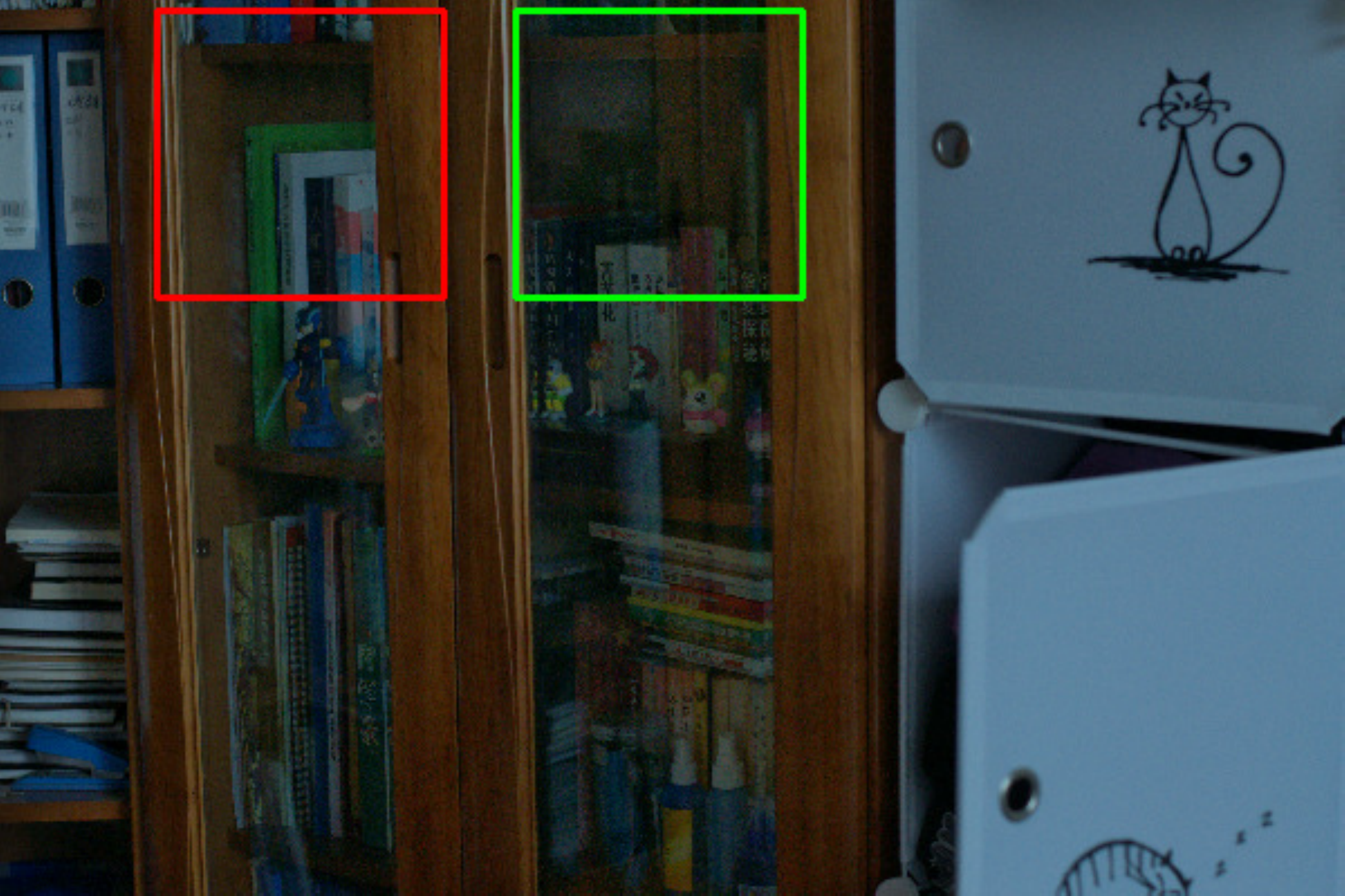}   		\put(70,0){\includegraphics[scale=.28]%
				{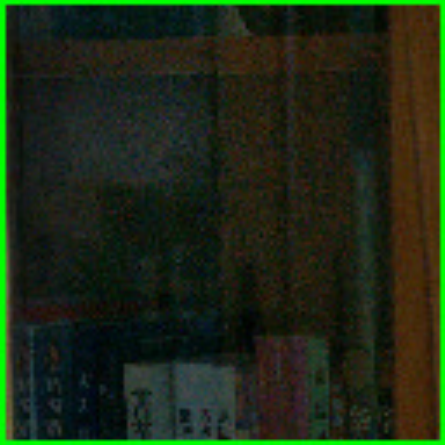}} 
		\end{overpic}
	}%
	\subfigure[MF \cite{fu2016fusion}]{
		\begin{overpic}[scale=.205]{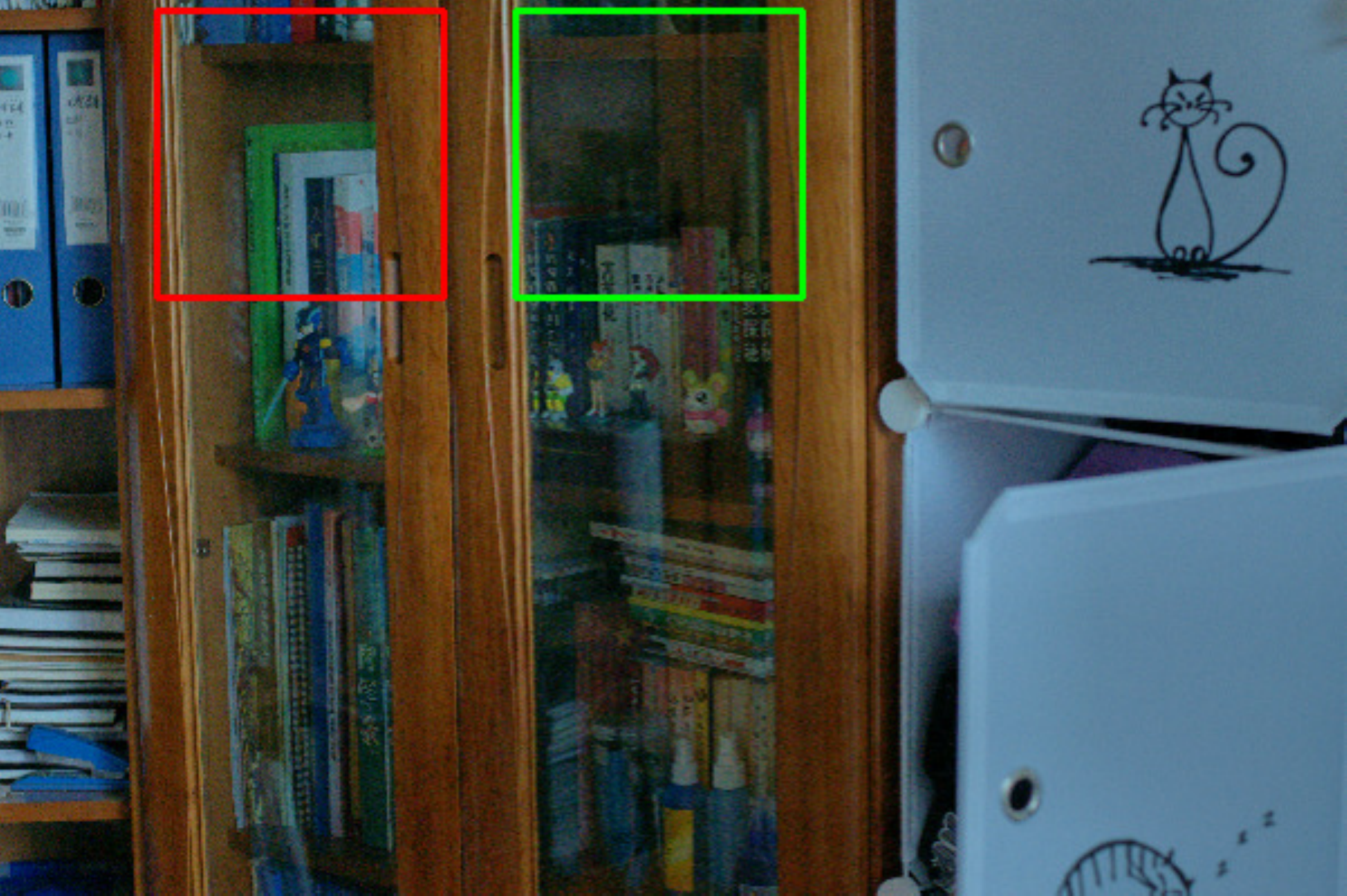}   		\put(70,0){\includegraphics[scale=.28]%
				{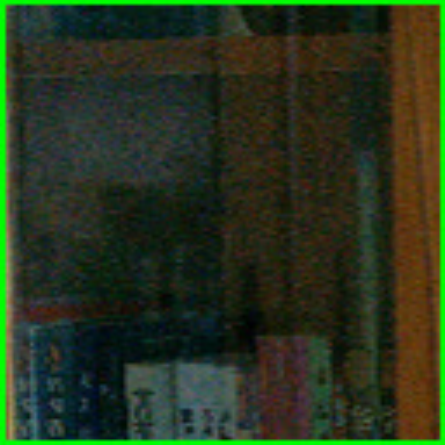}} 
		\end{overpic}
	}%
	\subfigure[NPE \cite{wang2013naturalness}]{
		\begin{overpic}[scale=.205]{pic/1/Dong-eps-converted-to.pdf}   		\put(70,0){\includegraphics[scale=.28]%
				{pic/1/Dong_1-eps-converted-to.pdf}} 
		\end{overpic}
	}%
	
	\subfigure[RRM \cite{li2018structure}]{
		\begin{overpic}[scale=.205]{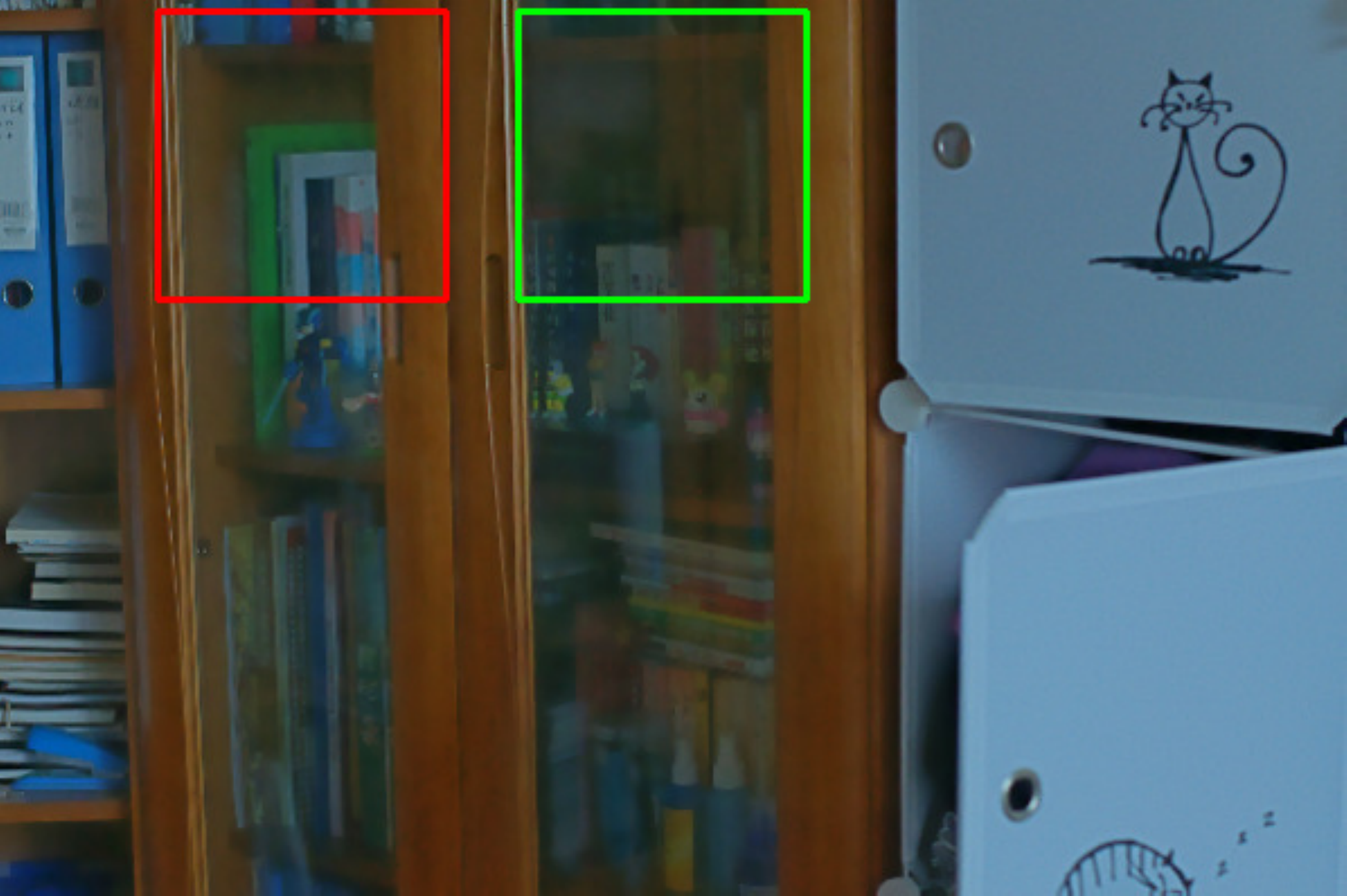}   		\put(70,0){\includegraphics[scale=.28]%
				{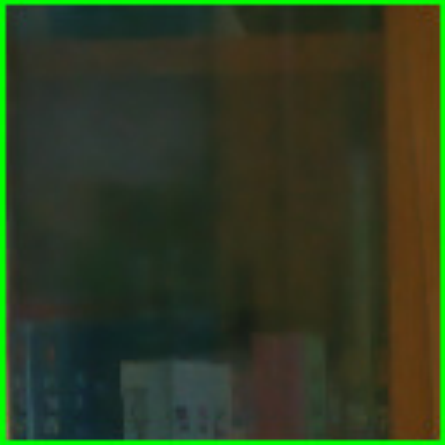}} 
		\end{overpic}
	}%
	\subfigure[MBLLEN \cite{lv2018mbllen}]{
		\begin{overpic}[scale=.205]{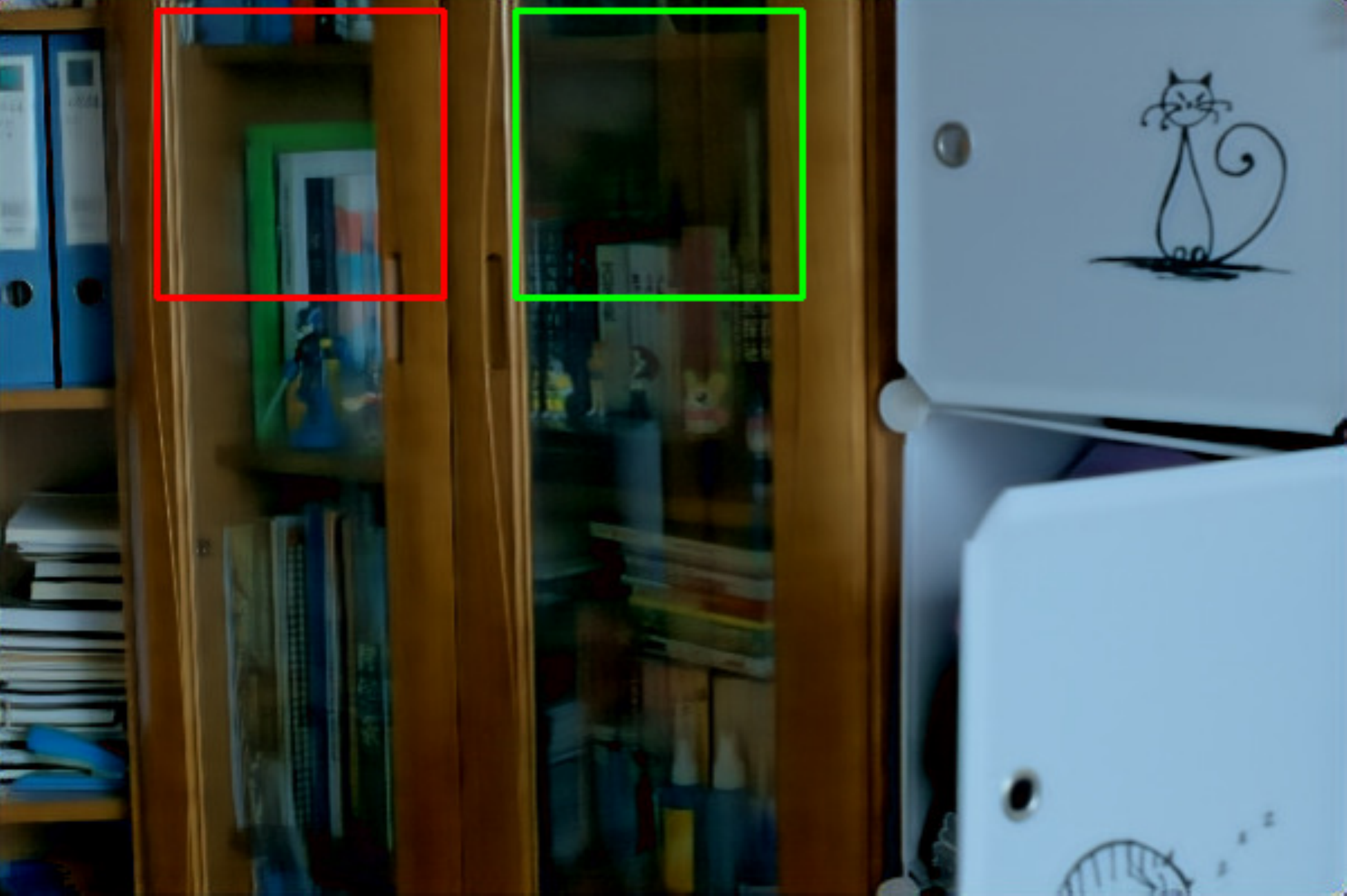}   		\put(70,0){\includegraphics[scale=.28]%
				{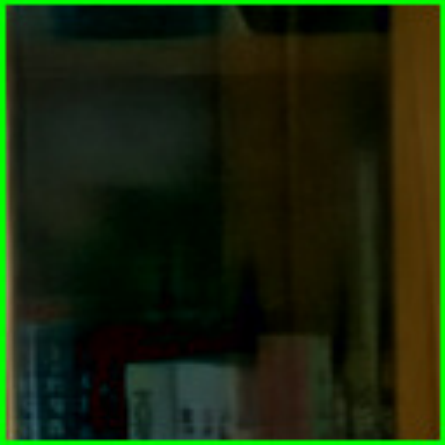}} 
		\end{overpic}
	}%
	\subfigure[RetinexNet \cite{wei2018deep}]{
		\begin{overpic}[scale=.205]{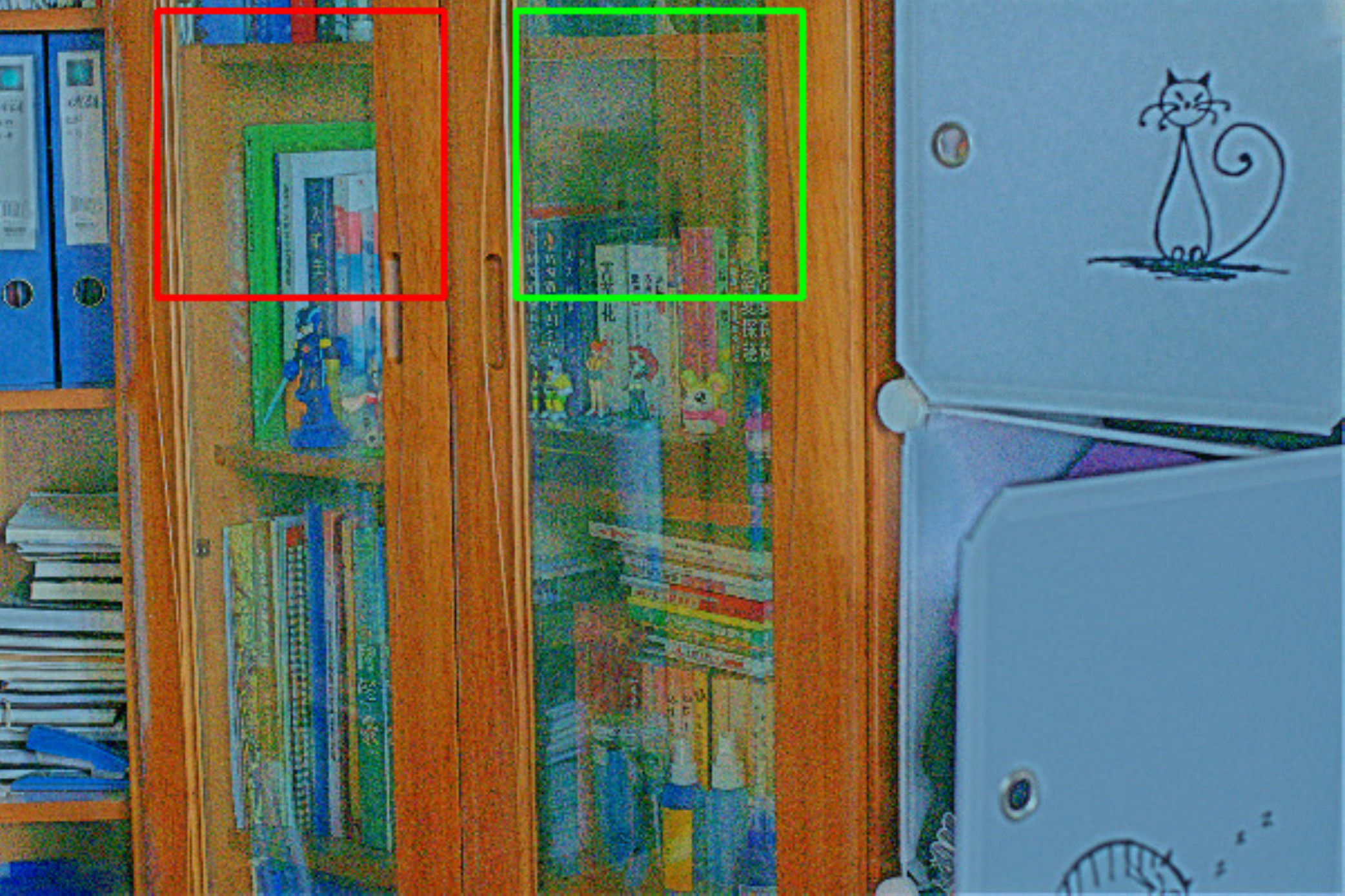}   		\put(70,0){\includegraphics[scale=.28]%
				{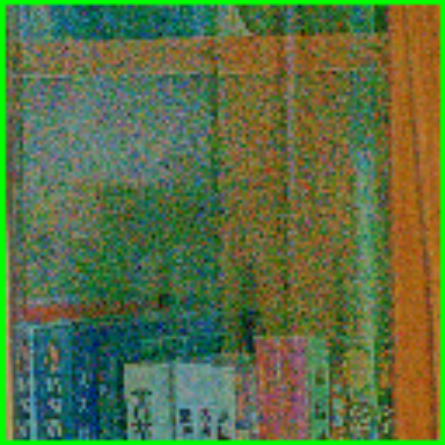}} 
		\end{overpic}
	}%
	\subfigure[GLAD \cite{wang2018gladnet}]{
		\begin{overpic}[scale=.205]{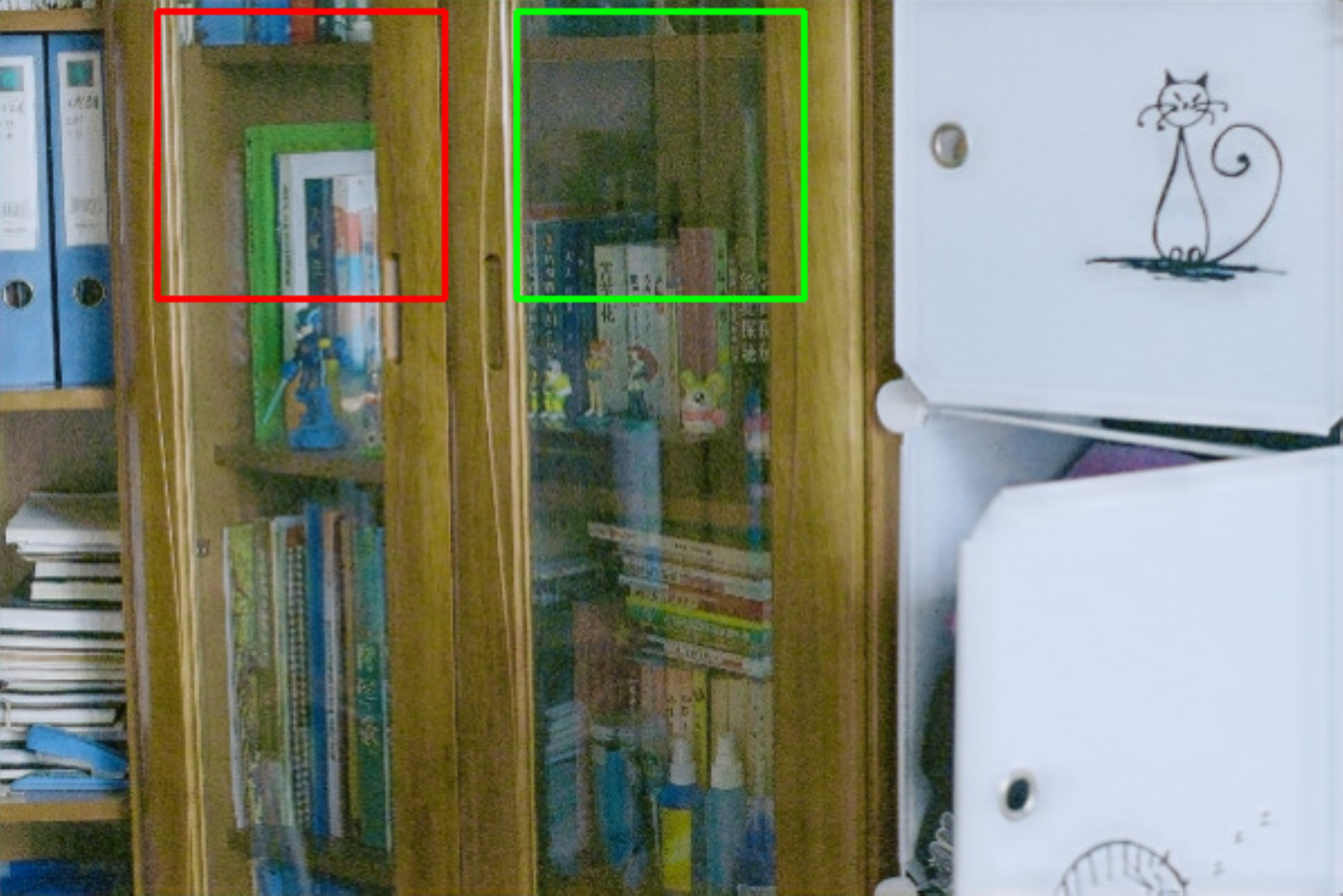}   		\put(70,0){\includegraphics[scale=.28]%
				{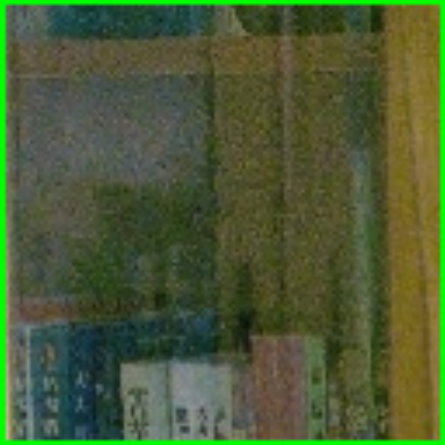}} 
		\end{overpic}
	}%
	
	\subfigure[EnlightenGan \cite{jiang2021enlightengan}]{
		\begin{overpic}[scale=.205]{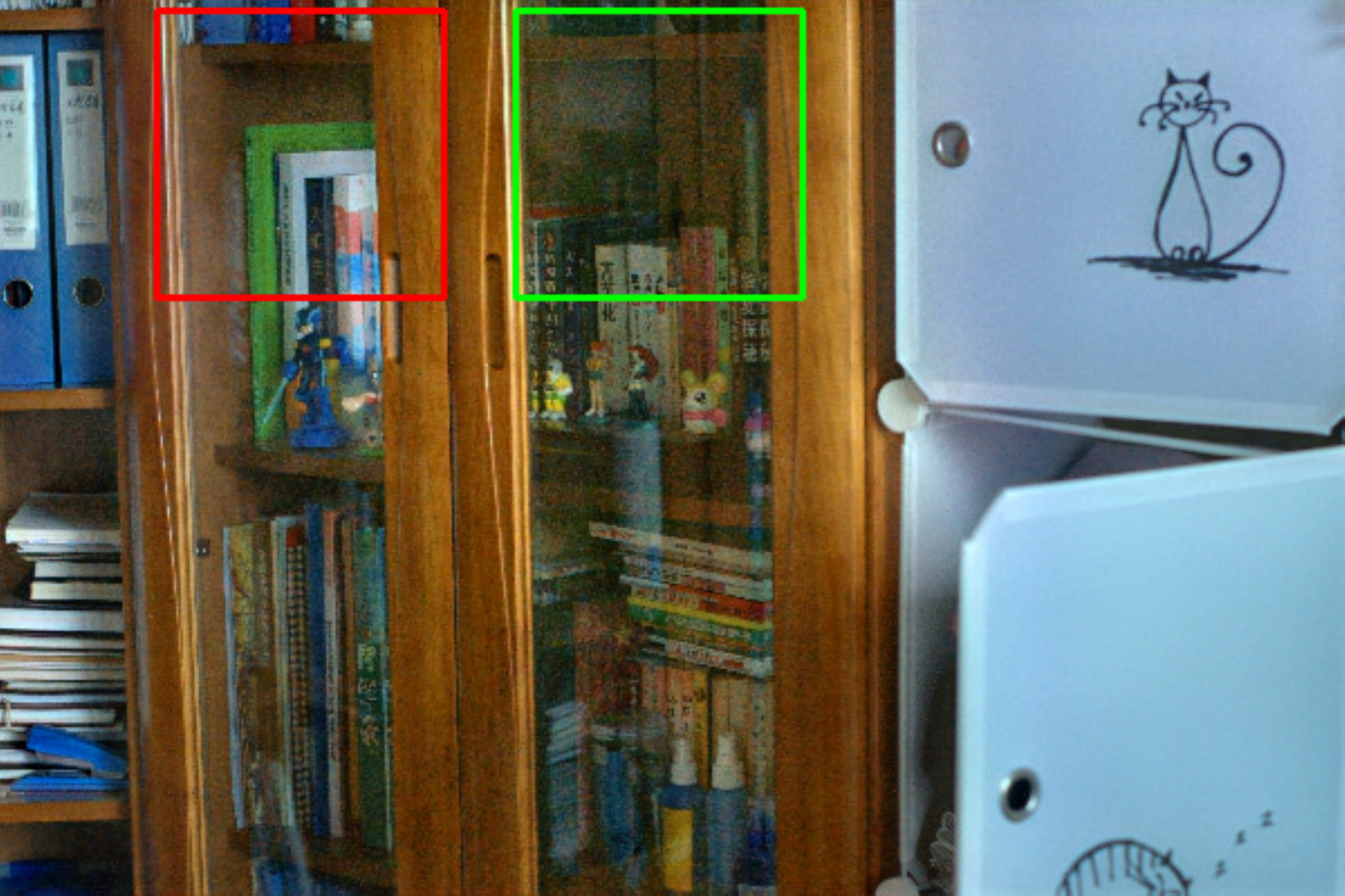}   		\put(70,0){\includegraphics[scale=.28]%
				{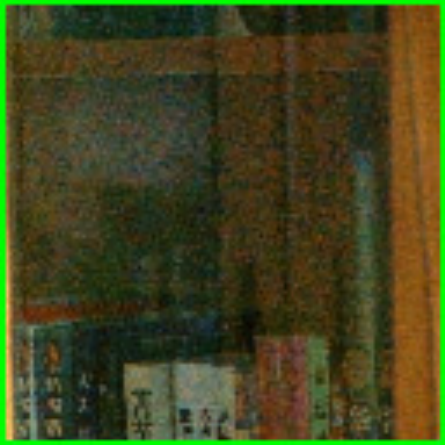}} 
		\end{overpic}
	}%
	\subfigure[Zero-DCE \cite{guo2020zero}]{
		\begin{overpic}[scale=.205]{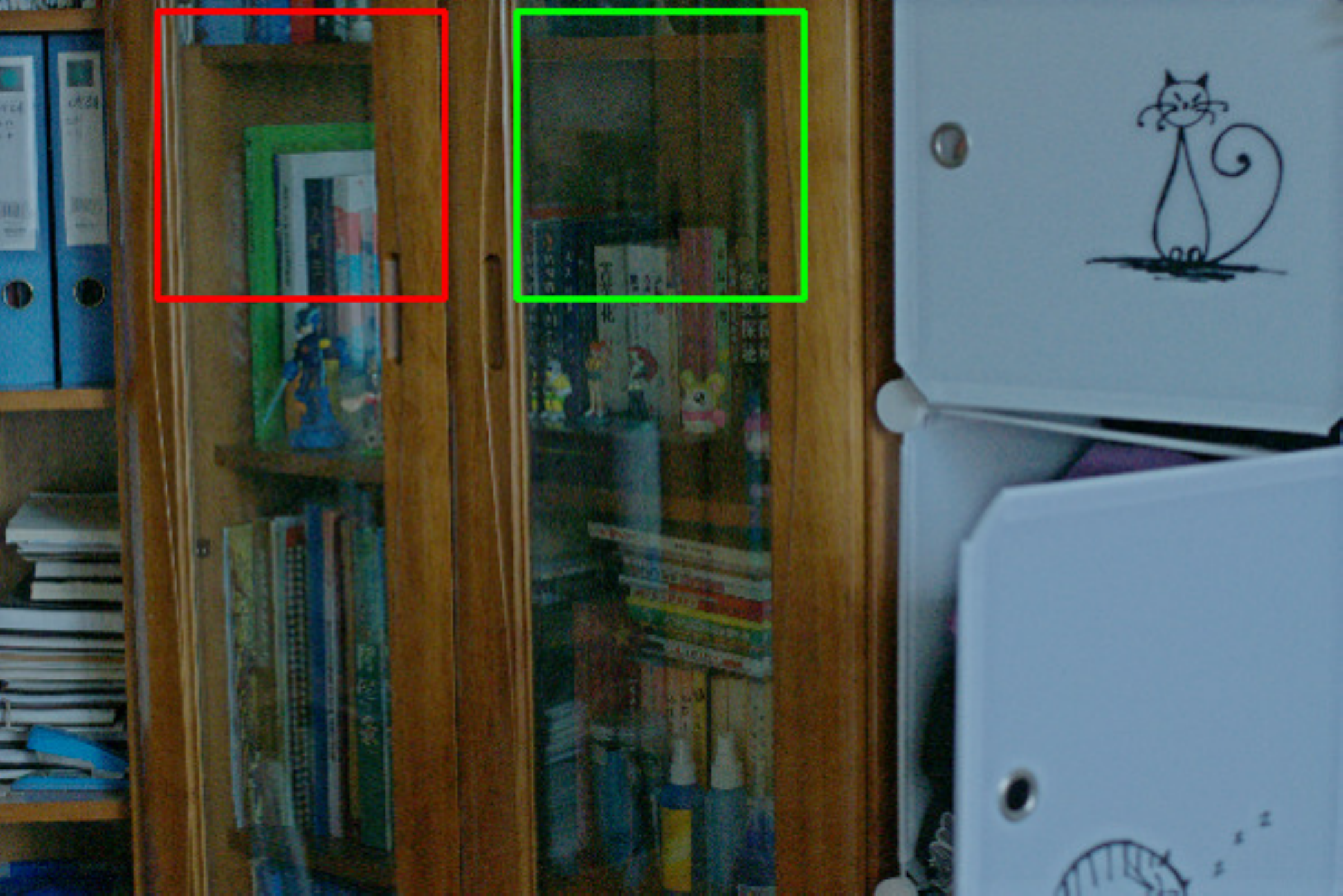}   		\put(70,0){\includegraphics[scale=.28]%
				{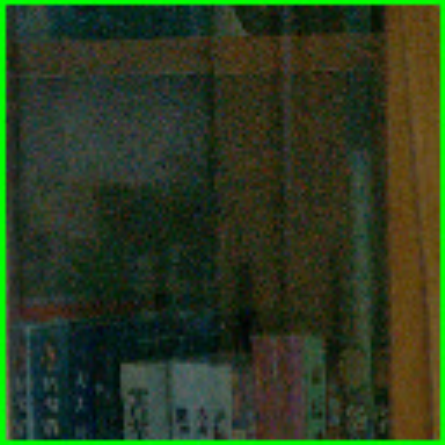}} 
		\end{overpic}
	}%
	\subfigure[Ours]{
		\begin{overpic}[scale=.205]{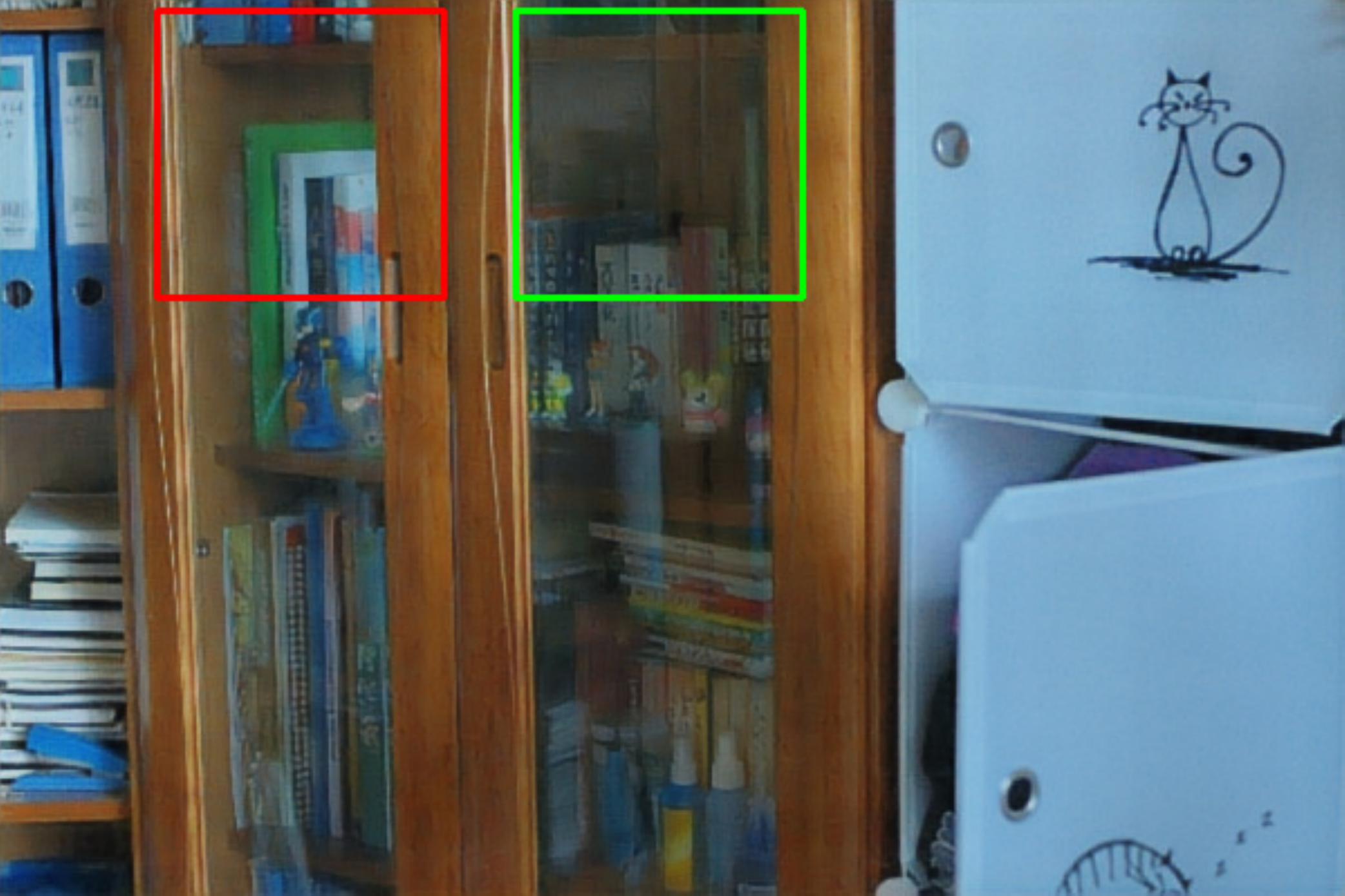}   		\put(70,0){\includegraphics[scale=.28]%
				{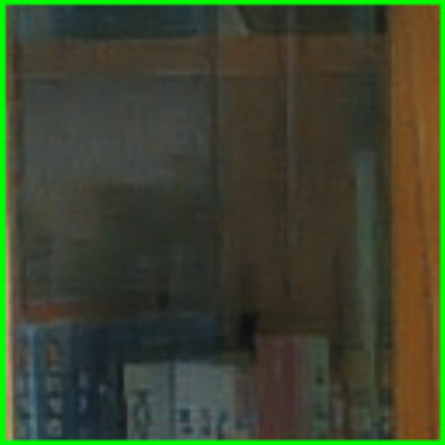}} 
		\end{overpic}
	}%
	\subfigure[GroundTruth]{
		\begin{overpic}[scale=.205]{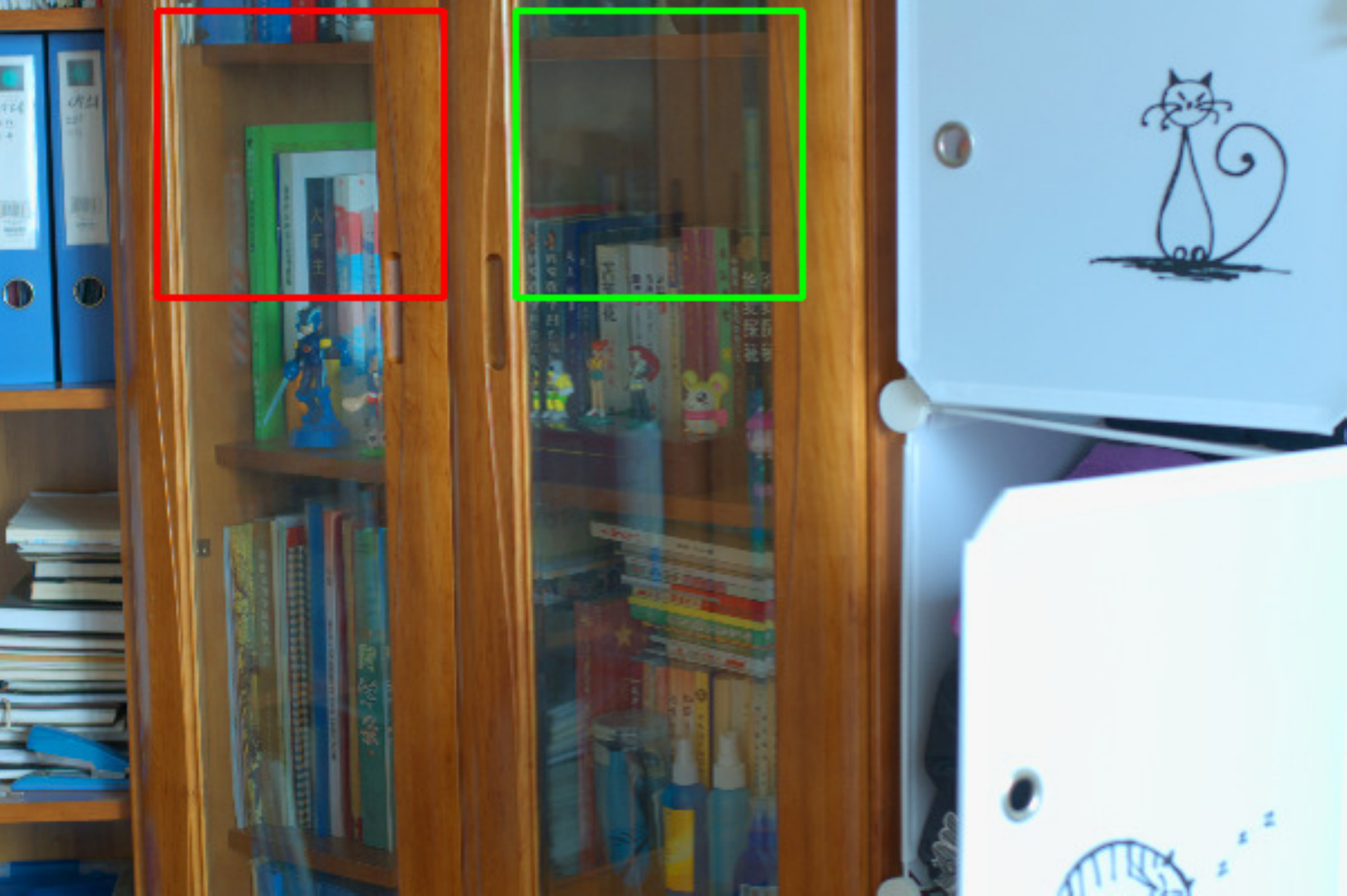}   		\put(70,0){\includegraphics[scale=.28]%
				{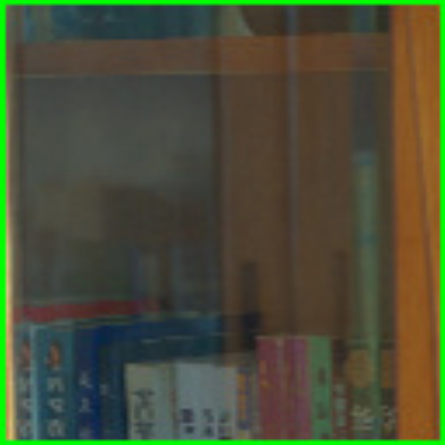}} 
		\end{overpic}
	}%
	
	\flushleft
	\caption{Visual comparison with other state-of-the-art methods on LOL real-world validation dataset, where the degradation is hidden in darkness.}
	\label{book}
\end{figure*}

\begin{figure*}
	\flushleft
	
	
	\subfigure[Input]{
		\begin{minipage}[b]{0.155\textwidth}
			\includegraphics[width=2.8cm]{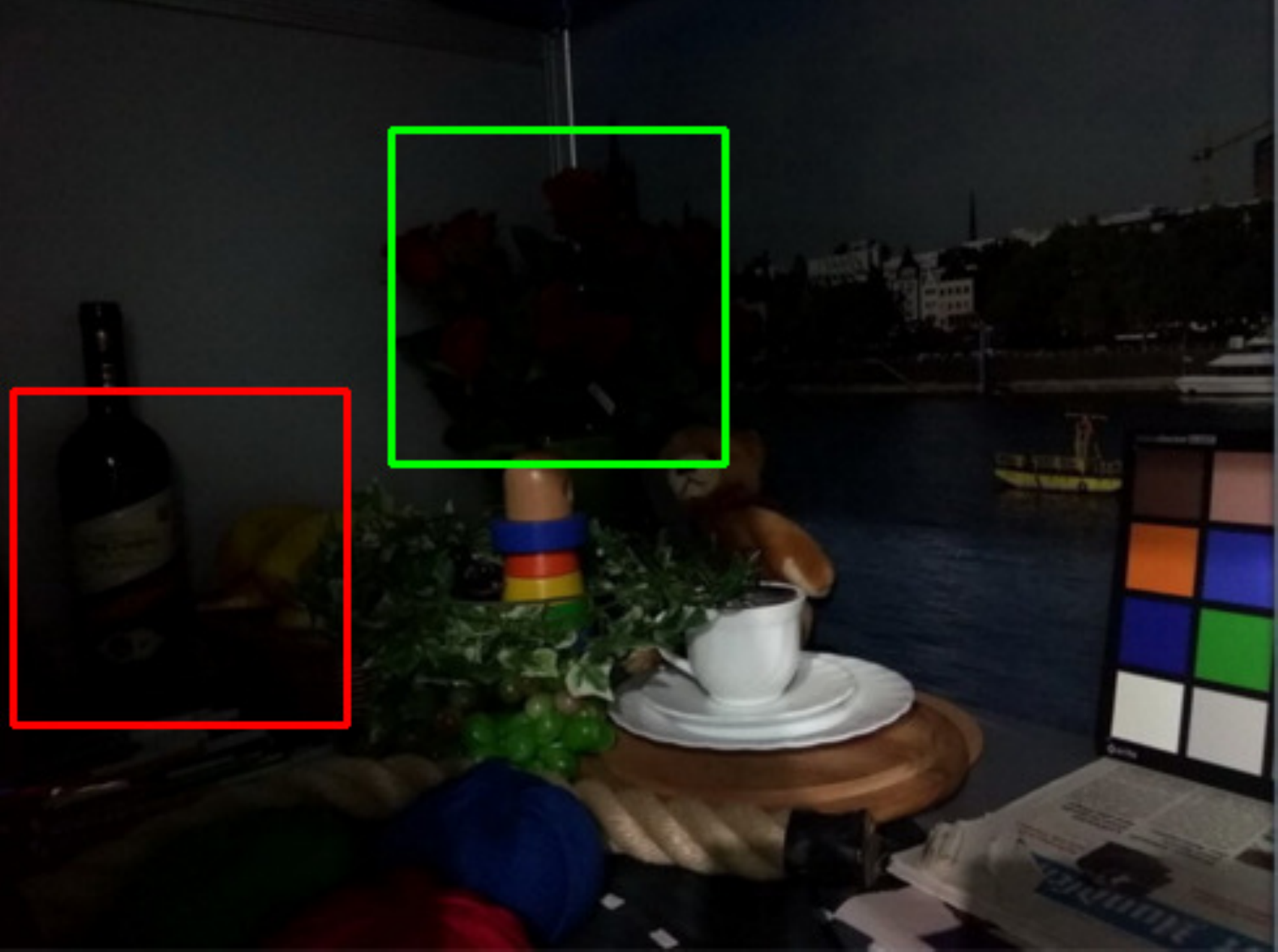}\vspace{1pt} \\
			\includegraphics[width=1.35cm]{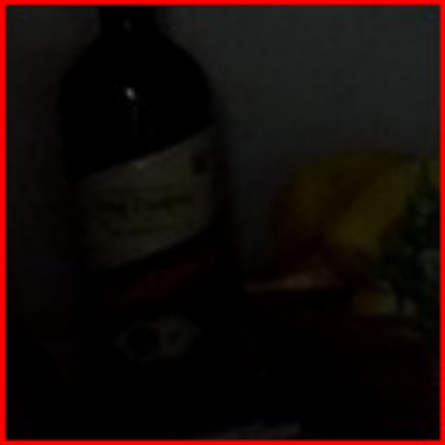}
			\includegraphics[width=1.35cm]{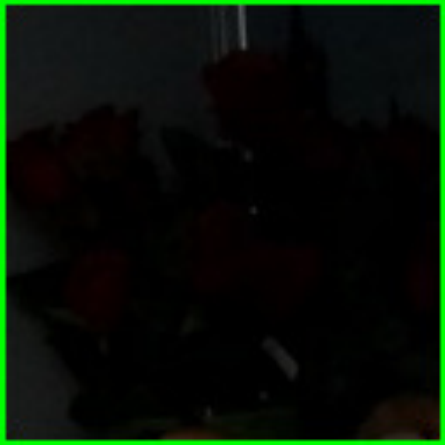}\vspace{5pt}
			\includegraphics[width=2.8cm]{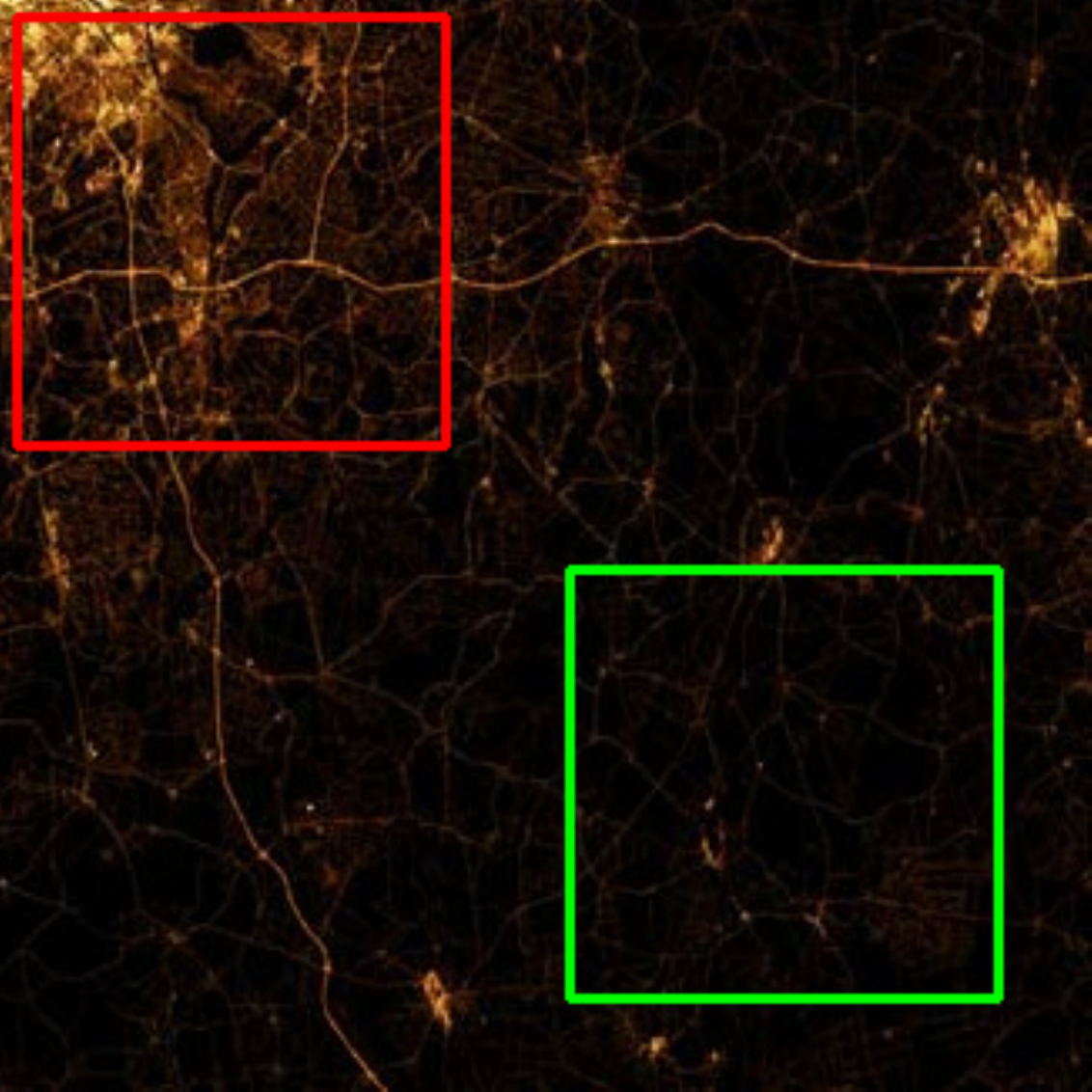}\vspace{1.5pt} \\
			\includegraphics[width=1.35cm]{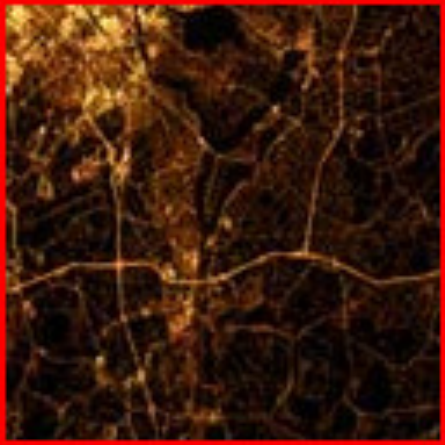}
			\includegraphics[width=1.35cm]{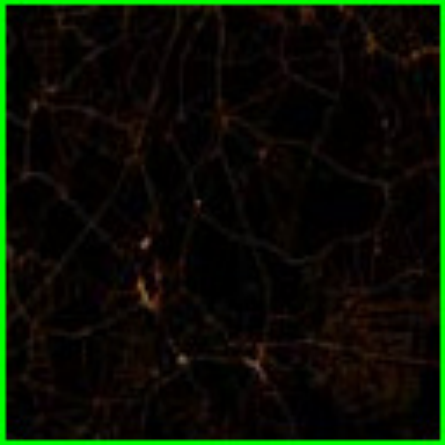}
		\end{minipage}
	}\hspace{-5pt}
	\subfigure[MSRCR\cite{jobson1997multiscale}]{
		\begin{minipage}[b]{0.155\textwidth}
			\includegraphics[width=2.8cm]{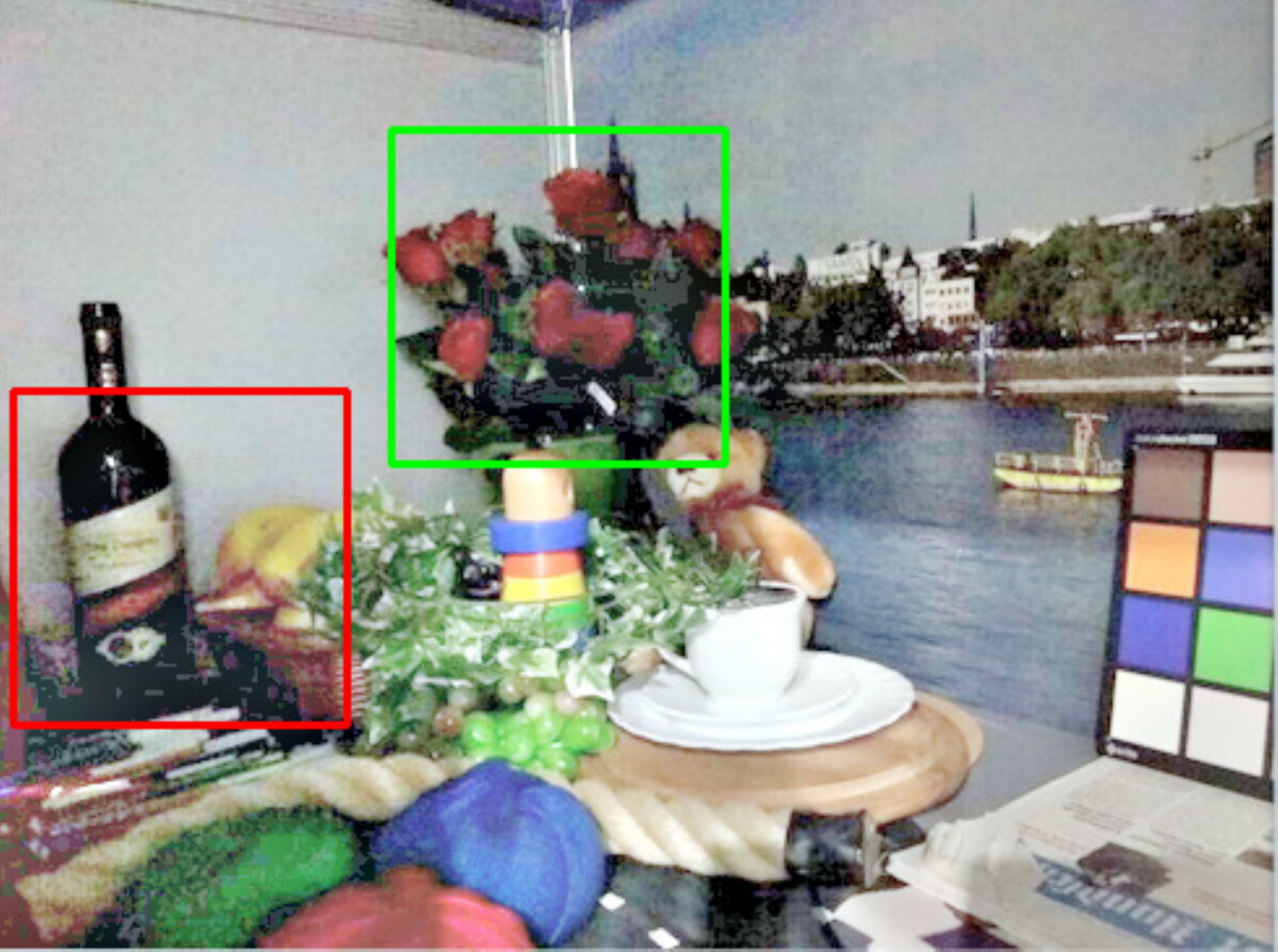}\vspace{1pt} \\
			\includegraphics[width=1.35cm]{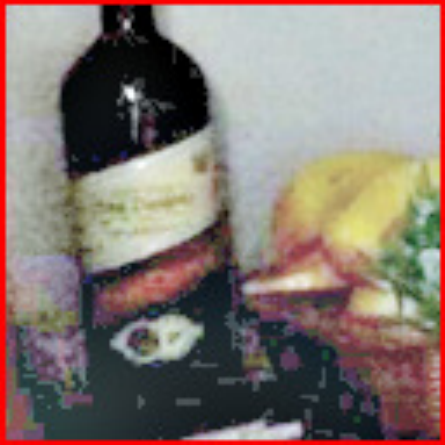}
			\includegraphics[width=1.35cm]{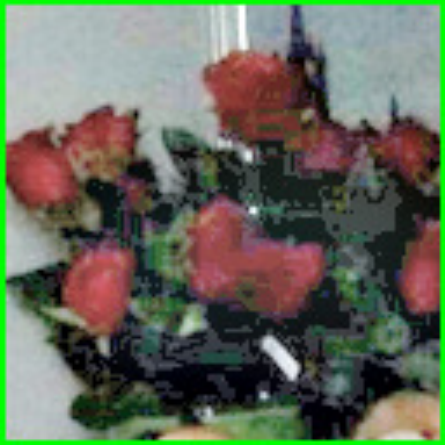}\vspace{5pt}
			\includegraphics[width=2.8cm]{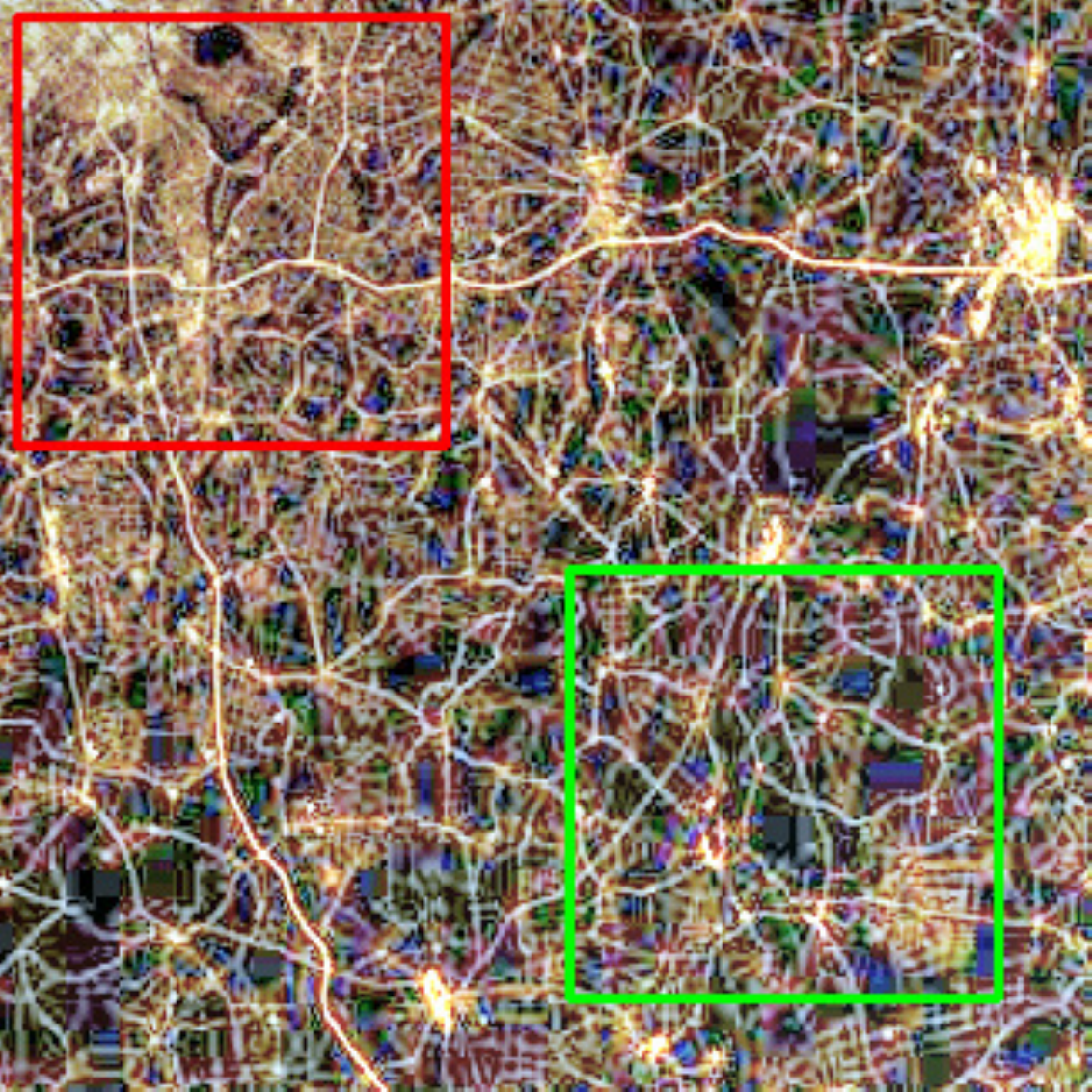}\vspace{1.5pt} \\
			\includegraphics[width=1.35cm]{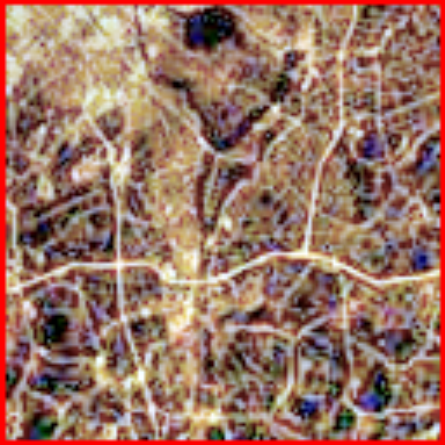}
			\includegraphics[width=1.35cm]{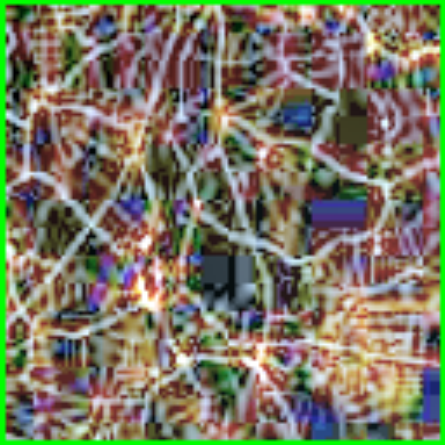}
		\end{minipage}
	}\hspace{-5pt}
	\subfigure[BIMEF\cite{ying2017bio}]{
		\begin{minipage}[b]{0.155\textwidth}
			\includegraphics[width=2.8cm]{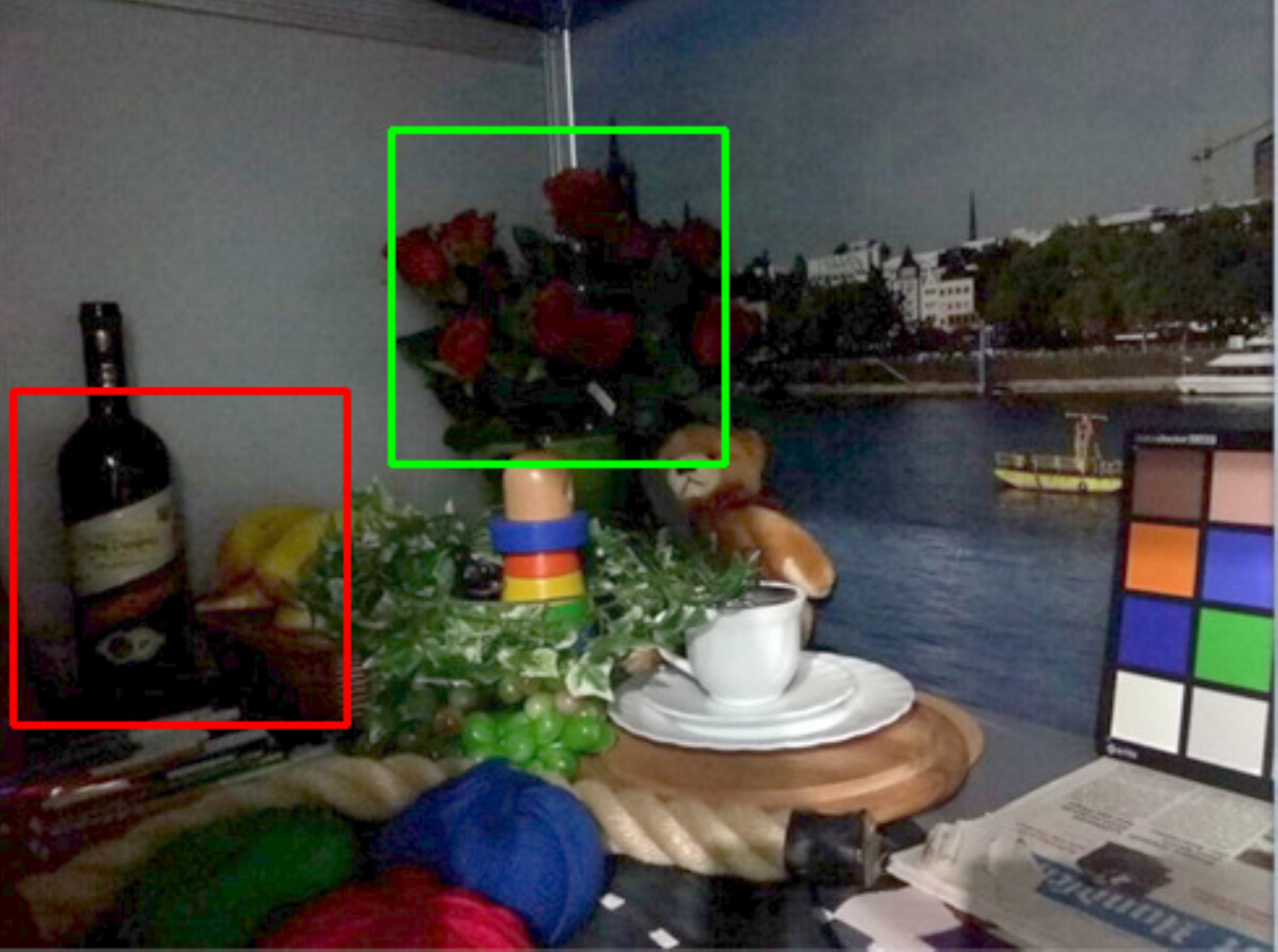}\vspace{1pt} \\
			\includegraphics[width=1.35cm]{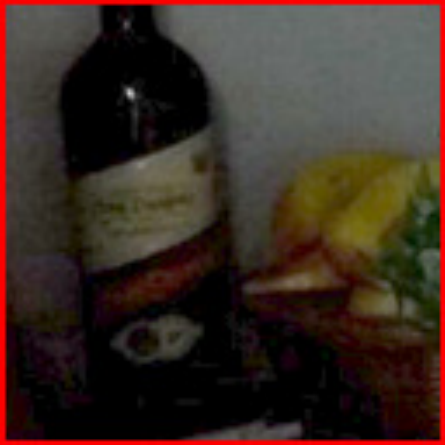}
			\includegraphics[width=1.35cm]{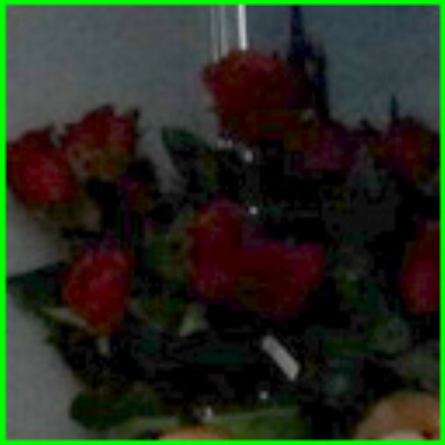}\vspace{5pt}
			\includegraphics[width=2.8cm]{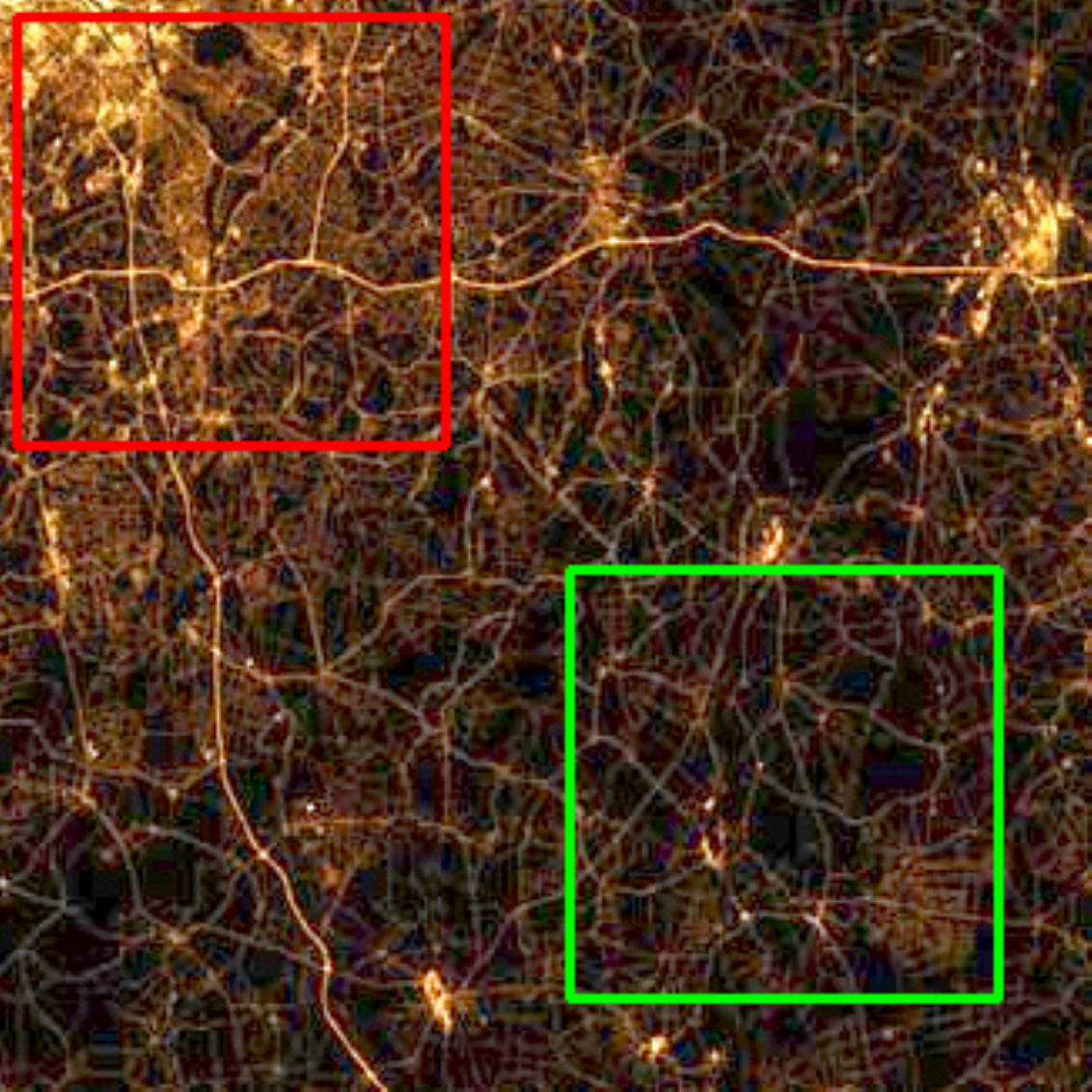}\vspace{1.5pt} \\
			\includegraphics[width=1.35cm]{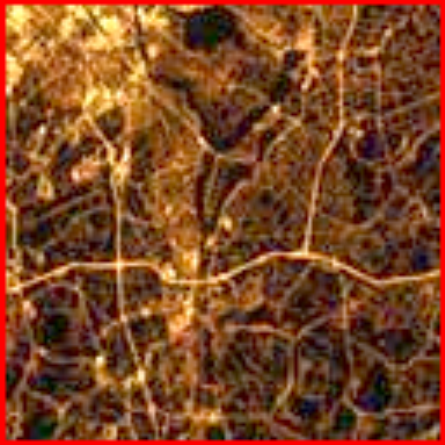}
			\includegraphics[width=1.35cm]{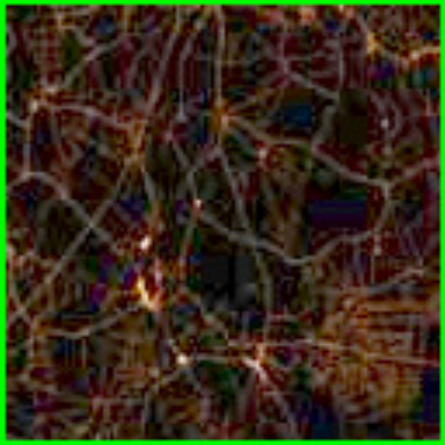}
		\end{minipage}
	}\hspace{-5pt}
	\subfigure[Dong\cite{dong2011fast}]{
		\begin{minipage}[b]{0.155\textwidth}
			\includegraphics[width=2.8cm]{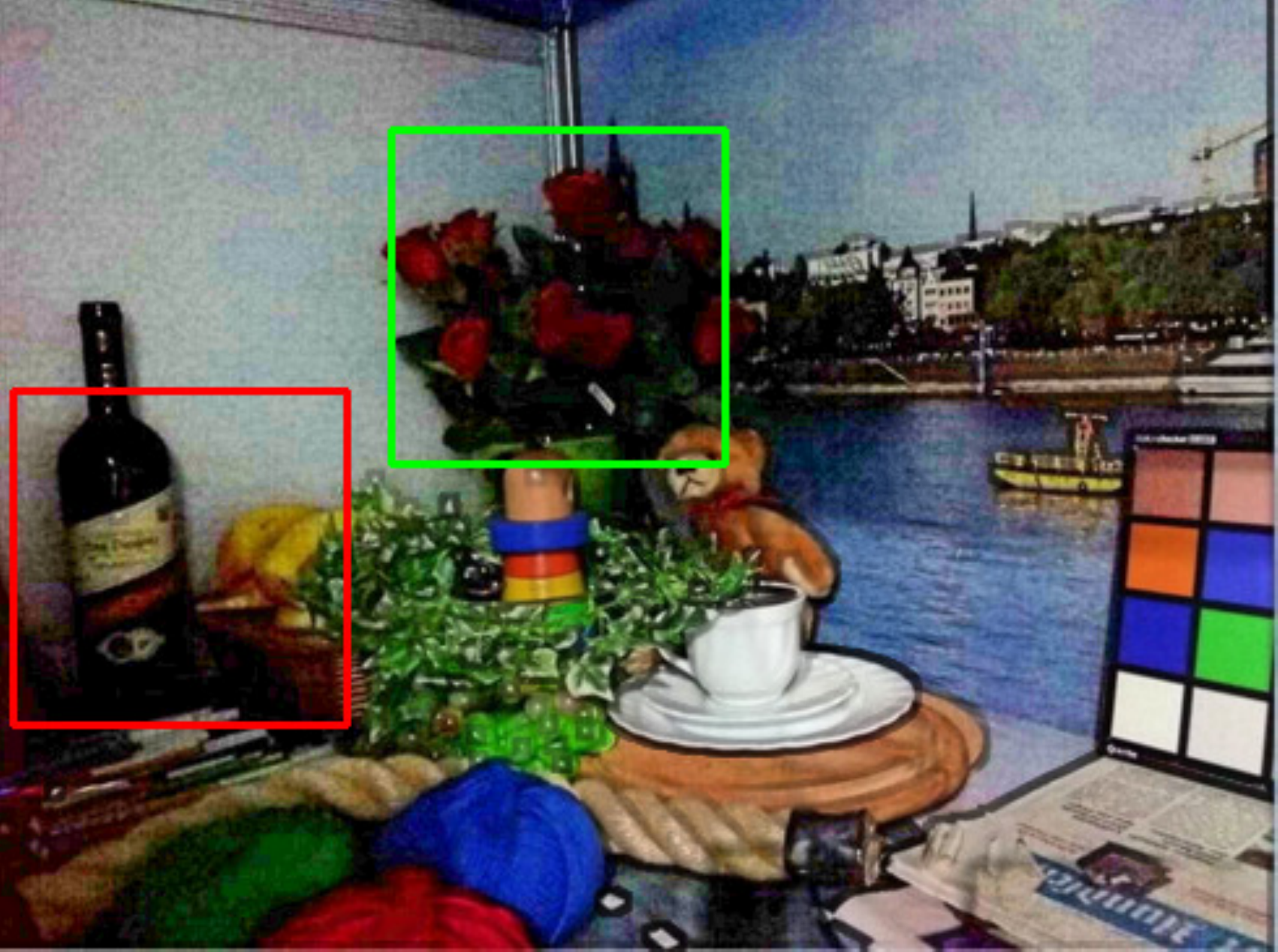}\vspace{1pt} \\
			\includegraphics[width=1.35cm]{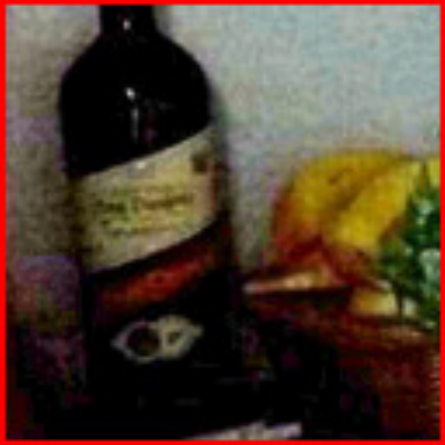}
			\includegraphics[width=1.35cm]{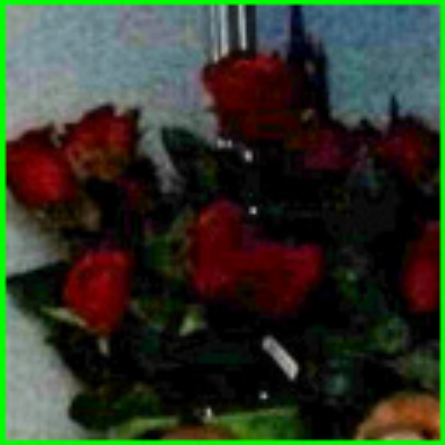}\vspace{5pt}
			\includegraphics[width=2.8cm]{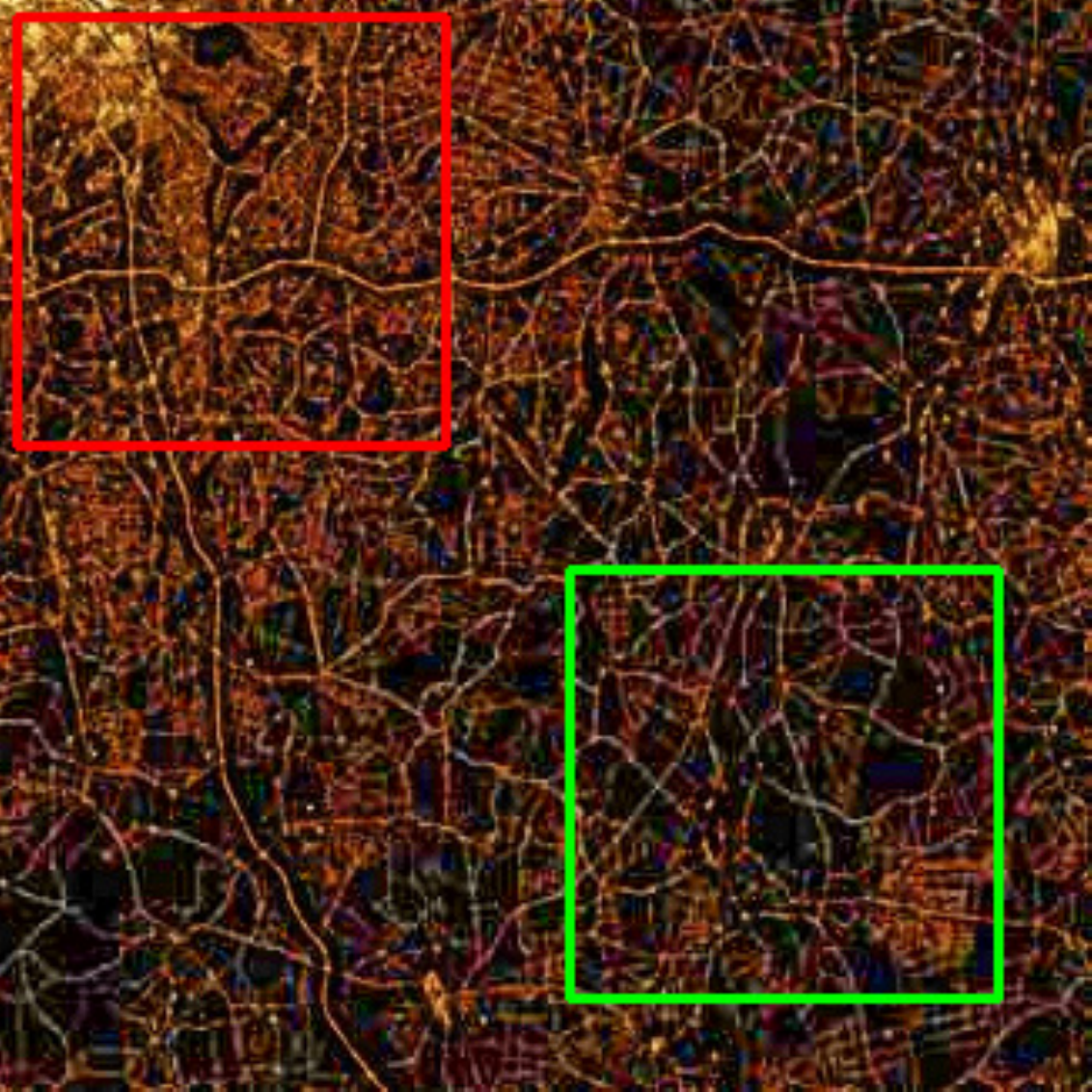}\vspace{1.5pt} \\
			\includegraphics[width=1.35cm]{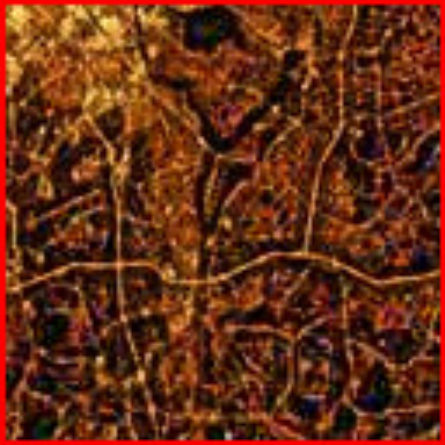}
			\includegraphics[width=1.35cm]{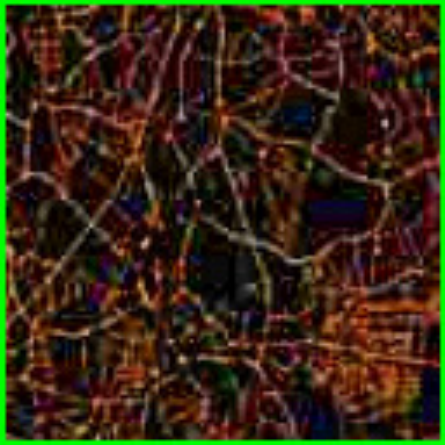}
		\end{minipage}
	}\hspace{-5pt}
	\subfigure[LECARM\cite{ren2018lecarm}]{
		\begin{minipage}[b]{0.155\textwidth}
			\includegraphics[width=2.8cm]{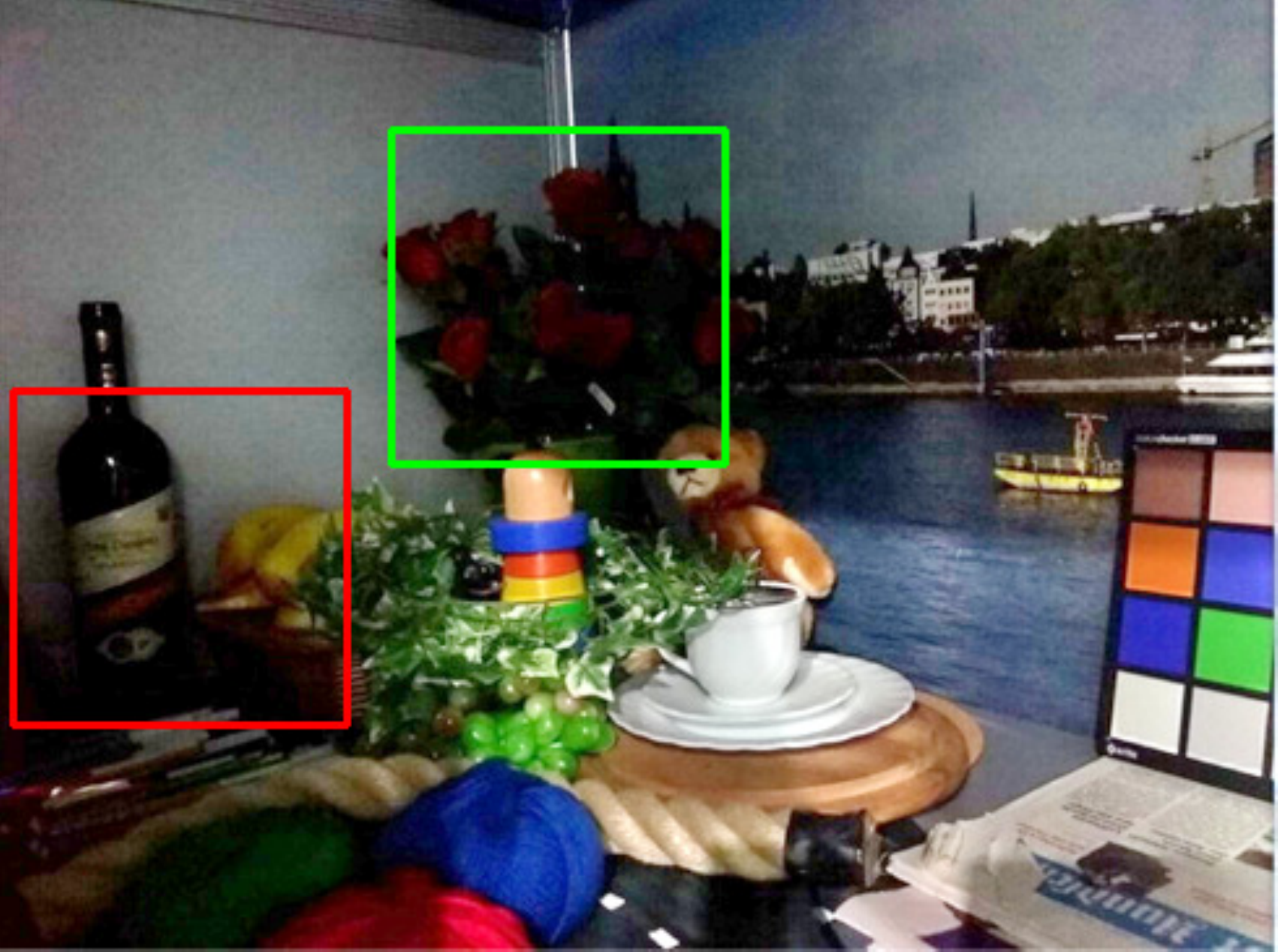}\vspace{1pt} \\
			\includegraphics[width=1.35cm]{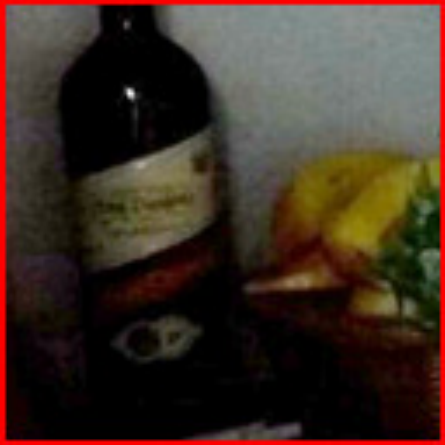}
			\includegraphics[width=1.35cm]{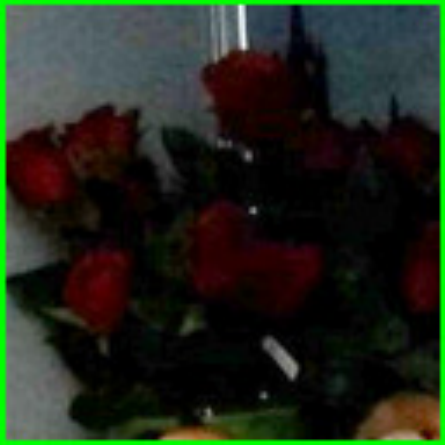}\vspace{5pt}
			\includegraphics[width=2.8cm]{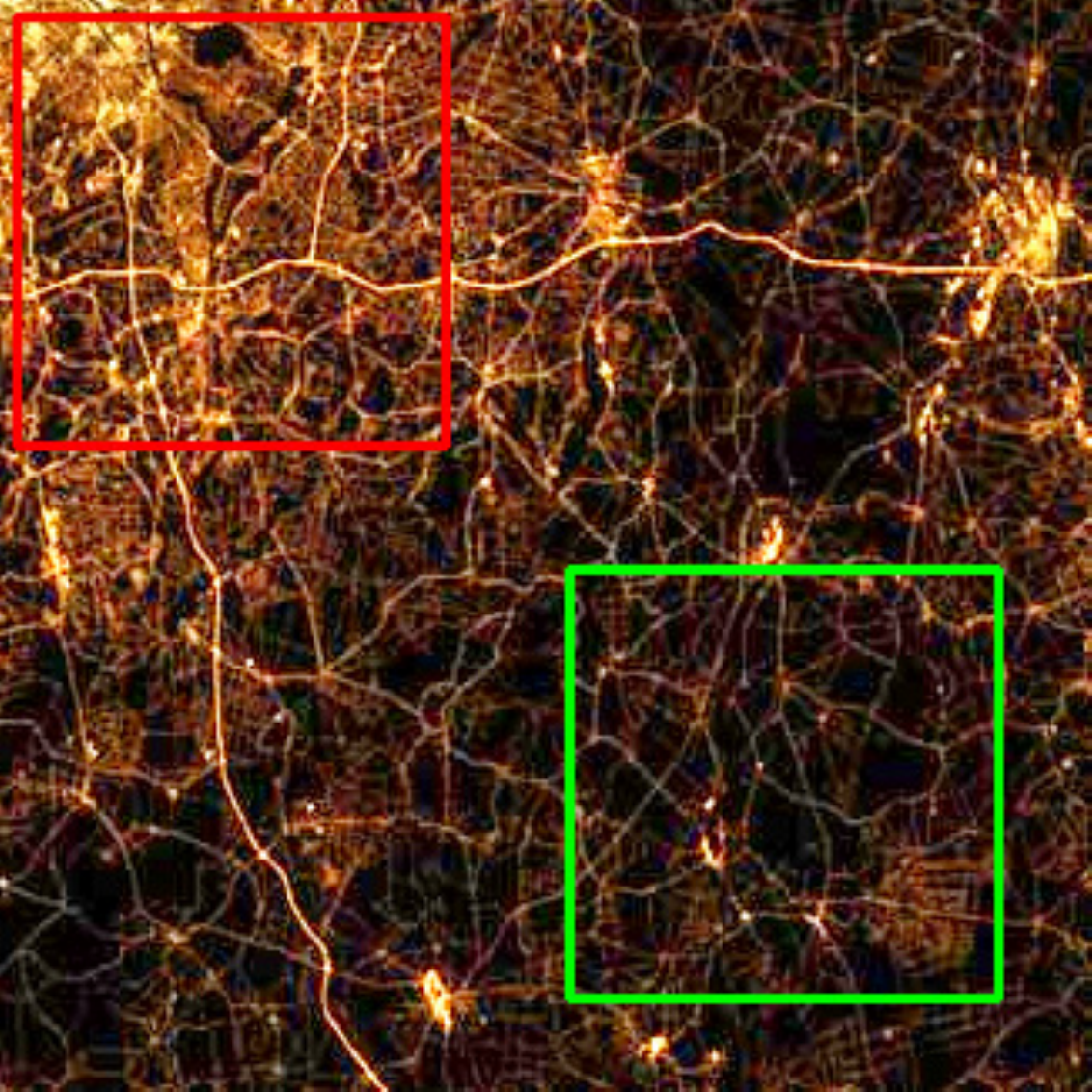}\vspace{1.5pt} \\
			\includegraphics[width=1.35cm]{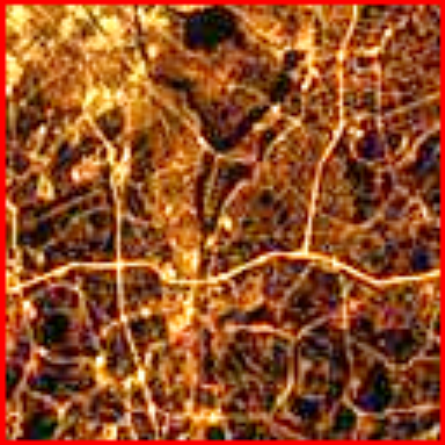}
			\includegraphics[width=1.35cm]{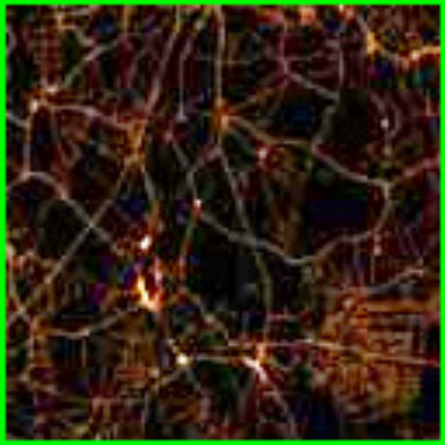}
		\end{minipage}
	}\hspace{-5pt}
	\subfigure[Ours]{
		\begin{minipage}[b]{0.155\textwidth}
			\includegraphics[width=2.8cm]{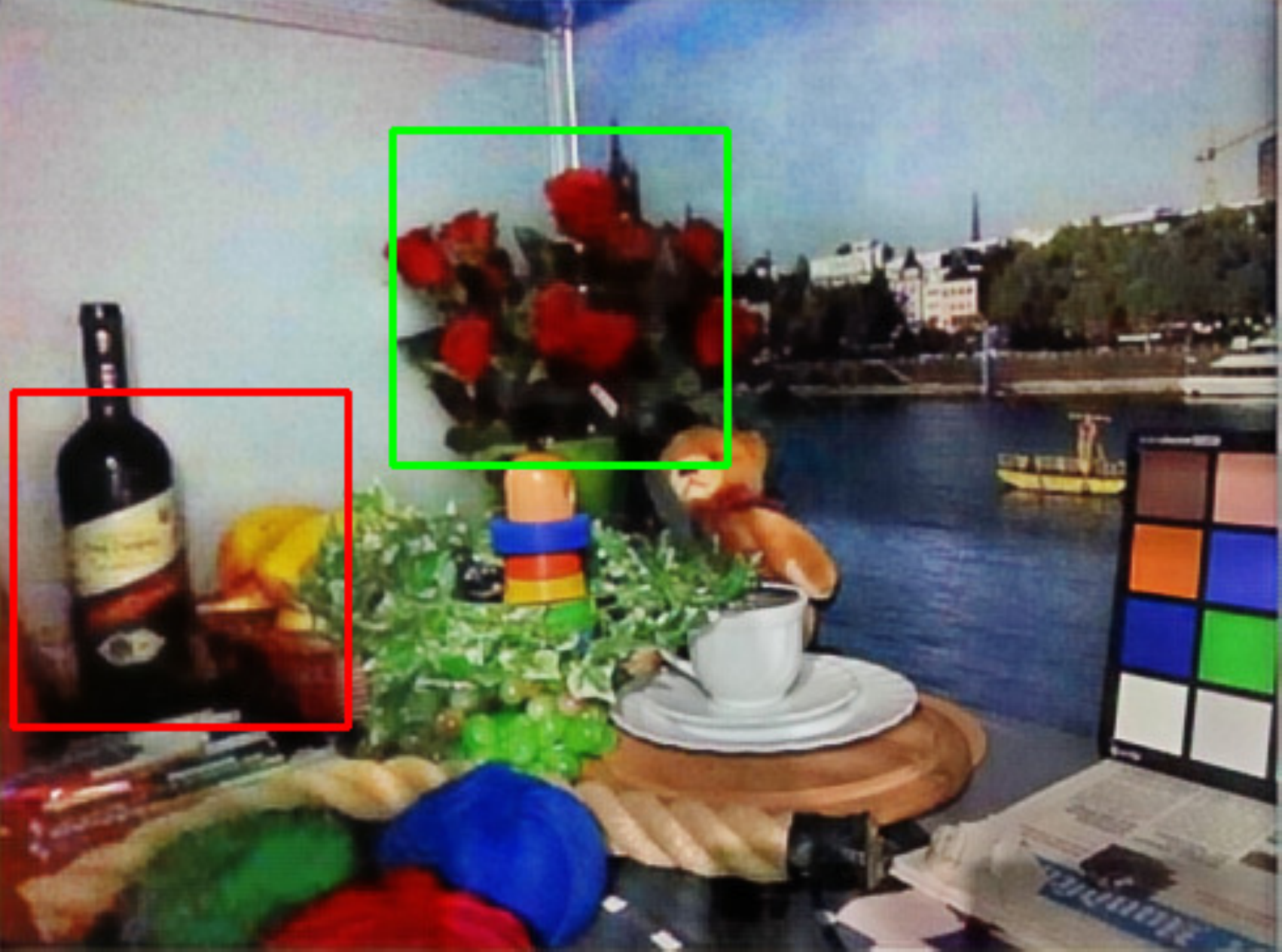}\vspace{1pt} \\
			\includegraphics[width=1.35cm]{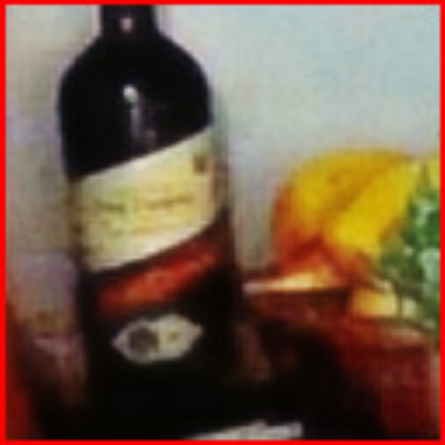}
			\includegraphics[width=1.35cm]{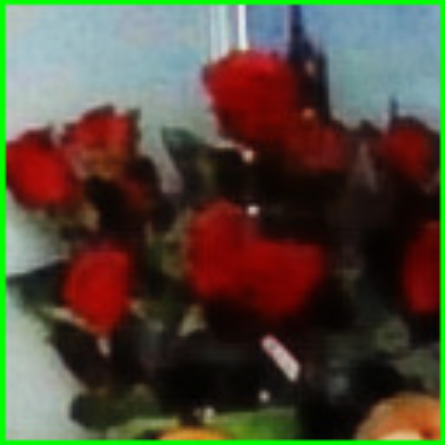}\vspace{5pt}
			\includegraphics[width=2.8cm]{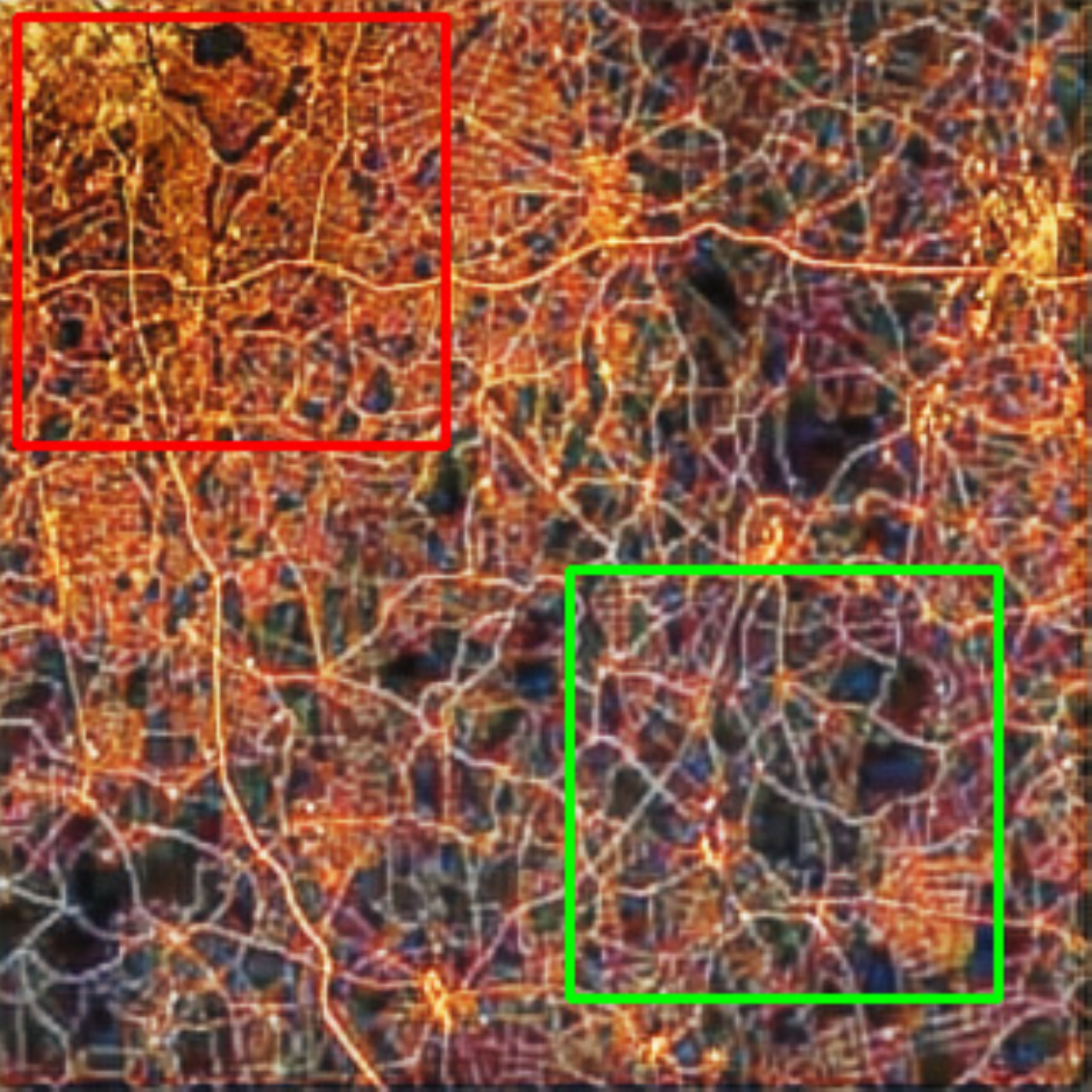}\vspace{1.5pt} \\
			\includegraphics[width=1.35cm]{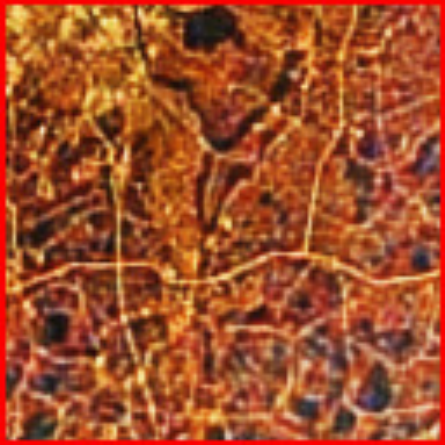}
			\includegraphics[width=1.35cm]{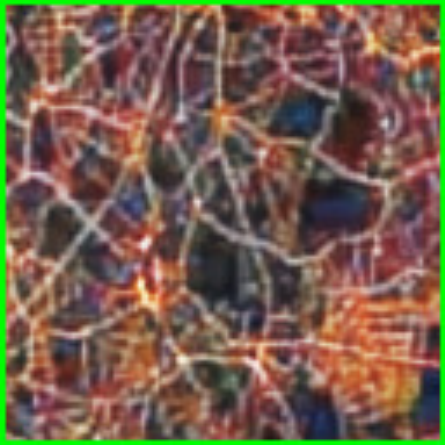}
		\end{minipage}
	}
	\caption{Visual comparison with other state-of-the-art methods on LIME dataset. There is no GroundTruth in LIME dataset.}
	\label{LIME}
\end{figure*}

\begin{figure*}
	\flushleft
	
	
	\subfigure[Input]{
		\begin{minipage}[b]{0.155\textwidth}
			\includegraphics[width=2.8cm]{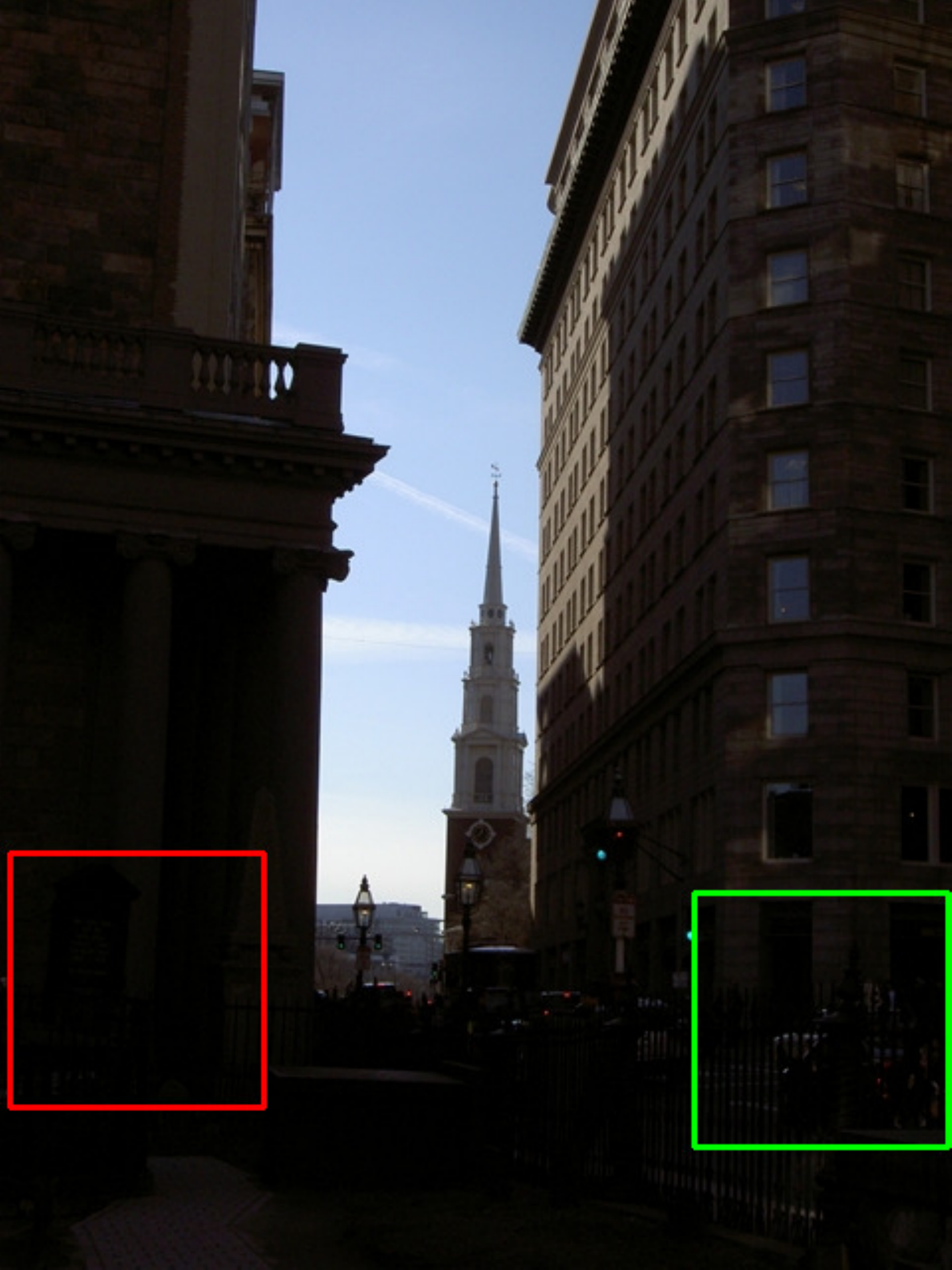}\vspace{1pt} \\
			\includegraphics[width=1.35cm]{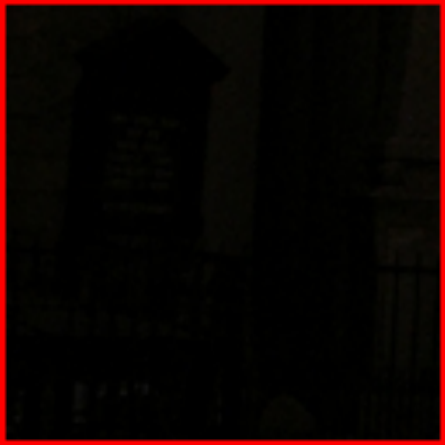}
			\includegraphics[width=1.35cm]{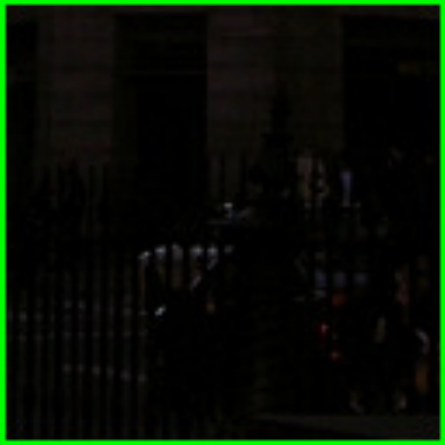}\vspace{5pt}
			\includegraphics[width=2.8cm]{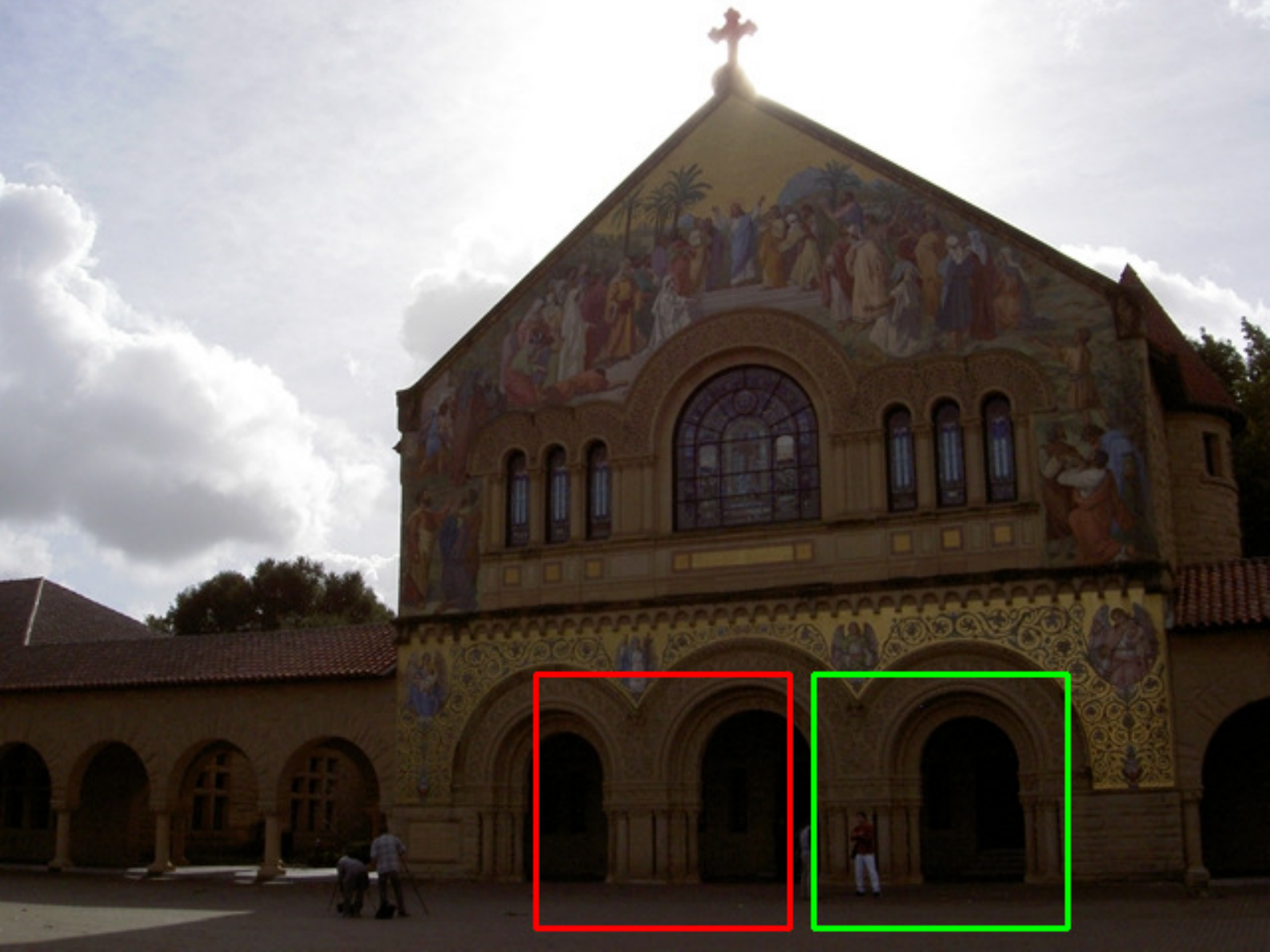}\vspace{1.5pt} \\
			\includegraphics[width=1.35cm]{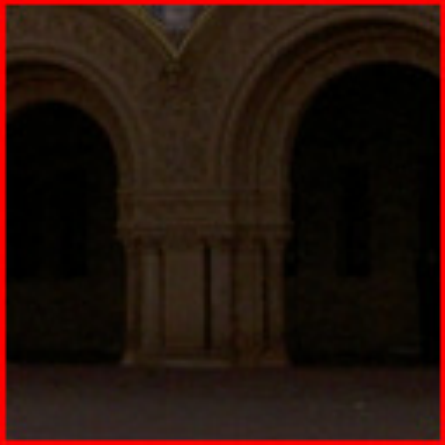}
			\includegraphics[width=1.35cm]{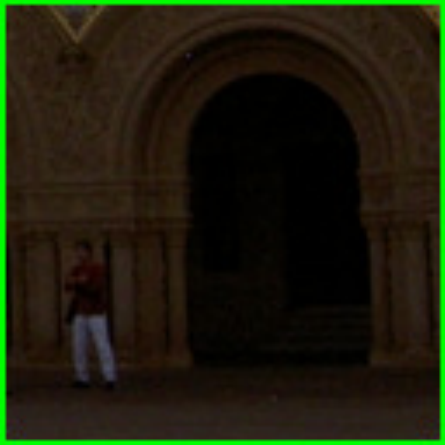}\vspace{5pt}
			\includegraphics[width=2.8cm]{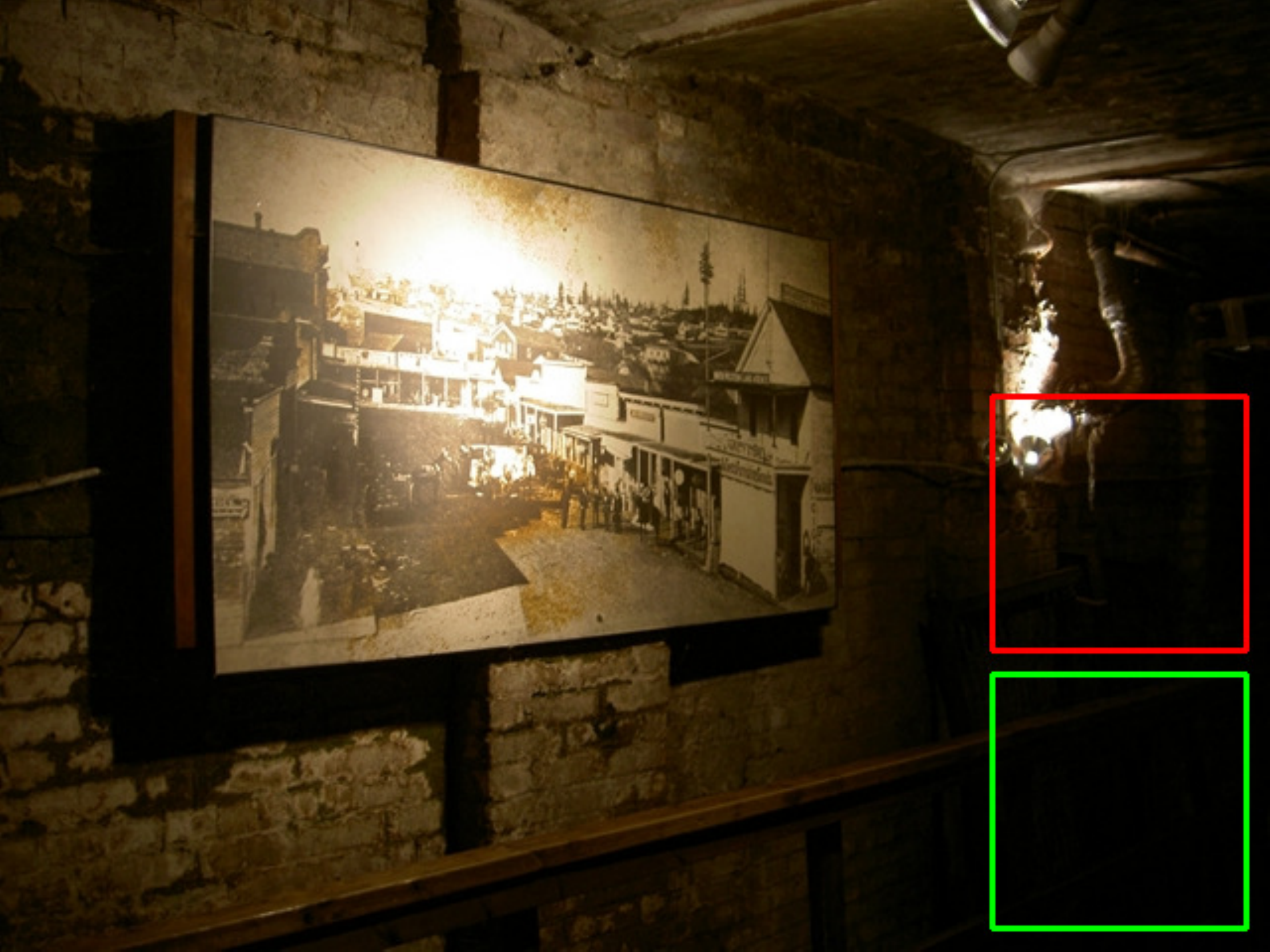}\vspace{1pt} \\
			\includegraphics[width=1.35cm]{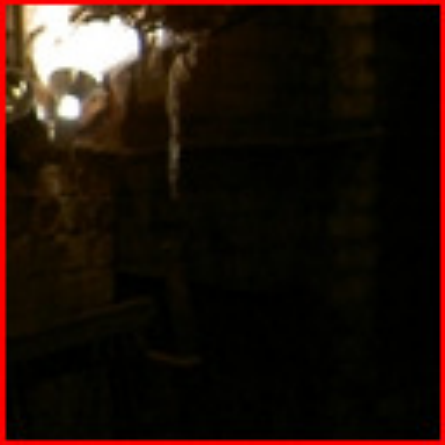}
			\includegraphics[width=1.35cm]{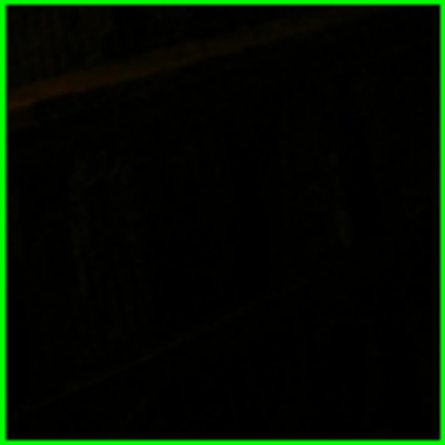}
		\end{minipage}
	}\hspace{-5pt}
	\subfigure[NPE\cite{wang2013naturalness}]{
		\begin{minipage}[b]{0.155\textwidth}
			\includegraphics[width=2.8cm]{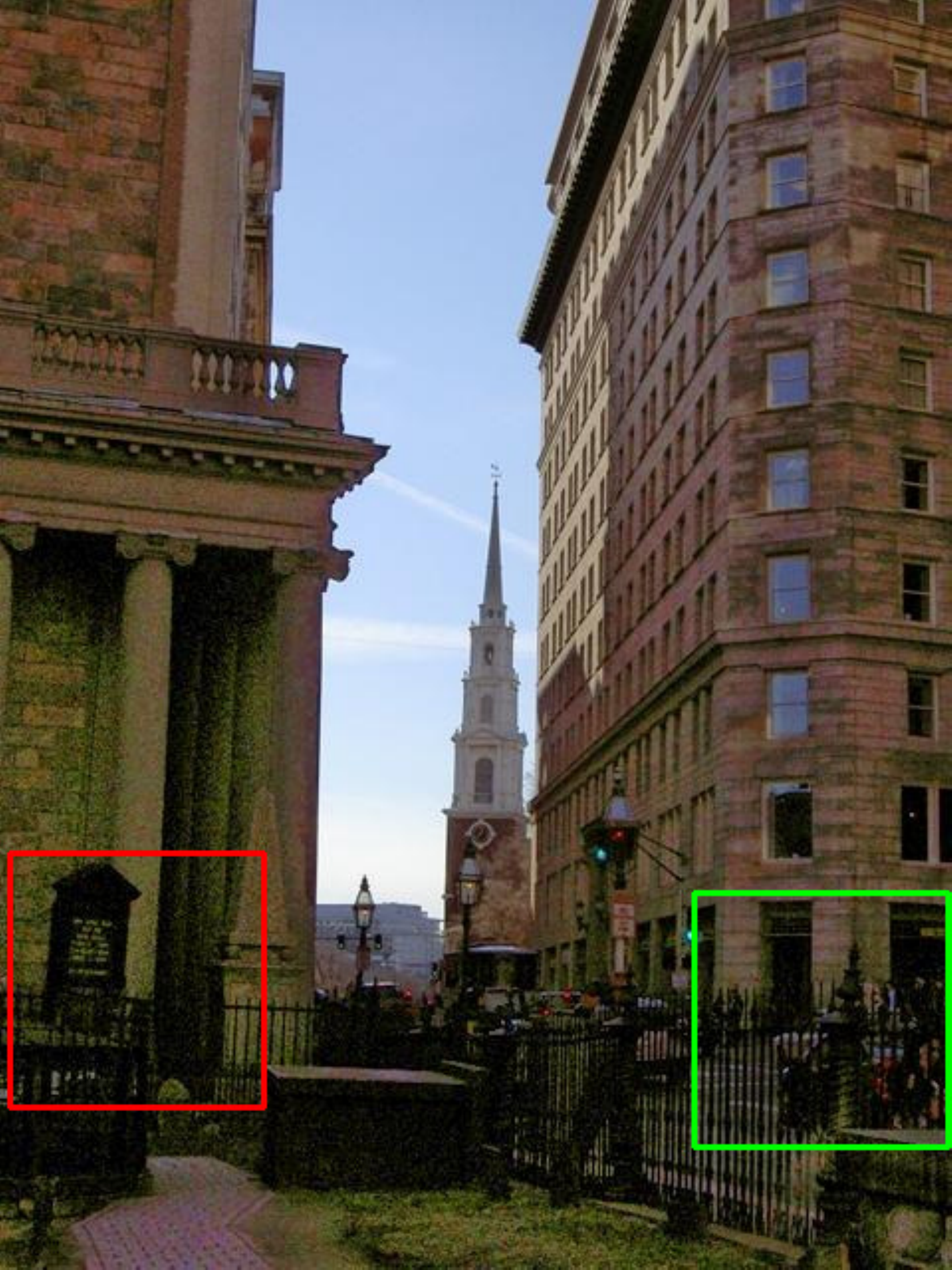}\vspace{1pt} \\
			\includegraphics[width=1.35cm]{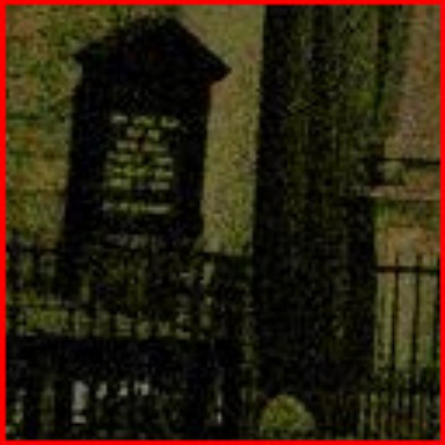}
			\includegraphics[width=1.35cm]{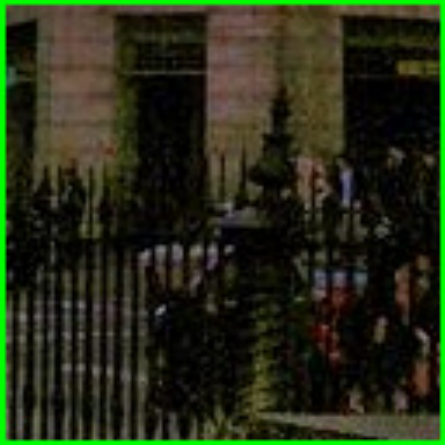}\vspace{5pt}
			\includegraphics[width=2.8cm]{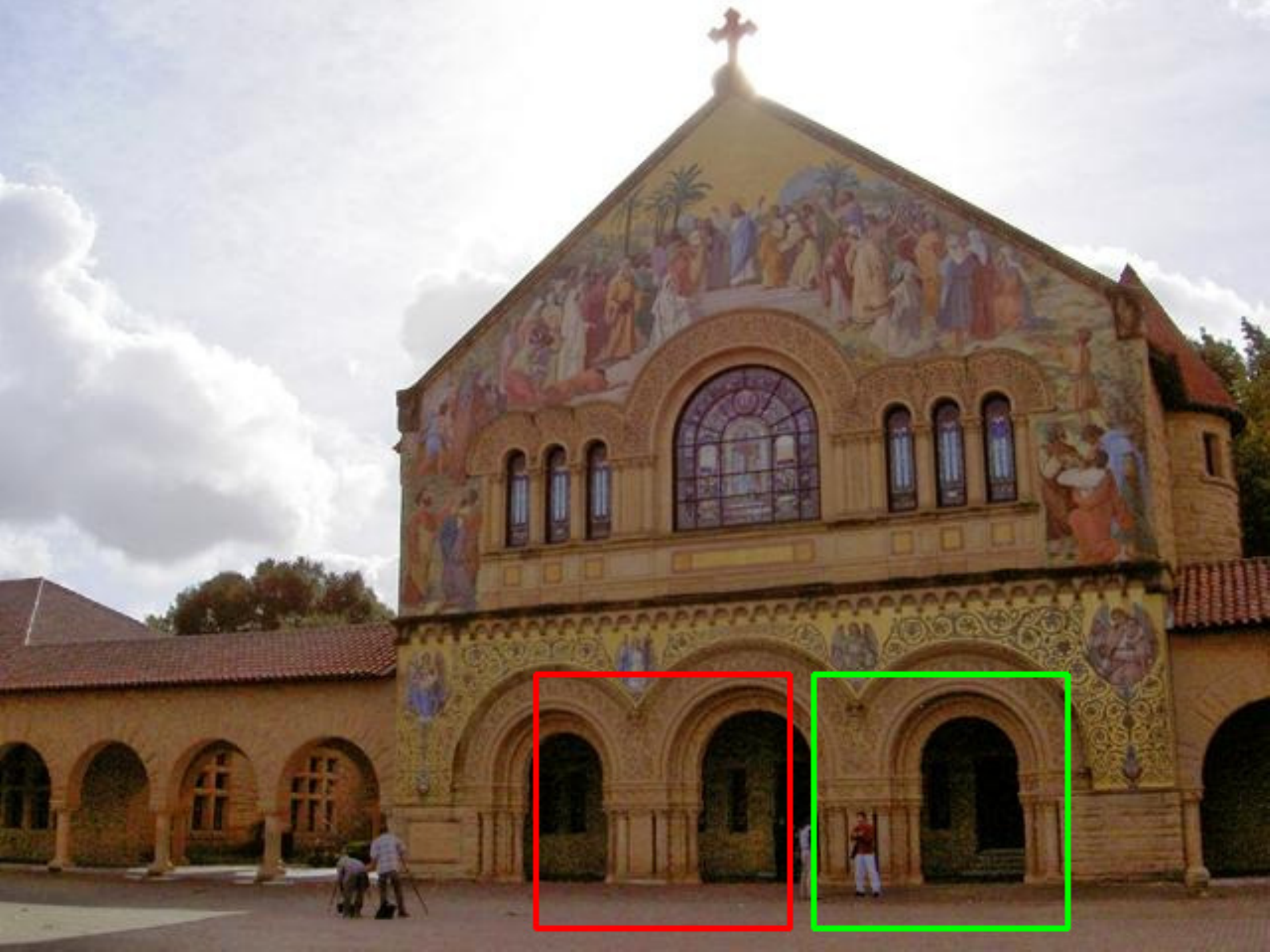}\vspace{1.5pt} \\
			\includegraphics[width=1.35cm]{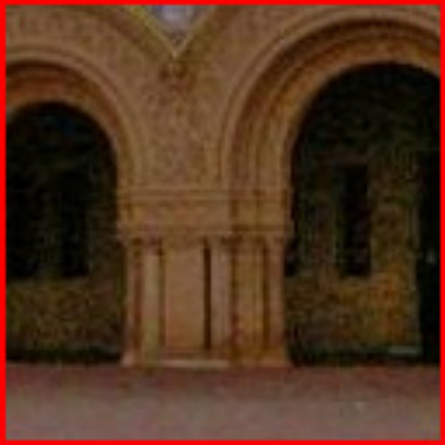}
			\includegraphics[width=1.35cm]{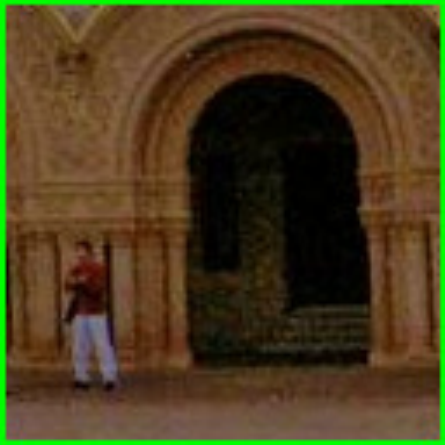}\vspace{5pt}
			\includegraphics[width=2.8cm]{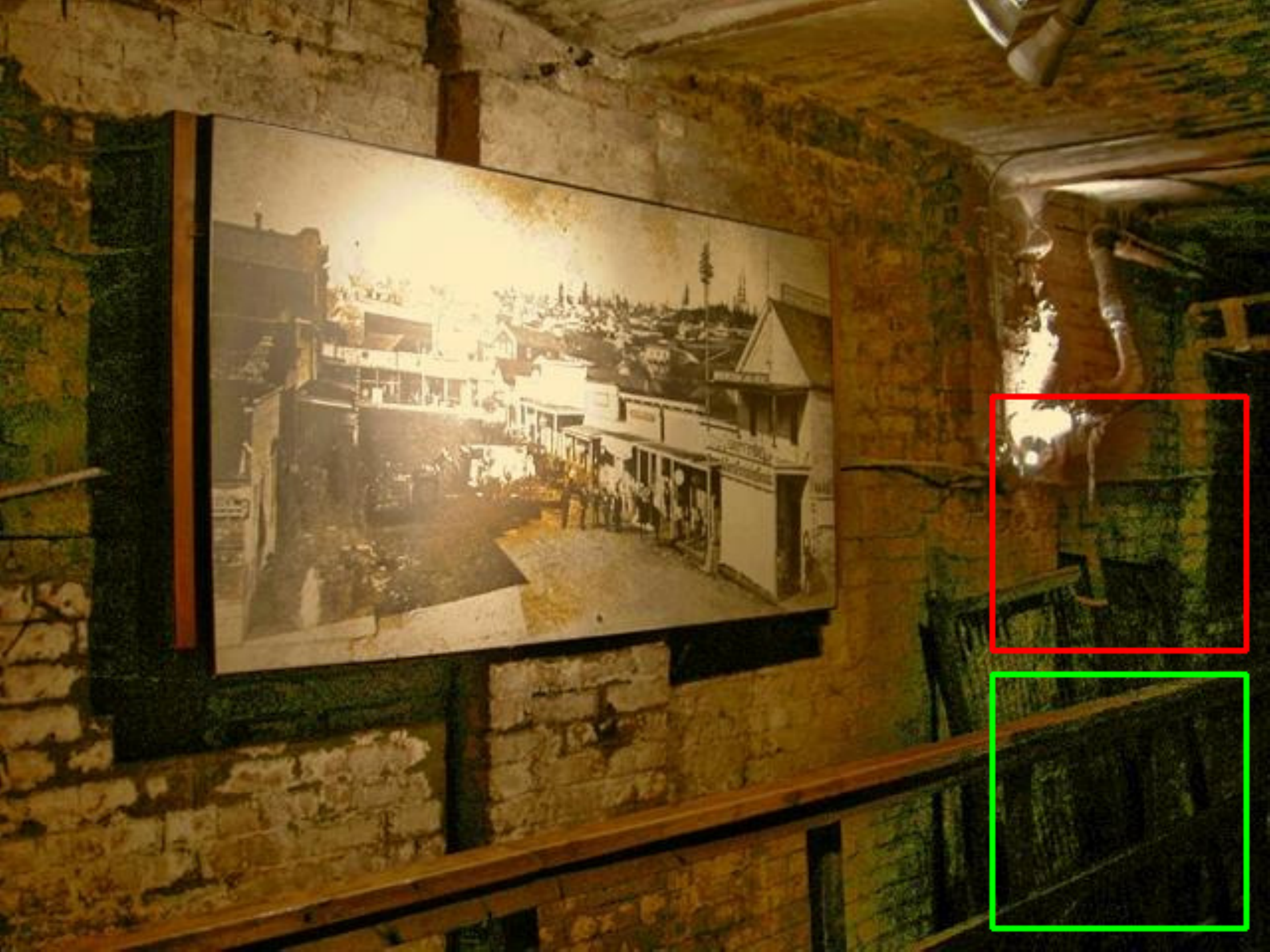}\vspace{1pt} \\
			\includegraphics[width=1.35cm]{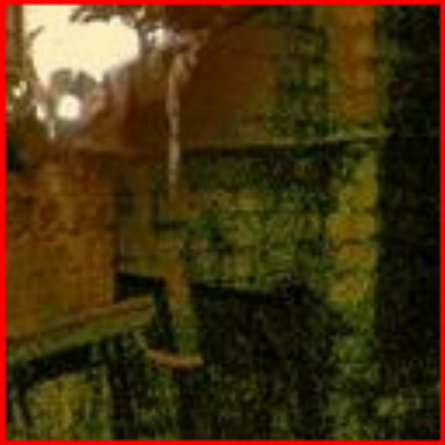}
			\includegraphics[width=1.35cm]{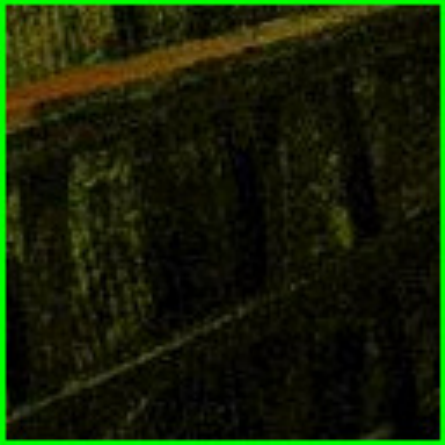}
		\end{minipage}
	}\hspace{-5pt}
	\subfigure[GLAD\cite{wang2018gladnet}]{
		\begin{minipage}[b]{0.155\textwidth}
			\includegraphics[width=2.8cm]{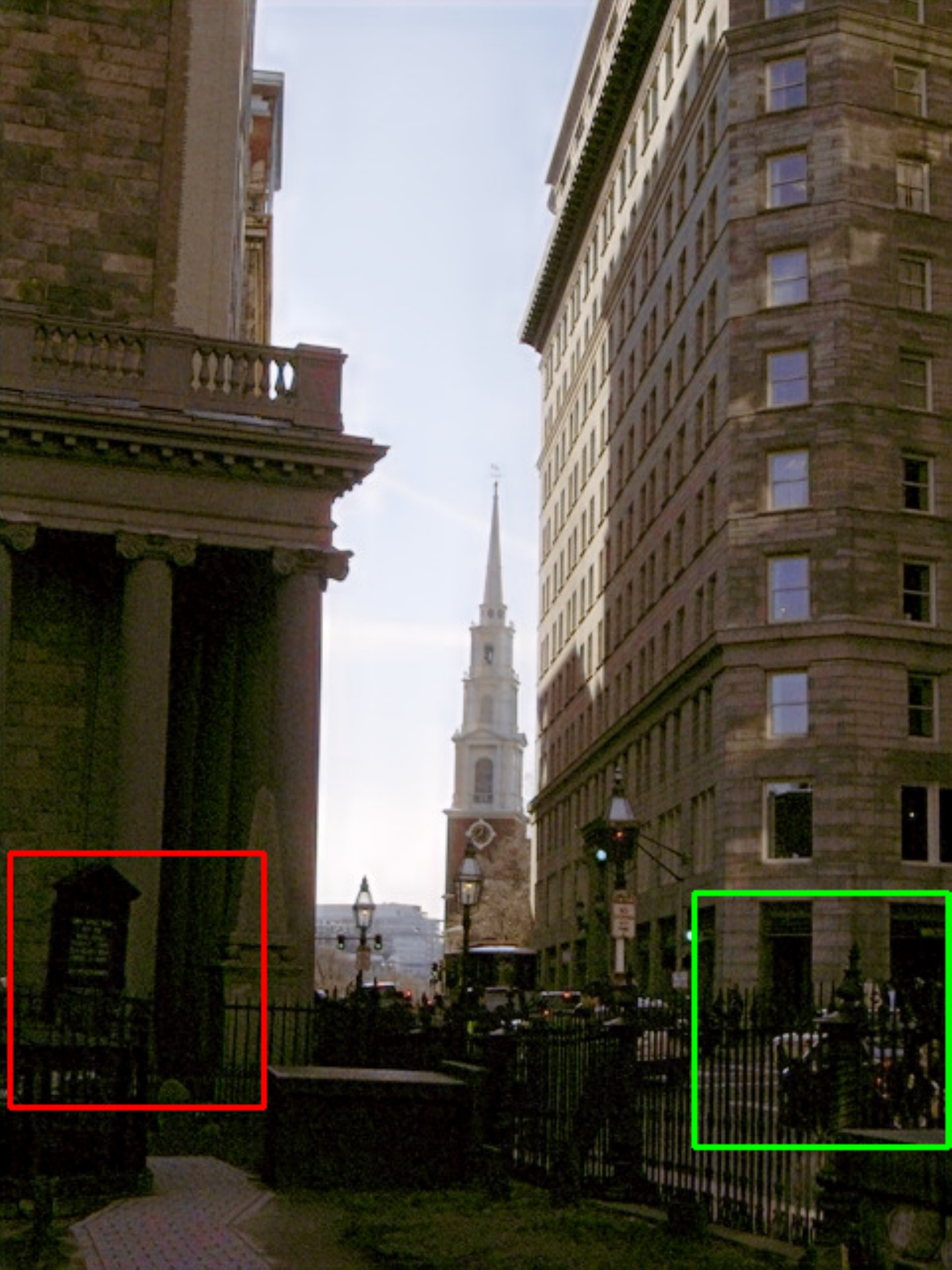}\vspace{1pt} \\
			\includegraphics[width=1.35cm]{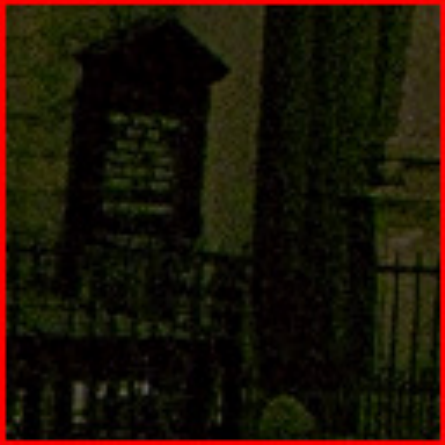}
			\includegraphics[width=1.35cm]{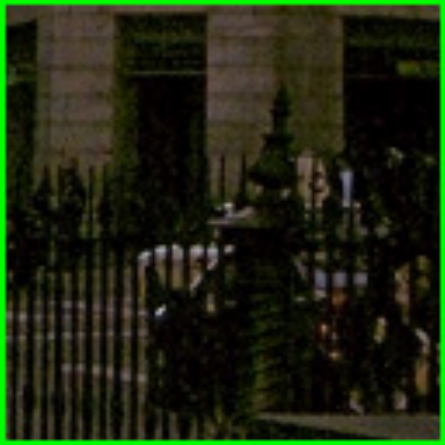}\vspace{5pt}
			\includegraphics[width=2.8cm]{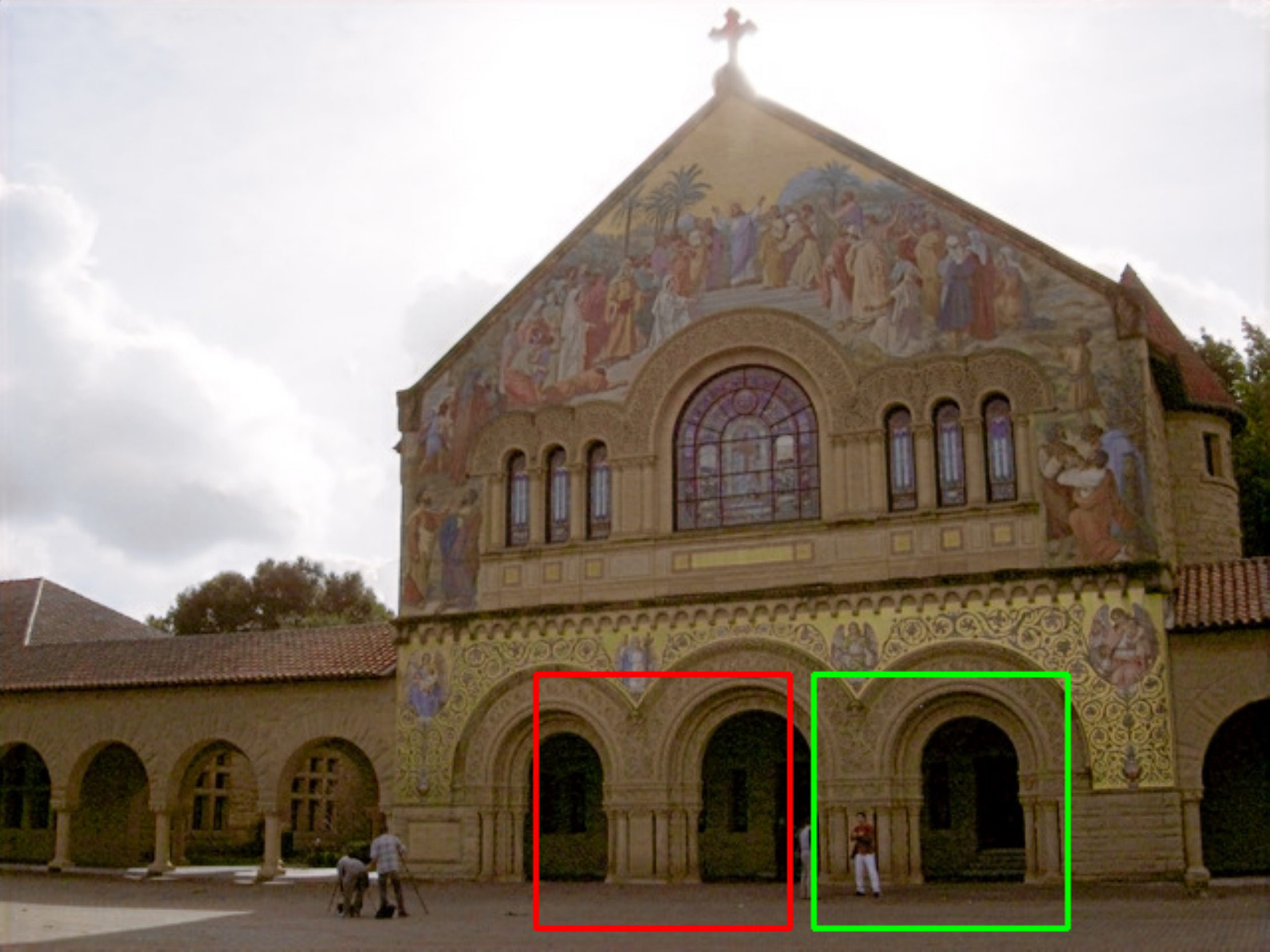}\vspace{1.5pt} \\
			\includegraphics[width=1.35cm]{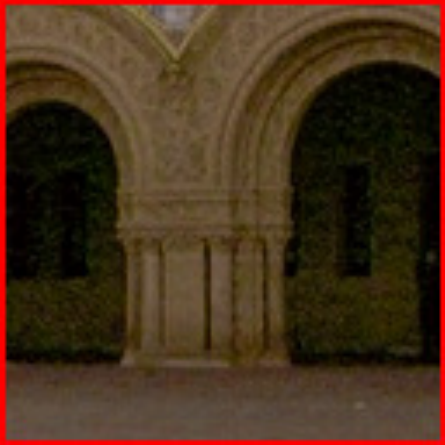}
			\includegraphics[width=1.35cm]{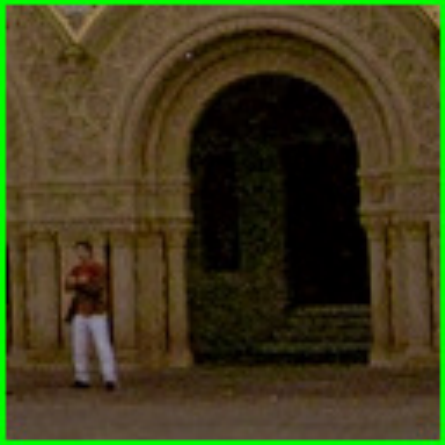}\vspace{5pt}
			\includegraphics[width=2.8cm]{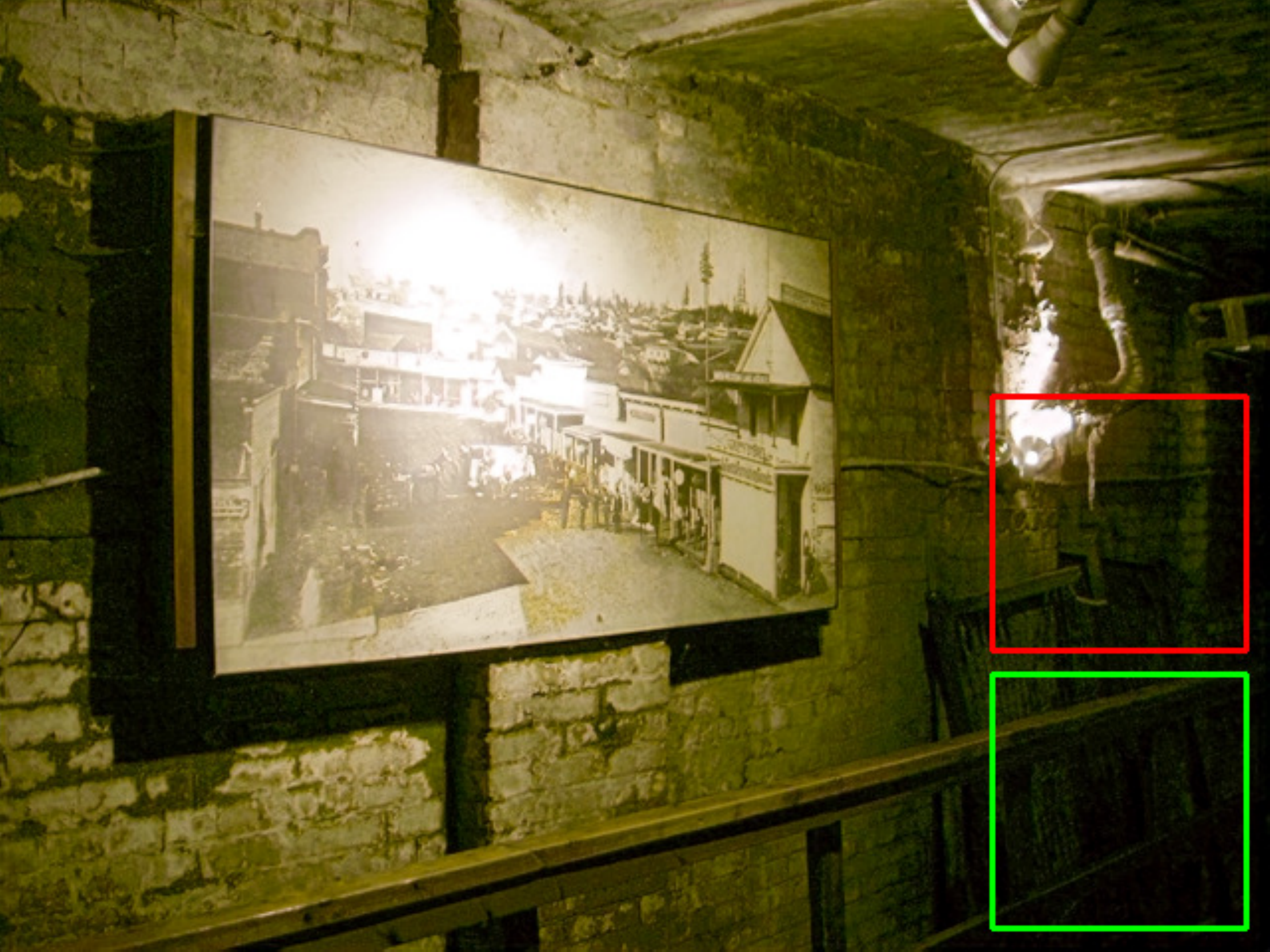}\vspace{1pt} \\
			\includegraphics[width=1.35cm]{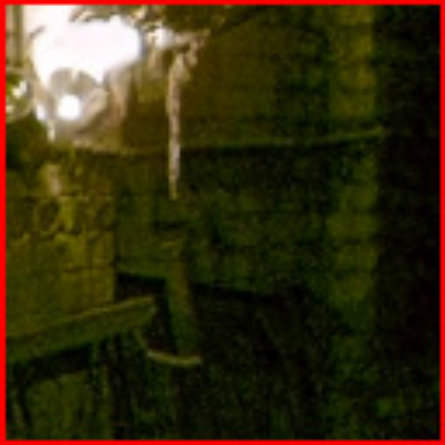}
			\includegraphics[width=1.35cm]{pic/test/DICM03_crop/low_1-eps-converted-to.pdf}
		\end{minipage}
	}\hspace{-5pt}
	\subfigure[RetinexNet\cite{wei2018deep}]{
		\begin{minipage}[b]{0.155\textwidth}
			\includegraphics[width=2.8cm]{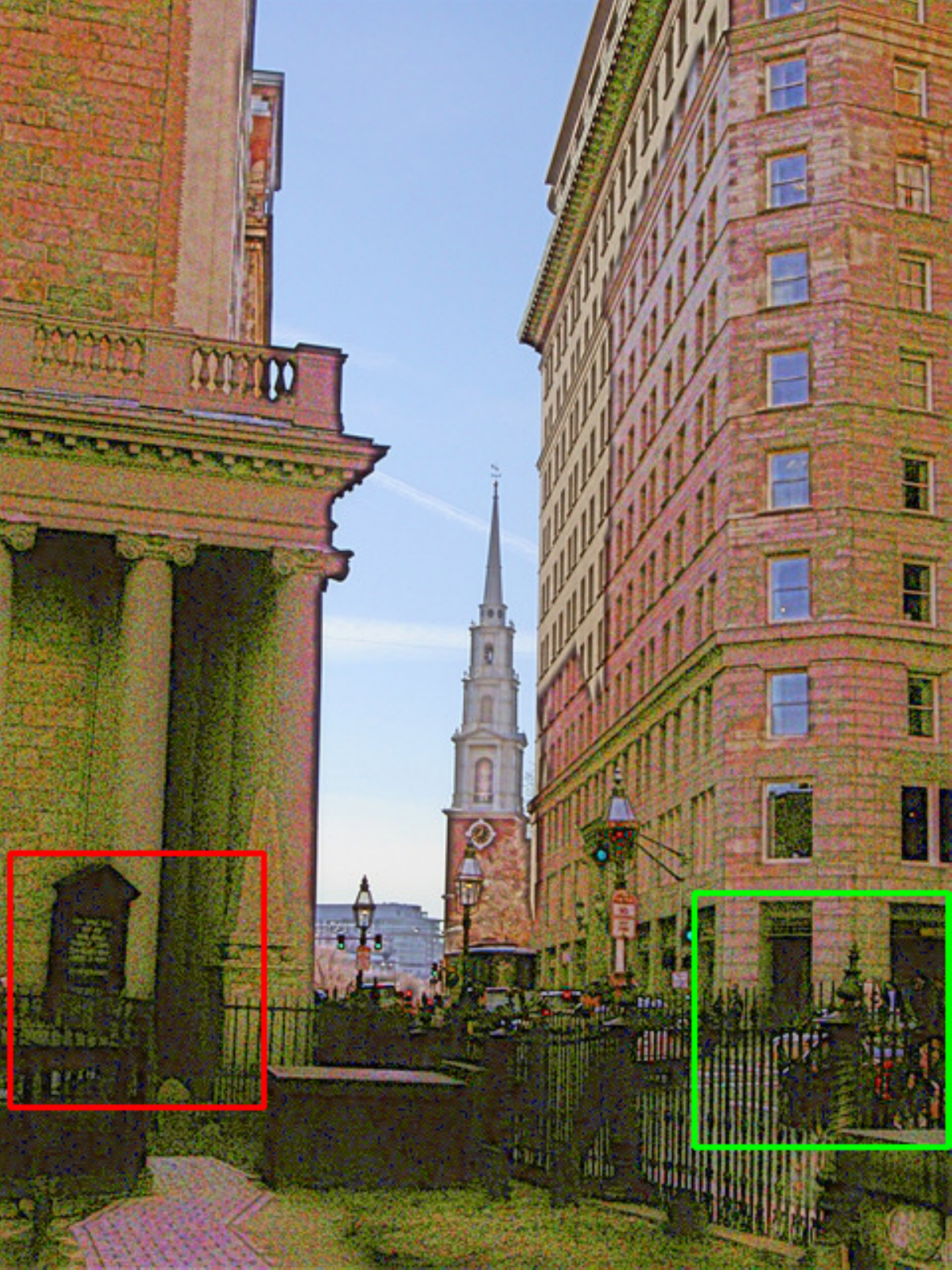}\vspace{1pt} \\
			\includegraphics[width=1.35cm]{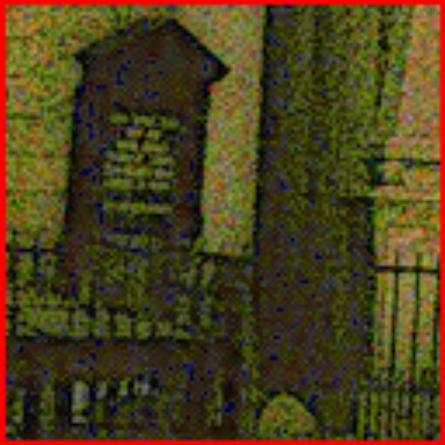}
			\includegraphics[width=1.35cm]{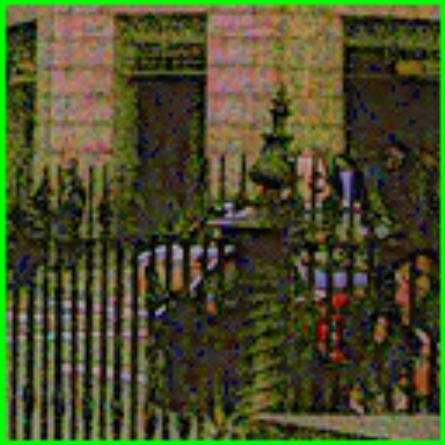}\vspace{5pt}
			\includegraphics[width=2.8cm]{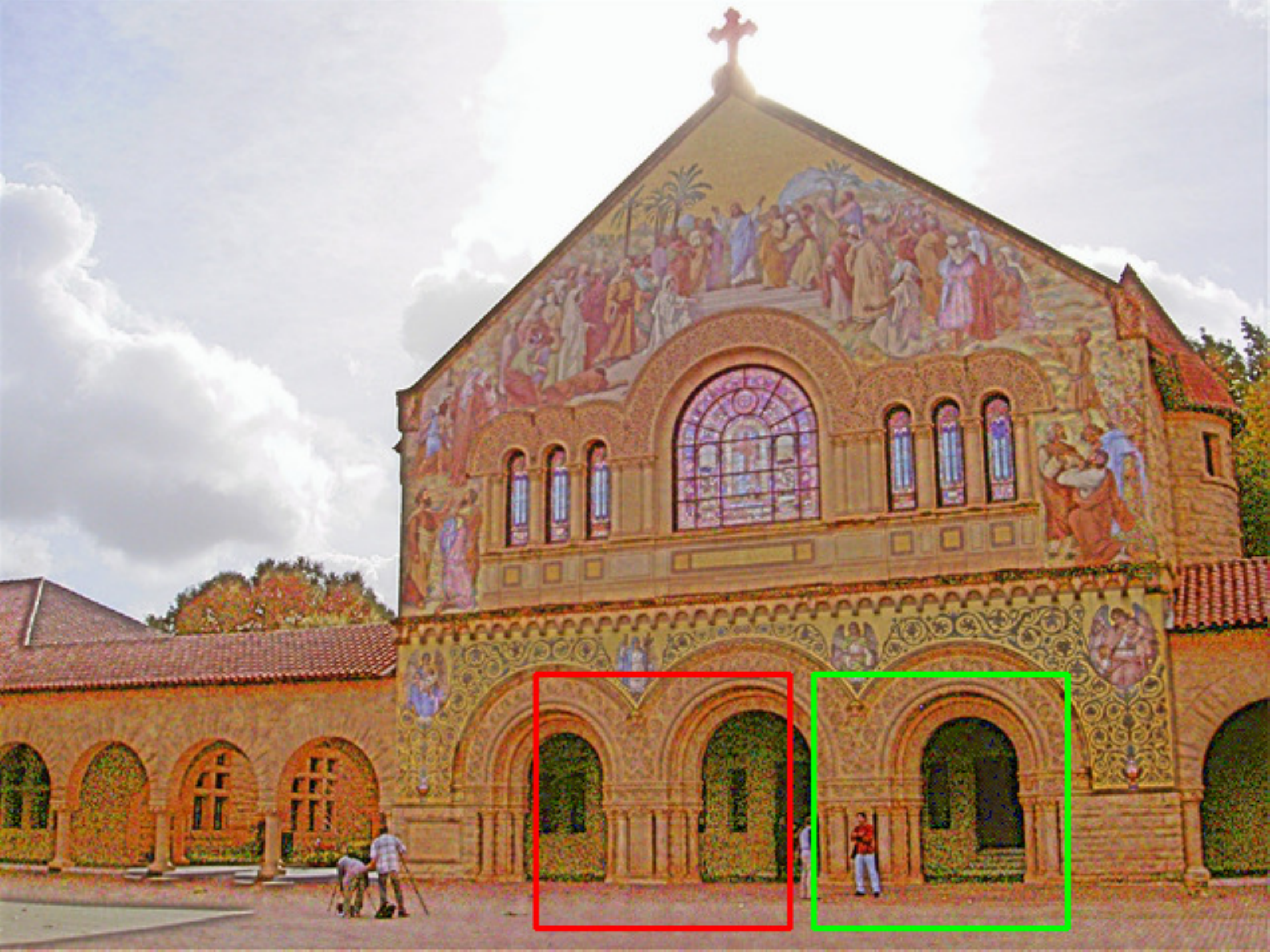}\vspace{1.5pt} \\
			\includegraphics[width=1.35cm]{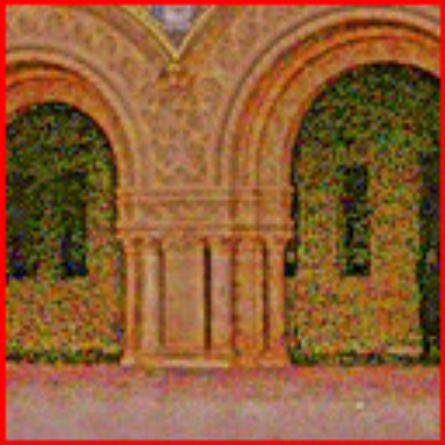}
			\includegraphics[width=1.35cm]{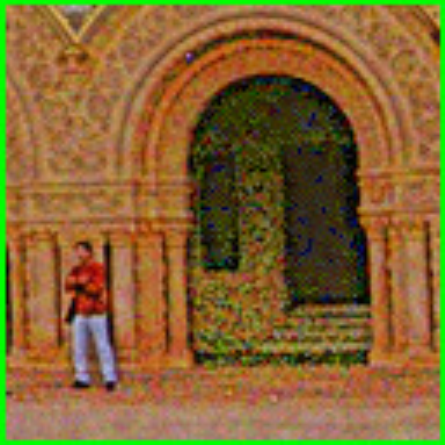}\vspace{5pt}
			\includegraphics[width=2.8cm]{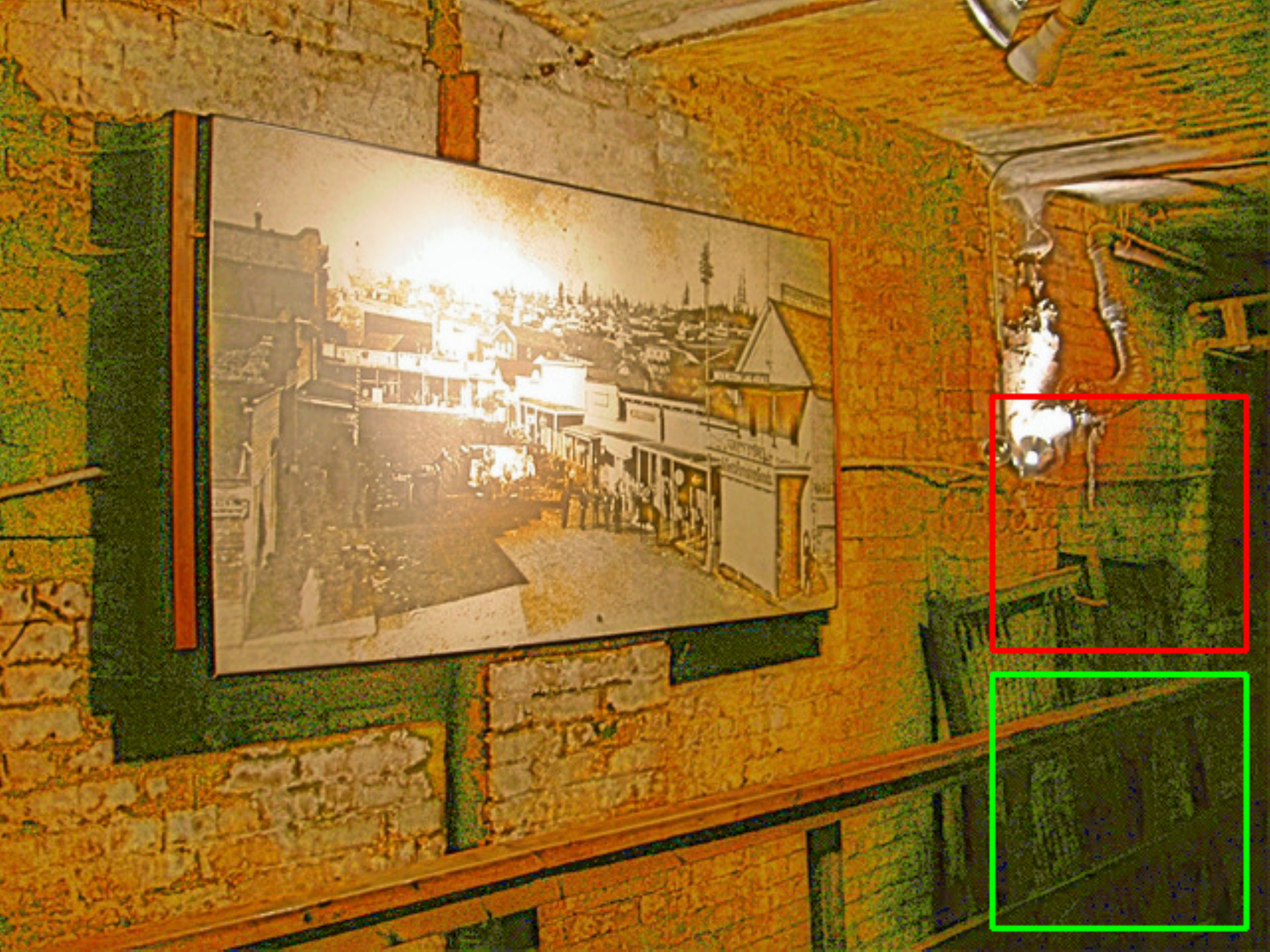}\vspace{1pt} \\
			\includegraphics[width=1.35cm]{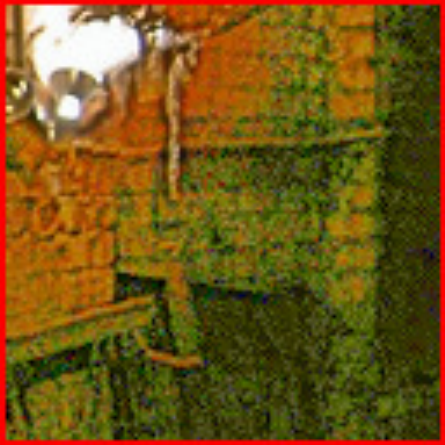}
			\includegraphics[width=1.35cm]{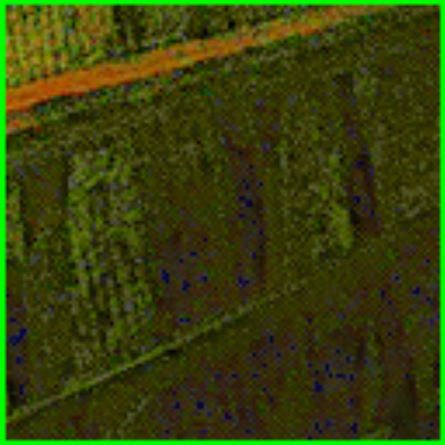}
		\end{minipage}
	}\hspace{-5pt}
	\subfigure[MBLLEN\cite{lv2018mbllen}]{
		\begin{minipage}[b]{0.155\textwidth}
			\includegraphics[width=2.8cm]{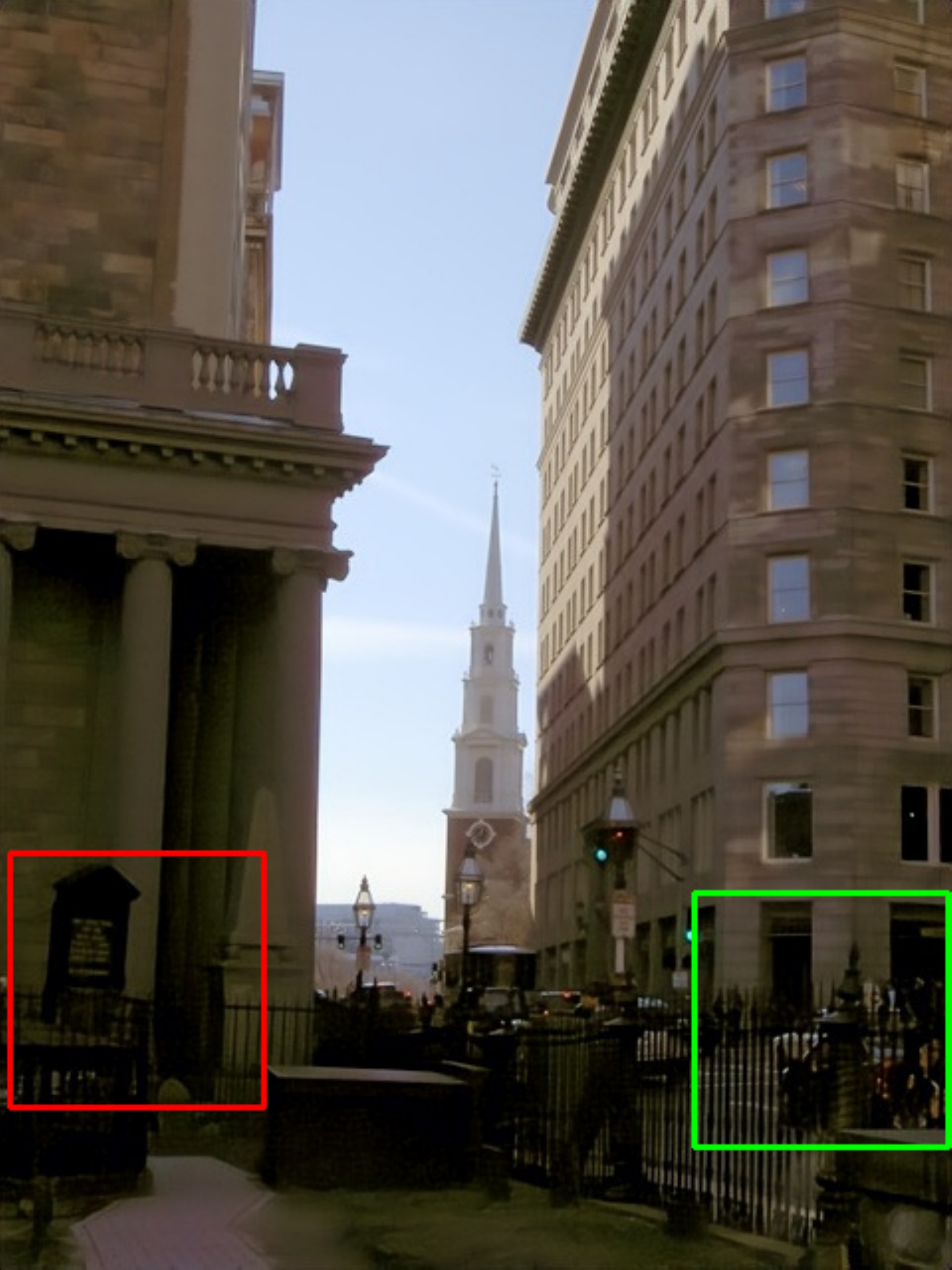}\vspace{1pt} \\
			\includegraphics[width=1.35cm]{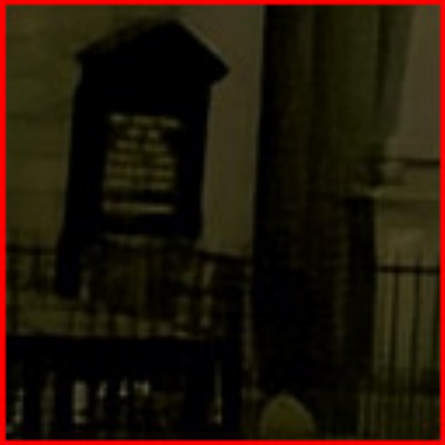}
			\includegraphics[width=1.35cm]{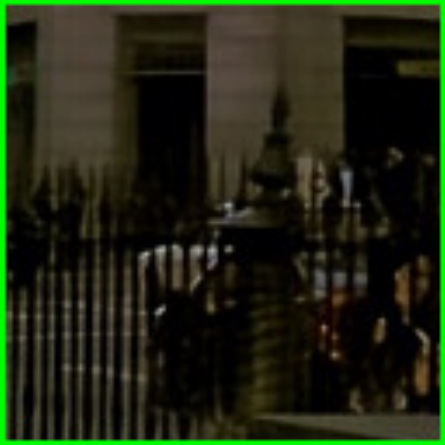}\vspace{5pt}
			\includegraphics[width=2.8cm]{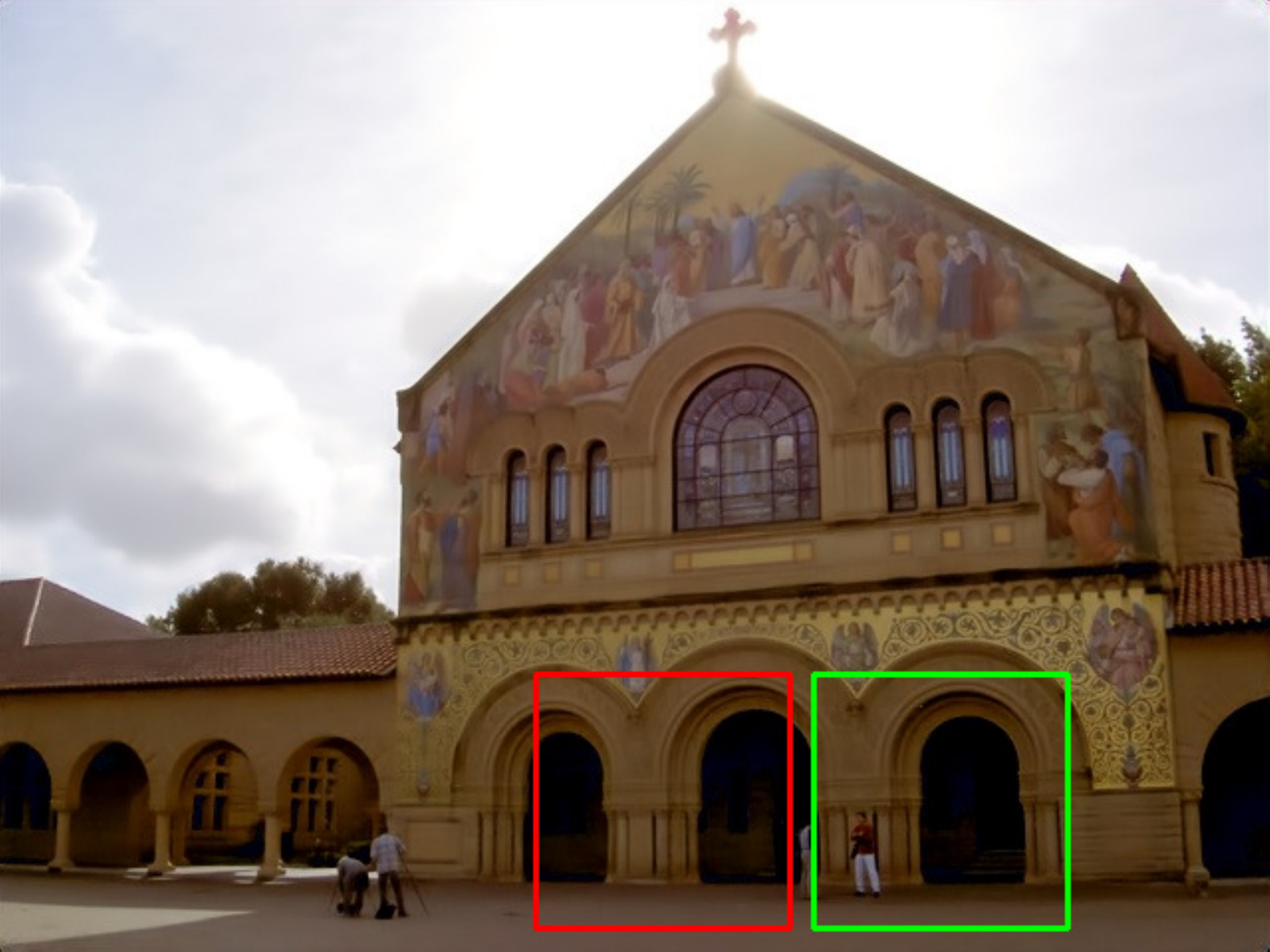}\vspace{1.5pt} \\
			\includegraphics[width=1.35cm]{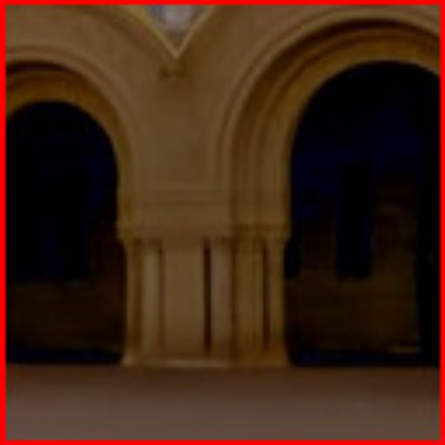}
			\includegraphics[width=1.35cm]{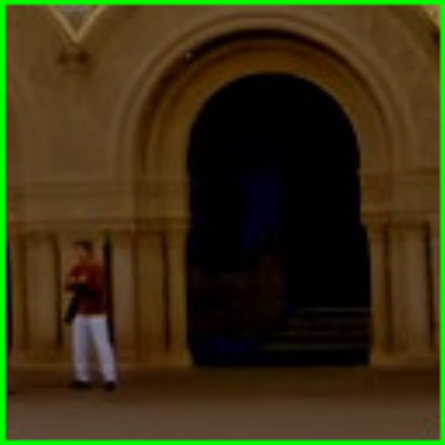}\vspace{5pt}
			\includegraphics[width=2.8cm]{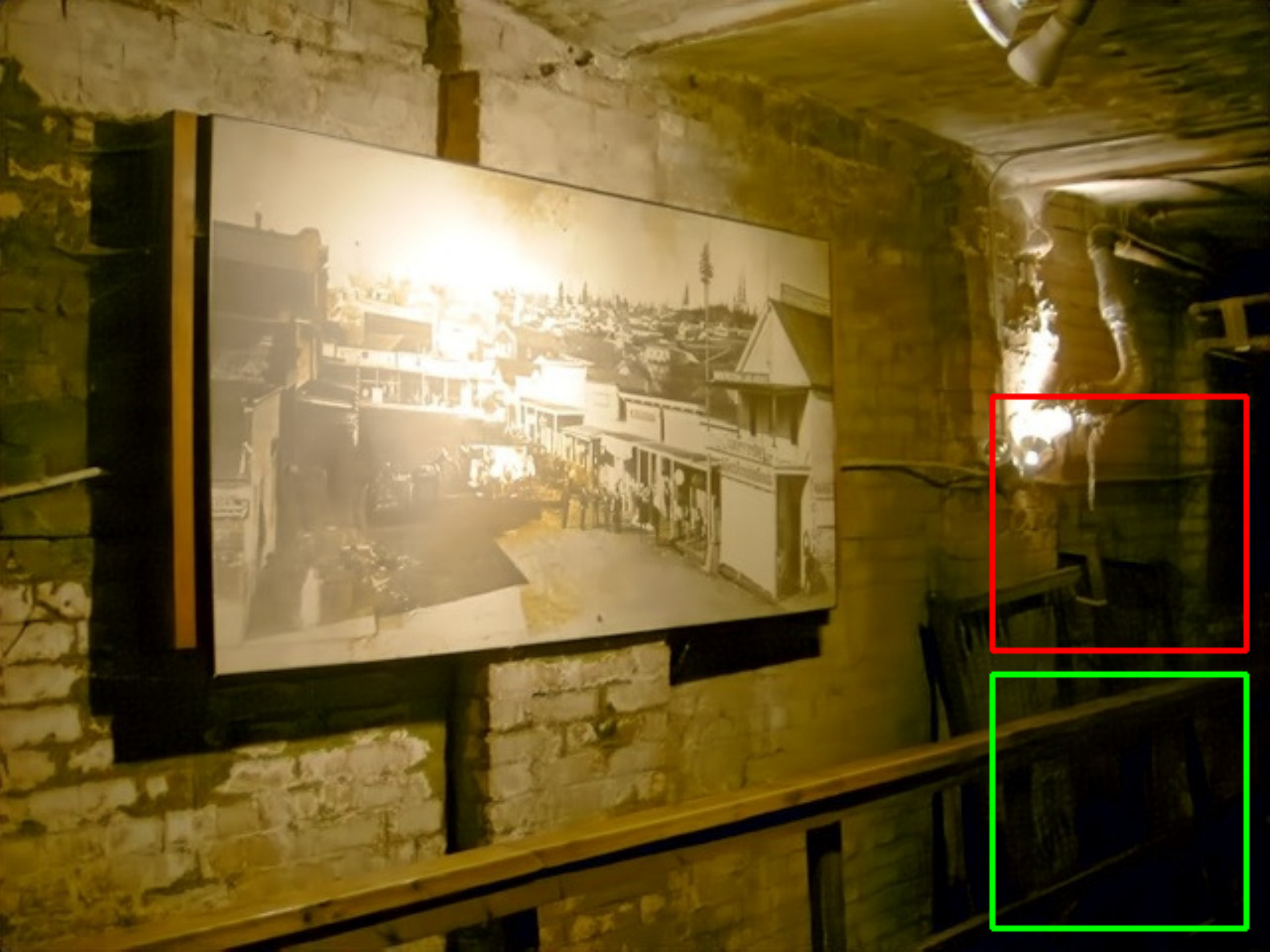}\vspace{1pt} \\
			\includegraphics[width=1.35cm]{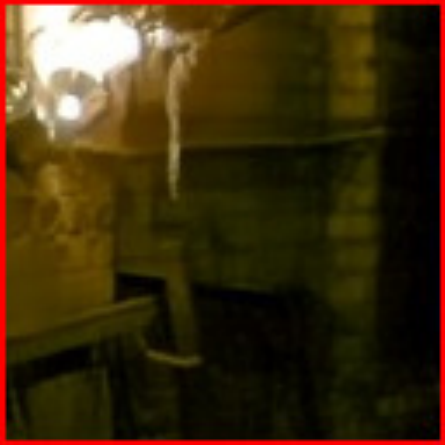}
			\includegraphics[width=1.35cm]{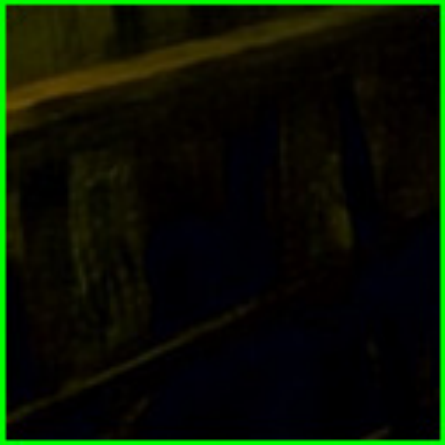}
		\end{minipage}
	}\hspace{-5pt}
	\subfigure[Ours]{
		\begin{minipage}[b]{0.155\textwidth}
			\includegraphics[width=2.8cm]{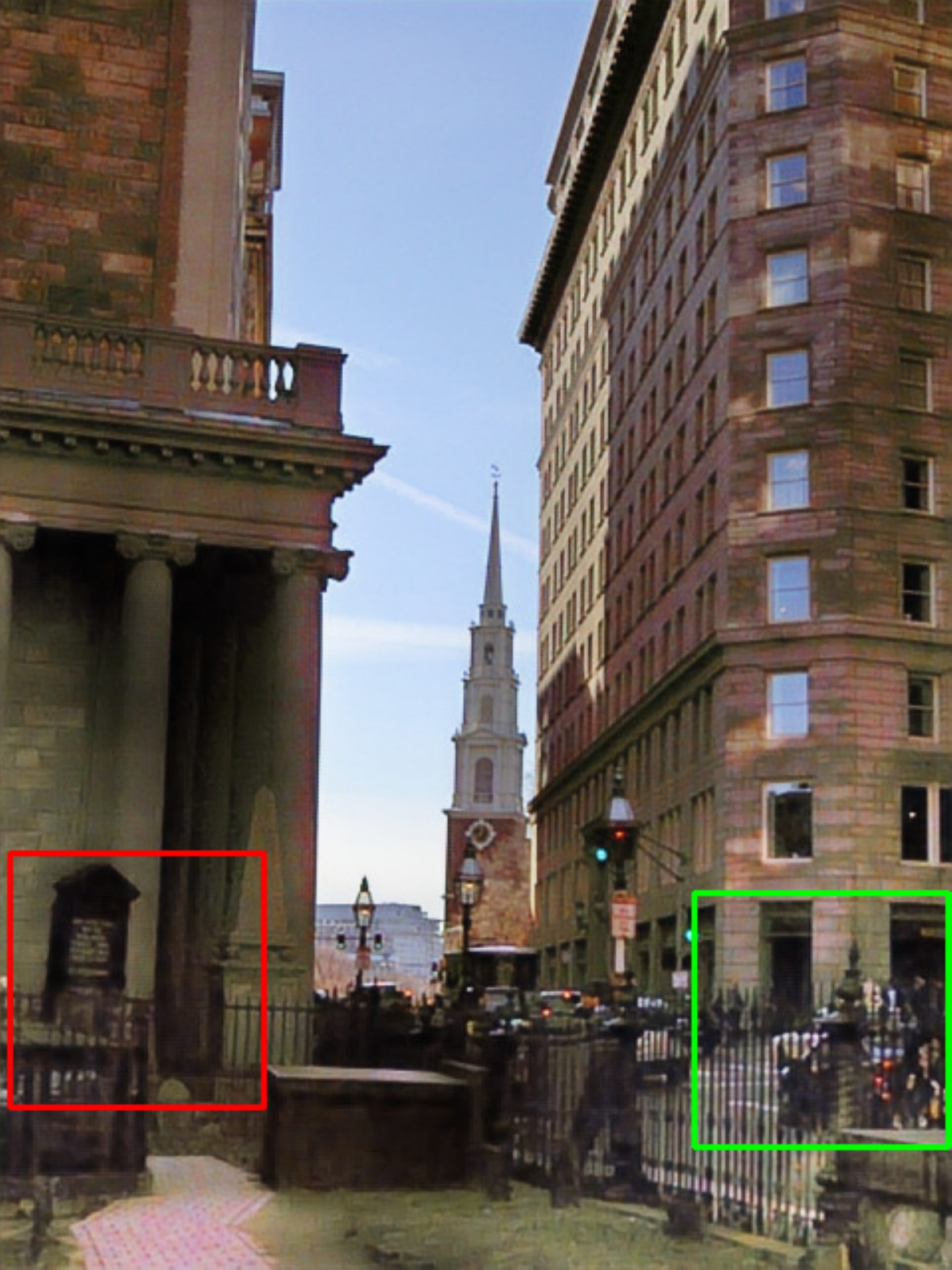}\vspace{1pt} \\
			\includegraphics[width=1.35cm]{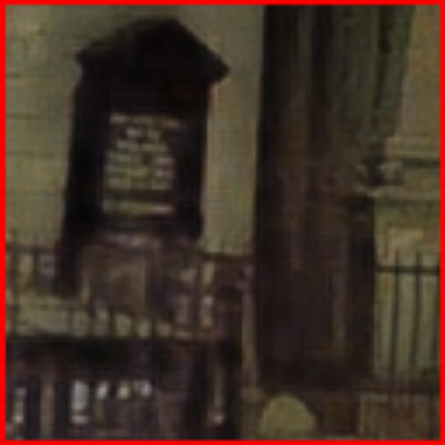}
			\includegraphics[width=1.35cm]{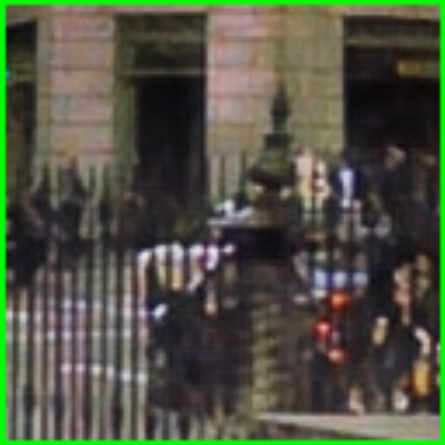}\vspace{5pt}
			\includegraphics[width=2.8cm]{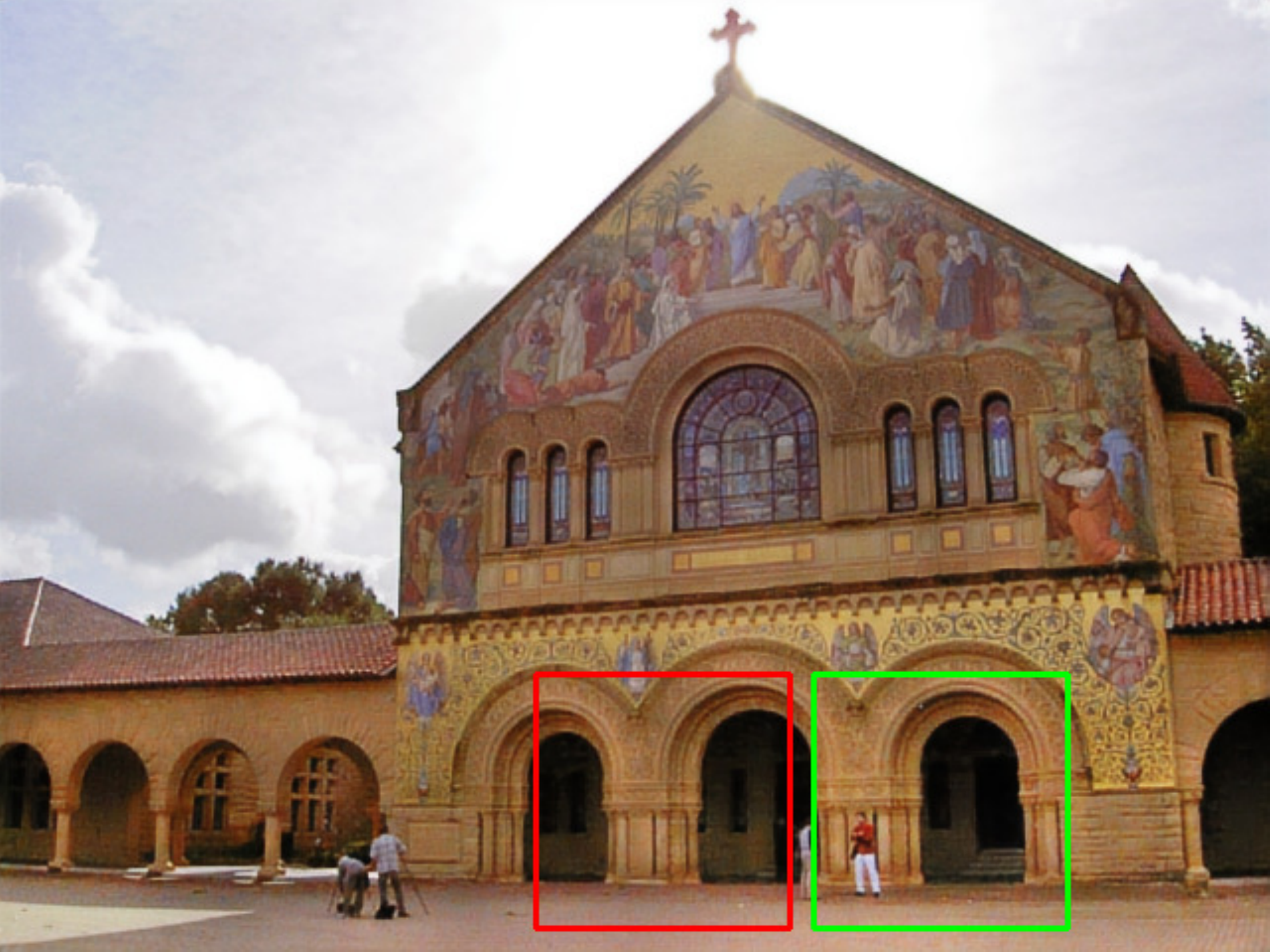}\vspace{1.5pt} \\
			\includegraphics[width=1.35cm]{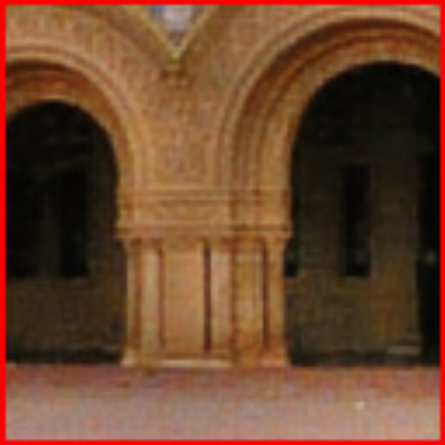}
			\includegraphics[width=1.35cm]{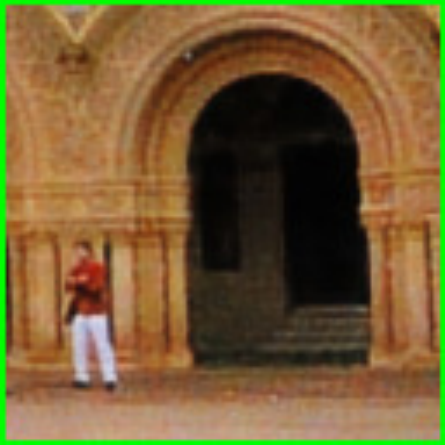}\vspace{5pt}
			\includegraphics[width=2.8cm]{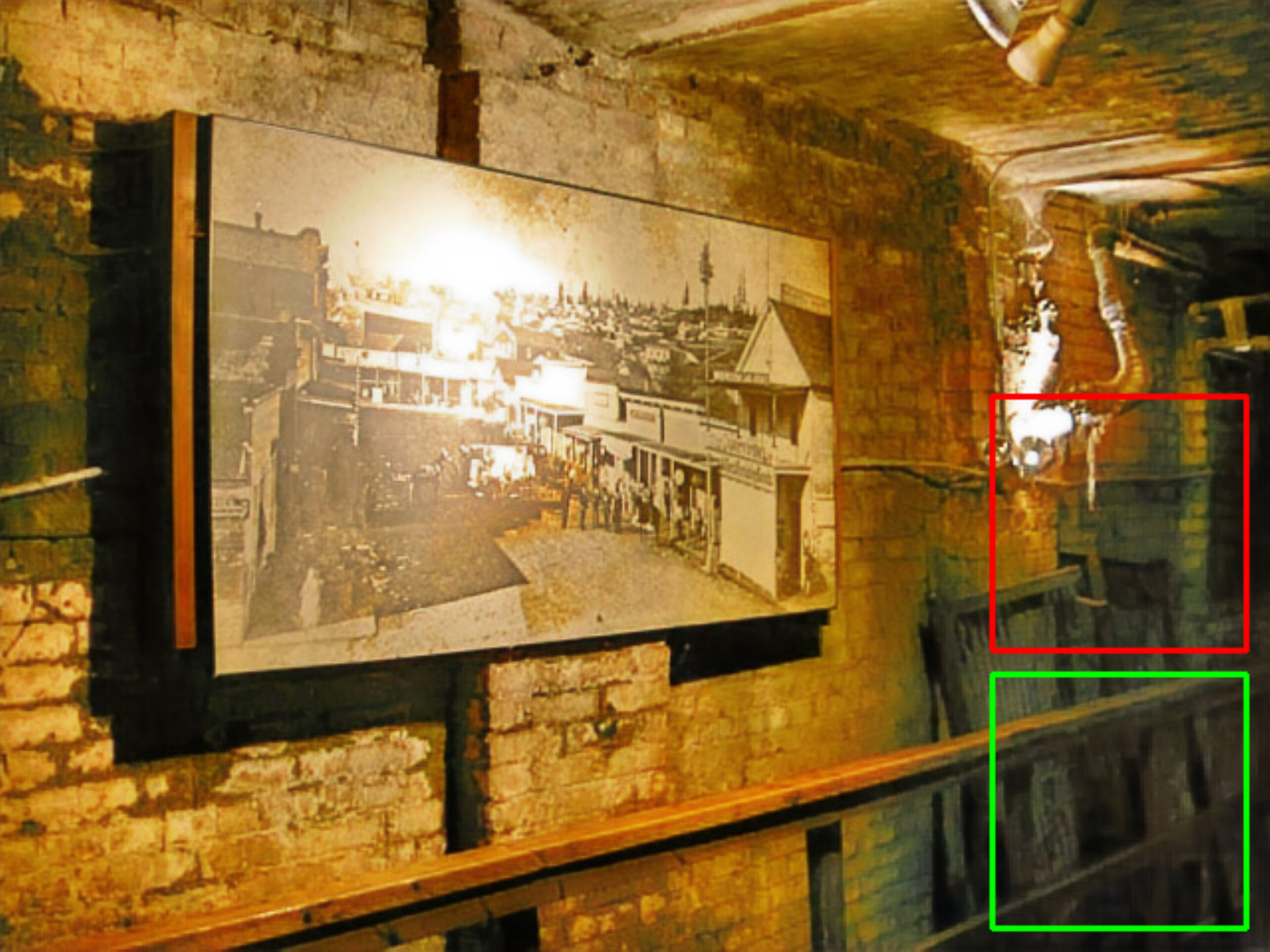}\vspace{1pt} \\
			\includegraphics[width=1.35cm]{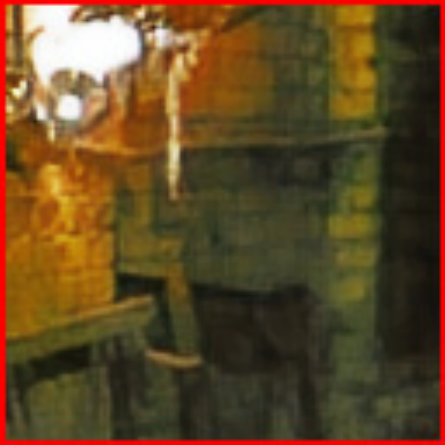}
			\includegraphics[width=1.35cm]{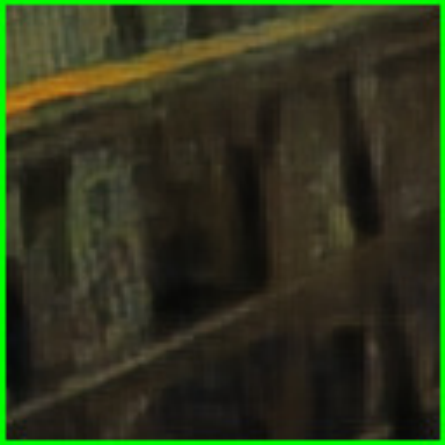}
		\end{minipage}
	}
	\caption{Visual comparison with other state-of-the-art methods on DICM dataset. There is no GroundTruth in DICM dataset.}
	\label{DICM}
\end{figure*}

\begin{figure*}
	\flushleft
	
	
	\subfigure[Input]{
		\begin{minipage}[b]{0.155\textwidth}
			\includegraphics[width=2.8cm]{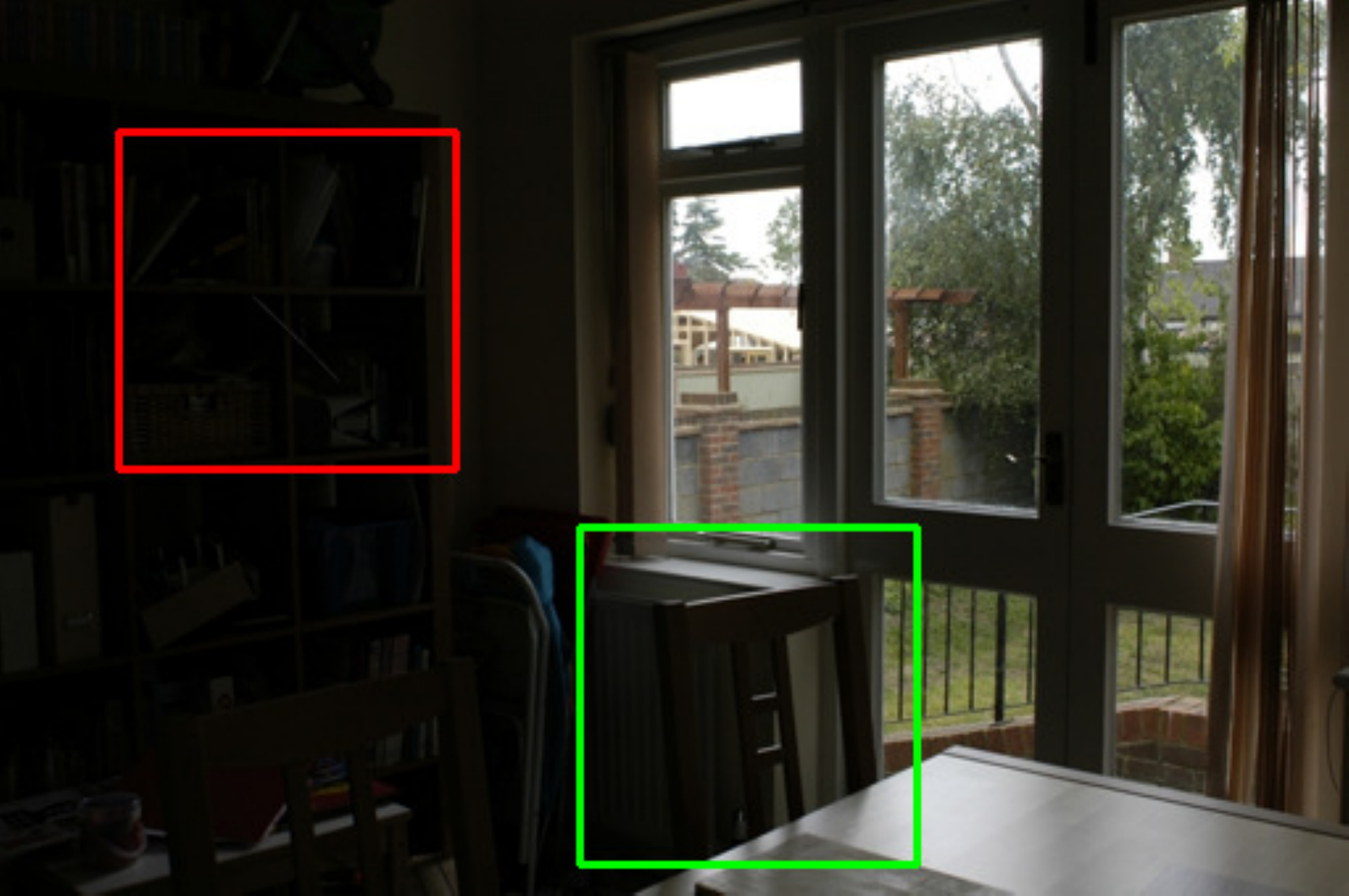}\vspace{1pt} \\
			\includegraphics[width=1.35cm]{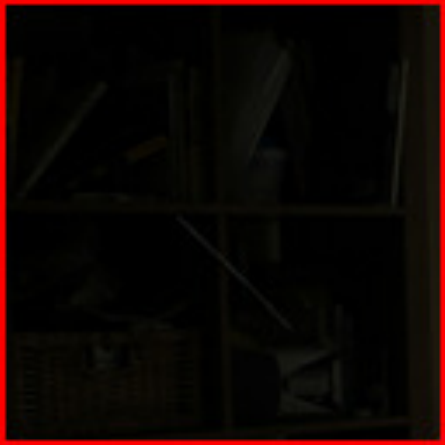}
			\includegraphics[width=1.35cm]{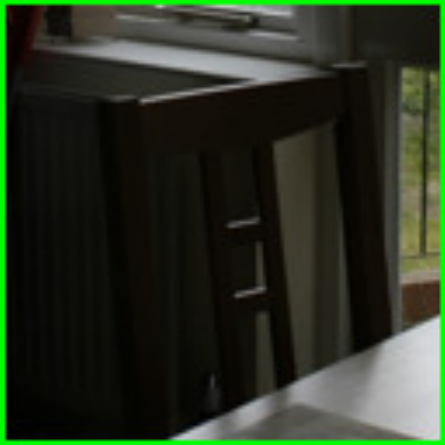}\vspace{5pt}
			\includegraphics[width=2.8cm]{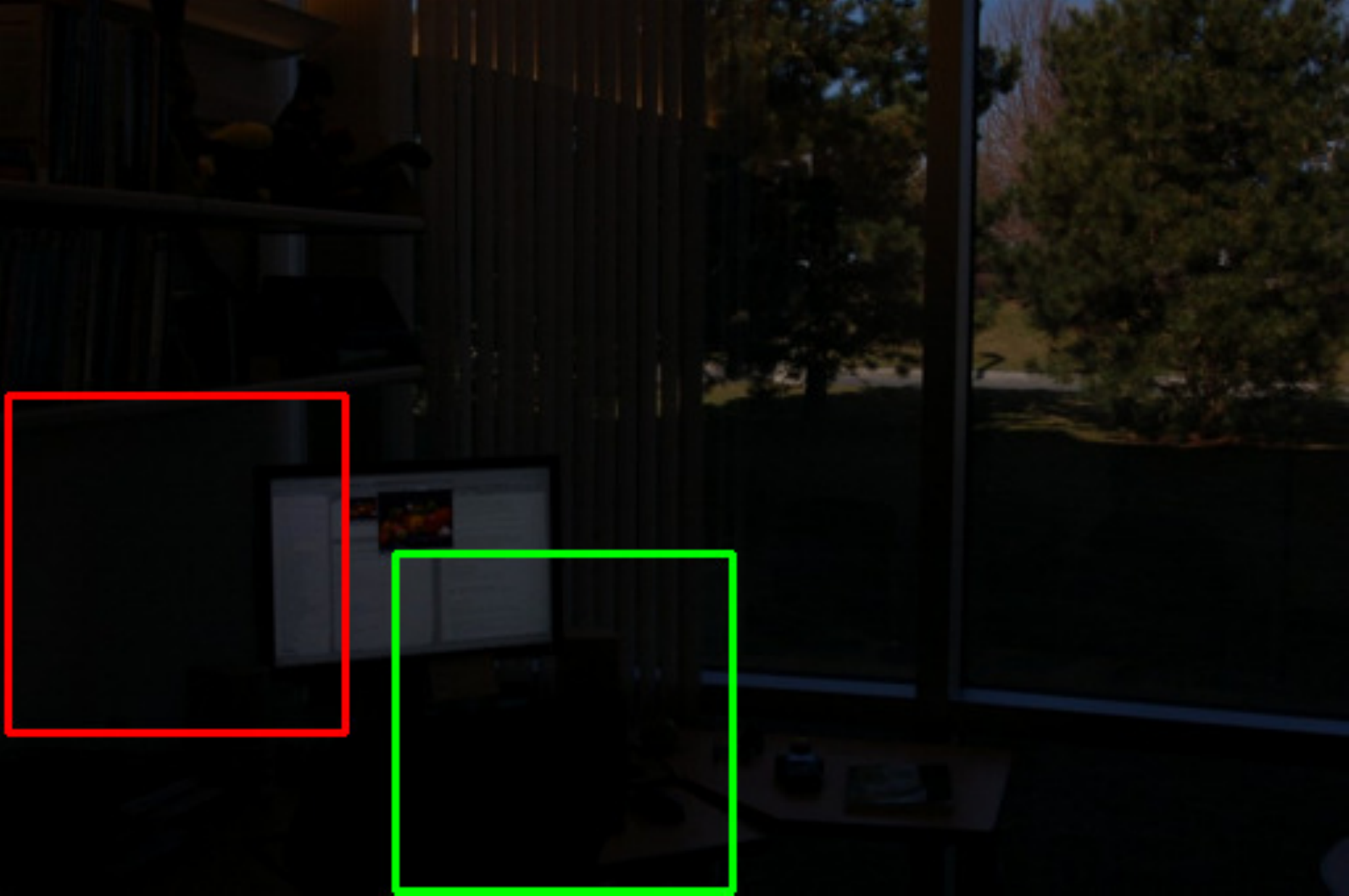}\vspace{1.5pt} \\
			\includegraphics[width=1.35cm]{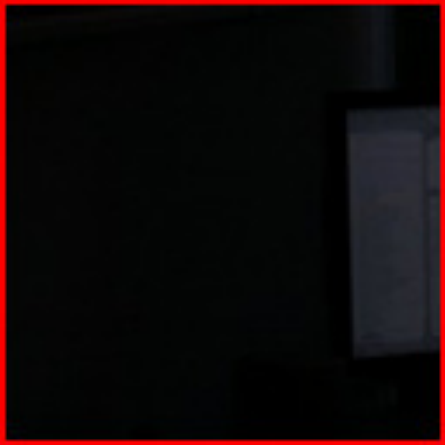}
			\includegraphics[width=1.35cm]{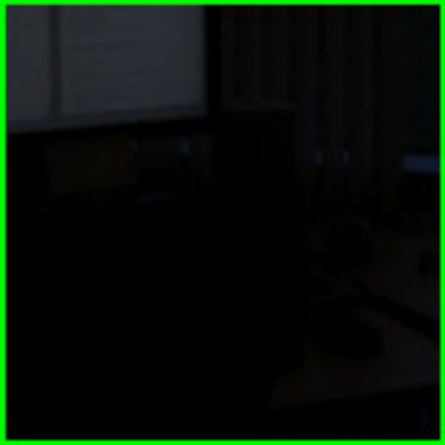}
		\end{minipage}
	}\hspace{-5pt}
	\subfigure[NPE\cite{wang2013naturalness}]{
		\begin{minipage}[b]{0.155\textwidth}
			\includegraphics[width=2.8cm]{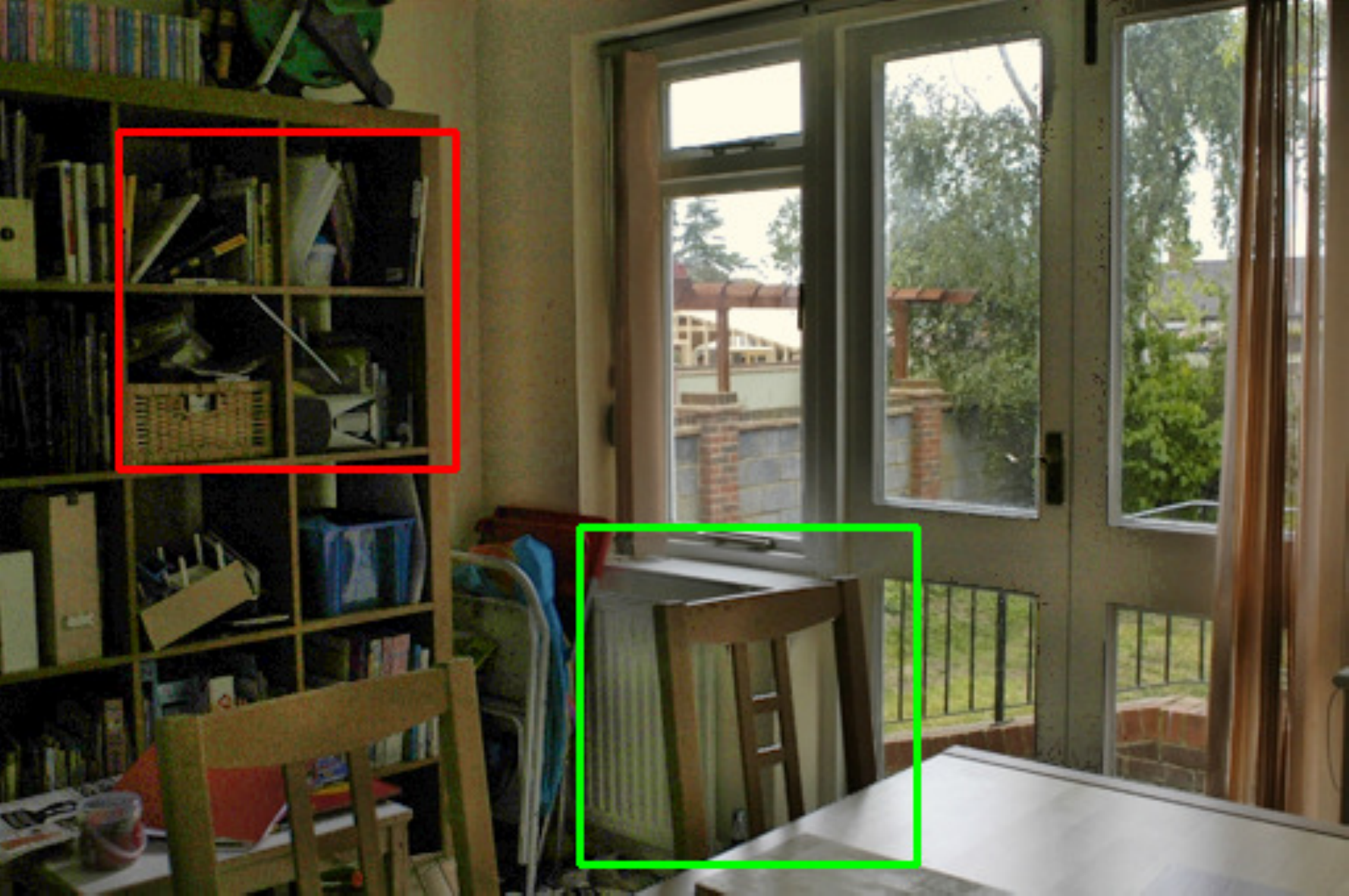}\vspace{1pt} \\
			\includegraphics[width=1.35cm]{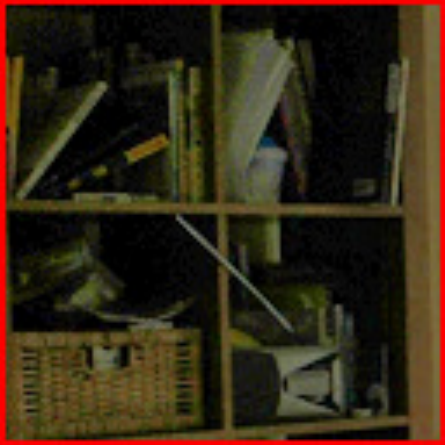}
			\includegraphics[width=1.35cm]{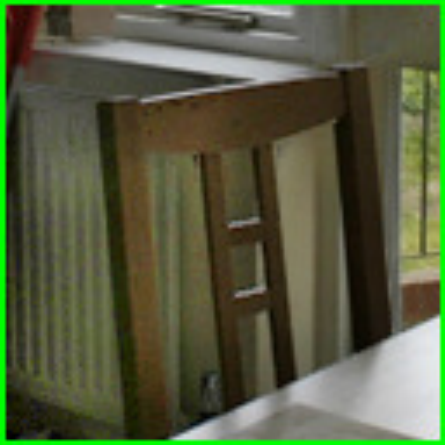}\vspace{5pt}
			\includegraphics[width=2.8cm]{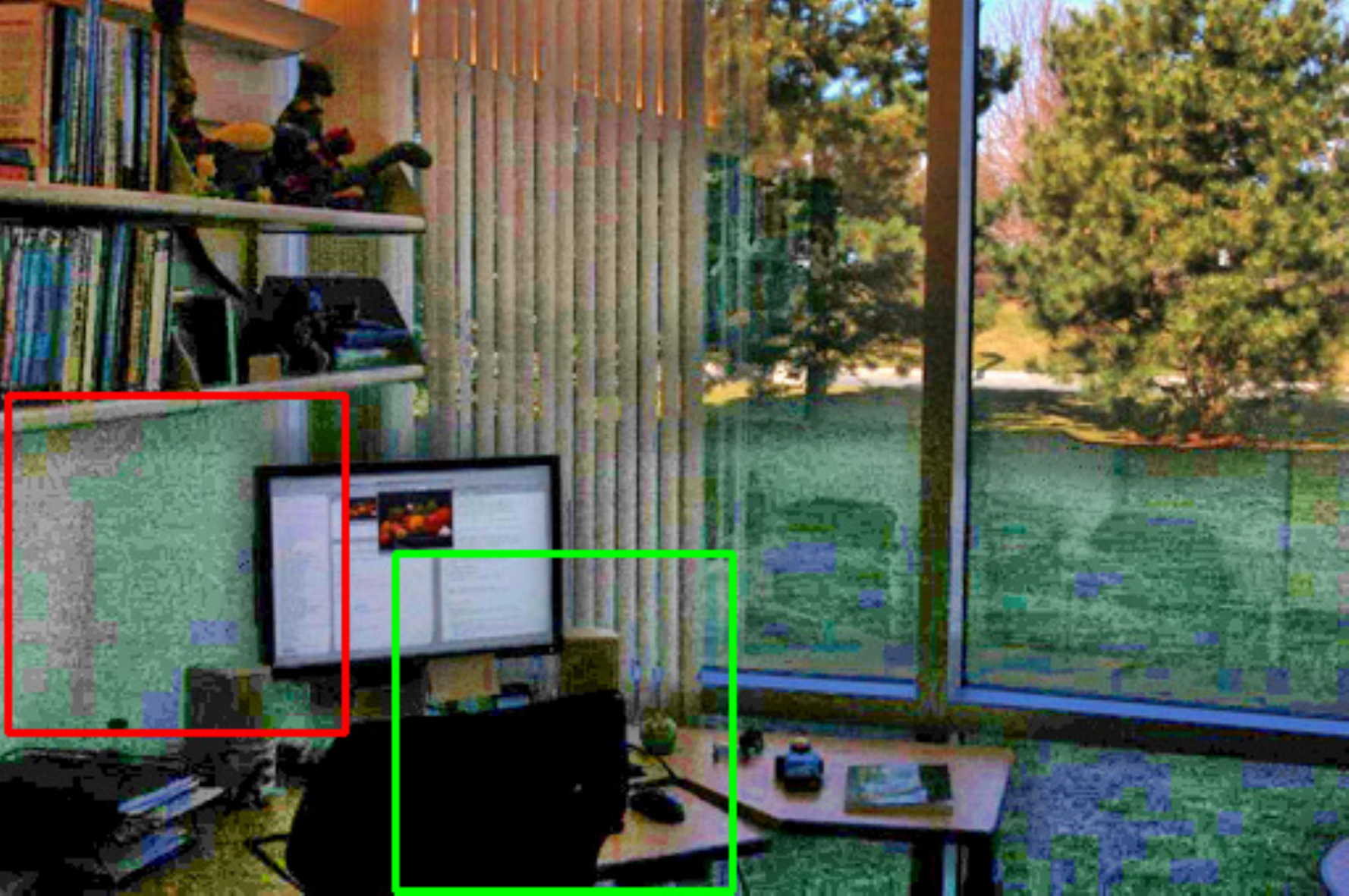}\vspace{1.5pt} \\
			\includegraphics[width=1.35cm]{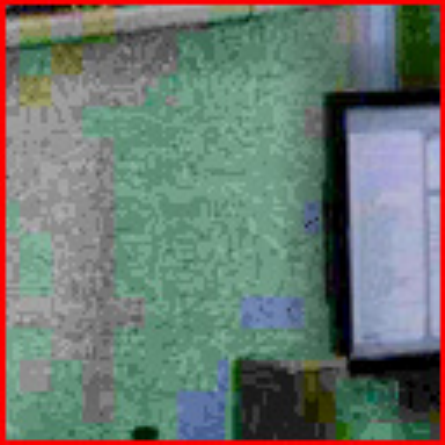}
			\includegraphics[width=1.35cm]{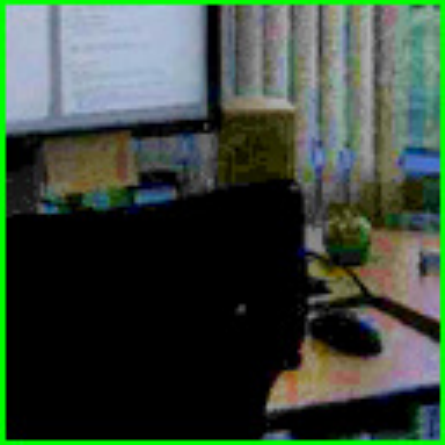}
		\end{minipage}
	}\hspace{-5pt}
	\subfigure[MF\cite{fu2016fusion}]{
		\begin{minipage}[b]{0.155\textwidth}
			\includegraphics[width=2.8cm]{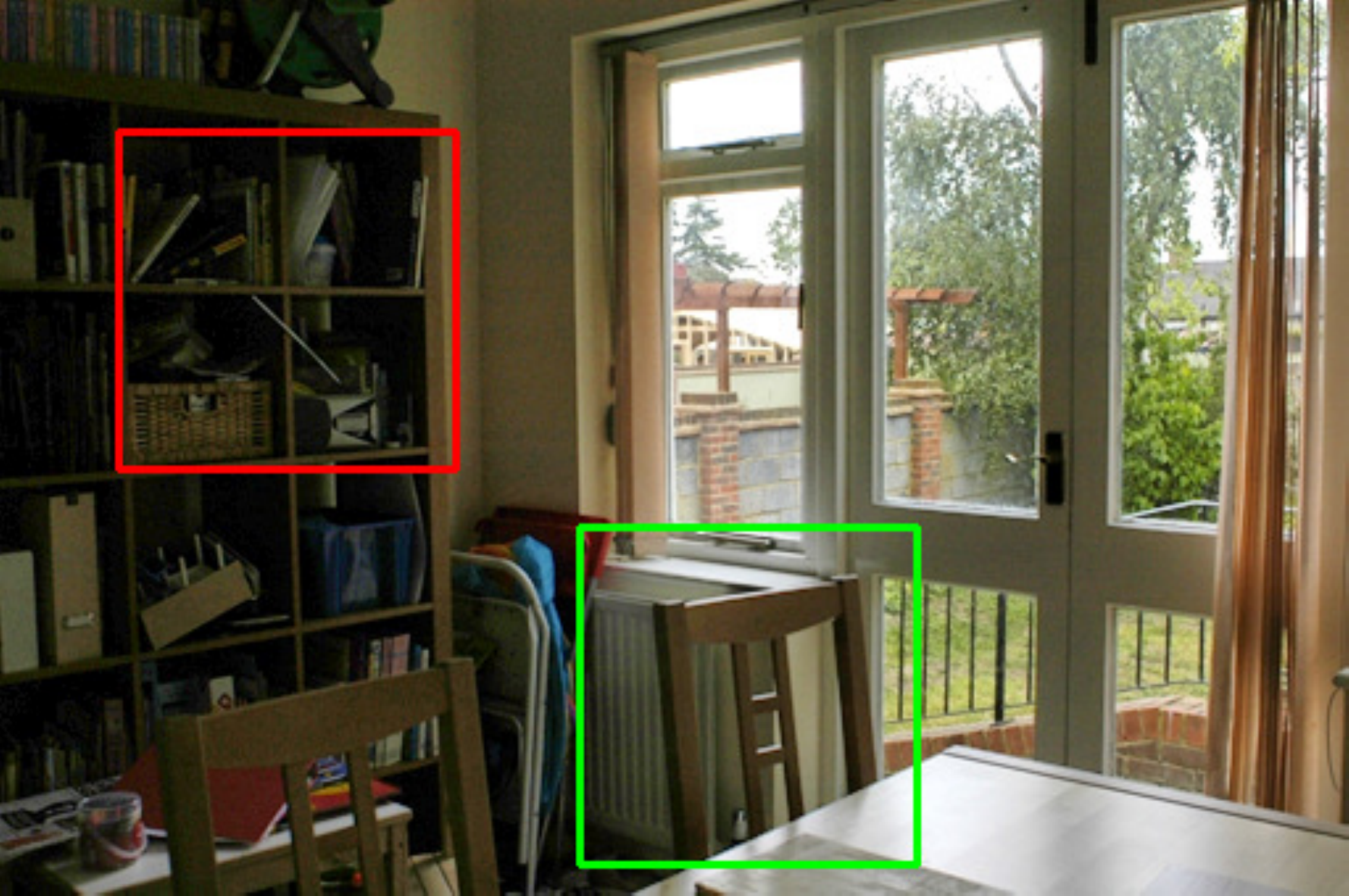}\vspace{1pt} \\
			\includegraphics[width=1.35cm]{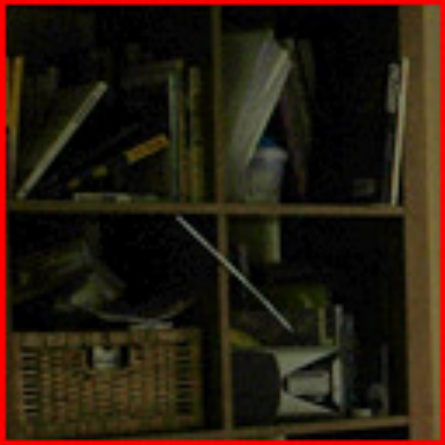}
			\includegraphics[width=1.35cm]{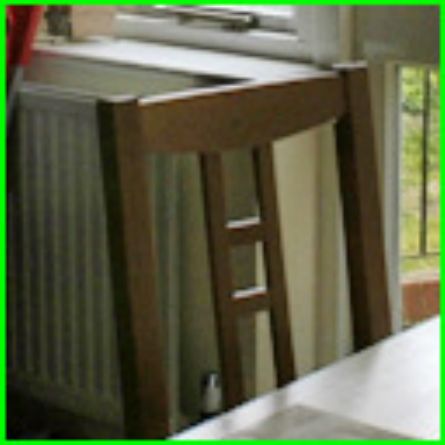}\vspace{5pt}
			\includegraphics[width=2.8cm]{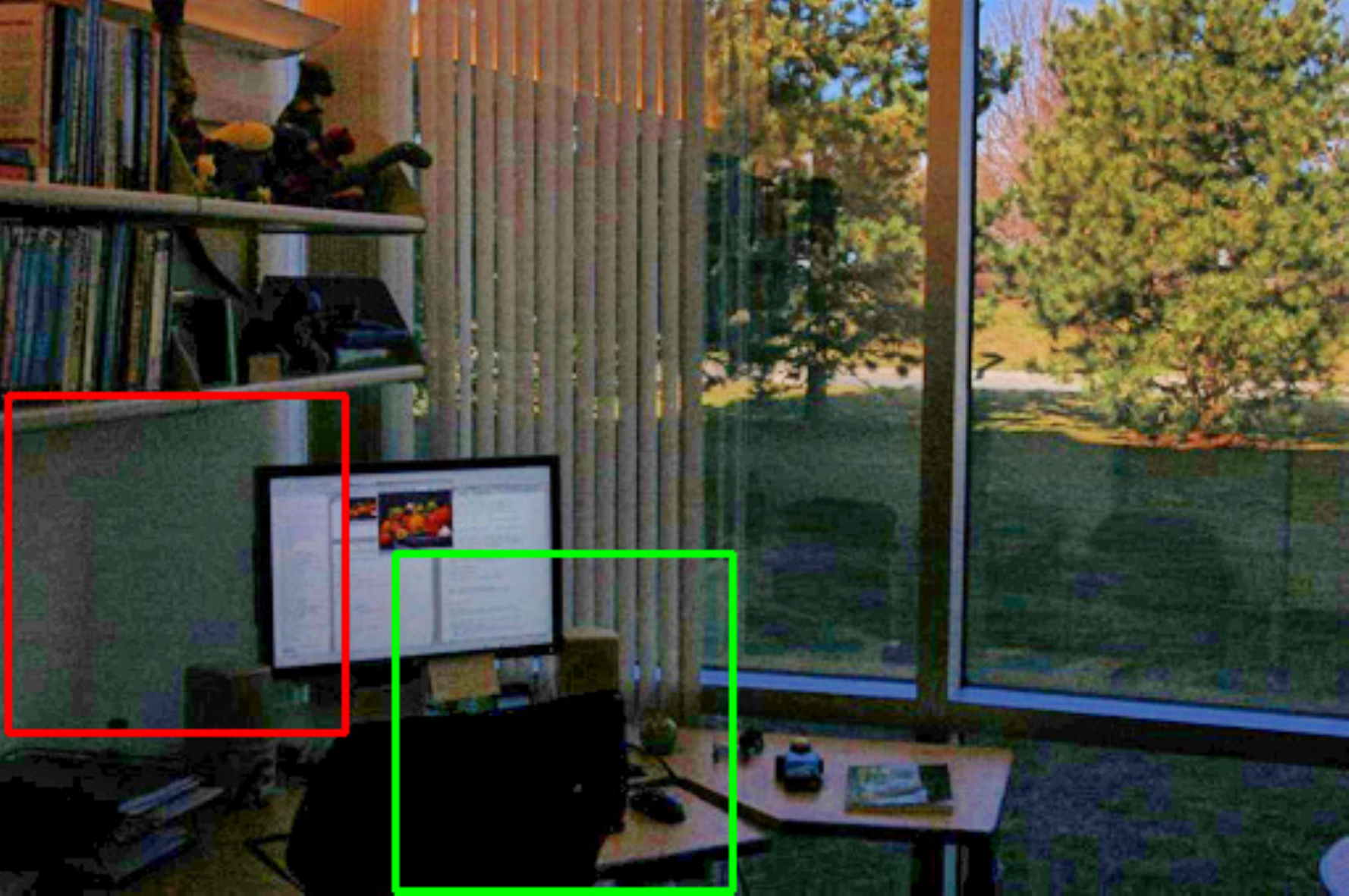}\vspace{1.5pt} \\
			\includegraphics[width=1.35cm]{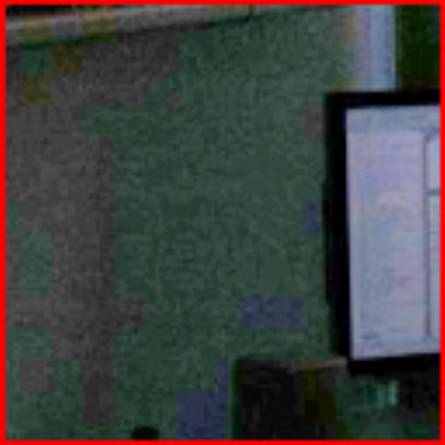}
			\includegraphics[width=1.35cm]{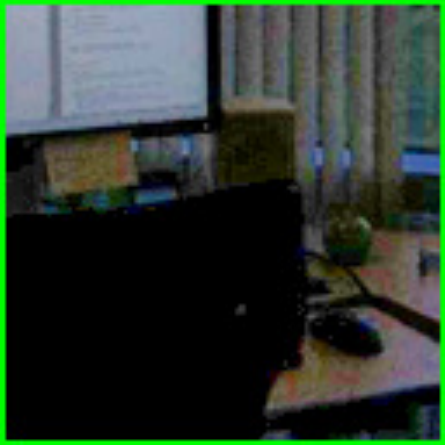}
		\end{minipage}
	}\hspace{-5pt}
	\subfigure[Zero-DCE\cite{guo2020zero}]{
		\begin{minipage}[b]{0.155\textwidth}
			\includegraphics[width=2.8cm]{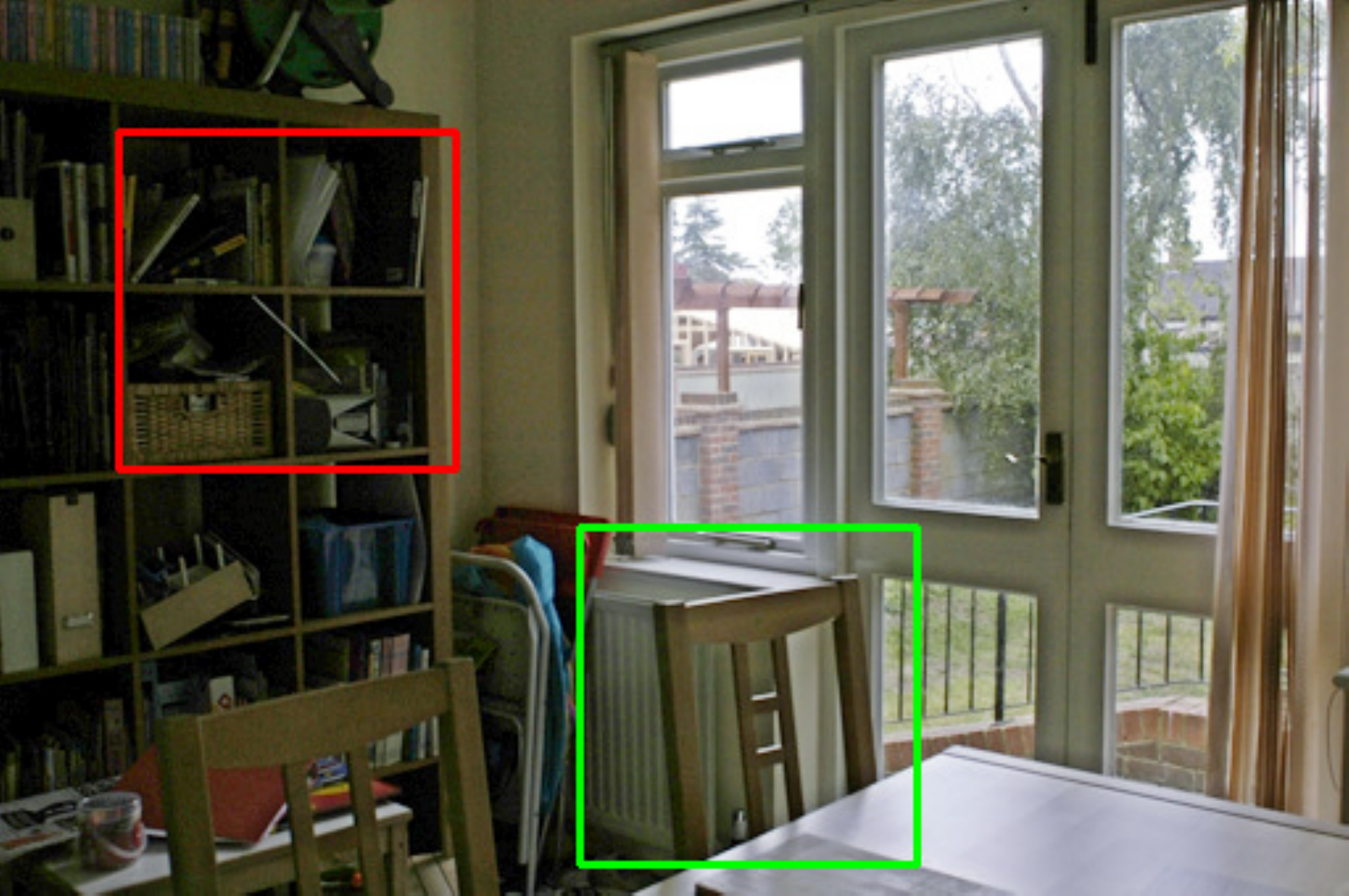}\vspace{1pt} \\
			\includegraphics[width=1.35cm]{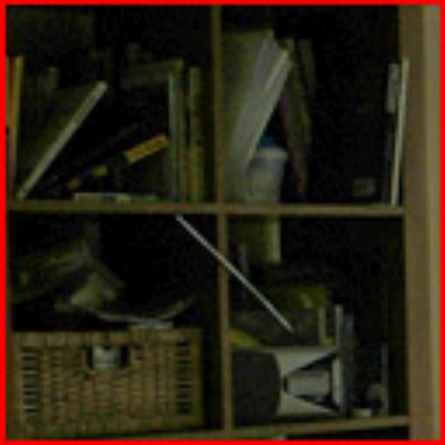}
			\includegraphics[width=1.35cm]{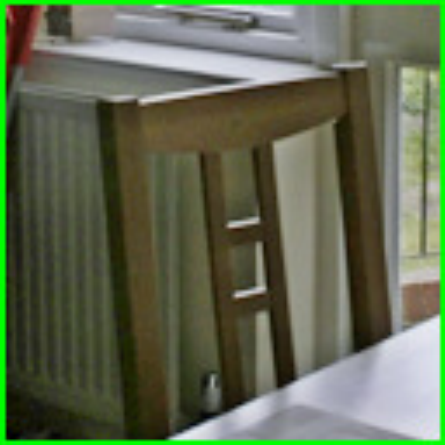}\vspace{5pt}
			\includegraphics[width=2.8cm]{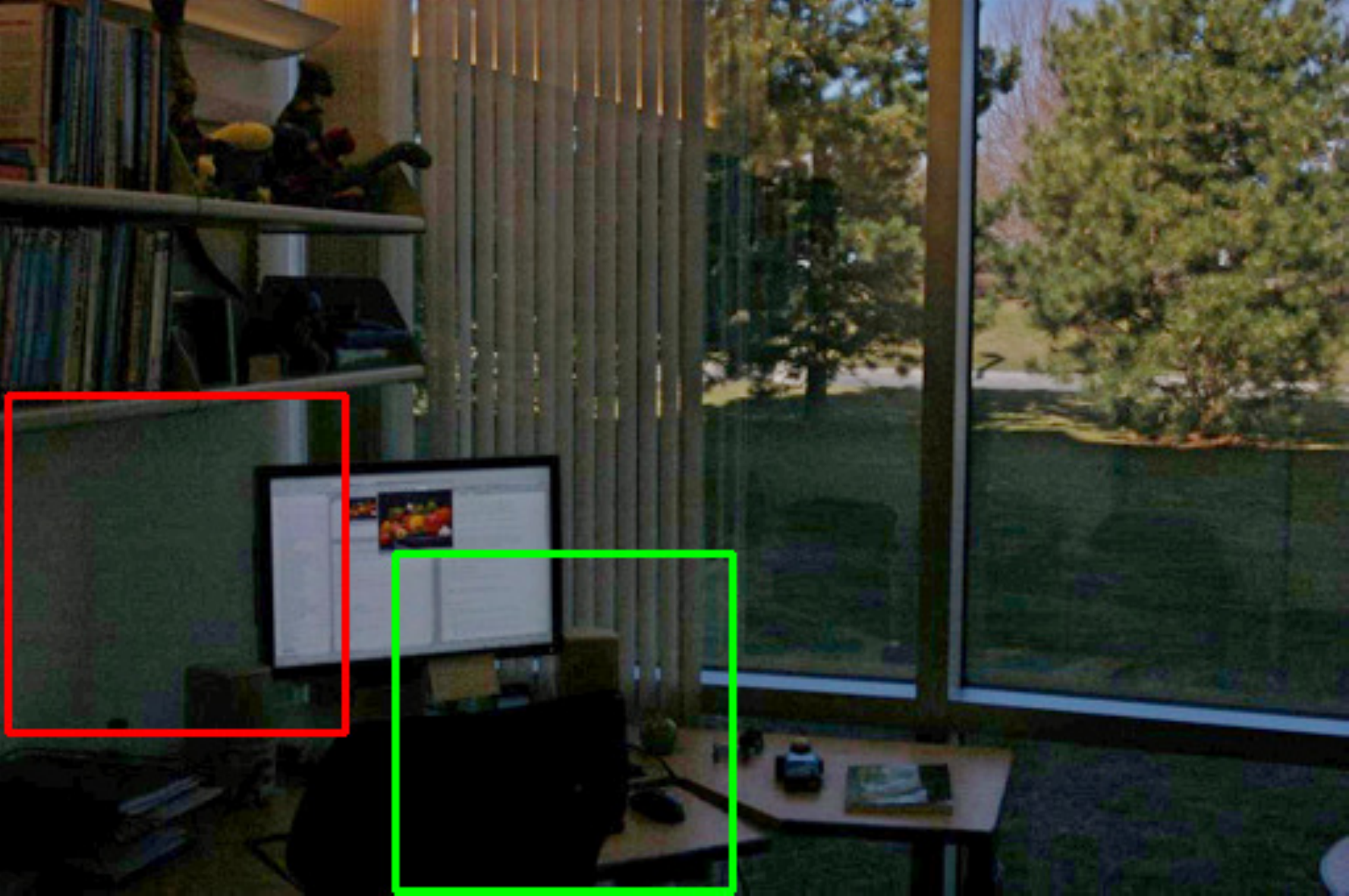}\vspace{1.5pt} \\
			\includegraphics[width=1.35cm]{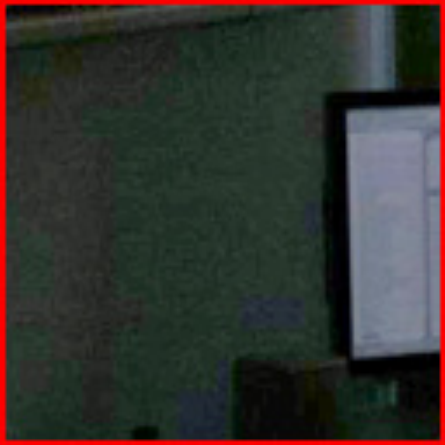}
			\includegraphics[width=1.35cm]{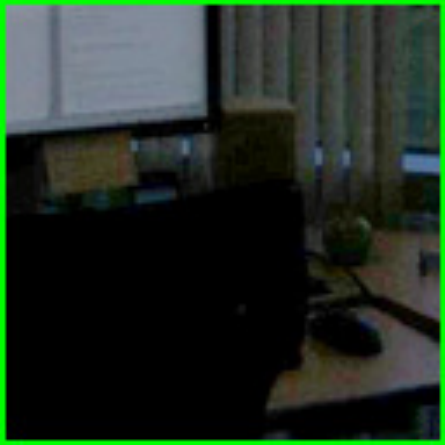}
		\end{minipage}
	}\hspace{-5pt}
	\subfigure[SRIE\cite{fu2016weighted}]{
		\begin{minipage}[b]{0.155\textwidth}
			\includegraphics[width=2.8cm]{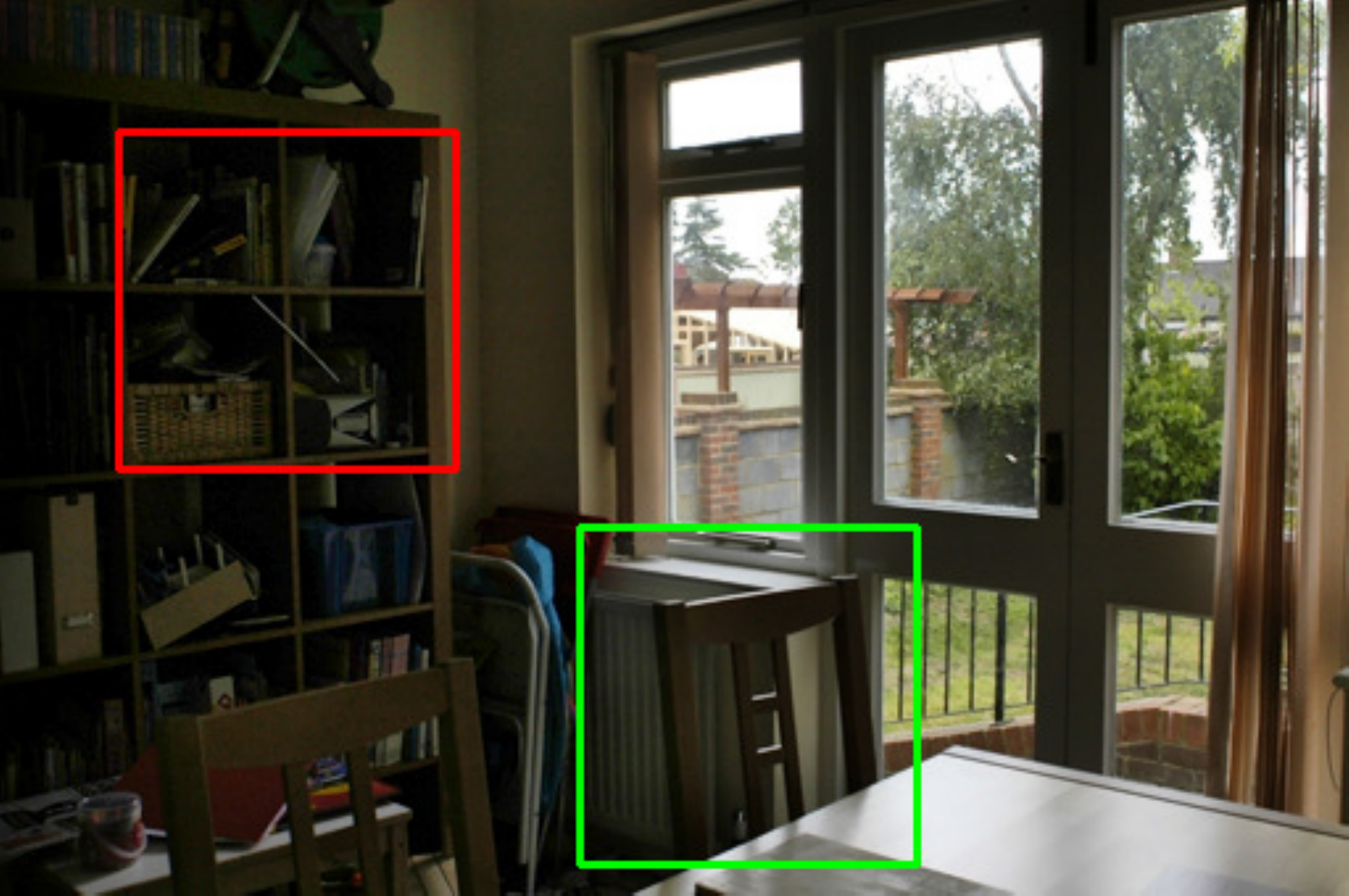}\vspace{1pt} \\
			\includegraphics[width=1.35cm]{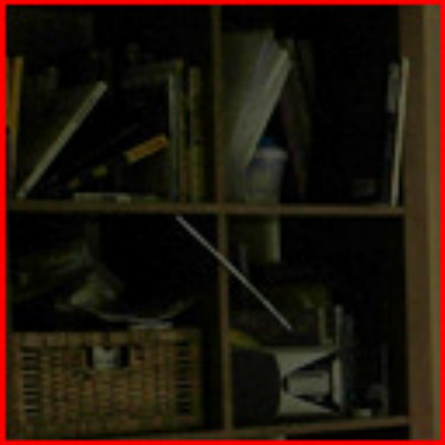}
			\includegraphics[width=1.35cm]{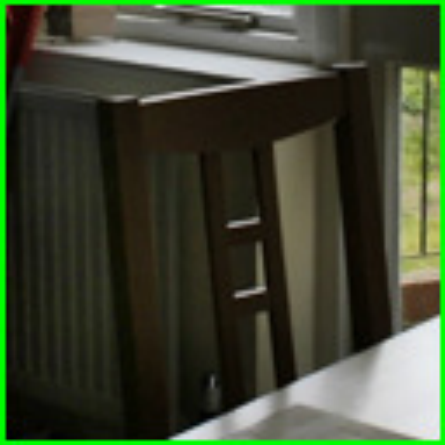}\vspace{5pt}
			\includegraphics[width=2.8cm]{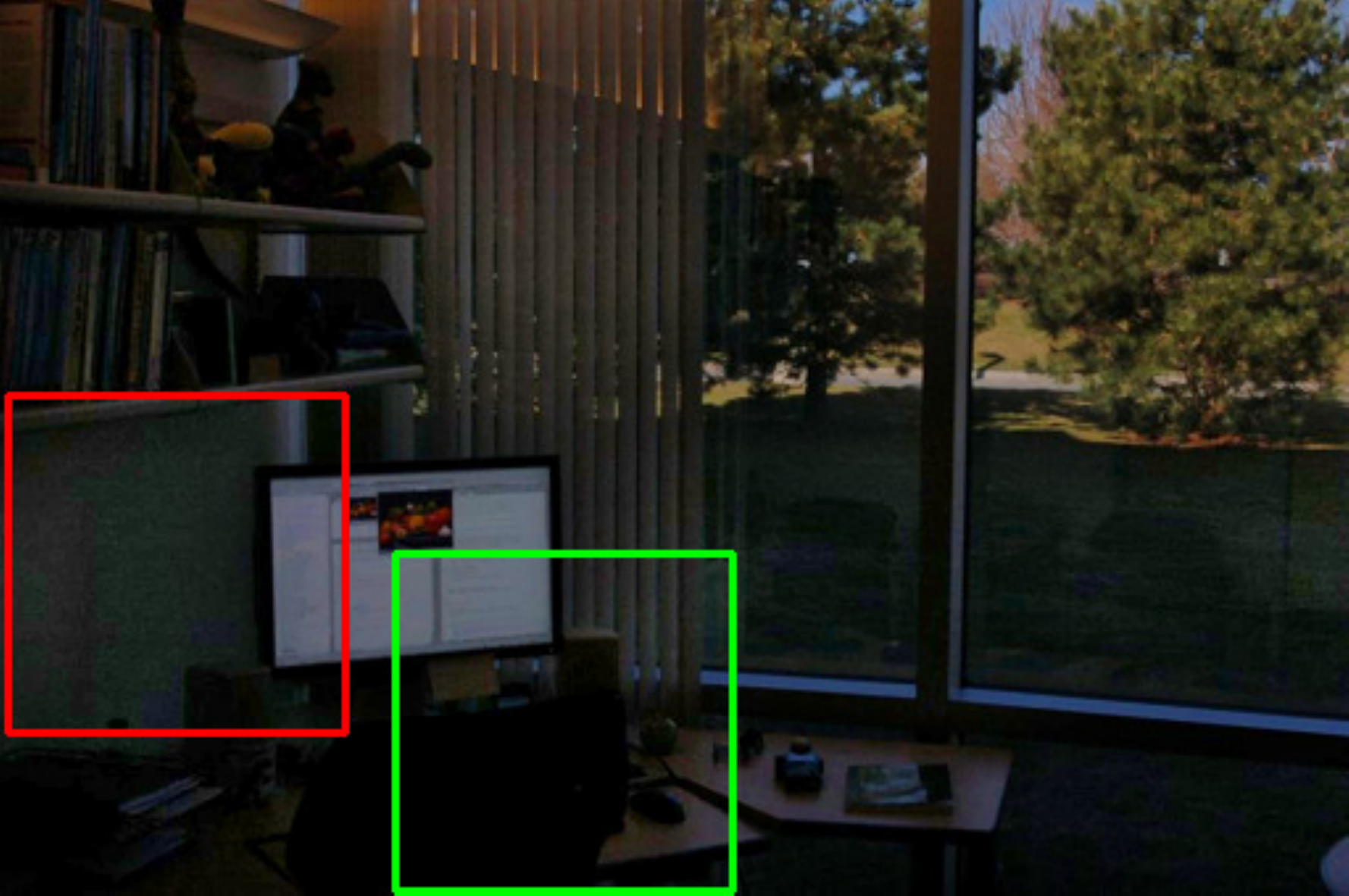}\vspace{1.5pt} \\
			\includegraphics[width=1.35cm]{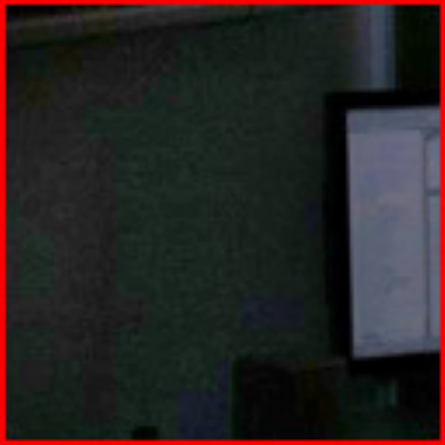}
			\includegraphics[width=1.35cm]{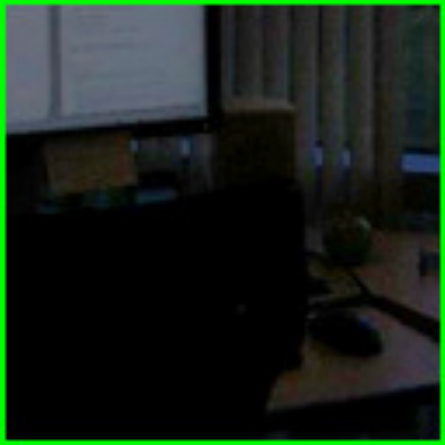}
		\end{minipage}
	}\hspace{-5pt}
	\subfigure[Ours]{
		\begin{minipage}[b]{0.155\textwidth}
			\includegraphics[width=2.8cm]{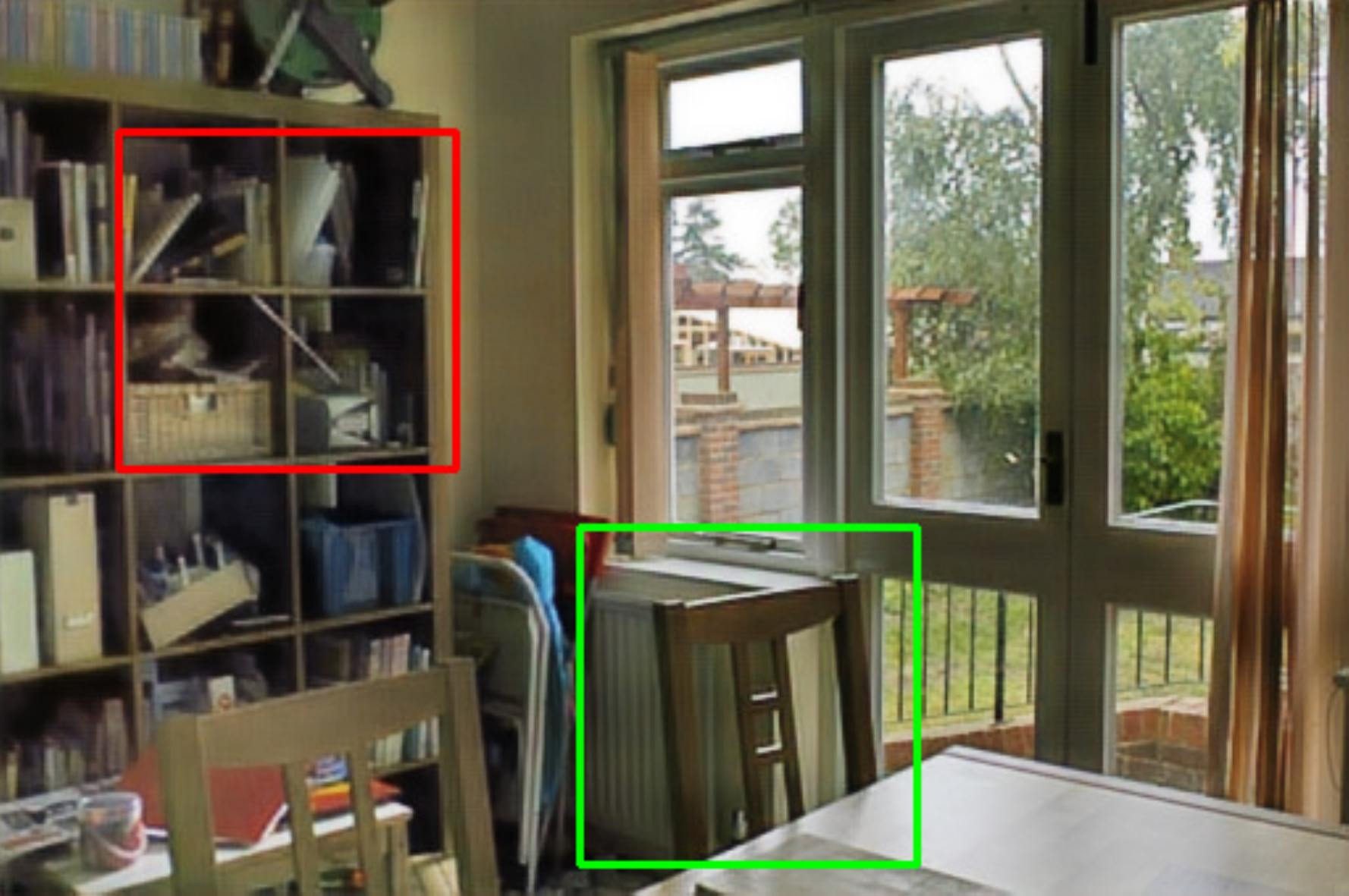}\vspace{1pt} \\
			\includegraphics[width=1.35cm]{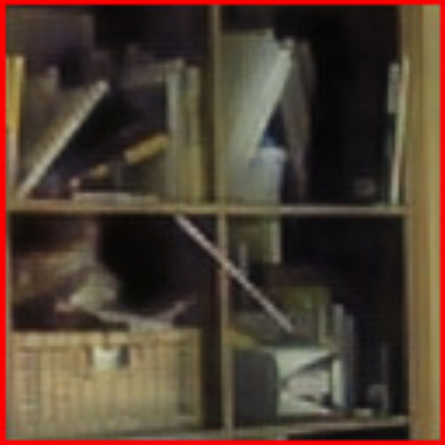}
			\includegraphics[width=1.35cm]{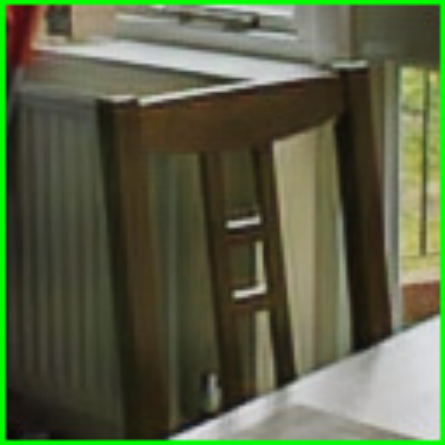}\vspace{5pt}
			\includegraphics[width=2.8cm]{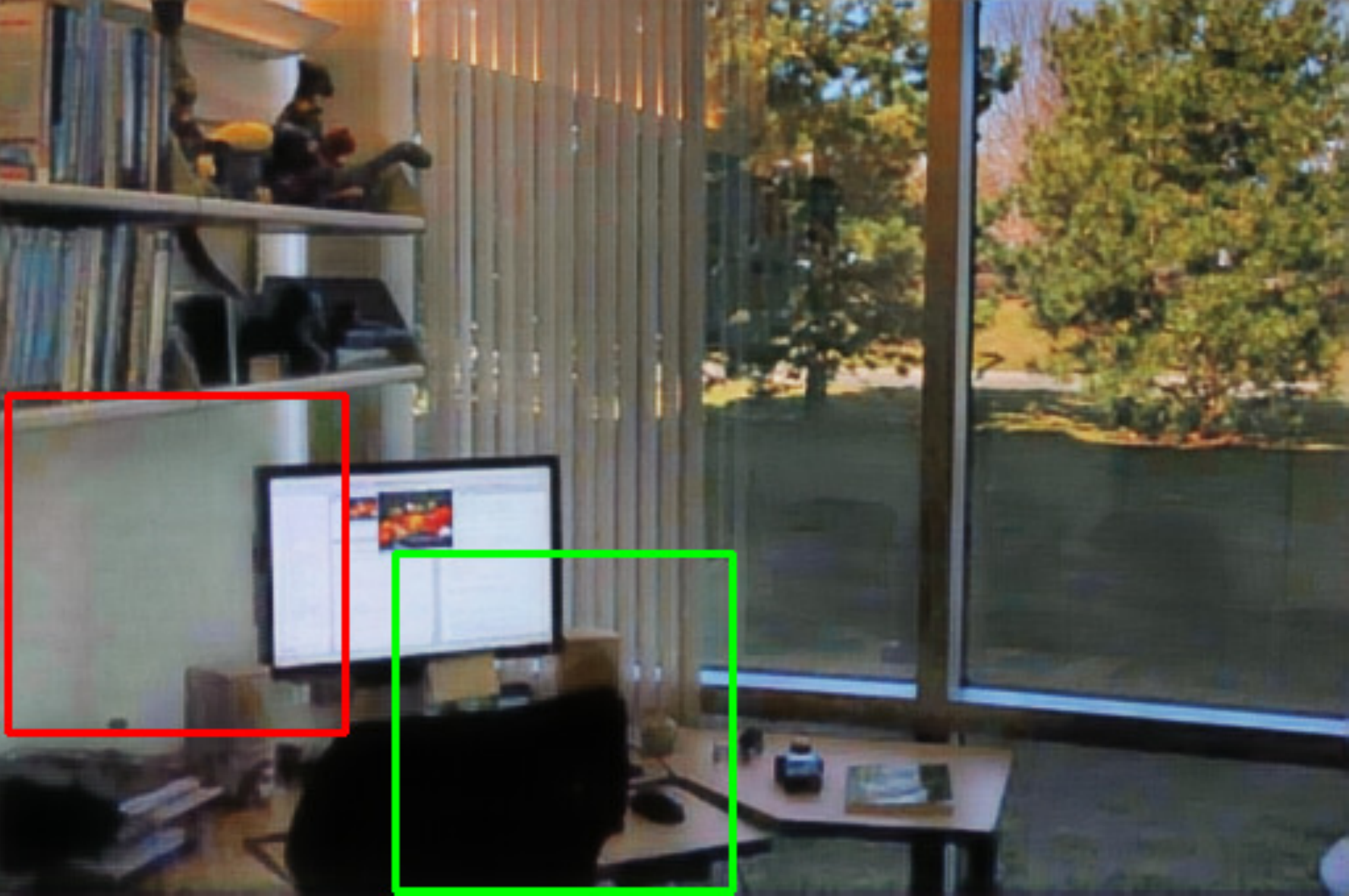}\vspace{1.5pt} \\
			\includegraphics[width=1.35cm]{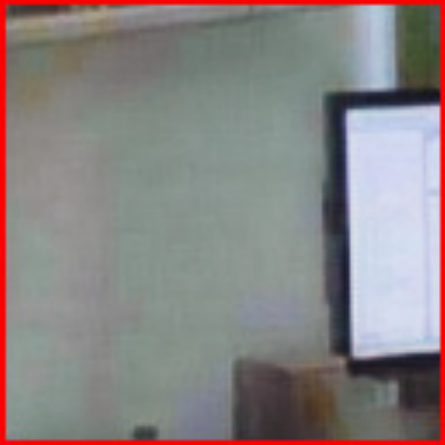}
			\includegraphics[width=1.35cm]{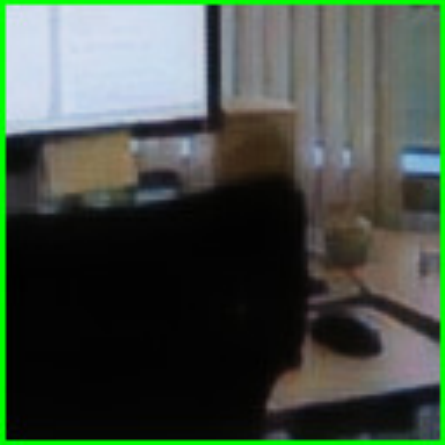}
		\end{minipage}
	}
	\caption{Visual comparison with other state-of-the-art methods on MEF dataset. There is no GroundTruth in MEF dataset.}
	\label{MEF}
\end{figure*}

\subsection{Visual Performance Analysis}
As shown in Fig. \ref{court} and Fig. \ref{book}, the low-light images from the LOL dataset are real-world images with very low brightness. The images enhanced by other methods contain serious noise pollution and color bias, low brightness and local over-exposure and under-exposure degradations. In particular, for RetinexNet\cite{wei2018deep}, there is so much amplified noise and color deviation in the images that they no longer look like real-world images. In contrast, our method can effectively solve these problems. It can be seen that the images enhanced by our method have almost no noise, the most abundant and reasonable color information and no over-exposure or under-exposure phenomena.\\
In addition, we test our model on LIME, DICM and MEF datasets. These three frequently used datasets have no GroundTruth. As shown in Fig. \ref{LIME}, Fig. \ref{DICM} and Fig \ref{MEF}, these methods have unpleasant problems, such as color bias, noise and low brightness. In contrast, the output results of our model can effectively suppress noise, reasonably improve brightness and correctly restore color information. Our results can also achieve a very good visual effect.
\subsection{Quantitative Performance Analysis}
Higher values of PSNR, SSIM, FSIM and UQI and lower value of LPIPS indicate better quality of images. We compared our method with other state-of-the-art methods in terms of these metrics, including traditional methods such as MSRCR \cite{jobson1997multiscale}, BIMEF \cite{ying2017bio}, LIME \cite{guo2016lime}, Dong \cite{dong2011fast}, SRIE \cite{fu2016weighted}, MF \cite{fu2016fusion}, NPE \cite{wang2013naturalness}, RRM \cite{li2018structure}, LECARM \cite{ren2018lecarm}, JED \cite{ren2018joint}, PLM \cite{yu2017low} and DIE \cite{zhang2019dual} and deep learning methods such as MBLLEN \cite{lv2018mbllen}, RetinexNet \cite{wei2018deep}, GLAD \cite{wang2018gladnet}, RDGAN\cite{wang2019rdgan}, Zero-DCE \cite{guo2020zero}, Zhang \cite{zhang2020self} and EnlightenGan \cite{jiang2021enlightengan}. As shown in Table \ref{lol_eval}, our method achieves the best performance in PSNR, SSIM, LPIPS, FSIM and UQI.\\
We also adopted Angular Error \cite{hordley2004re} as a quantitative index to measure the degree of color bias. Lower values of Angular Error indicate less color bias and better performance against color distortion. We compared four measures, i.e., deltaE \cite{sharma2005ciede2000}, the mean and median values of the Angular Error and the average value of them on the LOL validation dataset, which contains 15 real-world low/normal-light image pairs. As shown in Table \ref{lol_eval_color}, compared with other methods, our method achieves the best performance in all four measures.\\
In addition, we adopted Gray Entropy (GE) and Color Entropy (CE) as objective indicators to measure the amount of information in the image. CE calculates the information entropy of the three channels of the image and then adds them up. Image information entropy is also a common index for image quality evaluation. It reflects the richness of image information from the perspective of information theory. In general, larger entropy values for an image correspond to richer information and better quality. If the image brightness is too low, all kinds of information in the image will be hidden by darkness, so the calculated information entropy will be very low. We compare our method with other state-of-the-art deep learning low-light image enhancement methods in terms of GE and CE. As shown in Table \ref{lol_eval_entropy}, our results have the highest entropy of information, meaning our results contain the most information, such as color, structure, texure and detail information.
\begin{figure}[htbp]
	\flushleft
	\includegraphics[width=9cm]{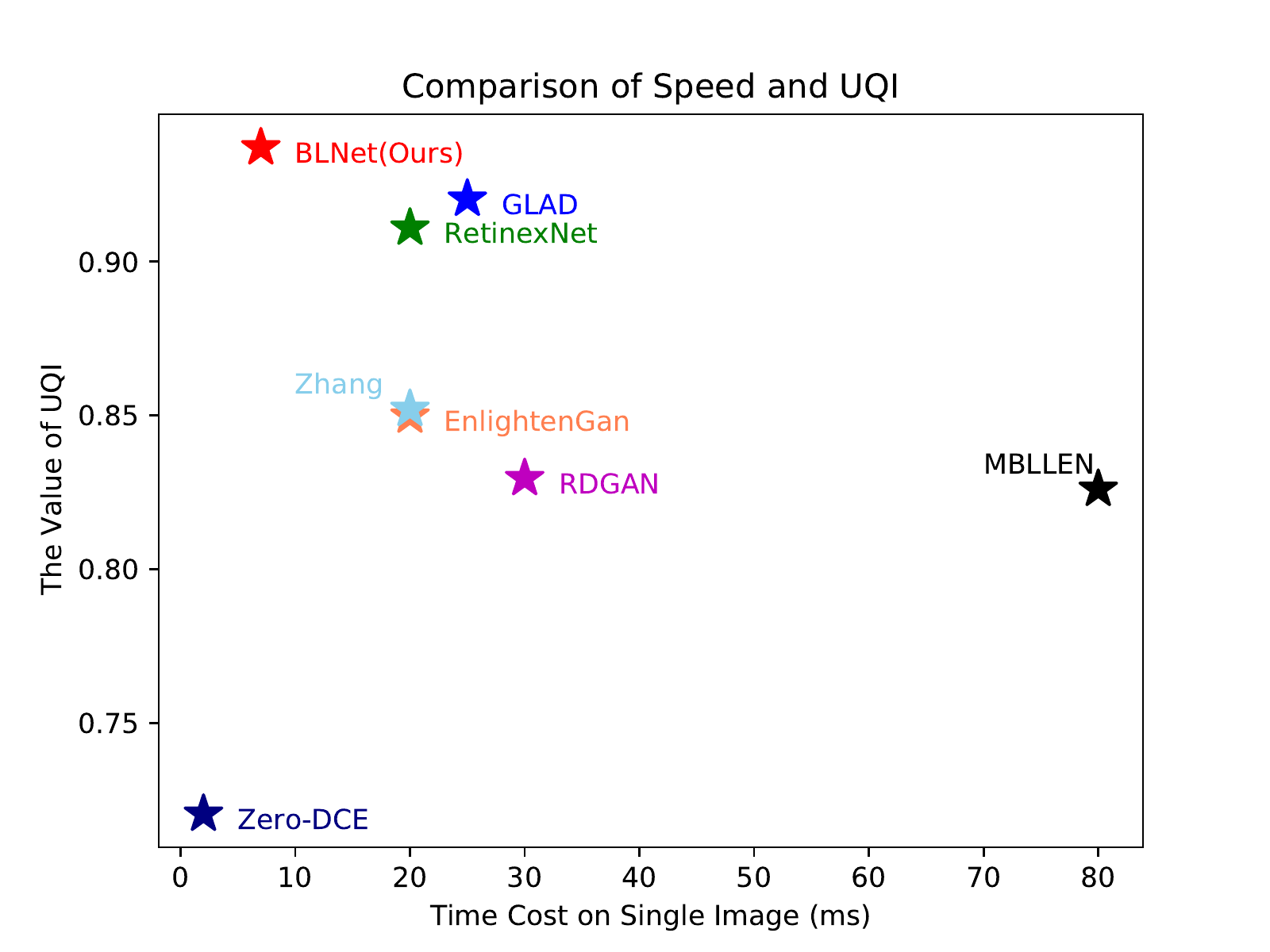}
	\caption{Runtime cost and performance comparison of our method and other state-of-the-art deep learning methods on the LOL dataset.}
	\label{speed}
\end{figure}

The inference speed of our model is very fast. As shown in Fig. \ref{speed}, we tested all of the deep learning models on a TITAN XP GPU with 12G RAM and compared the inference speed and UQI index of our model with those of other deep learning methods. The closer the method is to the upper left corner in the figure, the faster the method is and the higher the UQI is. As shown in the last colume of Table \ref{lol_eval_entropy}, our model only takes 7 ms to process a real-world low-light image with a resolution of 600x400 from the LOL dataset, and introducing our NCBC Module does not slow down our framework. Zero-DCE \cite{guo2020zero} is the fastest in terms of the inference speed, taking only 2 ms for a 600x400 resolution image. The speed of our model ranks second among all of the compared methods. This is because Zero-DCE adopts a very lightweight DCE-Net to estimate a set of best-fitting Light-Enhancement curves that iteratively enhance a given input image, so it can be very fast in terms of inference speed. Our model adopts two relatively deep U-Nets and a convolutional module, but the decomposition task is simple, so the reasoning can be achieved very quickly. In the brightness enhancement network, we only need to enhance the brightness of the illumination map and generate a single-channel illumination map correctly, and this is also a very simple task that can be accomplished quickly. Denoising and color bias correction are the most time-consuming steps. Our NCBC Module only calculates the loss functions, which enables our model to achieve the functions of noise removal and color restoration without additional time consumption on these two steps. This is why our model can run at such a high speed.

\subsection{Failure Cases}
\begin{figure}
	\flushleft
	\subfigure{
		\begin{minipage}{10\linewidth} 
			\includegraphics[width=3.95cm]{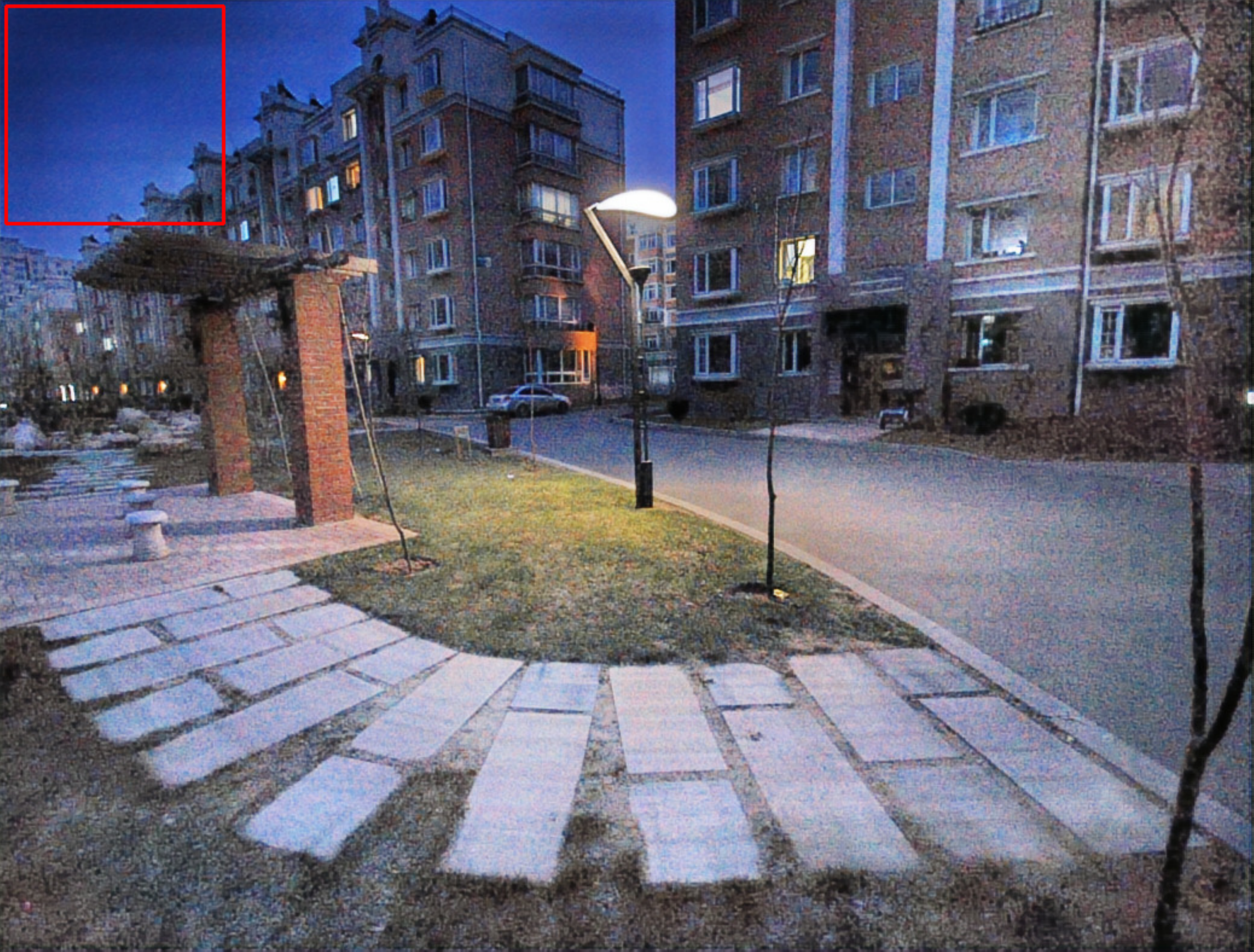}\vspace{-3pt}  
			\includegraphics[width=4.5cm]{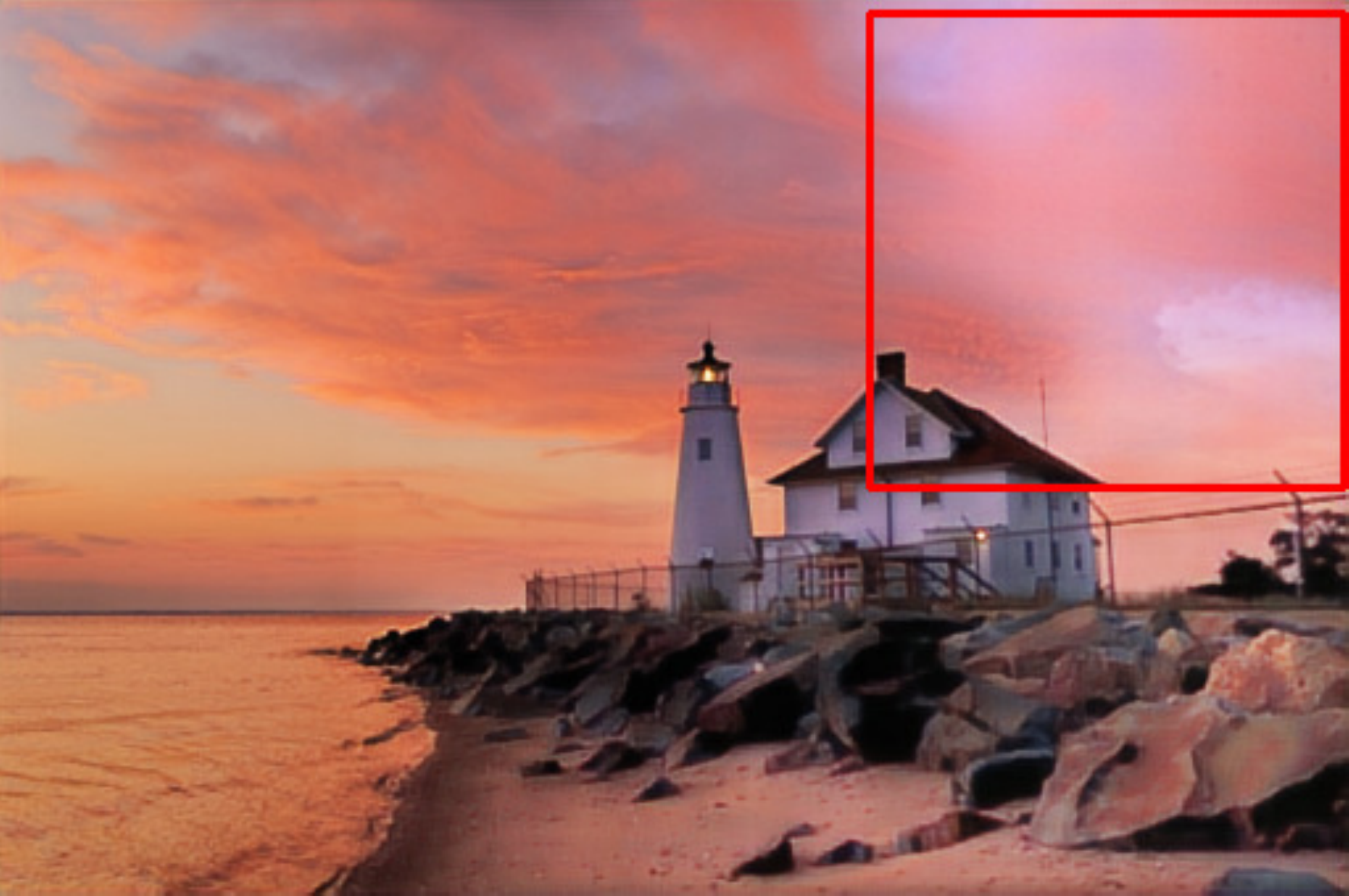}\vspace{-3pt} 
		\end{minipage}
	}
	\caption{Some failure cases of our results.}
	\label{failure}
\end{figure}

There are still some shortcomings in our method. If an image contains some areas that are too dark, unpleasant halo artifacts and local shadow will appear in such areas of the processed image when using our model. For example, as shown in Fig. \ref{failure}, our method may produce some failure cases with shadows and white blocks. 

\section*{Conclusion}

In this paper, inspired by the Retinex theory \cite{land1977retinex} and the idea of RetinexNet \cite{wei2018deep}, we propose a fast deep learning framework called \textbf{BLNet}, which consists of two U-Nets and a plain CNN with several stacked convolutional blocks. In the decomposition phase, we use the first U-Net to decouple images into reflectance and illumination. We also propose a CNN-based Noise and Color Bias Control Module (NCBC Module) to remove noise from the reflectance while controlling color bias and preserving details by transferring some of the details and color information to the illumination. In the enhancement phase, we use a similar U-Net to enhance the brightness of the illumination. Extensive experiments demonstrate the superiority and effectiveness of our method. In particular, our method is able to fix all of the degradations, such as low brightness, distorted contrast, real-world noise, color bias and detail loss, that occur in the process of low-light image enhancement. Another important advantage of our model is its fast computation speed, which is achieved as a result of the time-saving noise removal and color restoration contributed by our NCBC module.

\bibliographystyle{IEEEtran}
\bibliography{ref}

\end{document}